\documentclass{emulateapj}

\usepackage{amsmath,amssymb}
\usepackage{graphicx}
\usepackage[breaklinks,colorlinks,citecolor=blue,linkcolor=magenta]{hyperref} 
\usepackage[all]{hypcap}
\usepackage{aas_macros}

\slugcomment{Accepted for publication in The Astrophysical Journal}

\shorttitle{The EMPIRE survey}
\shortauthors{Jim\'enez-Donaire et al.} 

\begin{document}

%%---------------------------------------------------------------------
%                          Front Matter
%%---------------------------------------------------------------------

\title{EMPIRE: The IRAM 30-m Dense Gas Survey of Nearby Galaxies}

%\correspondingauthor{Mar\'ia Jes\'us Jim\'enez-Donaire}
%\email{mdonaire@cfa.harvard.edu}

\author{
Mar\'ia J. Jim\'enez-Donaire\altaffilmark{1,2,$\dagger$},
F. Bigiel\altaffilmark{3,2},
A. K. Leroy\altaffilmark{4},
A. Usero\altaffilmark{5},
D. Cormier\altaffilmark{6},
J. Puschnig\altaffilmark{3,2},
M. Gallagher\altaffilmark{4},
A. Kepley\altaffilmark{7},
A. D. Bolatto\altaffilmark{8},
S. Garc\'ia-Burillo\altaffilmark{5},
A. Hughes\altaffilmark{9,10},
C. Kramer\altaffilmark{11,12},
J. Pety\altaffilmark{12,13},
E. Schinnerer\altaffilmark{14},
A. Schruba\altaffilmark{15},
K. Schuster\altaffilmark{12},
F. Walter\altaffilmark{14}
}

\altaffiltext{1}{Harvard-Smithsonian Center for Astrophysics, 60 Garden Street, Cambridge, MA 02138, USA; \href{mailto:mdonaire@cfa.harvard.edu}{mdonaire@cfa.harvard.edu}}
\altaffiltext{2}{Institut f\"ur theoretische Astrophysik, Zentrum f\"ur Astronomie der Universit\"at Heidelberg, Albert-Ueberle Str. 2, 69120 Heidelberg, Germany}
\altaffiltext{3}{Argelander-Institut f\"ur Astronomie, Universit\"at Bonn, Auf dem H\"ugel 71, 53121 Bonn, Germany}
\altaffiltext{4}{Department of Astronomy, The Ohio State University, 140 W 18$^{\rm th}$ St, Columbus, OH 43210, USA}
\altaffiltext{5}{Observatorio Astron\'omico Nacional, Alfonso XII 3, 28014, Madrid, Spain}
\altaffiltext{6}{Laboratoire AIM, CEA/DSM-CNRS-Universit{\'e} Paris Diderot, Irfu/Service d'Astrophysique, CEA Saclay, F-91191 Gif-sur-Yvette, France}
\altaffiltext{7}{National Radio Astronomy Observatory, 520 Edgemont Road, Charlottesville, VA 22903, USA}
\altaffiltext{8}{Department of Astronomy and Laboratory for Millimeter-Wave Astronomy, University of Maryland, College Park, MD 20742, USA}
\altaffiltext{9}{CNRS, IRAP, 9 Av. colonel Roche, BP 44346, F-31028 Toulouse cedex 4, France}
\altaffiltext{10}{Universit\'{e} de Toulouse, UPS-OMP, IRAP, F-31028 Toulouse cedex 4, France}
\altaffiltext{11}{Instituto de Astrof\'isica de Andaluc\'ia IAA-CSIC, Glorieta de la Astronom\'ia s/n, E-18008, Granada, Spain}
\altaffiltext{12}{Institut de Radioastronomie Millim\'etrique (IRAM), 300 Rue de la Piscine, F-38406 Saint Martin d'H\`eres, France}
\altaffiltext{13}{Sorbonne Universit\'e, Observatoire de Paris, Universit\'e PSL, \'Ecole normale sup\'erieure, CNRS, LERMA, F-75005, Paris, France}
\altaffiltext{14}{Max-Planck-Institut f\"ur Astronomie, K\"onigstuhl 17, 69117 Heidelberg, Germany}
\altaffiltext{15}{Max-Planck-Institut f\"ur extraterrestrische Physik, Giessenbachstrasse 1, 85748 Garching, Germany}

\vspace{5mm}

\altaffiltext{$\dagger$}{Submillimeter Array Fellow}

\begin{abstract}
We present EMPIRE, an IRAM 30-m large program that mapped $\lambda = 3{-}4$~mm dense gas tracers at $\sim 1{-}2\,$kpc resolution across the whole star-forming disk of nine nearby, massive, spiral galaxies. We describe the EMPIRE observing and reduction strategies and show new whole-galaxy maps of HCN\,(1-0), HCO$^+$\,(1-0), HNC\,(1-0) and CO\,(1-0). We explore how the HCN-to-CO and IR-to-HCN ratios, observational proxies for the dense gas fraction and dense gas star formation efficiency, depend on host galaxy and local environment. We find that the fraction of dense gas correlates with stellar surface density, gas surface density, molecular-to-atomic gas ratio, and dynamical equilibrium pressure. In EMPIRE, the star formation rate per unit dense gas anti-correlates with these same environmental parameters. Thus, although dense gas appears abundant the central regions of many spiral galaxies, this gas appears relatively inefficient at forming stars. These results qualitatively agree with previous work on nearby galaxies and the Milky Way's Central Molecular Zone. To first order, EMPIRE demonstrates that the conditions in a galaxy disk set the gas density distribution and that the dense gas traced by HCN shows an environment-dependent relation to star formation. However, our results also show significant ($\pm 0.2$~dex) galaxy-to-galaxy variations. We suggest that gas structure below the scale of our observations and dynamical effects likely also play an important role.
\end{abstract}


%%----------------------------------------------------------------------
%                               Main Text
%%----------------------------------------------------------------------

\section{Introduction}

We present the ``EMIR Multiline Probe of the ISM Regulating Galaxy Evolution'' survey (EMPIRE; PI: F.~Bigiel). EMPIRE used the IRAM 30-m telescope to map multiple molecular lines in the 3-4\,mm atmospheric window across the whole star-forming area of nine nearby, massive spiral galaxies. The lines covered include the high critical density transitions HCN\,(1-0), HCO$^+$\,(1-0) and HNC\,(1-0), frequently referred to as ``dense gas tracers.'' Thanks to the wide bandwidth of the EMIR receiver, we simultaneously cover the CO isotopologues $^{13}$CO and C$^{18}$O as well as a number of fainter lines (e.g., the low-lying transitions of SiO, C$_2$H, N$_2$H$^+$).

The ratios among the lines mapped by EMPIRE constrain the density distribution and other physical conditions in the molecular gas. The faintness of these transitions at extragalactic distances has prevented previous large-scale mapping efforts. EMPIRE overcomes this obstacle by leveraging the wide bandwidth and excellent sensitivity of EMIR on the IRAM~30-m telescope. The result is the first resolved ($1-2\,$kpc resolution), wide area mapping survey of density-sensitive molecular lines in the 3-4\,mm atmospheric window. 

EMPIRE has two core goals. First, to constrain the density distribution within the molecular gas and to measure how the gas density distribution depends on galactic environment. Second, to measure how the star formation efficiency per unit molecular gas mass depends on the density distribution within the molecular gas and environment. More colloquially, EMPIRE aims to answer ``Where is gas dense in galaxies and how does dense gas relate to star formation?''

This paper describes the survey and addresses these two core questions. Here, we focus on the HCN-to-CO and IR-to-HCN line ratios as observational proxies for the dense gas fraction and dense gas star formation efficiency, respectively. We measure how these quantities depend on local conditions within galaxy disks. 

Our results build on previous observations: the pointed HCN survey by \citet{USERO15}, full-disk HCN mapping of M51 by \citet{BIGIEL16}, and the ALMA+IRAM study of four galaxies by \citet{GALLAGHER18}. J. Puschnig et al. (in preparation) will extend our analysis to leverage the full suite of EMPIRE line ratios, which we only discuss briefly here.

In addition to these studies, EMPIRE has already been used to study physical conditions in the molecular gas in a series of related papers. \citet{JIMENEZDONAIRE17} derived constraints on the optical depth of dense gas tracers by studying their less abundant isotopologues (H$^{13}$CN, H$^{13}$CO$^{+}$). \citet{JIMENEZDONAIRE17B} showed that the C$^{18}$O-to-$^{13}$CO line ratio increases systematically with radius in our targets. \citet{CORMIER18} measured the $^{13}$CO-to-$^{12}$CO ratio across our targets and showed how it depends on local physical conditions. They also calculated a spatially resolved $^{13}$CO-to-H$_2$ conversion factor, and found that $^{13}$CO may be a better tracer of the molecular gas mass than $^{12}$CO in galaxy centers. \cite{GALLAGHER18b} combined EMPIRE with higher resolution ALMA maps to show that, on average, the spectroscopic dense gas fraction, traced by HCN-to-CO, correlates with the cloud-scale molecular gas surface density.

We give the scientific background for the survey in Section \ref{sec:background}. We describe our IRAM 30-m observations, data reduction, and data processing in Section \ref{sec:data}. In Section \ref{sec:ancillary}, we summarize key supporting multi-wavelength data and in Section \ref{sec:phys_parameters} we explain how we convert these to physical quantities. Section \ref{sec:stacking} describes the stacking techniques that we use to improve the signal-to-noise ratio of our measurements. We present our results in Section \ref{sec:results}. In Sections \ref{sec:intensity}-\ref{sec:ratios}, we analyze the spatial extent of dense gas emission and compare it to the distribution of the CO emission. In Section \ref{sec:scaling} we compare our measurements to star formation scaling relations obtained from previous observations. We investigate the systematic variations of the star formation efficiencies and the dense gas fractions in Section \ref{sec:fdense_sfe}. Section \ref{sec:discussion} discusses our findings. We compare our observations to other recent results and discuss plausible physical drivers that could explain our observations. Finally, Section \ref{sec:summary} presents a summary of the survey and our key findings.

\subsection{Background}
\label{sec:background}

The low-$J$ $^{12}$CO emission lines have been used to map the molecular ISM in the Milky Way and many external galaxies. CO is the second most abundant molecule after H$_2$, and has been calibrated as a proxy to trace the distribution of H$_2$ mass \citep[e.g.,][]{BOLATTO13}. Given the low ``effective'' critical density required to excite the $J=1-0$ transition and its low excitation temperature, CO emission traces the bulk molecular medium. However, stars are thought to form preferentially in the densest regions of molecular clouds. Studies of the Milky Way \citep{HEIDERMAN10,LADA10,LADA12,EVANS14,VUTI16} and external galaxies \citep{GAO04,GARCIABURILLO12} have highlighted the role of dense gas as the immediate site of star formation. Thus, knowing the prevalence and star-forming ability of this dense gas is crucial to understand how gas is converted to stars in a galactic context. 

Line emission from molecules with higher dipole moments than that of CO, such as HCN or HCO$^+$, has a higher critical density than CO. This critical density represents the density for which the total radiative decay rate between an upper and lower rotational levels equals the rate of collisional de-excitation out of the upper level \citep[see e.g.,][]{SHIRLEY15}. For an optically thin line, the emissivity \citep[line emission per unit mass, as defined in][]{LEROY17} of the gas reaches a maximum at this value. In reality, radiative trapping effects can lead to a lower ``effective critical density.'' Also, because the emissivity of gas below the critical density is low but not zero, large masses of low density gas can produce significant emission even from high dipole moment molecules \cite[e.g.,][]{SHIRLEY15,LEROY17}. Despite these important caveats, the effective mean densities probed by low-$J$ HCN and HCO$^+$ lines are still notably higher than those accesses by low-$J$ CO lines. As a result, we expect these lines to trace gas more closely linked to star formation.

In a seminal paper, \citet{GAO04} observed the ground-state transition of HCN emission from 53 entire galaxies and bright galaxy centers across a large range of galaxy types, from normal spirals to (ultra)luminous infrared galaxies (L$_\textrm{TIR} \geq 10^{11}\,\textrm{L}_\odot$, hereafter (U)LIRGs). They observed a strong {\it linear} relationship between the recent star formation rate (SFR), as traced by the total infrared emission, and the HCN luminosity. Such a linear relationship does not hold for CO, because IR-bright, starburst galaxies, LIRGs and ULIRGs show a higher ratio of IR to CO emission than normal galaxies. Similar results were found by several subsequent studies of nearby galaxies \citep[e.g.,][]{GARCIACARPIO06,JUNEAU09,GARCIABURILLO12}. If the conversion from line luminosities to gas masses is the same for all galaxies, these results imply that the star formation efficiency of the molecular gas as traced by CO (SFE$_\textrm{mol} \equiv \textrm{SFR}/M_\textrm{mol}$) is higher in more luminous systems, while the star formation efficiency of the dense molecular gas (SFE$_\textrm{dense} \equiv \textrm{SFR}/M_\textrm{dense}$) as traced by HCN is approximately constant.

Observations isolating clouds and star-forming clumps in the Milky Way \citep[e.g.,][]{WU05,HEIDERMAN10,LADA10,LADA12,EVANS14} have extended the extragalactic IR-to-HCN correlation down to individual molecular clouds and dense cores. This suggests that the SFR per unit dense gas mass is nearly constant across many scales. These studies also found a good correspondence between the SFR in individual clouds (by counting young stellar objects) and the dense gas mass (by using extinction measurements). This suggests that dense gas mass is a strong predictor of how much star formation is occurring in a cloud.

As a result of these studies, a constant dense gas star formation efficiency, $\textrm{SFE}_{\rm dense}$, above some critical surface density, $\Sigma_\textrm{dense}$, has been hypothesized \citep[e.g.,][]{LADA10,LADA12,EVANS14}. In such ``density threshold models'' for star formation, there is a constant SFE of dense molecular gas and the overall star formation rate would then be regulated by the amount of dense gas available above this threshold. \citet{LADA13} and \citet{EVANS14} discussed in detail the limitations of this column density threshold idea and its applicability to Galactic molecular clouds. In particular \citet{LADA13} argue that a Kennicutt-Schmidt-type scaling relation is not enough to completely describe star formation in a cloud and, as a consequence, the observed scaling relation in unresolved galaxies is likely a consequence of unresolved observations of individual clouds, \citep[an idea also explored in][]{BIGIEL08,LEROY08}.

By contrast, turbulence regulated ``whole-cloud models'' for star formation postulate that the global properties of turbulent clouds set their density distribution and star formation efficiency \citep[e.g.,][]{PADOAN02,KRUMHOLZ07,FEDERRATH12}. In such a scenario, the fraction of star-forming dense gas, $f_\textrm{dense}=M_\textrm{dense}/M_\textrm{mol}$, and its efficiency, SFE$_\textrm{dense}$ depend on cloud parameters such as cloud mean density, virial parameter, and Mach number.

Observations of nearby galaxies, however, suggest that a constant SFE$_\textrm{dense}$ for star-forming regions across different galactic environments may be insufficient to explain the observations. \citet{GARCIABURILLO12} used the IRAM-30m telescope to observe a sample of 19 LIRGs in the $J=1-0$ lines of CO, HCN and HCO$^+$. Combined with literature data, they assembled a sample of $\sim$100 normal and (U)LIRG galaxies. Their observations, averaged across entire galaxies, largely obeyed the IR-to-dense gas power-law correlation found in previous Galactic and extragalactic work. However, the sample of LIRGs and ULIRGs deviates from this power-law. They measured $L_\textrm{IR}$-to-$L_\textrm{HCN}$ ratios as a proxy for SFE$_\textrm{dense}$ and found that these luminosity ratios are a factor of $2-3$ higher in LIRGs and ULIRGs than those measured in normal galaxies. These variations in the efficiency of star formation in dense gas suggest that real physical effects are still at play in different galactic environments and agree better with turbulence-regulated models.

Observations of entire galaxy disks at $\sim$kpc scales are  bridging the gap between cloud-scale studies and galaxy-scale surveys, which provide large number of systems but at too low resolution to connect to local ISM physics. Such observations are revealing systematic variations in the overall linear correlation between dense gas tracers and star formation rate tracers seen in global measurements of entire galaxies. \citet{USERO15} used the IRAM 30-m telescope to survey HCN\,(1-0) emission from 62 regions across 29 nearby star-forming galaxies. Their achieved resolution ($\sim 1-2\,$kpc) allowed for the investigation of the properties of the dense gas as a function of local conditions in galaxy disks. Their results show that the dense gas fraction ($f_\textrm{dense}$), as traced by the HCN/CO ratio, depends strongly on location in the disk, increasing with stellar surface densities ($\Sigma_*$) and molecular-to-atomic gas ratios ($R_\textrm{mol}$). On the other hand, they found that the star formation efficiency of dense molecular gas (SFE$_\textrm{dense}$), as traced by the IR/HCN ratio, anti-correlates systematically with those same parameters: it is $\sim 6-8$ times lower near galaxy centers than in the outer regions of the galaxy disks. 

Similar results have been found by \citet{BIGIEL16} and \citet{CHEN15} across the full disk of NGC\,5194 (M51): while there is an overall correlation between star formation rate tracers and HCN emission at $\sim$kpc resolution, the efficiency of dense gas to form stars drops at small galactocentric radii (taking the observables at face value). \citet{GALLAGHER18} presented new ALMA dense gas observations combined with IRAM 30-m short spacing, mapping the inner $\sim 3-5$~kpc of four local galaxies (NGC\,3351, NGC\,3627, NGC\,4254 and NGC\,4321). They found the same correlations between dense gas fraction, star formation rate, and local environment, and expressed them in terms of the dynamical equilibrium pressure needed to support the weight of gas disk in a galaxy region \citep[e.g.,][]{ELMEGREEN89,HELFER97,WONG02,BLITZ06}. \citet{QUEREJETA19} find similar results in resolved regions of M51's spiral arms, with high angular resolution observations from IRAM/NOEMA. Recent findings by \citet{BEMIS19} in the Antennae galaxy system also resemble these results. The two nuclei, NGC\,4038 and NGC\,4039, show the largest dense gas fractions, but the lowest SFE per unit dense gas mass.

Correspondingly, multiple observations in our own Milky Way have revealed that the SFR in the inner $\sim500\,$pc of the Galaxy (Central Molecular Zone, CMZ) appears strongly suppressed relative to its dense gas content \citep[e.g.,][]{JONES12,LONGMORE13,BARNES17,MILLS17}. This reinforces the trends discussed above. Dense gas in regions with high mean density appears to be inefficient at forming stars. Interpreting the observables at face value, these results may suggest changing density distributions and changing star formation rates per unit dense gas. The latter would be at odds with ``fixed density threshold'' models. Exploring these trends further across whole galaxy disks to understand the role of dense gas in galaxy scale star formation requires wide field mapping of the main dense gas tracers across full galaxy disks.

\section{Observations \& Data Processing}
\label{sec:data}

\subsection{Sample selection}

Table \ref{table:sample} lists the EMPIRE targets. All targets are nearby ($d \le 15\,$Mpc), face on ($i \le 65^{\circ}$) spiral galaxies that are also large on the sky ($\geq 2'$). 

We chose our targets from the HERA CO-Line Extragalactic Survey \citep[HERACLES,][]{LEROY09}. Because HERACLES builds on SINGS \citep{KENNICUTT03}, THINGS \citep{WALTER08}, and KINGFISH \citep{KENNICUTT11}, this ensures high quality and homogeneous multiwavelength data. We picked our targets to be CO-bright and actively star-forming, allowing us to detect the faint dense-gas tracing lines. We also required that they be relatively face on and close enough so that the IRAM 30-m beam ($\sim 30\arcsec$ at $\sim$90\,GHz) translates to physical scales of $\sim$1-2\,kpc. 

Finally, we aimed to cover a range of morphological and dynamical features. The sample contains galaxies which show a strong spiral arm structure (NGC\,628, NGC\,3184, and NGC\,5194). It also covers strongly barred galaxies (NGC\,2903, NGC\,3627), flocculent disks (NGC\,5055, NGC\,6946), strong nuclear bursts (NGC\,2903, NGC\,4321, NGC\,6946), Virgo cluster members (NGC\,4254, NGC\,4321), and interacting galaxies (NGC\,3627, NGC\,5194).

\begin{table*}
	\caption{\label{table:sample}EMPIRE galaxy sample.}
	\centering
	\begin{tabular}{lccrrcrrccrc}
		\hline\hline
Galaxy & RA & DEC & $i$ & P.A. & $r_{25}$ & $D$ & $V_\textnormal{hel}$ & Metal. & Morph. & $\langle \Sigma_\textnormal{SFR} \rangle$ & log$_{10}$($M_*$)\\
  & (EQ 2000) & (EQ 2000) & & & & & & & & & \\
  & hh mm ss.s & dd mm ss & ($^o$) & ($^o$) & (') & (Mpc) & (km s$^{-1}$) & $12+$log(O/H) &  & ($M_\odot\,\textnormal{yr}^{-1} \textnormal{kpc}^{-2}$) & log$_{10}$($M_\odot$)\\
(1) & (2) & (3) & (4) & (5) & (6) & (7) & (8) & (9) & (10) & (11) & (12)\\
\hline
NGC 628 & 01:36:41.8 & 15:47:00 & 7 & 20 & 4.9 & 9.0 & 659.1 & 8.35 & SAc & 4.0$\times 10^{-3}$ & 10.0\\
NGC 2903 & 09:32:10.1 & 21:30:03 & 65 & 204 & 5.9 & 8.5 & 556.6 & 8.68 & SABbc & 5.7$\times 10^{-3}$ & 10.1\\
NGC 3184 & 10:18:17.0 & 41:25:28 & 16 & 179 & 3.7 & 13.0 & 593.3 & 8.51 & SABcd & 2.8$\times 10^{-3}$ & 10.2\\
NGC 3627 & 11:20:15.0 & 12:59:30 & 62 & 173 & 5.1 & 9.4 & 717.3 & 8.34 & SABb & 7.7$\times 10^{-3}$ & 10.5\\
NGC 4254 & 12:18:50.0 & 14:24:59 & 32 & 55 & 2.5 & 16.8 & 2407.0 & 8.45 & SAc & 18$\times 10^{-3}$ & 10.5\\
NGC 4321 & 12:22:55.0 & 15:49:19 & 30 & 153 & 3.0 & 15.2 & 1571.0 & 8.50 & SABbc & 9.0$\times 10^{-3}$ & 10.6\\
NGC 5055 & 13:15:49.2 & 42:01:45 & 59 & 102 & 5.9 & 8.9 & 499.3 & 8.40 & SAbc & 4.1$\times 10^{-3}$ & 10.5\\
NGC 5194 & 13:29:52.7 & 47:11:43 & 20 & 172 & 3.9 & 8.4 & 456.2 & 8.55 & SAbc & 20$\times 10^{-3}$ & 10.5\\
NGC 6946 & 20:34:52.2 & 60:09:14 & 33 & 243 & 5.7 & 7.0 & 42.4 & 8.40 & SABcd & 21$\times 10^{-3}$ & 10.5\\
		\hline
	\end{tabular}
    	\\ \flushleft{{\bf Notes:} Galaxy names (1), adopted centers (2-3) and morphological types (10) are taken as listed in NED, the NASA Extragalactic Database. The orientation parameters: inclinations (4), position angles (5) and radius of the $B$-band 25th magnitude isophote (6) are taken from the HyperLeda database \citep{MAKAROV14}. Distances (7) are adopted from the Extragalactic Distance Database \citep[EDD,][]{TULLY09}. Heliocentric central velocities (8) are taken from \citet{WALTER08}. Globally averaged metallicities (9) from \citet{MOUSTAKAS10}, except for NGC\,2903 \citep{ENGELBRACHT08}. Average star formation rate surface density (11) inside 0.75 $r_{25}$, adopted from \citet{LEROY13}. Integrated stellar mass (12) of the entire galaxies based on $3.6\mu$m emission from \citet{DALE07} and \citet{DALE09}.}
\end{table*}

\subsection{IRAM 30-m observations}

\begin{table}
	\caption{Main spectral lines covered by our EMIR setups.}
	\label{table:lines}
	\centering
	\begin{tabular}{lcccc}
		\hline\hline
		Species & $\nu_\textrm{rest}$ & $\textrm{E}_\textrm{up}$ $^a$ & $n_\textrm{crit}$ $^a$ & Beam size$^b$\\
		& (GHz) & (K) & (cm$^{-3}$) & ($\arcsec$)\\
		\hline
		SiO 2-1$^c$ & 86.45 & 6.25 & 1$\times 10^5$ & 34.04\\
		C$_2$H 1-0 & 87.32 & 4.19 & 1$\times 10^5$ & 33.86\\
		HNCO 4-3 & 87.93 & 10.55 & 1$\times 10^4$ & 33.63\\
		HCN 1-0 & 88.63 & 4.25 & 2$\times 10^5$ & 33.36\\
		HCO$^{+}$ 1-0 & 89.19 & 4.28 & 3$\times 10^4$ & 33.15\\
		HNC 1-0 & 90.66 & 4.35 & 1$\times 10^5$ & 32.61\\
	%	HC$_3$N 10-9 & 90.98 & 24.0 & 9$\times 10^4$ & \\
		N$_2$H$^+$ 1-0 & 93.20 & 4.47 & 4$\times 10^{4}$ & 31.73\\
	%	SO 23-12 & 109.30 & & & \\
		C$^{18}$O 1-0$^d$ & 109.78 & 5.27 & 4$\times 10^2$ & 26.83\\
		HNCO 5-4 & 109.90 & 15.8 & 1$\times 10^7$ & 26.70\\
		$^{13}$CO 1-0$^d$ & 110.20 & 5.29 & 4$\times 10^2$ & 26.13\\
		$^{12}$CO 1-0 & 115.27 & 5.53 & 4$\times 10^2$ & 25.65\\
		\hline
	\end{tabular}
	\\ \flushleft{{\bf Notes:} $(a)$ The critical densities ($n_\textrm{crit}$) at 20\,K and energies of the upper level were calculated from the radiative and collisional coefficients taken from the Leiden LAMDA database, \citep{TAK07}. $(b)$ The beam size in this table refers to the resolution of our final cubes, not the telescope native resolution. $(c)$ Only available for NGC\,5194. $(d)$ Not available for NGC\,5194.}
\end{table}

The observations for EMPIRE were carried out at the IRAM 30m telescope located at Pico Veleta, Spain. Most of the data were taken from December 2014 through December 2016 for $\sim$440\,h over the course of 16 runs. We refer the reader to the EMPIRE survey website\footnote{\url{https://empiresurvey.webstarts.com}} for additional information. A link to the official IRAM repositories containing the data products will be available in the website. We used the 3\,mm band (E090) of the dual-polarization EMIR receiver \citep{CARTER12}, which yields an instantaneous bandwidth of 15.6\,GHz per polarization. The data were recorded using the Fast Fourier Transform Spectrometers (FTS), with a spectral resolution of \mbox{195 kHz}, corresponding to \mbox{$\sim 0.5$ km s$^{-1}$} for the E090 band. We tuned EMIR with a local oscillator frequency of $\sim 98.6\,$GHz. This allowed us to simultaneously observe the bright high critical density tracers HCN\,(1-0), HCO$^+$\,(1-0), and HNC\,(1-0) as well as the optically thin molecular column tracers $^{13}$CO\,(1-0) and C$^{18}$O\,(1-0). In addition, many fainter transitions of other molecules are also present in the band (see Table \ref{table:lines}). These are mostly not detected in individual lines of sight. In future work we will explore if these are accessible by means of spectral stacking.

For the remainder of the paper we refer to HCN\,(1-0) emission simply as HCN, and proceed analogously for the other molecular lines.

\begin{figure}
\includegraphics[scale=0.15]{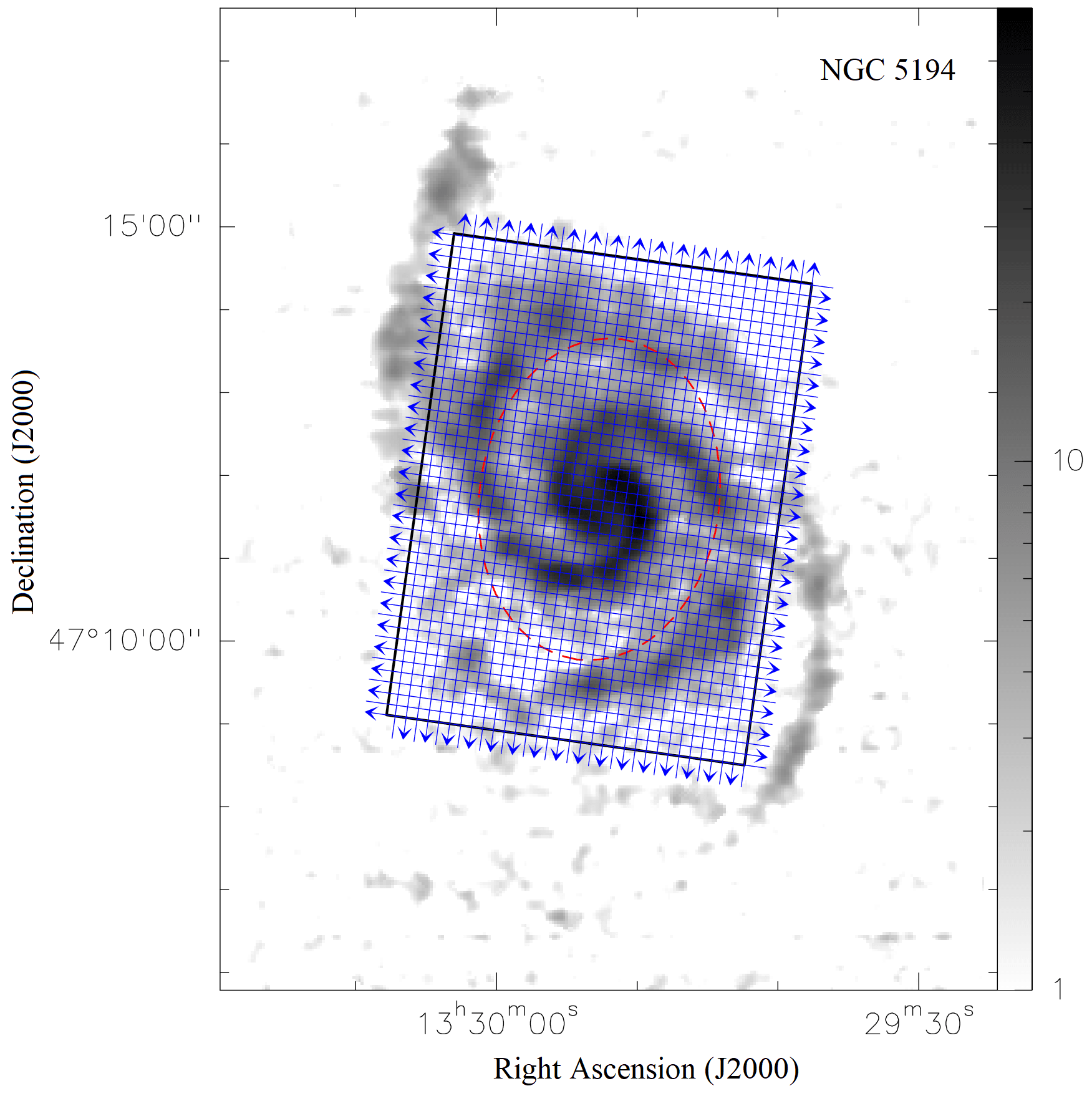}
\caption{On-the-fly mapping strategy for NGC\,5194. The gray scale map shows the $^{12}$CO\,(2-1) integrated intensity map in units of K\,km\,s$^{-1}$ from HERACLES. Arrows indicate the length and orientation of individual scan legs which fully sample the area of interest. The red ellipse shows the $0.5\times r_{25}$ radius of the galaxy}
\label{fig:otf}
\end{figure}

In every target galaxy, we defined a rectangular field that encompassed the area where $^{12}$CO\,(2-1) emission is detected in the HERACLES maps (Table \ref{obsdates}). We mapped these fields using the on-the-fly (OTF) mapping mode with emission-free reference positions close to the galaxies. We scanned each galaxy at $8\arcsec$ per second in multiple paths offset by $8\arcsec$, parallel to the major axis of the scanned field to cover the entire molecular disk. While scanning, we read out one dump every $4\arcsec$ (every 0.5 seconds) to ensure Nyquist sampling. To avoid remnant scan patterns in the final data products, we also scanned the same area in perpendicular (minor axis) direction, with the orientation of the cross-hatched pattern set by the position angle of the galaxy. See Figure \ref{fig:otf} for an example. Additionally, we shifted the grid center by $N\times \sqrt{2}\arcsec$ with $N=0,2,4,6$ along the diagonal of the grid cell to end up with a finer $2\arcsec$-grid (Figure \ref{fig:otf} shows the case of $N=0$). The observing dates for the individual galaxies, area covered, and the orientation of the OTF scans are shown in Table \ref{obsdates}. 

NGC\,5194 (M51) was observed in July and August 2012, as a precursor program to EMPIRE. For this galaxy a slightly different E090 tuning was used, where the local oscillator frequency was set to $\sim 88.7\,$GHz. This configuration allowed to capture the isotopologues from the main dense gas tracers, H$^{13}$CN, H$^{13}$CO$^+$ and HN$^{13}$C \citep[see][]{JIMENEZDONAIRE17}, leaving $^{13}$CO and C$^{18}$O unobserved for this galaxy. The $^{13}$CO data for NGC\,5194, however, was observed as part of the PAWS survey \citep{SCHINNERER13}. We refer the reader to \citet{BIGIEL16} for an analysis of this galaxy and details of the observations.

At the beginning of each observing session the focus of the telescope was set using observations of planets or bright quasars and then observed and corrected again every $\sim 3$ hours, as well as at sunset and sunrise. The telescope pointing was corrected every $1-1.5\,$ hours using a point-like source (quasar or planet) close to the target galaxy. Chopper wheel calibrations were performed every $\sim 10-15$ minutes employing standard hot/cold-load absorber and sky measurements; these are used to perform the first basic calibration and to convert the data to corrected antenna temperature scale ($T_a^*$). Line calibrators were observed as part of EMPIRE once a day during the observing runs, and the measured velocity-integrated intensities varied by only $\sim 3-8\,\%$ between different runs (see Figure \ref{calibrators}), implying a stable relative calibration in EMPIRE.

\begin{table*}
	\caption{EMPIRE dense gas and complementary CO\,(1-0) observing dates and scan orientation.}
	\label{obsdates}
	\centering
	\begin{tabular}{l c c c c c c}
		\hline\hline
		Galaxy & Scanned Field & Scan P.A.$^{(a)}$ & Dates & On-source time & RMS$^{(b)}$ & $T_\textrm{sys}^{(c)}$\\
		& (arcmin$^{2}$) & ($^{\circ}$) & & (h) & (mK) & (K)\\
		\hline
		\multicolumn{7}{c}{{\bf EMPIRE dense gas observations}}\\
		\hline
		NGC~628 & $4.0 \times 4.0$ & 45& Jun 2015; Sep 2015; Oct 2015 & 44 & 2.8 & 110\\ %10
		NGC~2903 & $2.0 \times 3.5$ & 0 & Apr 2016; May 2016; Dec 2016 & 19 & 2.3 & 80\\ %10
		NGC~3184 & $3.0 \times 3.0$ & 0 & May 2016; Jun 2016; Jul 2016 & 24 & 2.2 & 80\\%10
		NGC~3627 & $2.5 \times 4.0$ & 0 & Nov 2015; Dec 2016 & 30 & 2.8 & 120\\%9
		NGC~4254 & $3.0 \times 3.0$ & 90 & Mar 2015; Apr 2015 & 27 & 1.8 & 80\\%18
		NGC~4321 & $4.0 \times 2.5$ & 115 & May 2015; Nov 2015; Jan 2016 & 30 & 2.1 & 90\\%10.
		NGC~5055 & $6.0 \times 3.0$ & 110& Aug 2015; Jan 2016; Jun 2016; Jul 2016 & 38 & 2.7 & 80\\%11
		NGC~5194 & $4.2 \times 5.7$ & 0 & Jul-Aug 2012 & 75 & 2.8 & 110\\%9
		NGC~6946 & $4.5 \times 6.5$ & 243 & Dec 2014; Apr 2015; Jun 2015; Aug 2015 & 52 & 2.7 & 100\\%12.5
		\hline
		\multicolumn{7}{c}{{\bf CO\,(1-0) complementary observations}}\\
		\hline
		NGC~628 & $8.0 \times 8.0$ & 20 & Aug 2016; Oct 2016 & 18 & 16.3 & 300\\
		NGC~2903 & $4.0 \times 7.0$ & 204 & Sep 2016 & 4 & 22.5 & 260\\
		NGC~3184 & $6.0 \times 5.5$ & 180 & Sep 2016 & 8 & 17.3 & 300\\
		NGC~3627 & $5.0 \times 2.5$ & 0 & Sep 2015; Sep 2016 & 6 & 16.2 & 300\\
		NGC~4254 & $7.0 \times 7.0$ & 35 & May 2015 & 8 & 20.2 & 200\\
		NGC~4321 & $6.0 \times 6.5$ & 30 & Aug 2016; Oct 2016 & 12 & 13.1 & 300\\
		NGC~5055 & $6.0 \times 3.2$ & 100 & Aug 2013 & 12 & 15.0 & 200\\		
		NGC~6946 & $10.5 \times 9.5$ & 243 & Aug 2016; Oct 2016 & 13 & 23.5 &300 \\
		\hline
	\end{tabular}
    	\\ \flushleft{{\bf Notes:} $(a)$ Orientation of major axis scans measured from North through East. (b) Typical RMS ($T_\textrm{MB}$) values after reduction for the HCN\,(1-0) and CO\,(1-0) data cubes, respectively, in 4 km s$^{-1}$ velocity channels and at $33\arcsec$ resolution. (c) Average system temperatures.}    
\end{table*}

\subsection{New $^{12}$CO\,(1-0) observations}
\label{sec:co}

A key goal of EMPIRE is to measure dense gas fraction variations and relate these to local ISM conditions. The ratio of high critical density ($>10^5\,$cm$^{-3}$) tracers like HCN\,(1-0), to tracers of total molecular material ($>10^2\,$cm$^{-3}$) such as CO\,(1-0), hereafter CO, is sensitive to gas density changes \citep{LEROY17}. Thus, e.g., the HCN-to-CO ratio is one immediately accessible observational diagnostic of the dense gas fraction. High-quality and uniform $^{12}$CO\,(1-0) data thus play a key role for such measurements.
Although ancillary CO\,(2-1) data exist for the EMPIRE galaxies \citep{LEROY09}, no uniform, high quality, single-dish CO\,(1-0) data set existed for all targets. Therefore, we also obtained new maps of the CO\,(1-0) line emission from each target using EMIR on the IRAM 30-m (PI Jim\'enez-Donaire, projects 061-15 and 059-16; PI Cormier, project D15-12; NGC\,5194 has 30m EMIR CO\,(1-0) data from PAWS). These ancillary data cover a larger region matched to the HERACLES coverage in $^{12}$CO\,(2-1). For these observations, the upper 8\,GHz sub-band of EMIR was set at 3mm to cover the $^{12}$CO\,(1-0) line and the isotopologues $^{13}$CO\,(1-0) and C$^{18}$O\,(1-0).  We centered the remaining 8\,GHz bandwidth at 212.98\,GHz, to capture the $J=2-1$ transition lines of $^{13}$CO and C$^{18}$O. The new CO data set was obtained and processed in the same way as the other EMPIRE data. 

\subsection{Data reduction and processing}

We use the Multichannel Imaging and Calibration Software for Receiver Arrays ({\tt MIRA}) to perform the basic calibration and data reduction. This software is part of the Grenoble Image and Line Data Analysis Software ({\tt GILDAS}) package\footnote{\url{http://www.iram.fr/IRAMFR/GILDAS}; see \citet{PETY05} for more detailed information}. We first convert each spectrum to the antenna temperature scale by combining it with the nearest ``chopper wheel'' calibration scan. Then we subtract the closest OFF measurement from the calibrated spectrum. After this basic calibration, the data for each observed line is written out using the Continuum and Line Analysis Single-dish Software ({\tt CLASS}) package. At every position where a data dump is taken, we extract a spectrum for each individual line of interest, subtract a zeroth order baseline from this spectrum, and regrid the spectrum to have channel width 4 km s$^{-1}$ across a 1500 km s$^{-1}$ bandpass. Then these spectra are written out into a FITS table.

After this, we read the spectra into a custom {\tt IDL} pipeline \citep[based on, but improved from, the HERACLES data reduction pipeline][]{LEROY09}. Here, we identify pathological data showing spikes or platforming effects (intensity offsets in the observed spectra) before fitting any baselines. In order to fit baselines to each spectrum we want to avoid the expected velocity range of the line, which we determine from CO~(1-0) emission. For every spectrum we define a window ranging between 50-300 km s$^{-1}$, centered around the galactic mean CO\,(1-0) velocity. After that, we define two additional windows adjacent to the central one and with the same width, which we use to fit a second order polynomial baseline. This fit is then subtracted from the entire spectrum.

In order to filter remaining pathological spectra, we sort all spectra according to their root-mean-square (RMS, calculated after subtracting the baselines on the line-free windows) relative to the expected value from the radiometer equation and reject the highest 10\%. In addition, we exclude spectra or parts of spectra in velocity, time or polarization, where careful inspection reveals remaining platforming or other issues.

Finally, the data for each spectral line are gridded into a cube, which is later convolved with a Gaussian kernel to a common working resolution of $33\arcsec$ for the purposes of this work. We employed forward and beam efficiencies available from the IRAM documentation\footnote{Online IRAM documentation: \url{http://www.iram.es/IRAMES/mainWiki/Iram30mEfficiencies}} in order to convert the temperature scales, $T^*_\textrm{A}$, to main beam temperature ($T_\textrm{MB}$):
\begin{equation}
T_\textrm{mb} = \frac{F_\textrm{eff}}{B_\textrm{eff}}\, T_\textrm{A}^*.
\end{equation}

For the 2013 campaign, the typical $T_\textrm{MB}/T^*_\textrm{A}$ ratios at 88.6 and 115.3\,GHz are 1.17 and 1.21, respectively. For the remainder of the paper, we work in units of $T_\textrm{MB}$.

\subsection{Final data products}

Because low-$J$ CO lines are much easier to excite, and generally much brighter than those from the dense gas tracers HCN, HCO$^+$ and HNC, and CO isotopologues, we use our new CO(1-0) data to construct masks for the dense gas tracers in position and velocity. For that we select those regions in each galaxy with a $^{12}$CO\,(1-0) signal-to-noise ratio (SNR)$>$4 in at least 2 coincident pixels, and we then expand these by incorporating adjacent pixels with SNR$>$2. In all cases, the dense gas emission appears well contained within these masks.

Maps of integrated intensity for each emission line are created by integrating the masked data cubes along every line of sight. We use the regions outside the bright CO~(1-0) mask (free of signal) to estimate the RMS noise in individual channels as the standard deviation in line-free parts of each spectrum in our datacubes. To generate uncertainty maps of the integrated intensity, we multiplied the derived RMS noise by the channel width (4\,km\,s$^{-1}$) in velocity and by the square-root of the number of channels used to compute the integrated intensity along the line of sight. We find an RMS noise range of 1.8-2.9\,mK ($T_\textrm{MB}$), with an average of $\sim$2.4 mK for HCN\,(1-0). These values are slightly higher in the case of our HCO$^+$\,(1-0) observations, where the typical RMS noise varies within 2.0-3.0\,mK, with an average of 2.6\,mK. We find our HNC\,(1-0) cubes to be the noisiest within the studied dense gas tracers, where the calculated RMS noise ranges from 2.7\,mK to 5.0\,mK, with an average of 3.5\,mK. Within our sample, we consistently find the datacubes for NGC\,628 and NGC\,5194 to be the noisiest, whereas NGC\,4254 shows the lowest RMS noise. Regarding our complementary CO\,(1-0) data taken at $\sim 115\,$GHz, we find typically a much higher RMS noise level of $\sim$18\,mK (within the range of 15-20\,mK) due to the much shorter integration times needed to detect this line (see Table \ref{obsdates}), although the SNR achieved is much higher than for the dense gas lines. This translates into uncertainties on the integrated intensities of the order 0.07 K km s$^{-1}$ for the dense gas tracers, about 0.09~K~km~s$^{-1}$ for the CO isotopologues ($^{13}$CO and C$^{18}$O) and 0.50~K~km~s$^{-1}$ for CO.

\section{Ancillary data}
\label{sec:ancillary}
The EMPIRE targets are some of the best-studied nearby galaxies. They all have existing data across the electromagnetic spectrum that provide an excellent characterization of the distribution of gas, stars, dust, and recent star formation.

The atomic gas content is measured using data from ``The H\,I Nearby Galaxy Survey'' \citep[THINGS,][]{WALTER08}. This VLA large program mapped 34 galaxies, including 7 of the 9 EMPIRE galaxies, in the 21 cm line with high angular ($\sim 10\arcsec$) and velocity ($\sim 5\,$km s$^{-1}$) resolution and sensitivity. NGC\,4254 and NGC\,4321 were not covered by THINGS, therefore we used archival VLA maps \citep[from][]{SCHRUBA11,LEROY13}. 
	
We employ broadband IR photometry in the $3.6-500\,\mu$m range, from the ``{\it Spitzer} Infrared Galaxies Survey'' \citep[SINGS,][]{KENNICUTT03} and the ``Key Insights on Nearby Galaxies: a Far-Infrared Survey with Herschel'' surveys \citep[KINGFISH,][]{KENNICUTT11}. This broadband IR emission is then used to estimate the total infrared emission following \citet{GALAMETZ13} and star formation rate surface densities ($\Sigma_\textrm{SFR}$), as described in Section \ref{sec:phys_parameters}. NGC\,2903 and NGC\,5194 lack KINGFISH coverage, therefore we use {\it Spitzer} 24$\mu$m emission maps \citep[Local Volume Legacy, LVL, from][]{DALE09} for NGC\,2903 and other {\it Herschel} data for NGC\,5194 \citep[Very Nearby Galaxy Survey, VNGS, from][]{BENDO12}. We use the {\it Spitzer} Survey of Stellar Structure in Galaxies (S4G) processing of the IRAC data \citep{SHETH10} and LVL processing of the MIPS data \citep{DALE09,LEE09} to compute stellar surface densities ($\Sigma_*$) in our targets.

\section{Estimating physical parameters and spectral stacking}
\label{sec:phys_parameters}
Converting observed intensities into physical quantities is subject to assumptions and hence is somewhat uncertain \citep[e.g.,][]{KENNICUTT12,BOLATTO13,SANDSTROM13,USERO15}. We therefore choose in the following to report our results using direct observables (e.g., intensities $I_\textrm{HCN}$) in addition to reporting physical quantities (e.g., $\Sigma_{\textrm{dense}}$). These are derived from linear transformations following \citet{USERO15} as detailed below.

\subsection{Molecular gas surface density} We estimate the mass surface density of molecular gas using our new maps of CO(1-0) line emission (Section \ref{sec:co}) to trace the molecular hydrogen (H$_2$) content. The molecular surface density can be derived as:
\begin{equation}
\Sigma_\textrm{mol}=\alpha_\textrm{CO}I_\textrm{CO}\textrm{cos}(i).
\end{equation}

The $\textrm{cos}(i)$ factor corrects for inclination and $\alpha_\textrm{CO}$ is the CO-to-H$_2$ conversion factor. We assume this value to be Milky Way-like throughout the sample ($\alpha_\textrm{CO}=$ 4.4 $M_\odot$ pc$^{-2}$ (K km s$^{-1}$)$^{-1}$, i.e. including the 1.36 factor for helium), which is commonly adopted for massive, solar metallicity galaxies \citep[see][]{BOLATTO13}. Although variations from galaxy to galaxy and within galaxies are present, the most up-to-date values calculated in the disks of nearby galaxies largely agree with the Galactic value \citep{SANDSTROM13,CORMIER18}. Galaxy centers show the largest differences, with systematically lower values and the scatter per radius is only a factor of 2 \citep{BOLATTO13,SANDSTROM13,CORMIER18}. Provided that our focus is on kpc-size regions and late-type, normal spirals, we do not expect large variations across the disks. In fact, the average disk metallicities only range from $12 + \log$O/H of 8.34 to 8.68 (about a factor of two) among our target galaxies (see Table \ref{table:sample}). Despite these observed variations of $\alpha_\textrm{CO}$ in galaxy centers, for simplicity and lacking a detailed physical understanding, we adopt the fixed Milky Way conversion factor and discuss its implications in Section \ref{sec:discussion}.

\subsection{Dense gas surface density} As for the dense gas surface densities, one can also define a conversion factor ($\alpha_\textrm{HCN}$) to calculate the mass surface density of dense molecular gas, $\Sigma_\textrm{dense}$, from the HCN\,(1-0) integrated intensity:
\begin{equation}
\Sigma_\textrm{dense}=\alpha_\textrm{HCN}I_\textrm{HCN}\textrm{cos}(i).
\end{equation}
\citet{GAO04} estimated $\alpha_\textrm{HCN}=$ 10 $M_\odot$ pc$^{-2}$ (K km s$^{-1}$)$^{-1}$ as a typical value for the disks of normal, star-forming galaxies based on virial theorem and radiative transfer arguments. For that, they assume self-gravitating dense gas clumps with typical $n\sim3\times10^4\,\textrm{cm}^{-3}$ and brightness temperatures of 35\,K. \citet{WU10} found larger values of $\sim 20$  $M_\odot$ pc$^{-2}$ (K km s$^{-1}$)$^{-1}$ with a 0.54~dex scatter in a more complete study of resolved dense clumps, for which the mass was determined through the virial method. However, this dense gas conversion factor is not as well characterized as $\alpha_\textrm{CO}$, and should thus be considered at least as uncertain as the latter (see Section \ref{sec:hcn_emission}). For consistency with previous extragalactic work, we use the conversion factor estimated in \citet{GAO04} to calculate dense gas surface densities.

There are a number of caveats associated with using HCN emission as a dense gas tracer. We review these in detail in Section \ref{sec:hcn_emission}, but also mention them briefly here. First, the mean density traced by any molecular line reflects the convolution of an underlying density distribution with a density-dependent emissivity. As a consequence, in the common case where low density gas is more abundant than high density gas, significant emission can also arise from gas density below the critical density. Second, the optical depth, and so strength of radiative trapping, associated with HCN is not strongly constrained. Nor is the HCN abundance or excitation perfectly known. While a detailed assessment of HCN emissivity and intensity, dense gas mass, and effective density with the data in hand is not possible, the strongly different effective critical densities of low-$J$ HCN and CO lines render, e.g., the HCN-to-CO ratio a good first-order proxy for changing mean gas density on kpc scales \citep[also supported by radiative transfer modeling,][]{LEROY17}. Several other processes such as UV, X-ray or cosmic ray heating can also alter the emissivity of HCN via chemistry. These issues are likewise impossible to address with only the EMPIRE data. We do indirectly address this issue by using different line ratios (e.g., HCO$^+$-to-CO), and we expect chemistry effects to average out at least to some degree on kpc-scales. However, because of these caveats we present and analyze our results using direct observables and recommend caution regarding their interpretation (see Section \ref{sec:discussion}).

\subsection{Atomic gas surface density} We calculate the atomic gas mass surface density, $\Sigma_\textrm{HI}$, from the 21 cm line integrated intensity maps obtained by THINGS \citep{WALTER08}, via:
\begin{equation}
\frac{\Sigma_\textrm{HI}}{M_\odot\,\textrm{pc}^{-2}}=0.020\,\frac{I_\textrm{21\,cm}}{\textrm{K km\,s}^{-1}}\,\textrm{cos}(i).
\end{equation}
This conversion assumes optically thin emission and takes any missing zero-spacing correction to be negligible, reasonable assumptions for the THINGS data set provided the good agreement found between interferometric-only and single dish measurements inside the THINGS $30'$ primary beam \citep[see][]{WALTER08}. In addition, it includes a factor of 1.36 to reflect the presence of helium.

\subsection{Stellar surface density} 
The stellar structure observed in galaxy disks can provide an interesting insight to the distribution of dense gas: gas follows the stellar gravitational potential and hence stellar distribution in the galaxy disk can be an important driver of the local dynamical equilibrium pressure. \citet{USERO15} and \citet{BIGIEL16} employed the {\it Spitzer} 3.6 $\mu$m maps \citep{DALE09} to derive the stellar surface density, $\Sigma_*$, since photospheric emission from old stars is responsible for most of the emission seen in the 3.6\,$\mu$m band. However there can be contamination from dust heated by young stellar populations, therefore we follow the approach used by \citet{GALLAGHER18}, and use contaminant-corrected maps from \citet{QUEREJETA15}. They employed re-processed 3.6\,$\mu$m and 4.5\,$\mu$m photometry as part of S$^4$G, and used the ``Independent Component Analysis'' (ICA) method presented in \citet{MEIDT12} to separate the contribution from the dust emission heated by young stars in the 3.6$\mu$m band (about $10{-}30$\%). Finally, we derive stellar surface densities by assuming a mass to light ratio of $352$~M$_\odot$~pc$^{-2}$~(MJy~sr$^{-1}$)$^{-1}$, which corresponds to approximately 0.5 M$_{\odot}$ per L$_{\odot}$ (\citealt{MEIDT14}).

\subsection{Total infrared intensity and star formation rate} Following the same approach as in our previous work \citep[e.g.,][]{USERO15,BIGIEL16,JIMENEZDONAIRE17B,CORMIER18}, we use the total infrared (TIR) surface brightness as a proxy for the local surface density of star formation. To estimate this, we combine $\lambda = 70$, 160, and 250\,$\mu$m maps from {\em Herschel} \citep[KINGFISH,][]{KENNICUTT11}. We convolve these to match the $33\arcsec$ beam of our EMPIRE data using the kernels from \citet{ANIANO11}, calculate the TIR surface brightness following \citet{GALAMETZ13}:
\begin{equation}
\label{eq:sir_galametz}
\Sigma_\textrm{TIR}= \sum\,c_i\,\Sigma_i,
\end{equation}

\noindent where $\Sigma_i$ refers to the surface brightness in a given {\it Herschel} band $i$. We then convert to star formation rate surface density using the prescription of \citet{MURPHY11}: 
\begin{equation}
\label{eq:sfr_murphy}
\frac{\Sigma_\textrm{SFR}}{M_\odot\,\textrm{yr}^{-1}\,\textrm{kpc}^{-2}}=1.48\times10^{-10}\frac{\Sigma_\textrm{TIR}}{L_\odot\,\textrm{kpc}^{-2}}.
\end{equation}

NGC~2903 lacks {\em Herschel} data, therefore we use {\it Spitzer} 24\,$\mu$m and 70\,$\mu$m \citep[from LVL,][]{DALE09} data to estimate the TIR surface brightness, following the same method. We motivate this choice and discuss alternative SFR tracers in Section \ref{appendix:sfr-tracers}. We find that our results are robust against the choice of SFR tracer, which was also the conclusion reached in an extensive similar analysis by \citet{GALLAGHER18}.

In the study of $^{13}$CO\,(1-0) emission from EMPIRE, \citet{CORMIER18} compared TIR estimates for all galaxies in our sample using both SED modeling and the prescriptions of \citet{GALAMETZ13}. They find differences between the two estimates on the order of 10\% when combining the MIPS, PACS and SPIRE bands, and about 20\% when using the MIPS bands only.

\subsection{Hydrostatic pressure of the ISM}
\label{sec:data_pressure}
The gravitational potential of a galaxy at any point in the disk is the sum of contributions from the ISM, stars and dark matter. In hydrostatic equilibrium, the midplane pressure of the gas in a galaxy disk will adjust to support its weight in this combined gravitational potential of gas and stars. We might expect this midplane, external pressure to be coupled to the mean internal pressure of molecular clouds \citep[e.g.,][]{OSTRIKER10,HUGHES13}, setting its individual pressure and (surface) densities {\em before star formation takes place}. These, in turn, may play a key role regulating the cloud density structure and subsequent star formation.

In this picture the hydrostatic pressure, $P_\textrm{h}$, increases with gas volume density, and it would determine not only the ability of the ISM to form molecular hydrogen \citep{ELMEGREEN89,ELMEGREEN94}, but also the initial ability of gas at any particular density to form stars \citep[e.g.,][]{HELFER97,USERO15,MEIDT18}. As $P_\textrm{h}$ rises, so would the mean density of the gas in the clouds, which would lead to higher observable dense gas fractions. Thus, a number of recent works have focused on this dynamical equilibrium pressure as a key parameter related to the fraction of dense gas and star formation across large parts of local galaxies \citep[e.g.,][]{GALLAGHER18}.

Following \citet{ELMEGREEN89}, \citet{WONG02}, and \citet{BLITZ06}, the hydrostatic pressure needed to balance the gravity on the gas in the disk can be expressed as:
\begin{equation}
\label{eq:pressure}
P_\textrm{h}=\frac{\pi}{2}\,G\,\Sigma_\textrm{gas}\left(\Sigma_\textrm{gas}+\frac{\sigma_\textrm{g}}{\sigma_{*,_\textrm{z}}}\,\Sigma_{*}\right),
\end{equation}
where $\Sigma_\textrm{gas}$ is the total atomic and molecular surface density, $\Sigma_\textrm{*}$ is the stellar surface density, $\sigma_\textrm{g}$ is the velocity dispersion of the gas and $\sigma_{*,_\textrm{z}}$ is the stellar velocity dispersion along the vertical direction. While the first term in the equation expresses the gas self-gravity, the second one reflects the weight of the gas in the stellar potential well. We neglect the contribution of dark matter to the mass volume density, which in the inner parts of galaxies is dominated by the stars near the disk midplane.

Since direct measurements of stellar velocity dispersion in nearby galaxy disks are rare, we adopt a series of assumptions to obtain $P_\textrm{h}$ as a function of more easily observable quantities. Following \citet{LEROY08}, we assume a self-gravitating stellar disk characterized by a scale height $h_*=\frac{1}{2}\,\sqrt{\frac{\sigma_{*,z}^2}{2\pi G \rho_*}}$, where $\rho_*$ is the stellar volume density. The scale height, $h_*$, is typically observed to be constant with radius across the star forming disks of spiral galaxies \citep[e.g.,][]{VANDERKRUIT88,KREGEL02,VANDERKRUIT11}. The stellar surface density and the midplane stellar volume density are then related: $\Sigma_*\approx 4\,\rho_*\,h_*$. We adopt $\sigma_\textrm{g} \approx 15$~km~s$^{-1}$, a value observed to be appropriate for large scales and high surface density regions of galaxy disks \citep[e.g.,][]{TAMBURRO09,CALDUPRIMO13,LEROY16,SUN18}. We refer the reader to \cite{LEROY08} and \cite{GALLAGHER18} for a more detailed description of the assumptions taken in the hydrostatic pressure derivation. We expect this pressure estimation to be a good representation of the time-averaged hydrostatic pressure needed to balance the galaxy disk against its own self-gravity and the stellar gravitational potential well. In the following, we will refer to it as dynamical-equilibrium pressure $P_{\rm DE}$.

\subsection{Spectral stacking technique}
\label{sec:stacking}
The emission coming from high critical density tracers such as HCN is faint for individual lines of sight, especially in the inter-arm regions and outer parts of galaxies. EMPIRE's wide coverage includes significant area where our target lines are not detected at high significance over individual lines of sight. To increase the signal-to-noise we thus also average independent spectra over extended regions (e.g., deriving radial profiles) using a spectral stacking technique that leverages our high signal-to-noise CO data as a prior \citep{SCHRUBA11,CALDUPRIMO13,JIMENEZDONAIRE17}. 

\begin{figure*}
	\centering
	\includegraphics[scale=0.13]{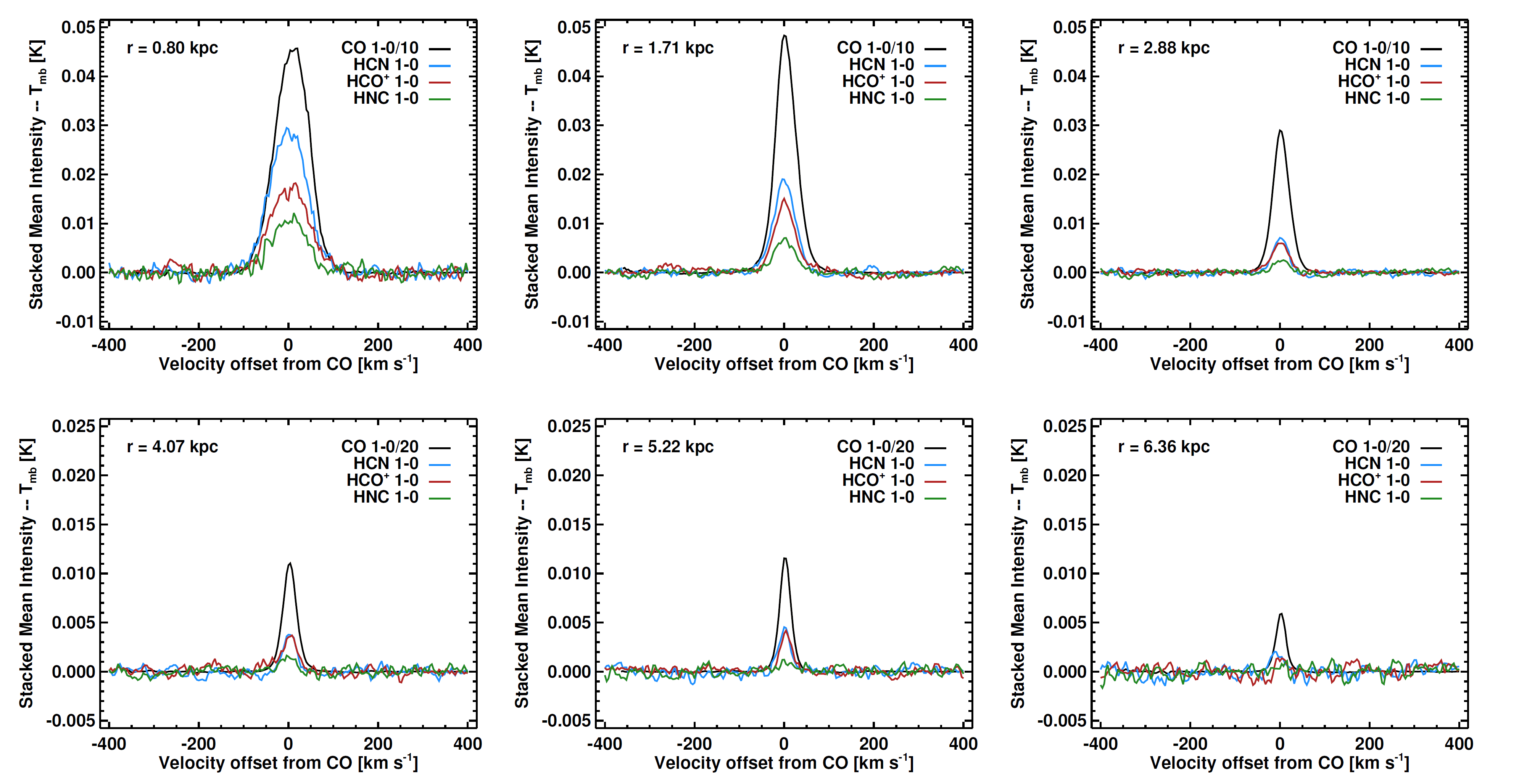}
	\caption{Example of stacked CO\,(1-0), HCN(1-0), HCO$^+$\,(1-0) and HNC(1-0) spectra in $30\arcsec$ ($\sim 1.5\,\textrm{kpc}$) radial bins for NGC\,5194. Galactocentric radii are shown in the upper left of each panel. For ease of comparison, CO spectra are scaled (see legend). }
	\label{fig:dif_stacks}
\end{figure*}

Specifically, we measure the mean velocity along each line of sight from the $^{12}$CO line, and assume that the dense gas tracer emission is distributed over similar velocities (as we do observe in HCN-bright regions). This value, which varies across galaxy disks due to rotation, is then used as a reference for the spectral stacking. The velocity axis of all our spectral lines is then aligned to the local CO mean velocity and the spectra is then subsequently stacked. In Figure \ref{fig:dif_stacks} and Figures \ref{fig:dif_stacks2}-\ref{fig:dif_stacks9} we provide an example of the resulting HCN spectral stacks (blue lines) in radial bins of $30\arcsec$ for one EMPIRE galaxy.

We fit the stacked spectrum of each molecular line with a single Gaussian profile or a double-horn profile. The latter profile is adapted to better describe some of the galaxy centers where the observed emission lines appear broad, with a flattened peak due to spatially unresolved gas motions that coincide with central bars or molecular rings. To perform the fit, we center a 100 km s$^{-1}$-wide window on the peak of emission to have an initial guess for the line width and use the {\tt MPFIT} function in {\tt IDL}. The free parameters we calculate from the fit are the line center velocity, the peak intensity, and the velocity dispersion. We compute the uncertainties on the integrated intensity as:
\begin{equation}
    \label{eq:unc}
    \Delta\,II = \sigma_\textrm{rms}\times \Delta\,v_\textrm{chan}\times \sqrt{\frac{\textrm{FWHM}_\textrm{line}}{\Delta\,v_\textrm{chan}}},
\end{equation}

\noindent where $\sigma_\textrm{rms}$ is the 1$\sigma$ RMS value of the noise in K, which is measured from the signal-free part of the spectrum, $\Delta\,v_\textrm{chan}$ is the width of each channel in units of km\,s$^{-1}$, and $\textrm{FWHM}_\textrm{line}$ is the full width at half maximum of the line derived from the fit, also in km\,s$^{-1}$. When the emission lines remain undetected (below 3$\sigma$ RMS of the noise), we compute 3$\sigma$ upper limits on the integrated intensity. These are derived integrating over a Gaussian profile with a peak set to the 3$\sigma$ RMS value of the noise, and a width set to the $\textrm{FWHM}_\textrm{line}$ found for the high signal-to-noise CO line, stacked over the same physical region.

\section{Results}
\label{sec:results}

\subsection{Distribution of dense gas emission}
\label{sec:intensity}

Figures \ref{fig:maps1}-\ref{fig:maps9} show integrated intensity maps and azimuthally averaged profiles of line intensities, line ratios, and physical conditions for each EMPIRE target. The top left panel shows the infrared dust continuum at 70\,$\mu$m tracing the location of recent star formation activity. The top right panel shows line-integrated $^{12}$CO\,(1-0) intensity, from our new maps. Grey contours in the top right panel show the HCN~(1-0) line integrated intensity. The middle row includes radial profiles for the brightest lines detected in the EMPIRE survey (left); and key quantities characterizing the galactic ISM structure (right). The bottom row shows the radial profiles of the ratios of the main dense gas tracers to CO\,(1-0), tracing molecular gas (HCN/CO, HCO$^+$/CO and HNC/CO, left panel), and among the dense gas tracers (HCO$^+$/HCN, HNC/HCN, HNC/HCO$^+$, right panel).

Generally, the distribution of HCN intensity matches the large-scale structure traced by CO and 70 $\mu$m emission. The HCN intensity peaks at the center of each target and then appears prominent along the spiral arms (e.g., NGC\,5055, NGC\,5194 and NGC\,6946) and central bars (e.g., NGC\,2903 and NGC\,3627). Globally, we find the brightest emission in NGC\,6946 and the weakest in NGC\,3184. 

Outside galaxy centers, we find that the HCN integrated intensity is $\sim 30-70$ times weaker than CO, on average. As a result, we only detect the brightest individual lines of sight at high signal-to-noise in HCN. The spectral stacking approach described in Section \ref{sec:stacking} still allows us to recover the line signal at good SNR after integrating over a larger area.

\begin{figure*}
	\begin{center}
		\includegraphics[scale=0.58]{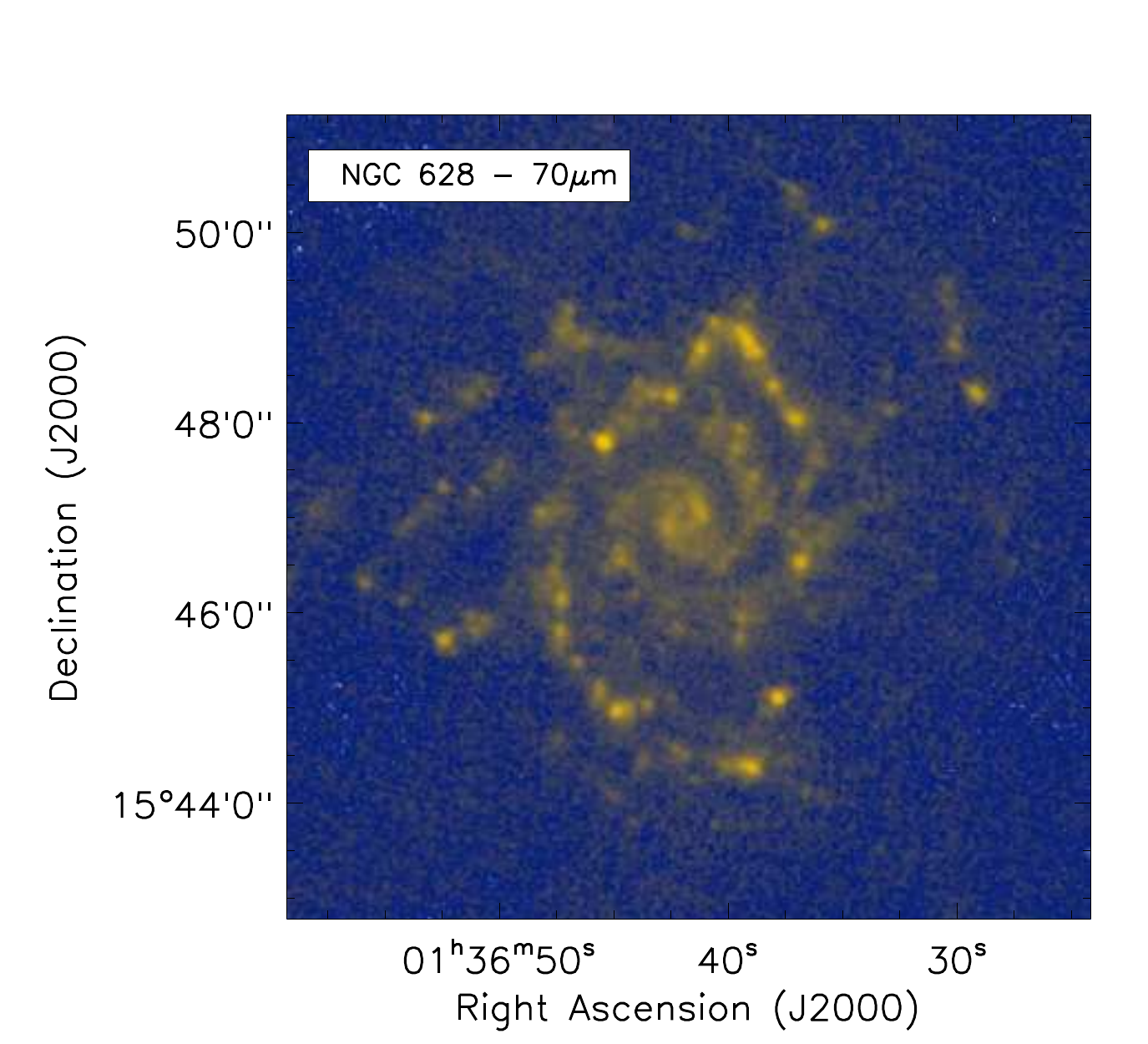}
		\includegraphics[scale=0.6]{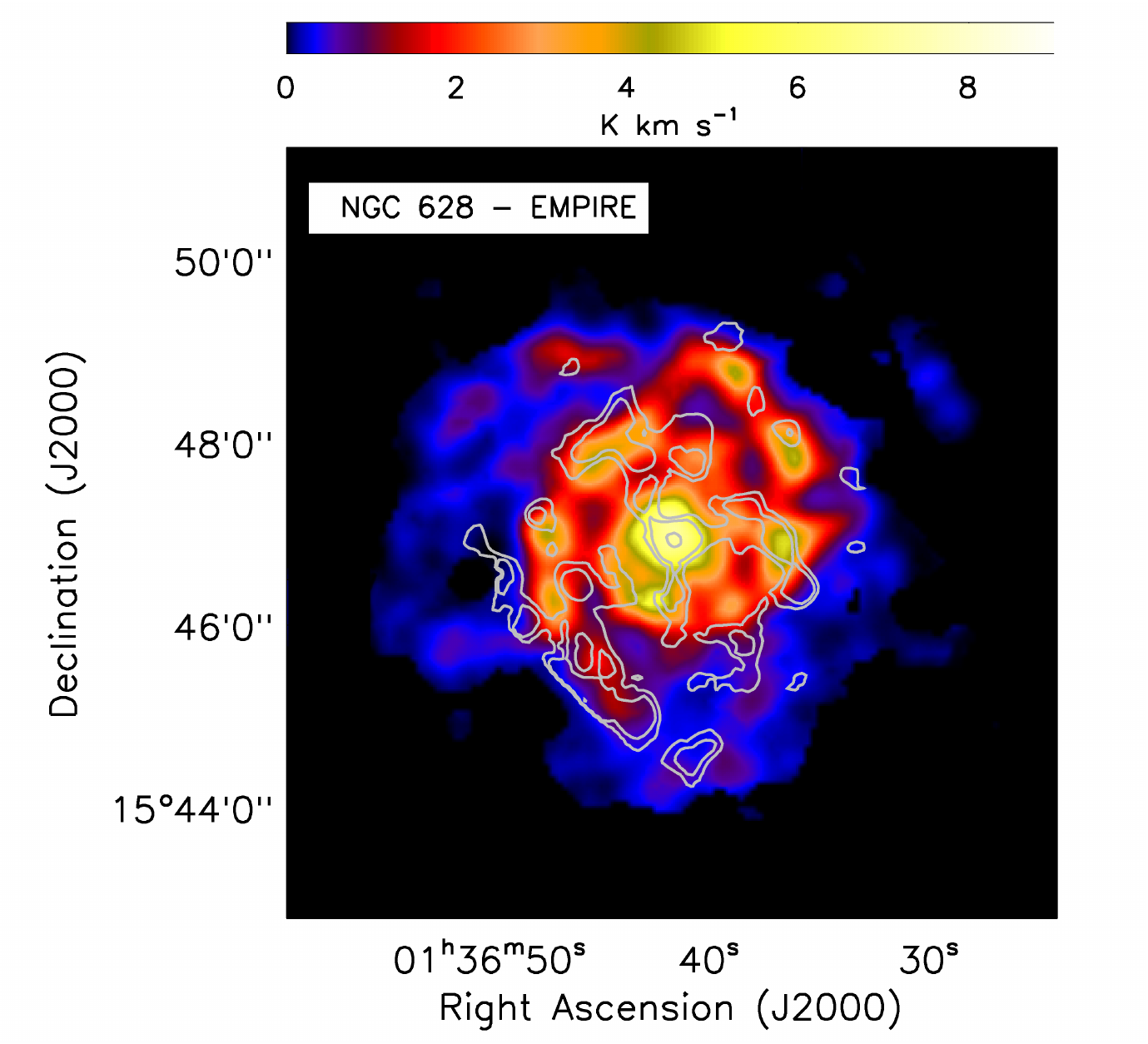}
		\includegraphics[scale=0.4]{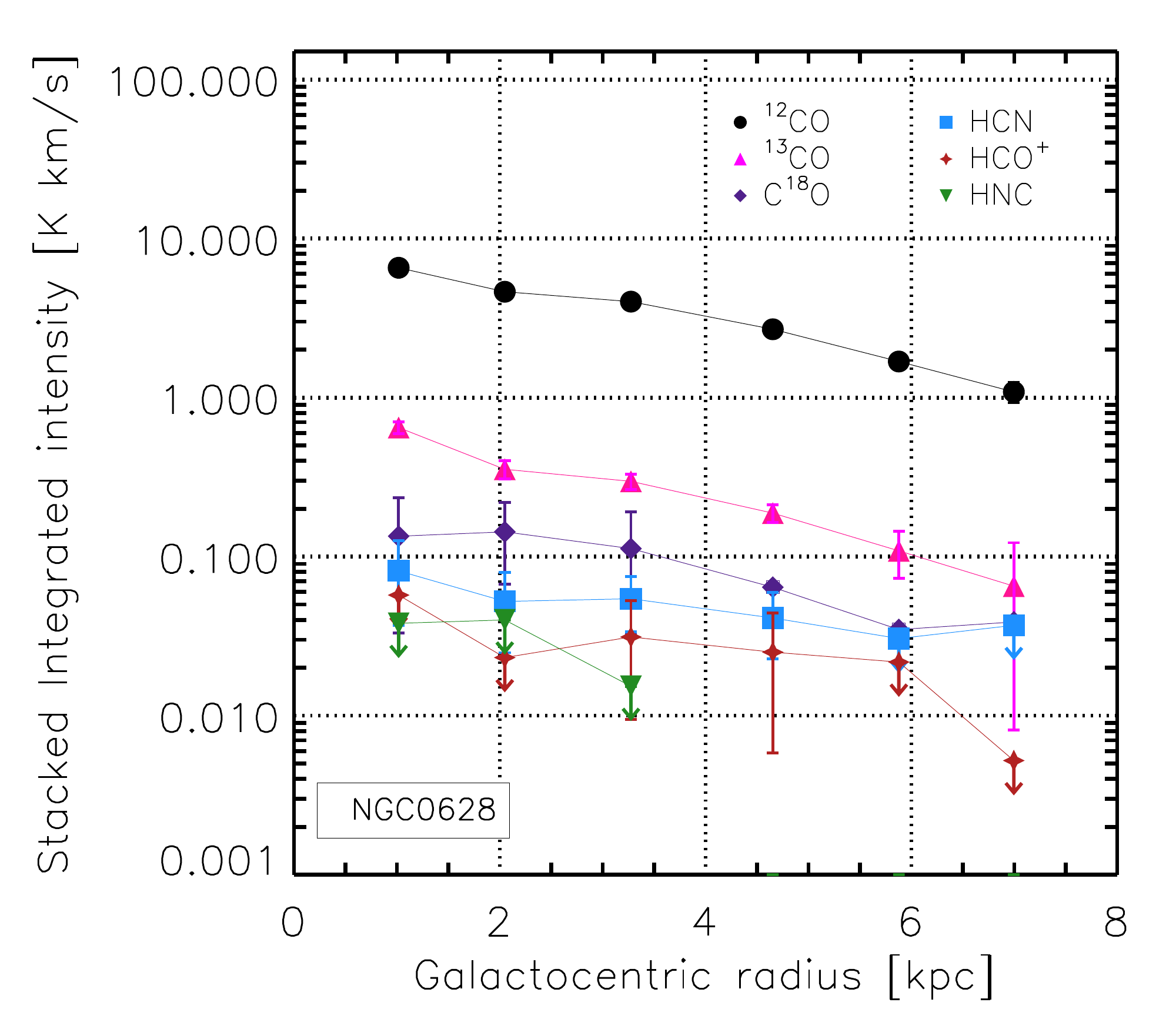}
		\includegraphics[scale=0.4]{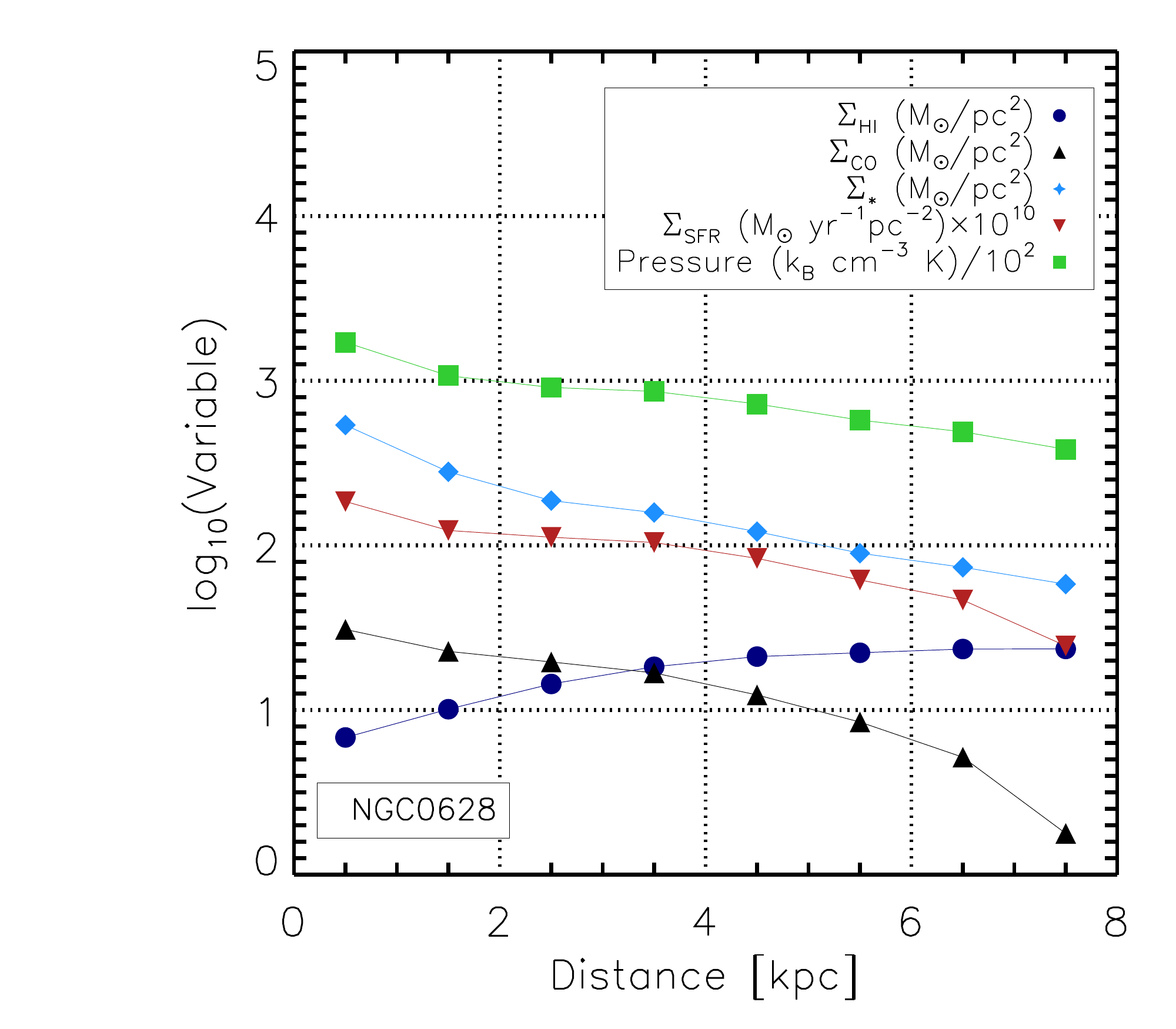}
		\includegraphics[scale=0.4]{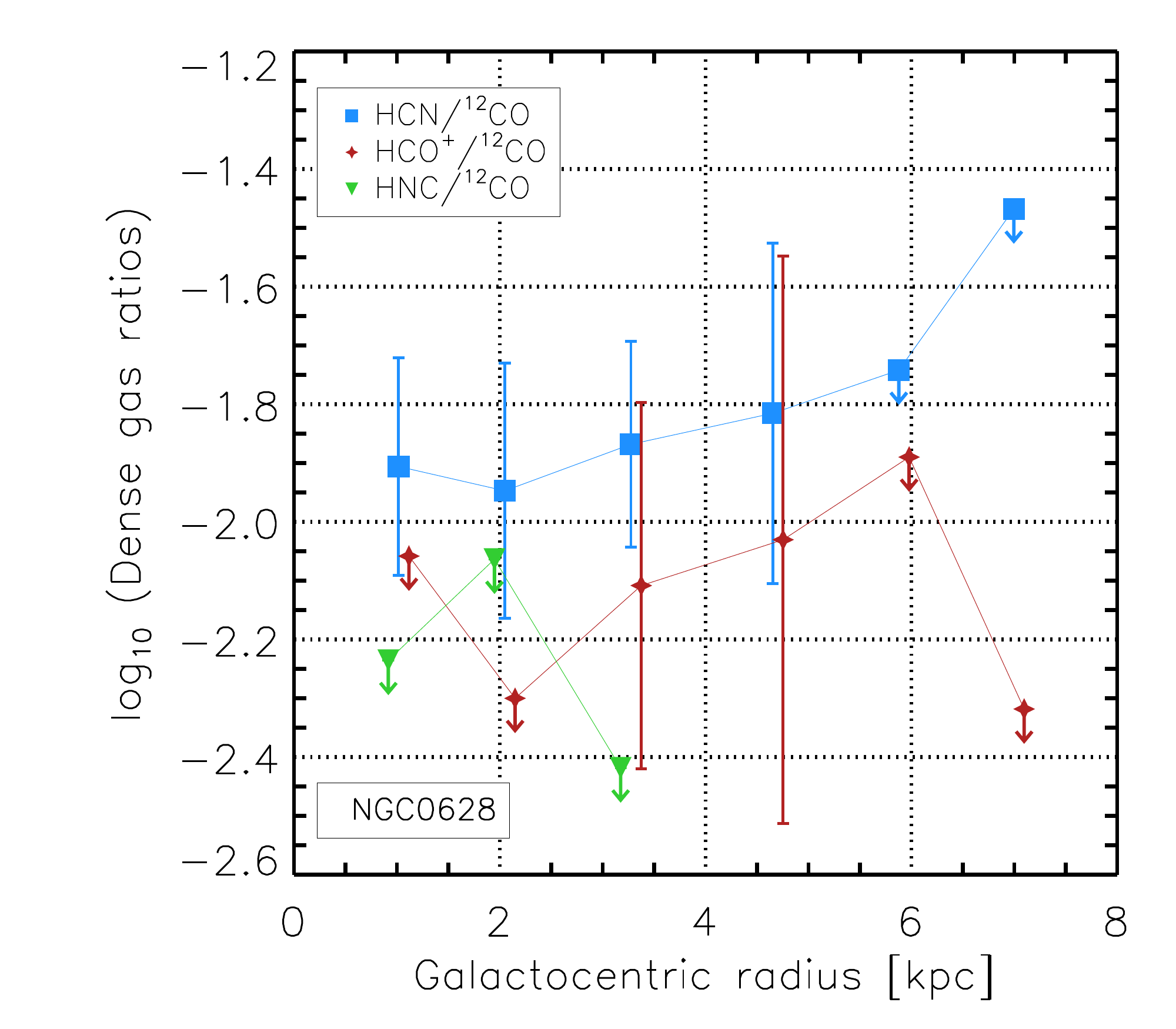}
		\includegraphics[scale=0.4]{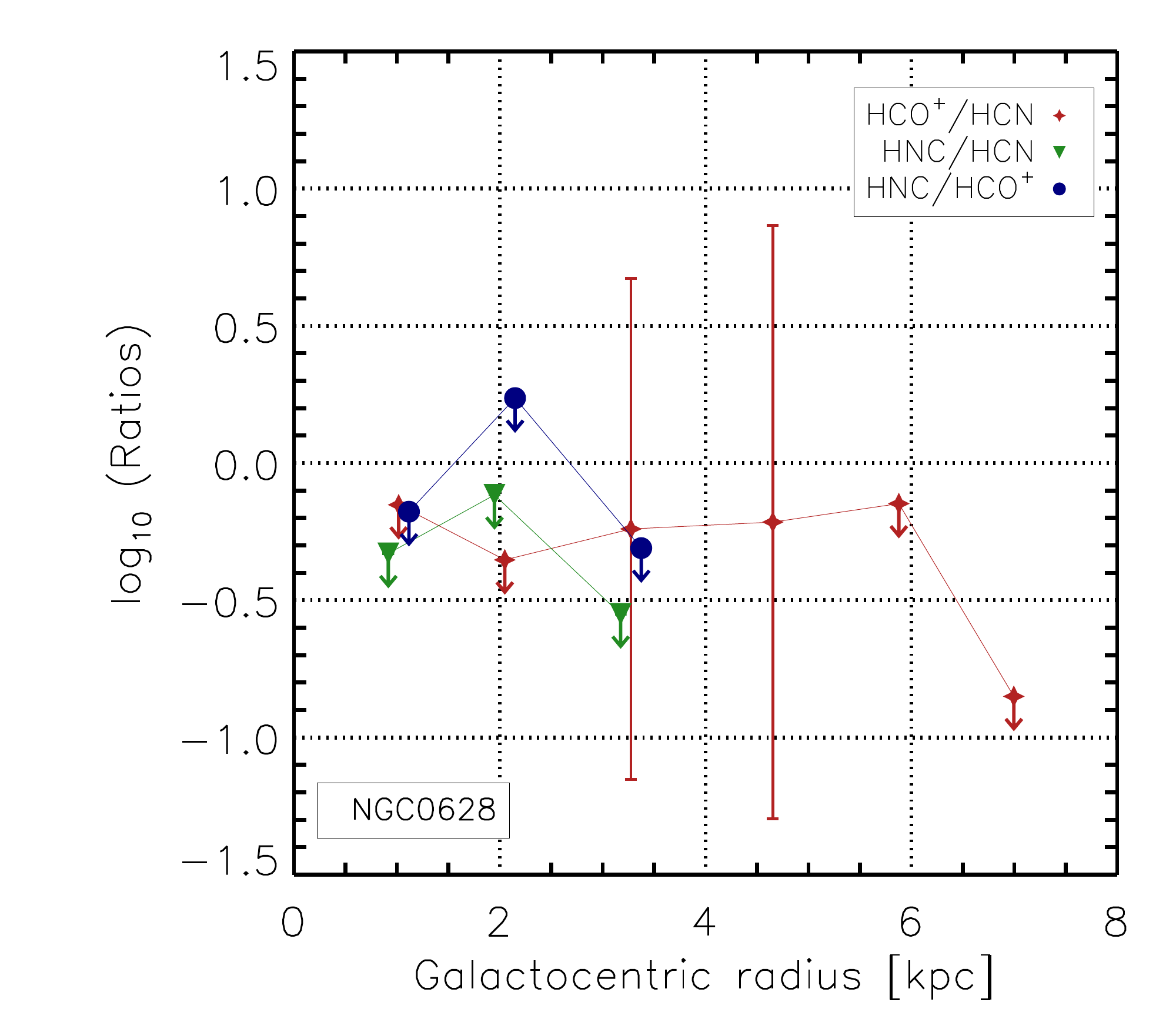}
	\end{center}
	\caption{Atlas of observations for the EMPIRE sources. {\it Top left: Herschel} 70$\mu$m map tracing star formation at its native resolution of $\sim 6\arcsec$. {\it Top right:} HCN\,(1-0) contours (0.5, 0.8 and 1.0\,K\,km\,s$^{-1}$, white) over $^{12}$CO\,(1-0) integrated intensity (K\,km\,s$^{-1}$) at $33\arcsec$ resolution. {\it Middle left}: Azimuthally stacked integrated intensity profiles for the main EMPIRE lines in $30\arcsec$ radial bins. The stacks span the entire galaxy disks, out to $\sim 8$\,kpc. Points show secure detections ($>3\sigma$) and arrows show 3$\sigma$ upper limits. {\it Middle right}: surface density profiles for tracers of atomic (H\,I), bulk molecular (CO), stellar ({\it Spitzer} 3.6$\mu$m), and SFR (TIR) surface density and dynamical equilibrium pressure. {\it Bottom}: Ratio (K\,km\,s$^{-1}$) of stacked integrated intensities of main dense gas tracers and CO integrated intensity (left) and among dense gas tracers (right) as a function of radius.}
	\label{fig:maps1}
\end{figure*}

\begin{figure*}
	\begin{center}
		\includegraphics[scale=0.6]{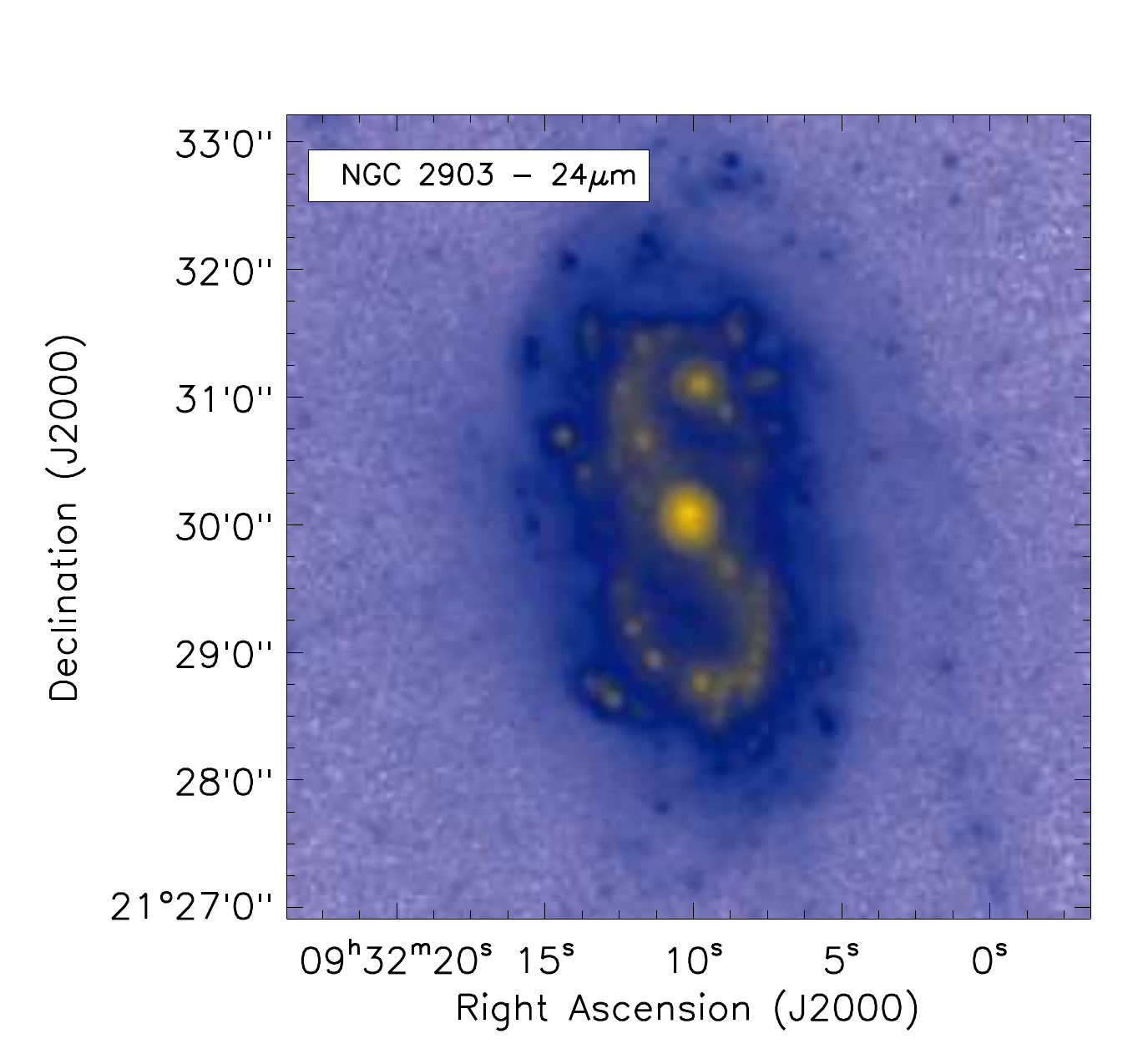}
		\includegraphics[scale=0.6]{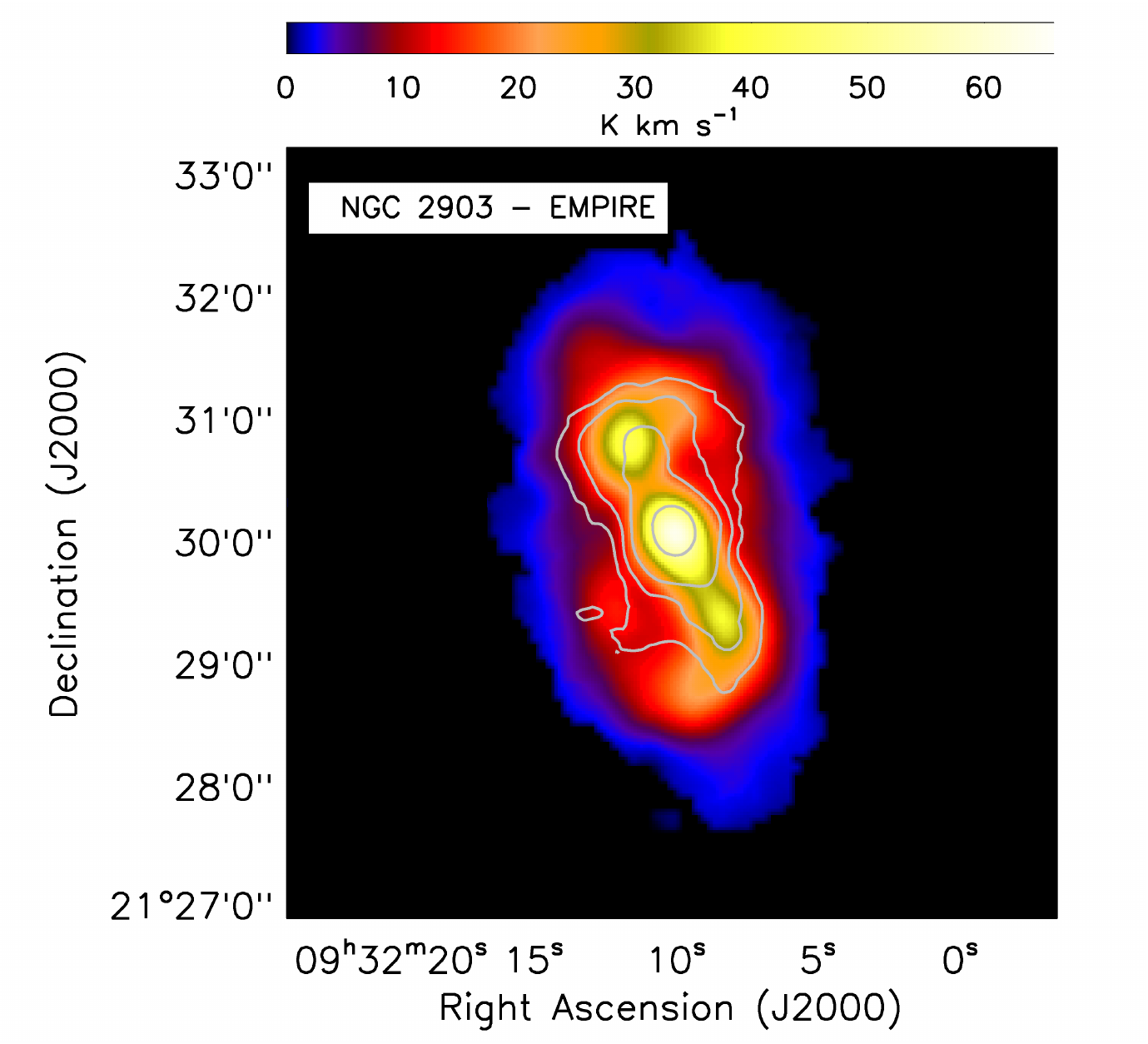}
		\includegraphics[scale=0.4]{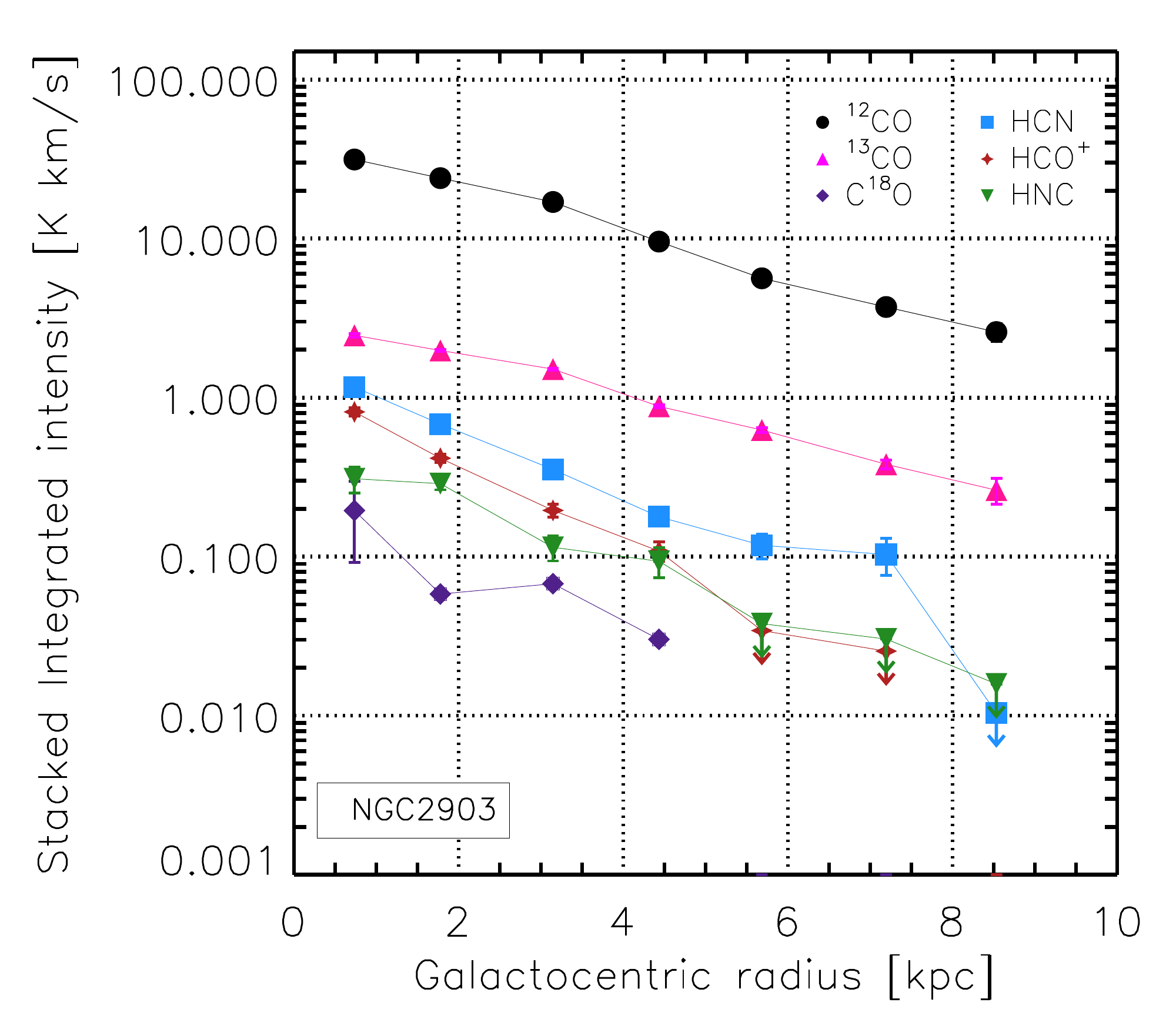}
		\includegraphics[scale=0.4]{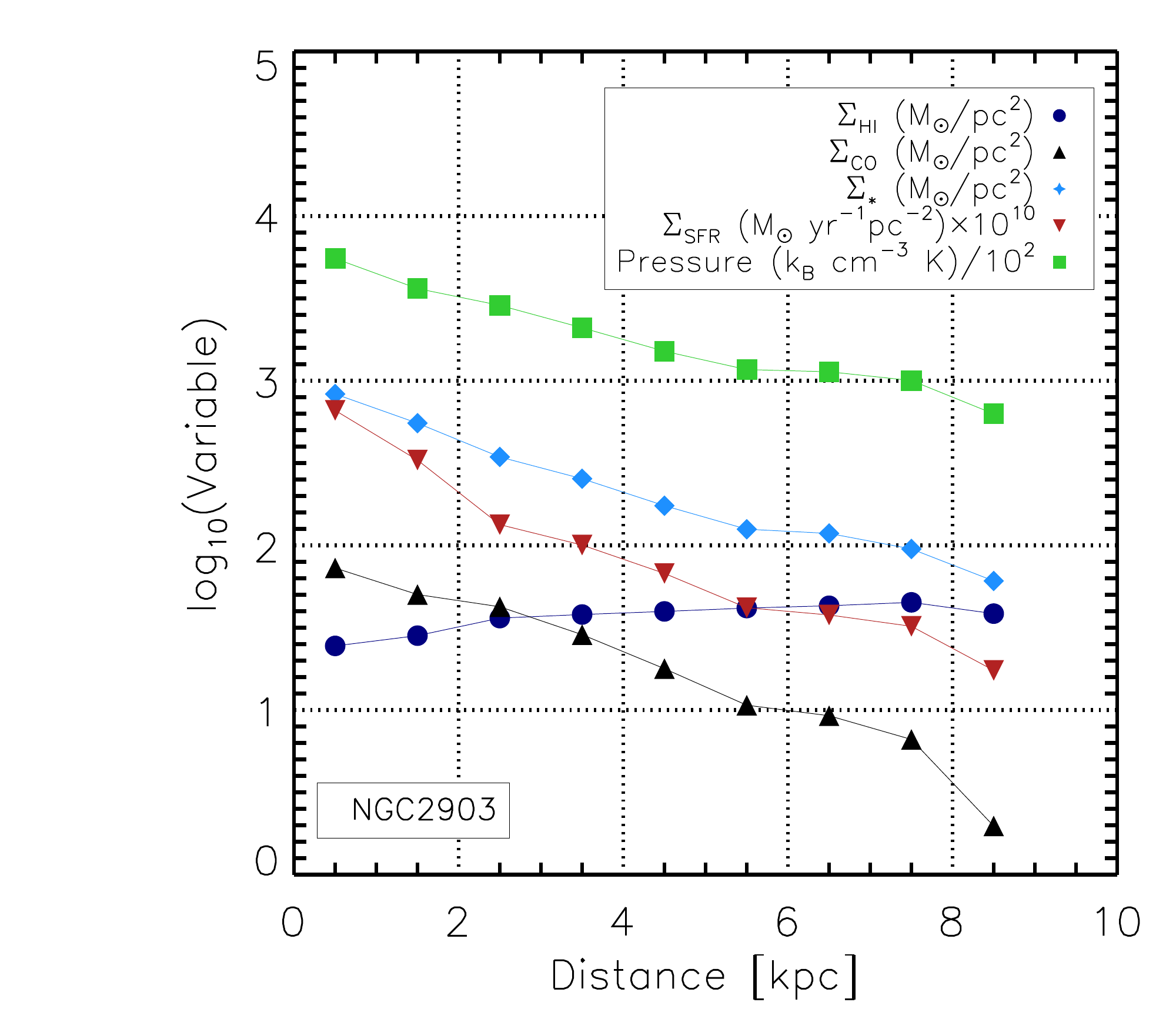}
		\includegraphics[scale=0.4]{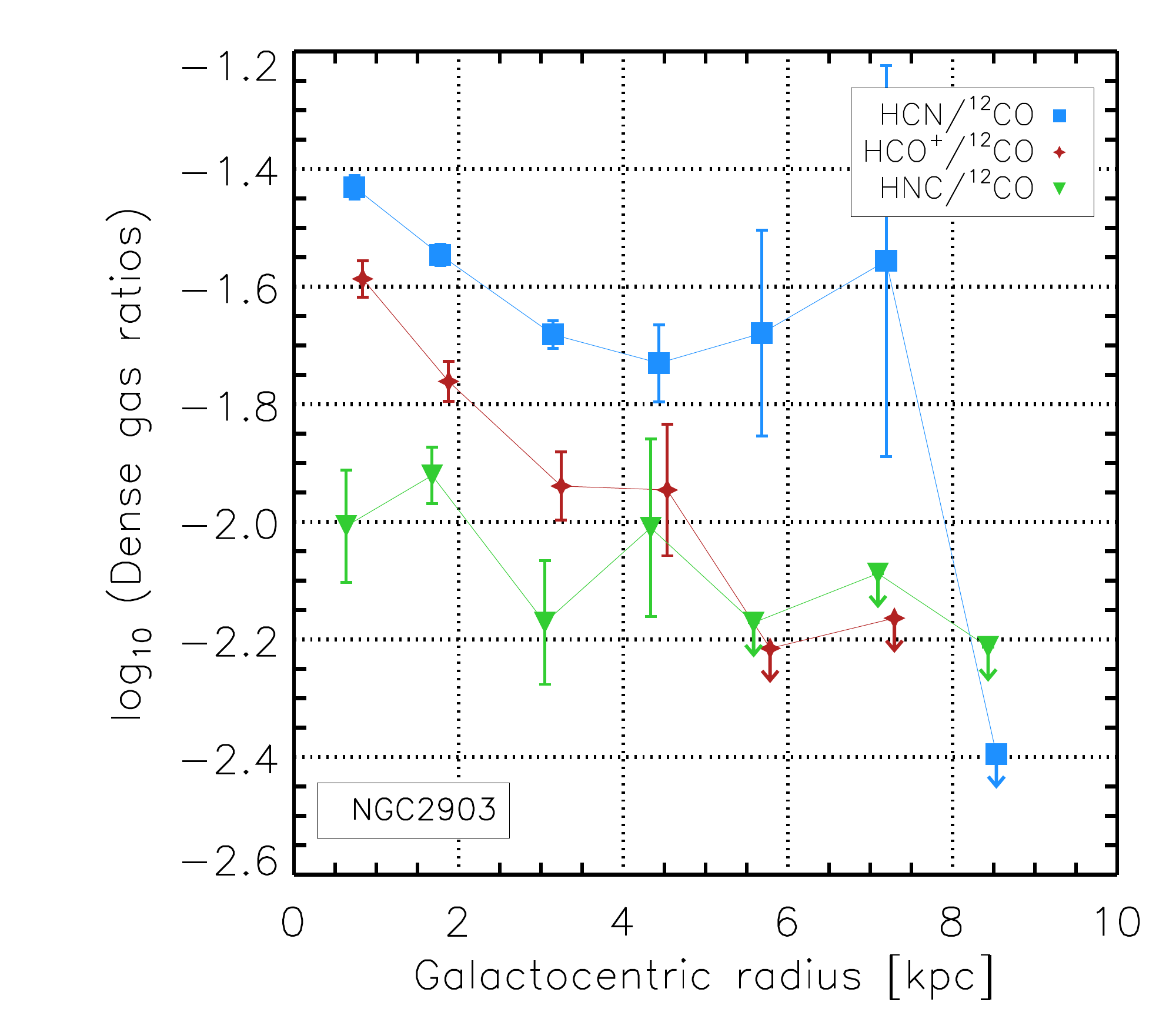}
		\includegraphics[scale=0.4]{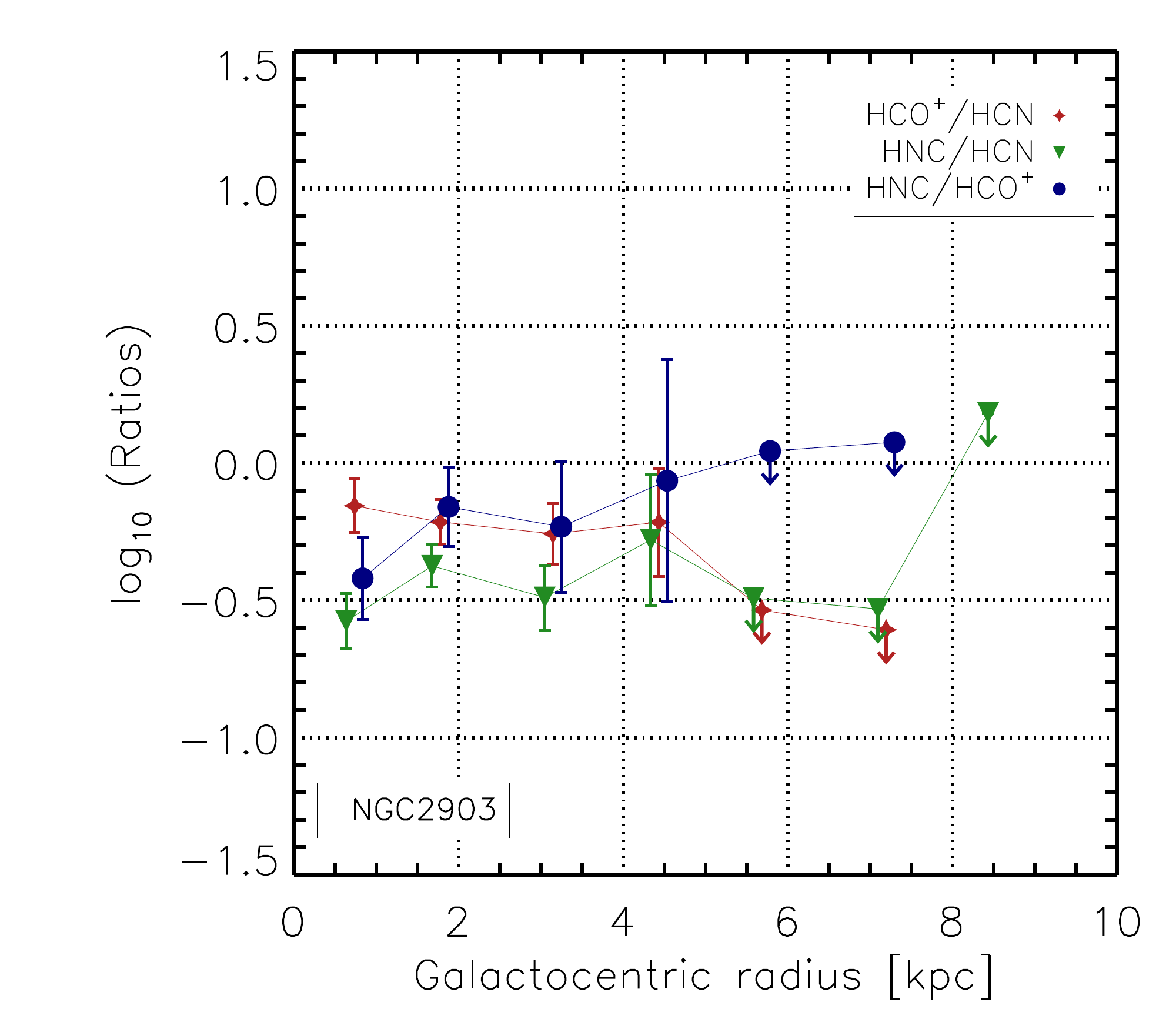}
	\end{center}
\caption{Continued for NGC~2903. {\it Spitzer} 24$\mu$m data are used in the case of NGC 2903 as a star formation rate tracer due to the lack of available Herschel data. The HCN\,(1-0) contours employed are 0.5, 0.8 and 1.7 K km s$^{-1}$.}
	\label{fig:maps2}
\end{figure*}

\begin{figure*}
	\begin{center}
		\includegraphics[scale=0.6]{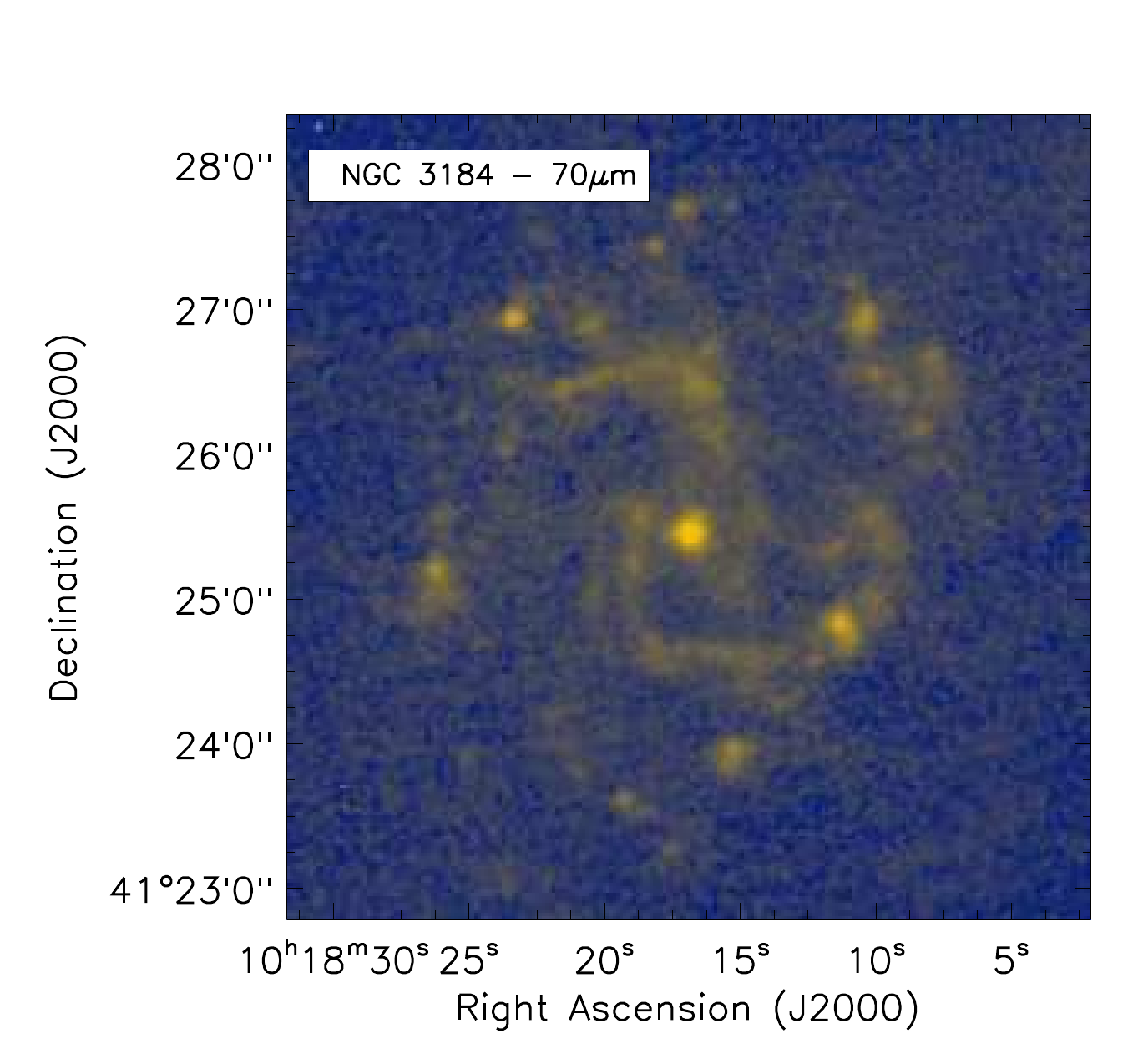}
		\includegraphics[scale=0.6]{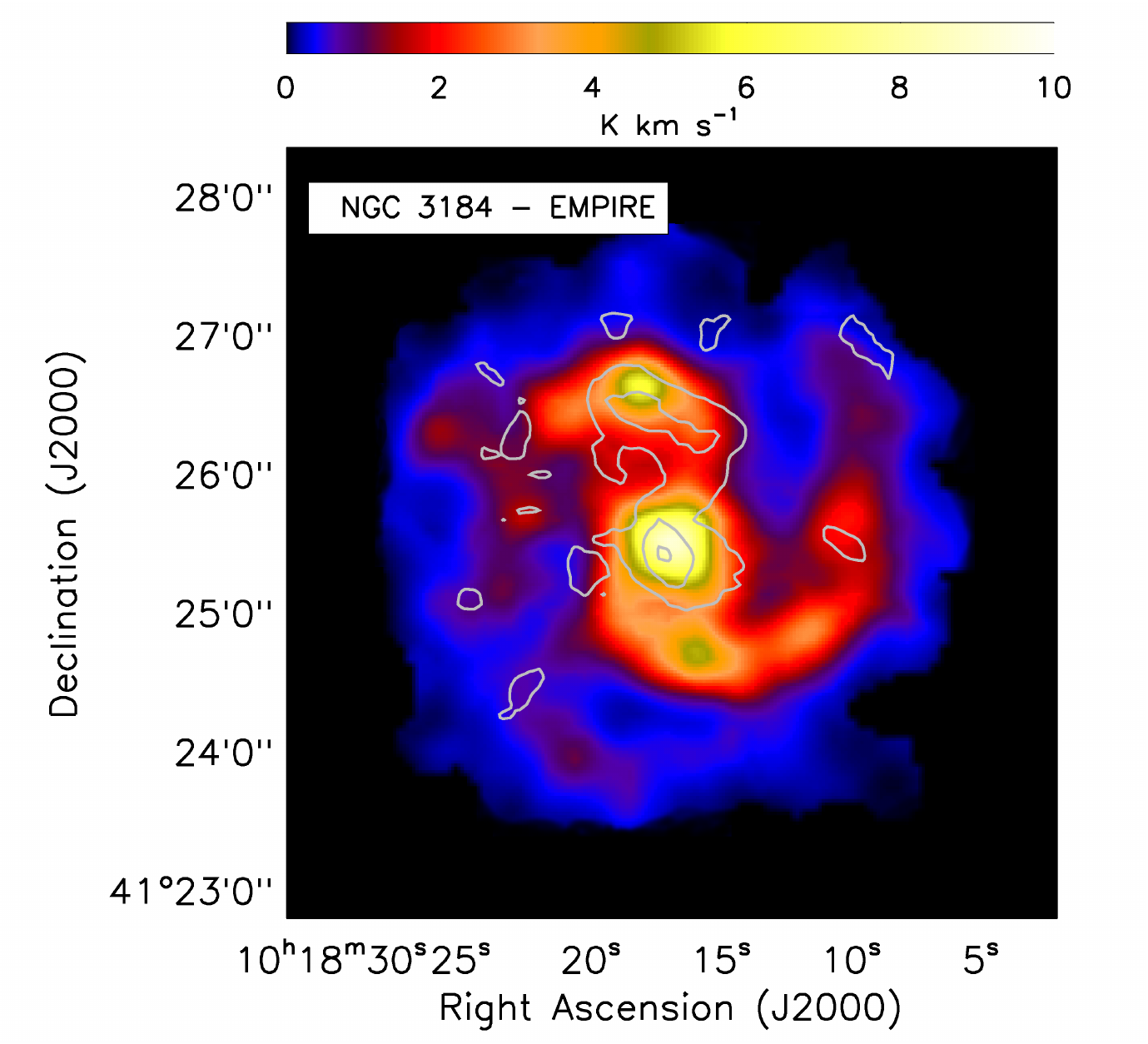}\\
		
		\includegraphics[scale=0.4]{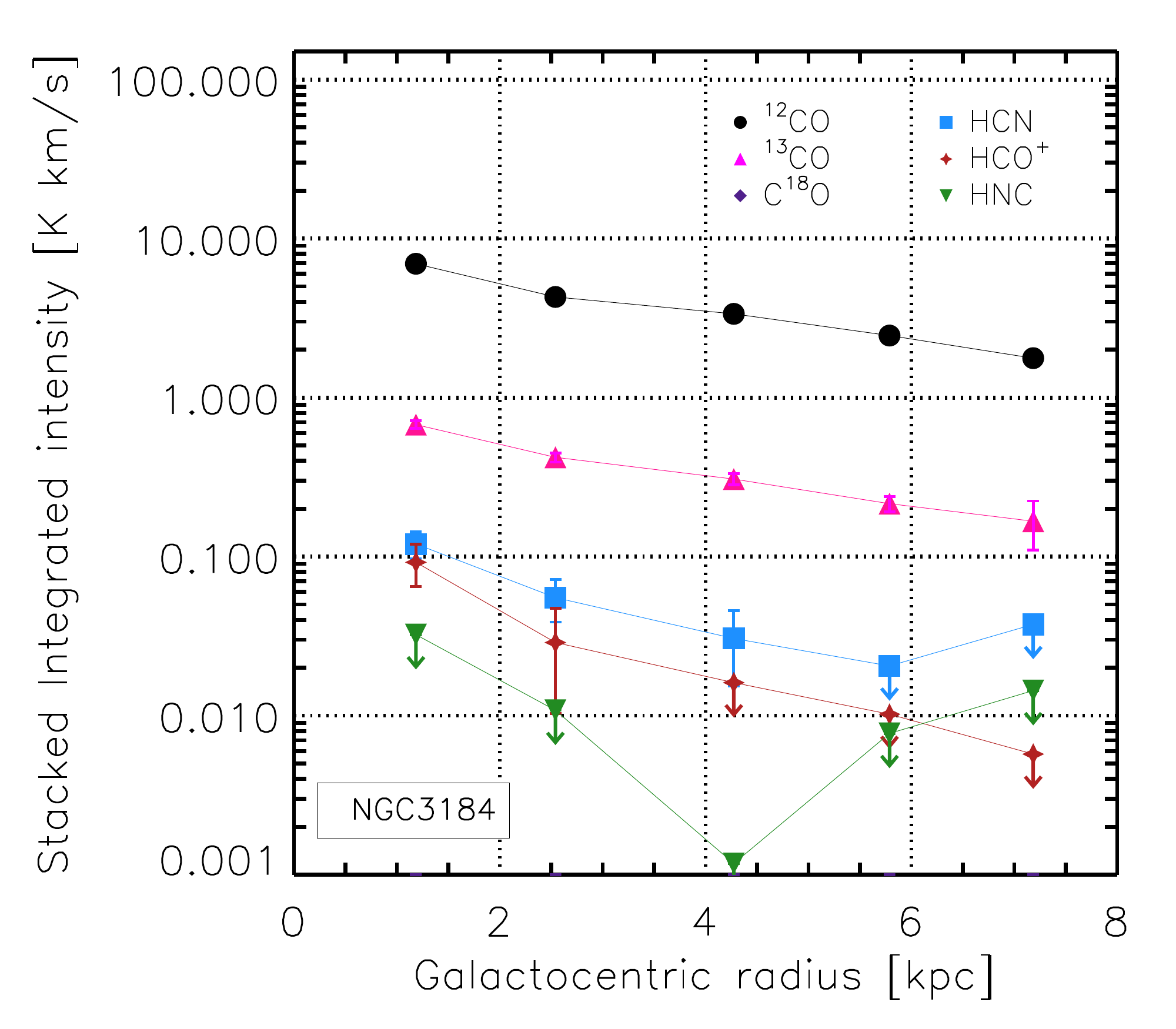}
		\includegraphics[scale=0.4]{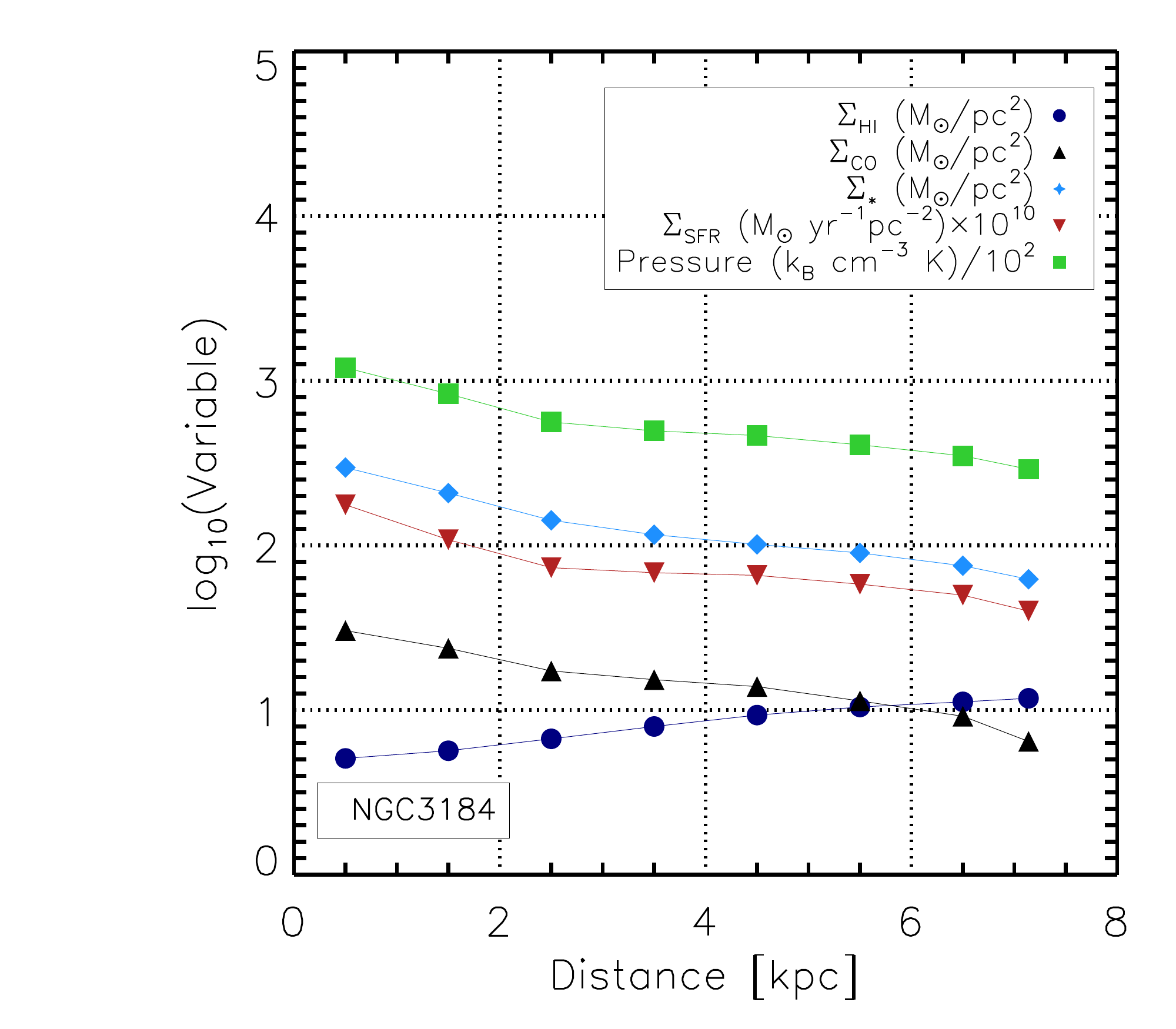}\\
		\includegraphics[scale=0.4]{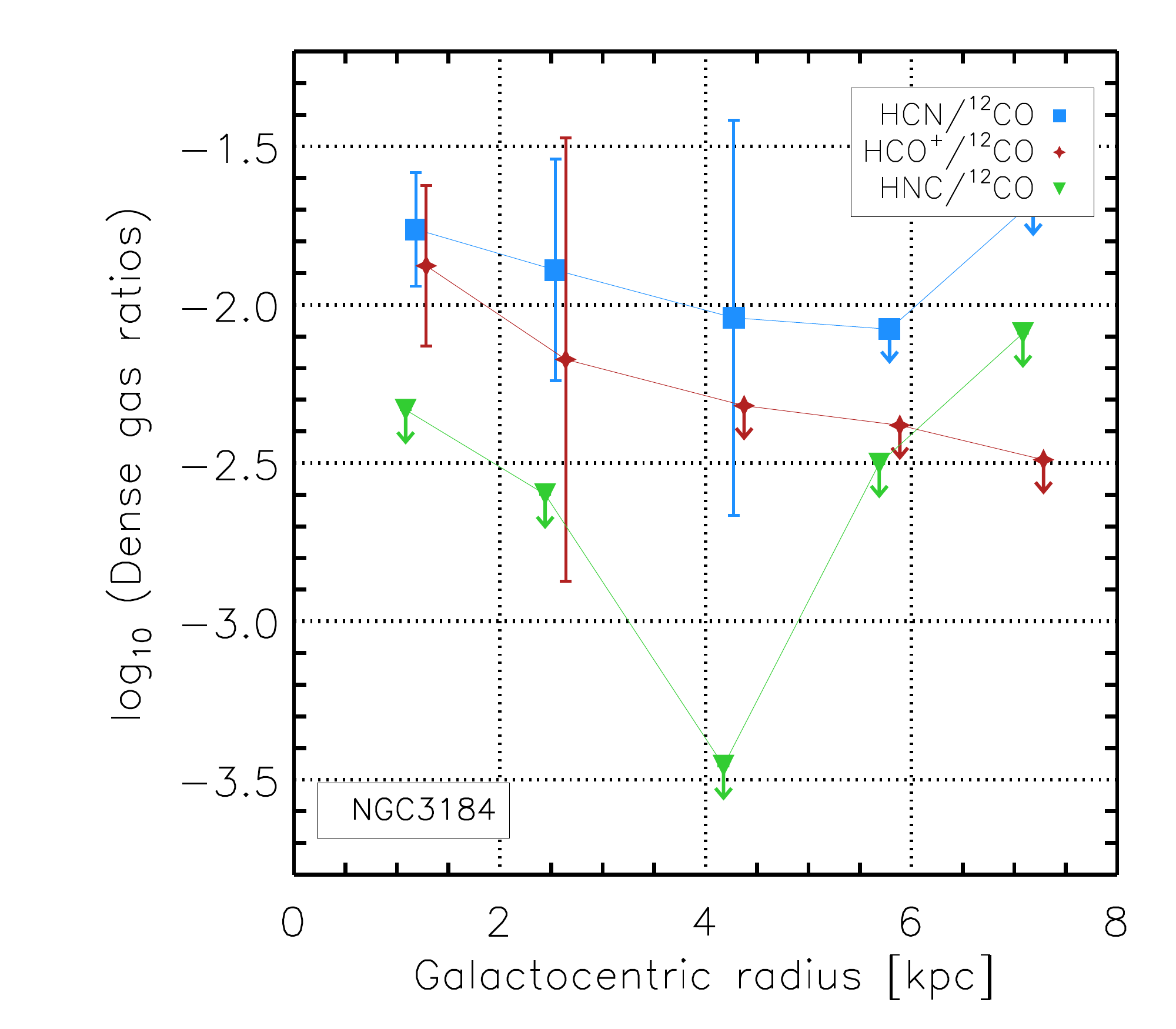}
		\includegraphics[scale=0.4]{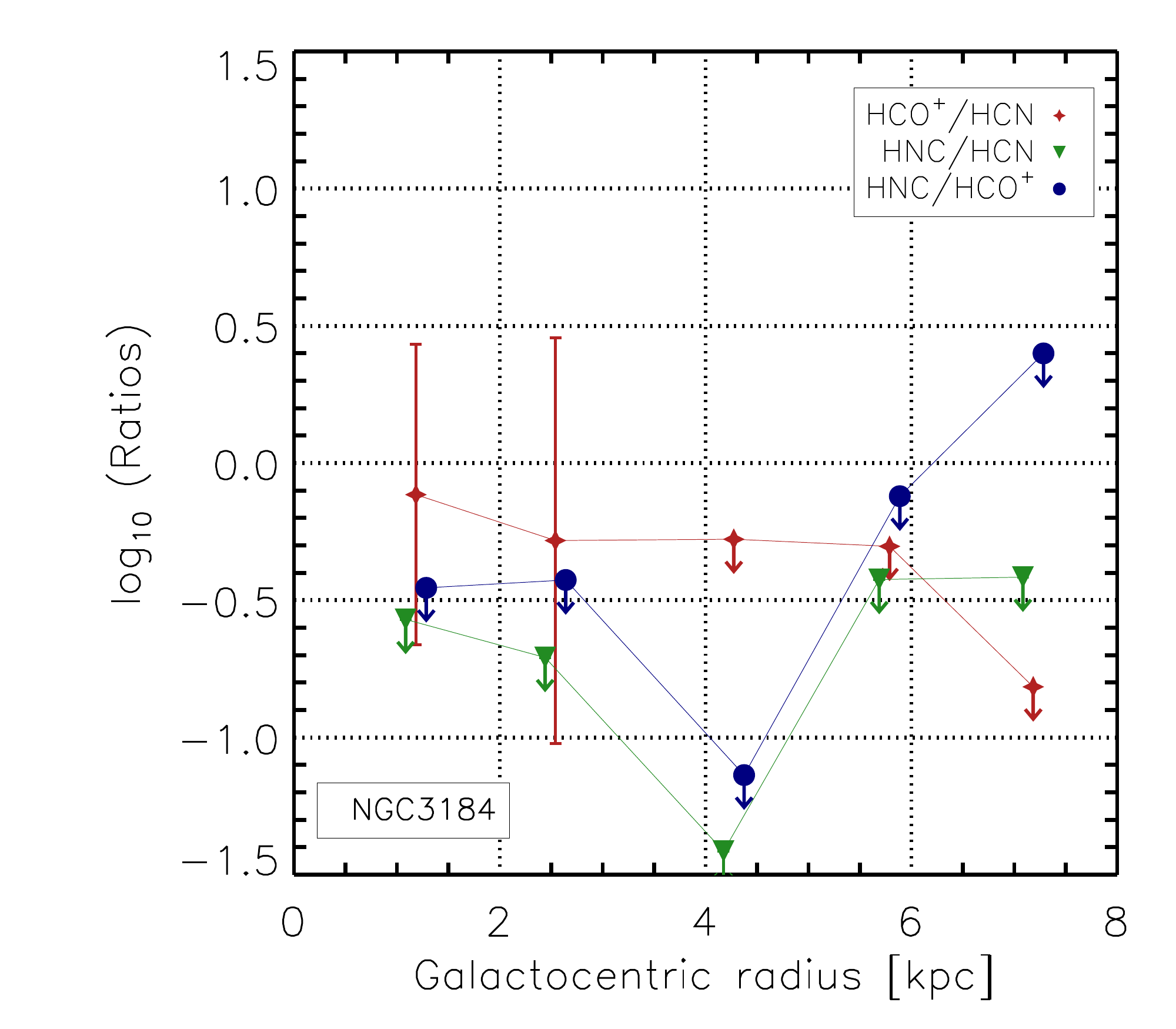}\\
	\end{center}
	\caption{Continued for NGC~3184. The HCN\,(1-0) contours employed are: 0.08, 0.16 and 0.22 K km s$^{-1}$.}
	\label{fig:maps3}
\end{figure*}

\begin{figure*}
	\begin{center}
		\includegraphics[scale=0.6]{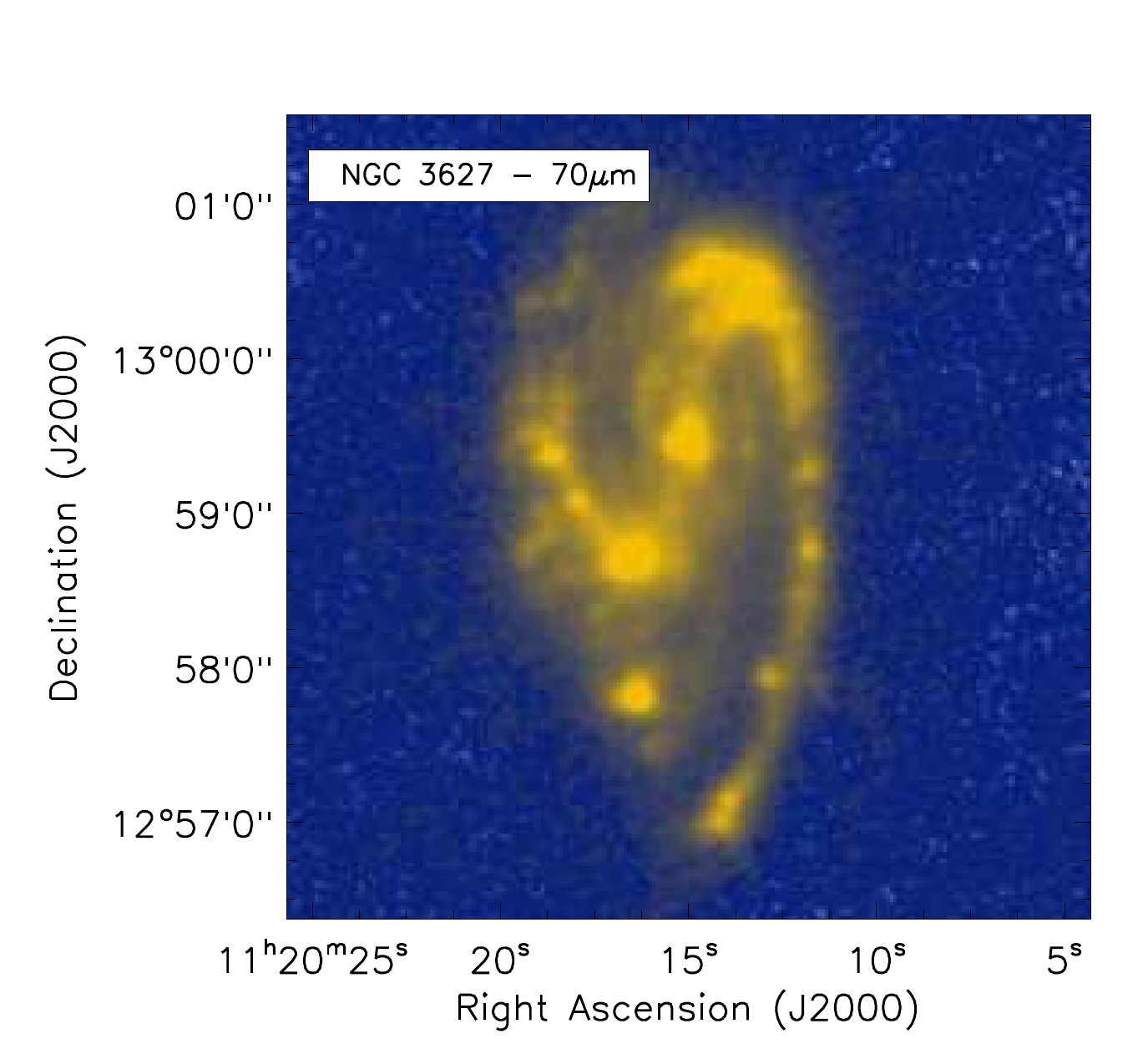}
		\includegraphics[scale=0.6]{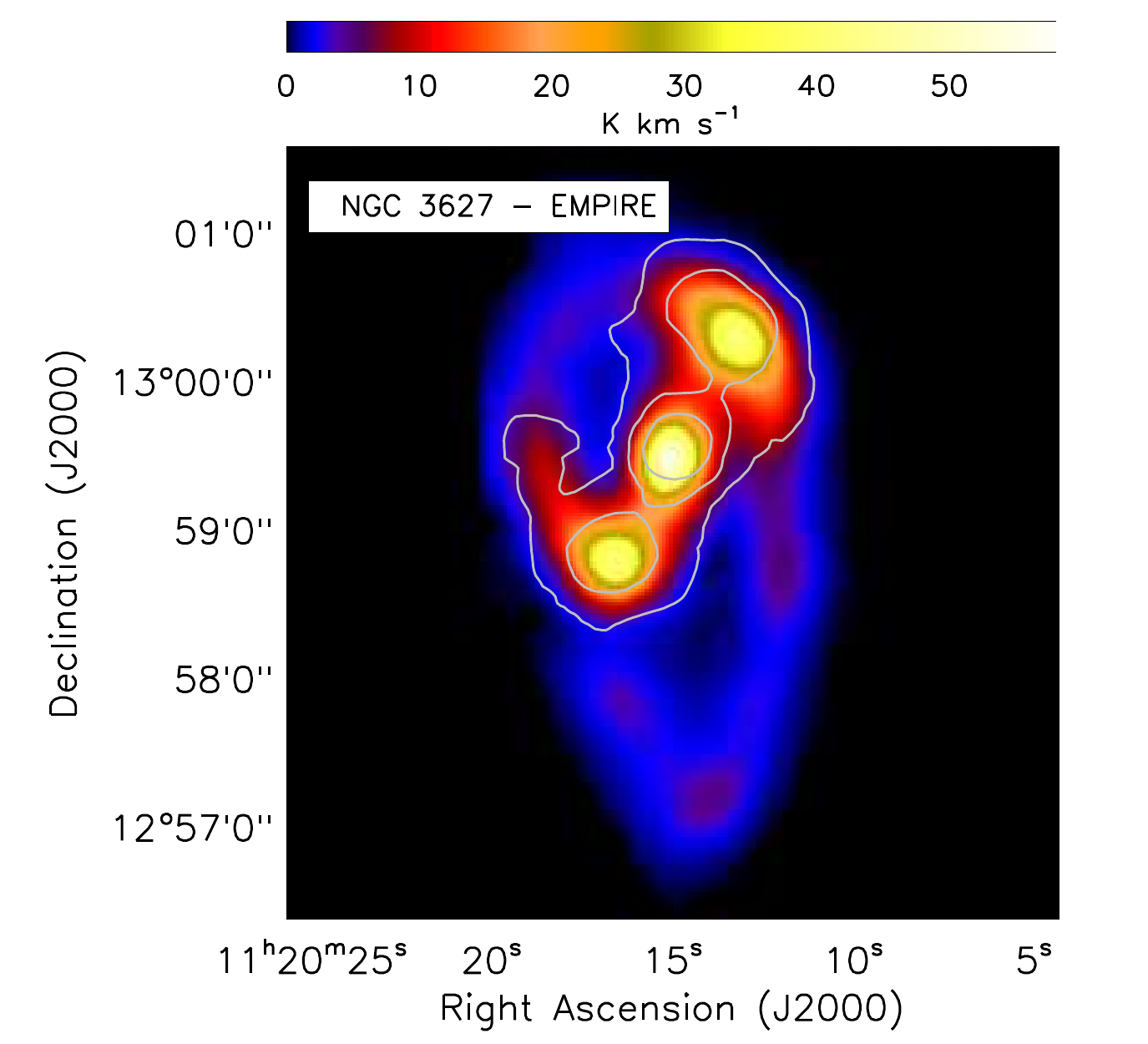}\\
		
		\includegraphics[scale=0.4]{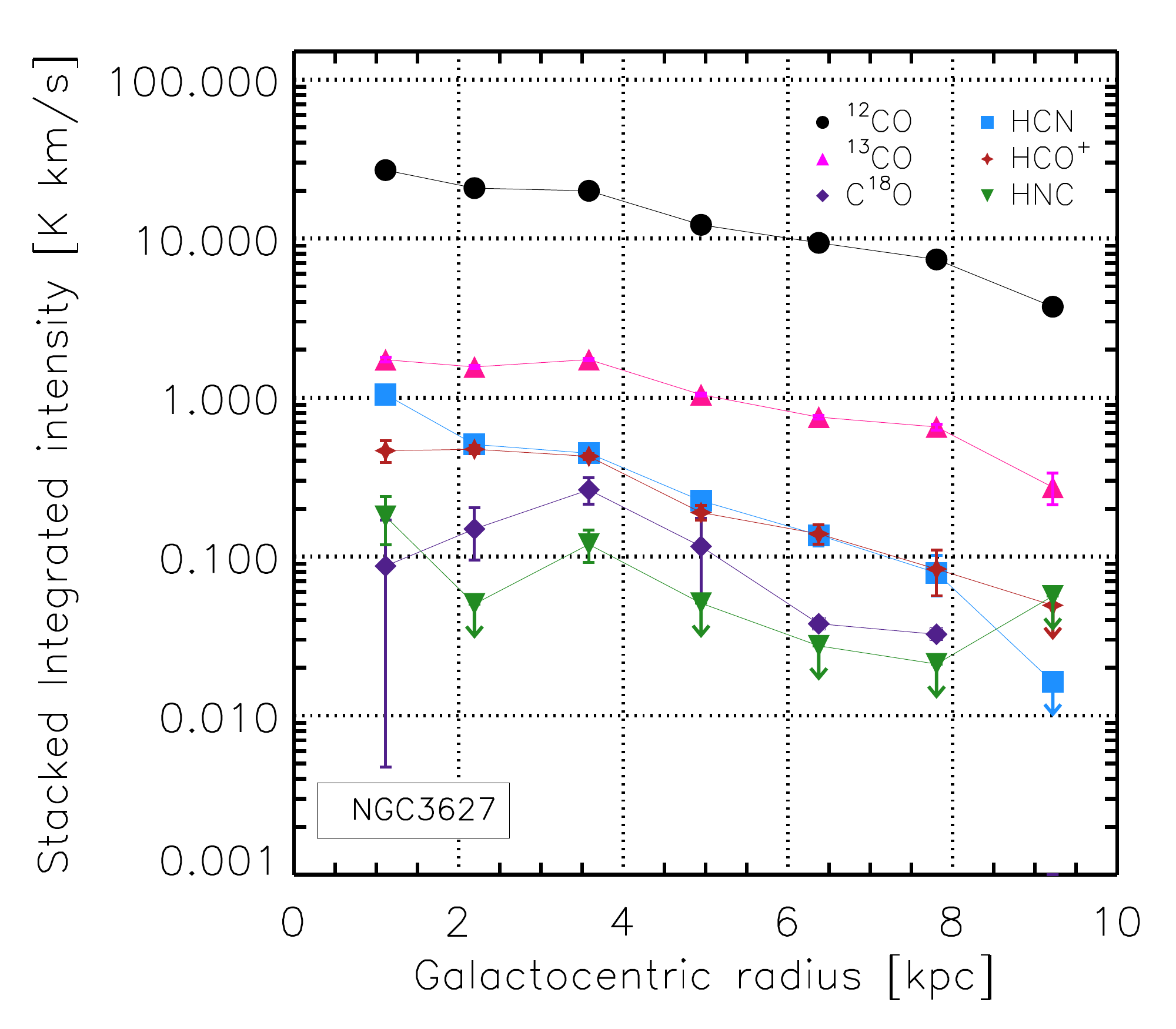}
		\includegraphics[scale=0.4]{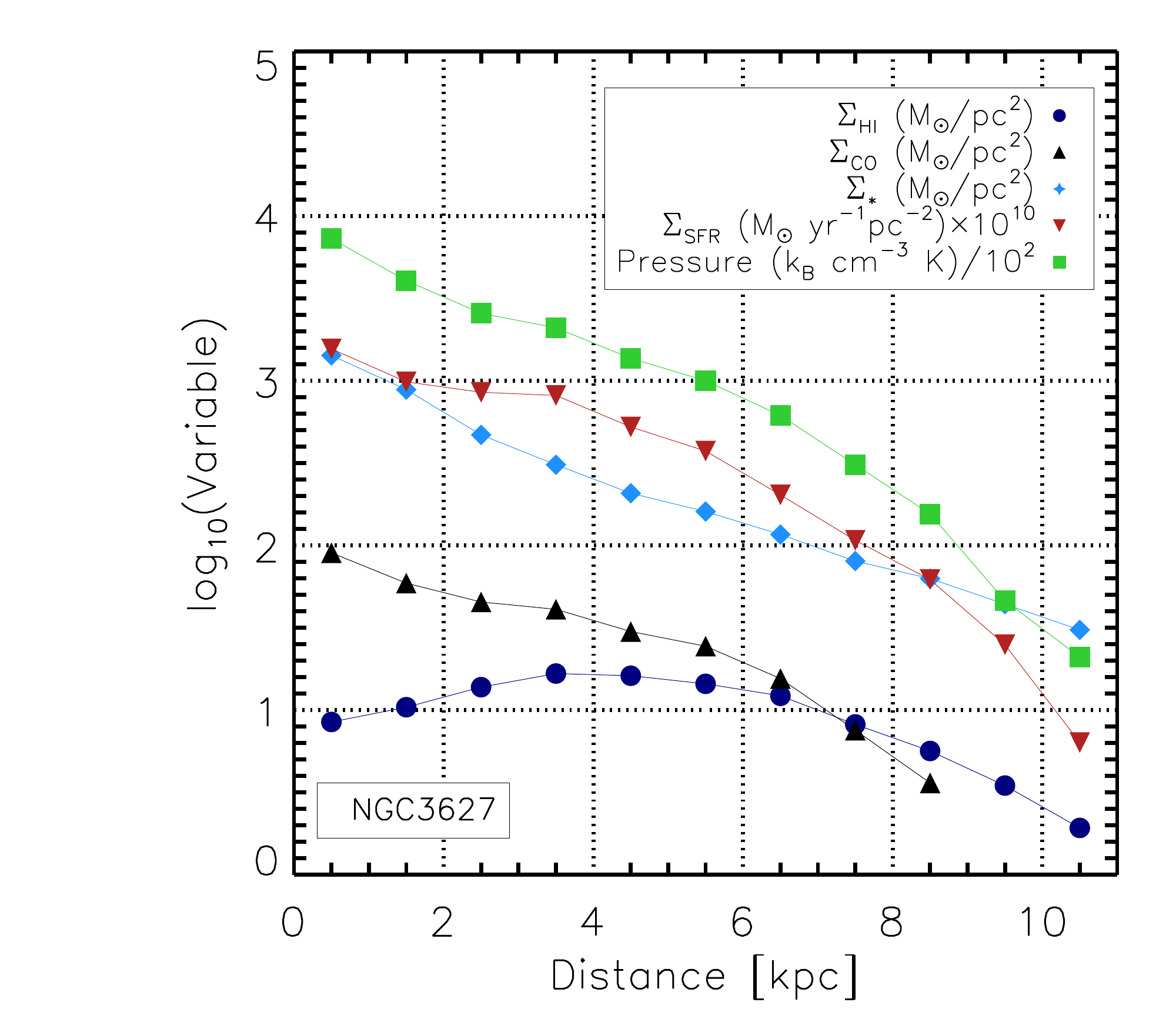}\\
		\includegraphics[scale=0.4]{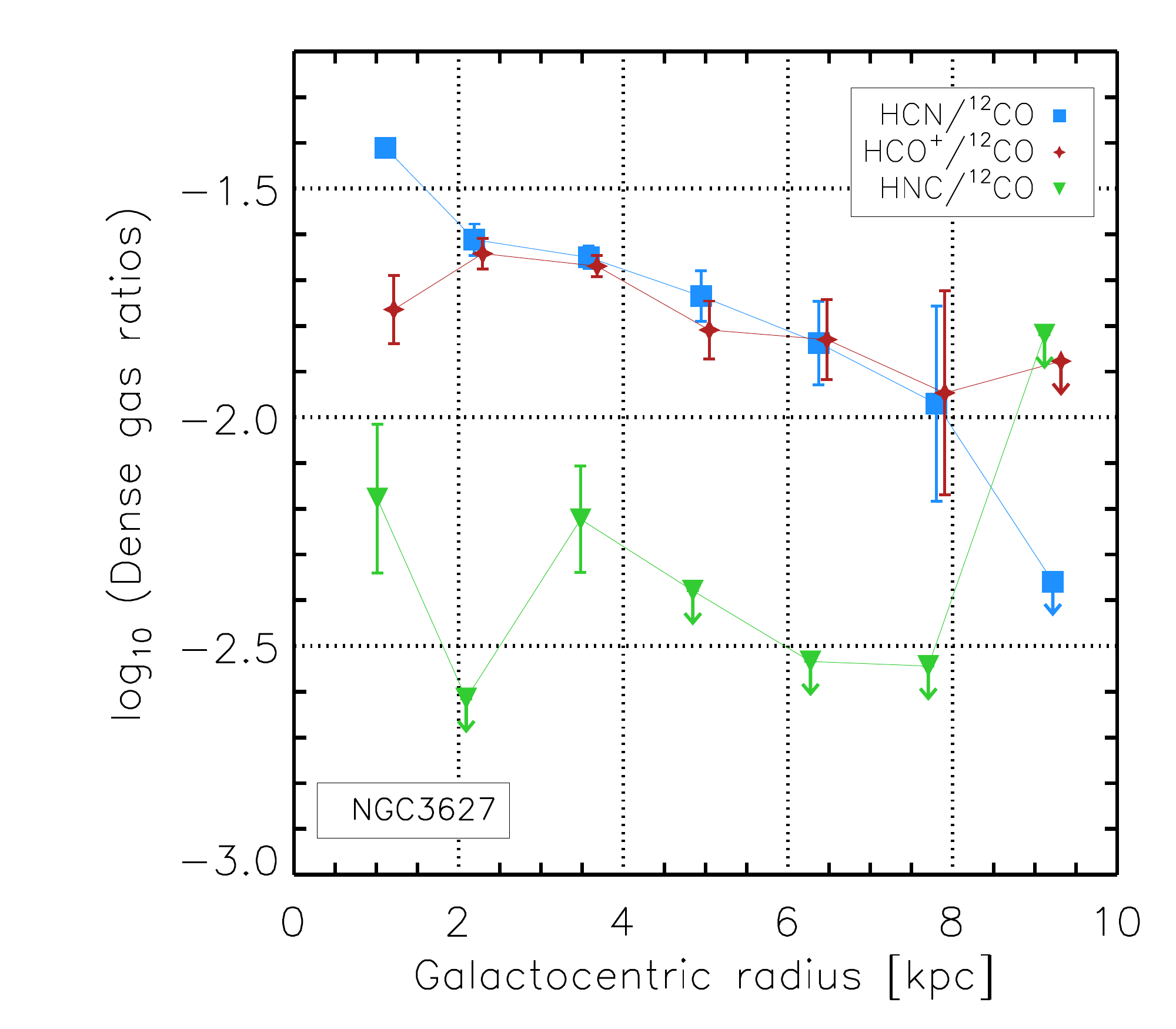}
		\includegraphics[scale=0.4]{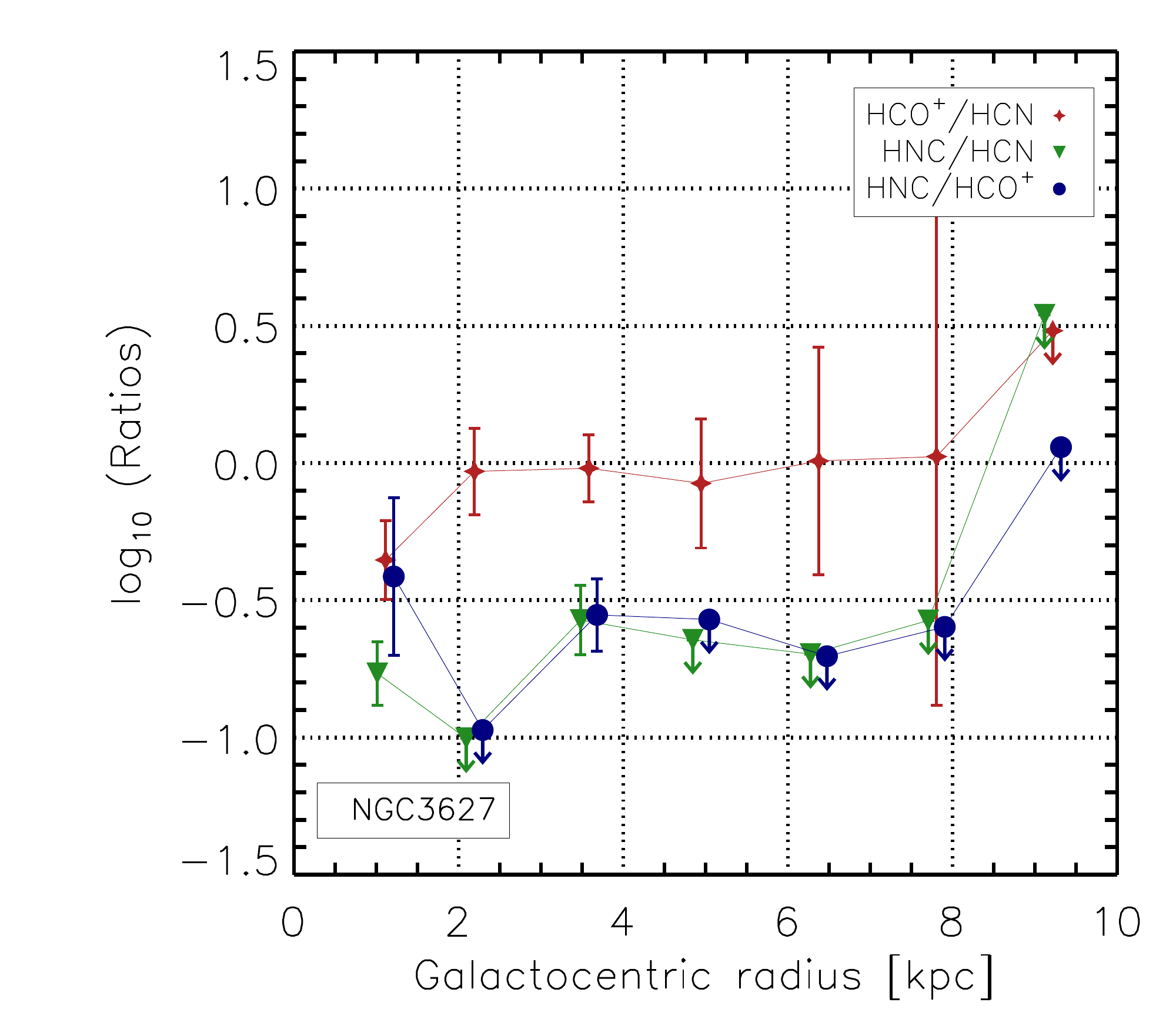}\\		
	\end{center}
	\caption{Continued for NGC~3627. The HCN\,(1-0) contours employed are: 0.4, 0.8 and 1.5 K km s$^{-1}$.}
	\label{fig:maps4}
\end{figure*}

\begin{figure*}
	\begin{center}
		\includegraphics[scale=0.6]{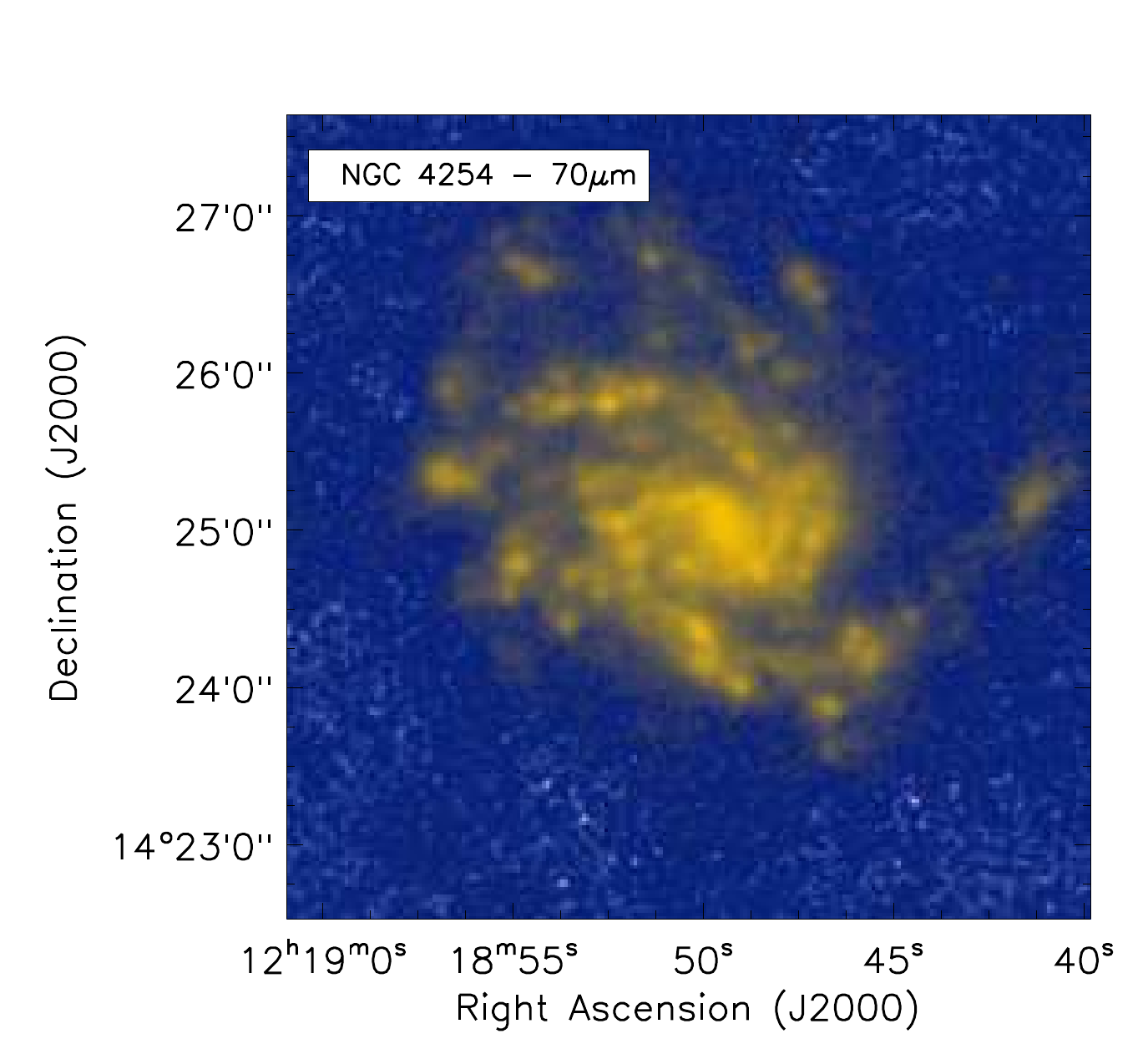}
		\includegraphics[scale=0.6]{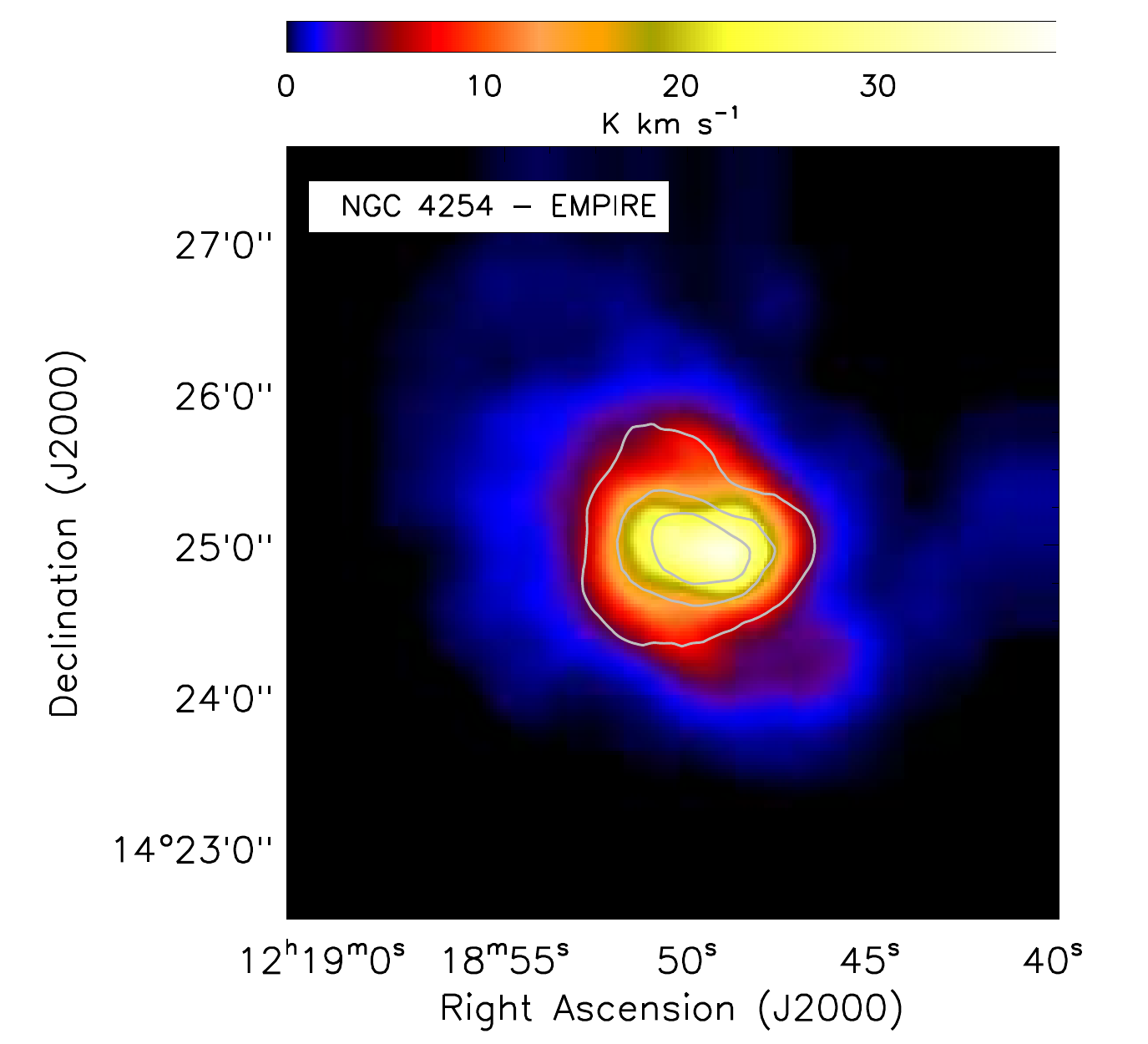}\\
		
		\includegraphics[scale=0.4]{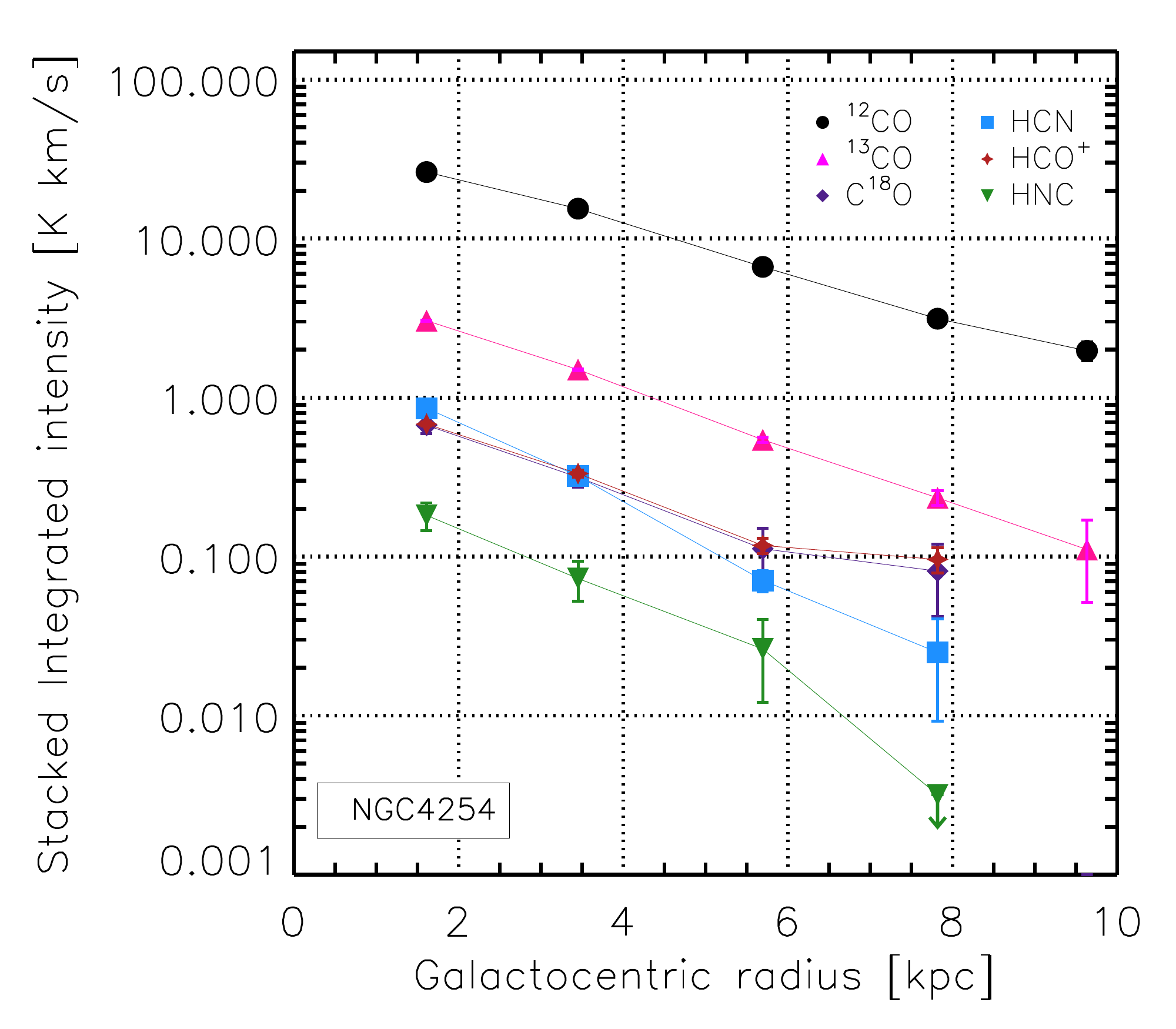}
		\includegraphics[scale=0.4]{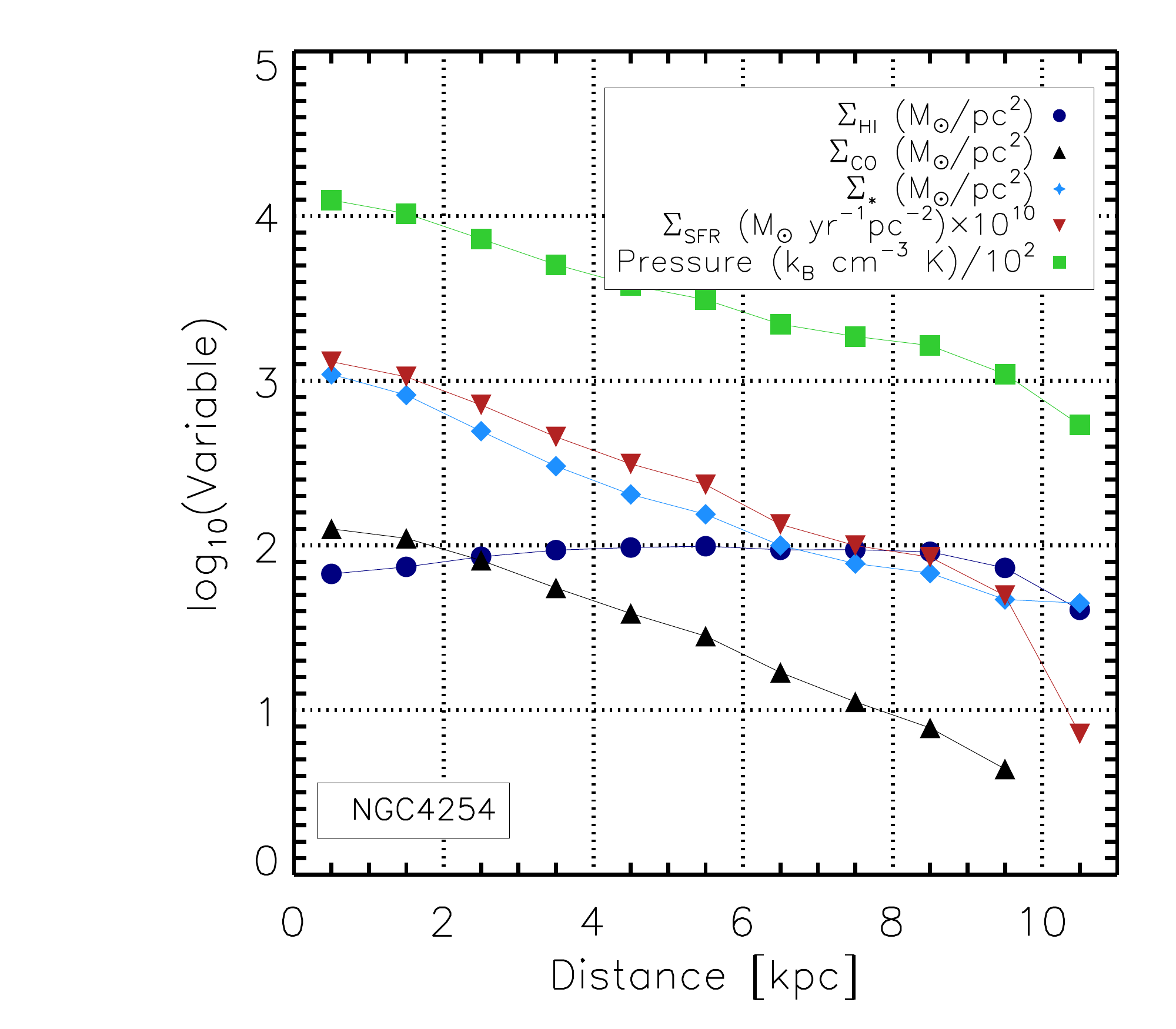}\\
		\includegraphics[scale=0.4]{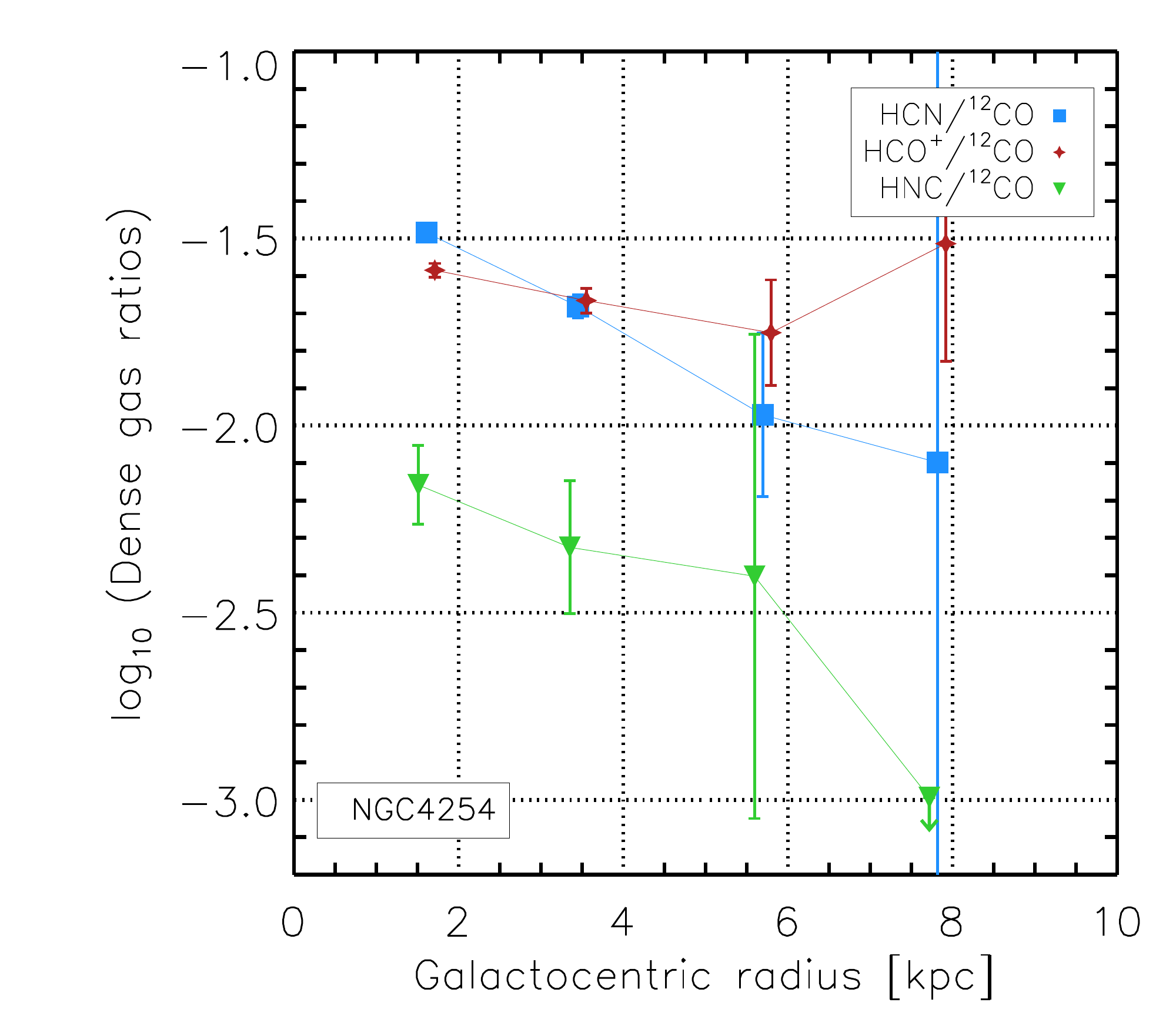}
		\includegraphics[scale=0.4]{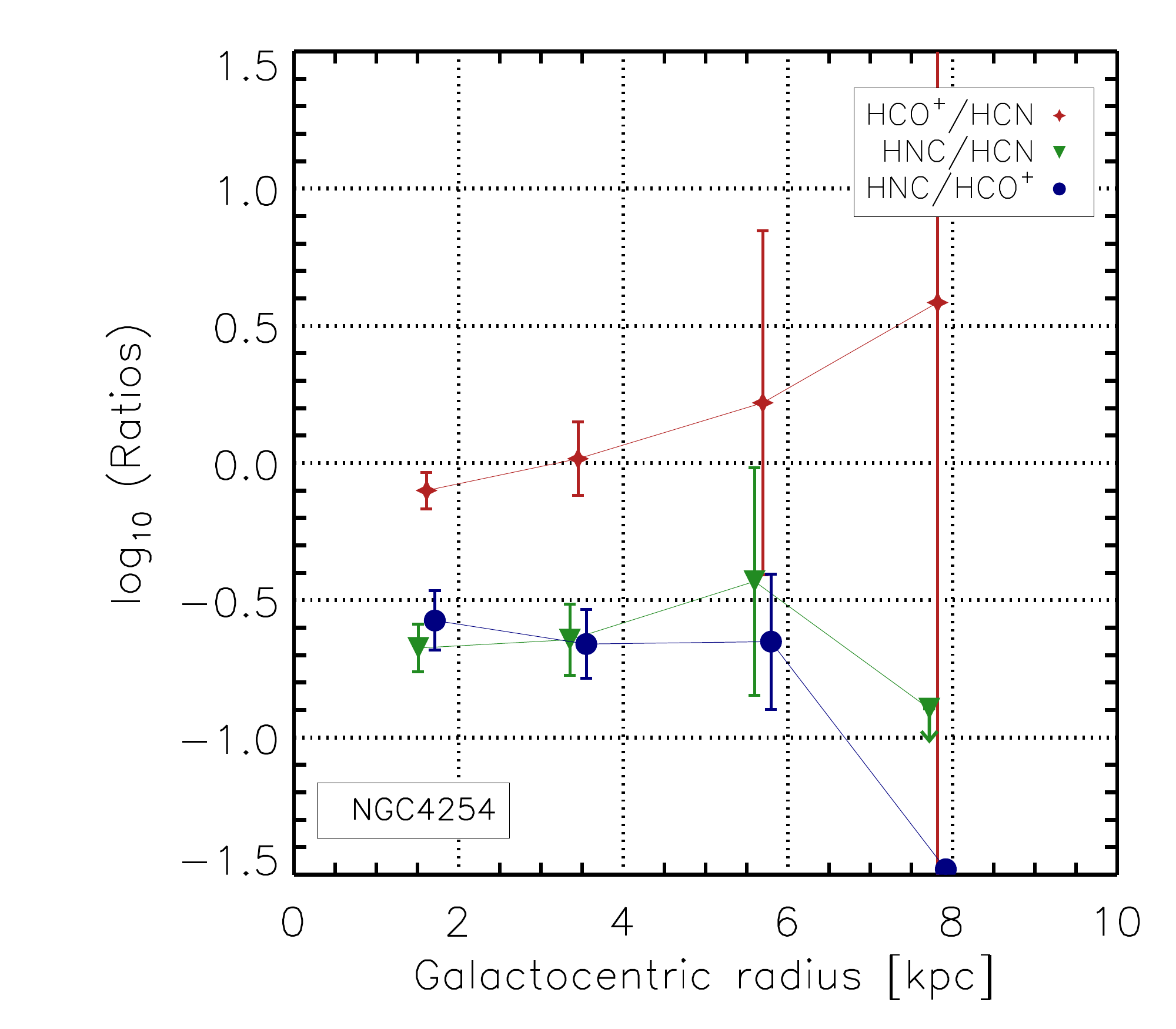}\\
	\end{center}
	\caption{Continued for NGC~4254. The HCN\,(1-0) contours employed are: 0.3, 0.7 and 1.0 K km s$^{-1}$.}
	\label{fig:maps5}
\end{figure*}

\begin{figure*}
	\begin{center}
		\includegraphics[scale=0.6]{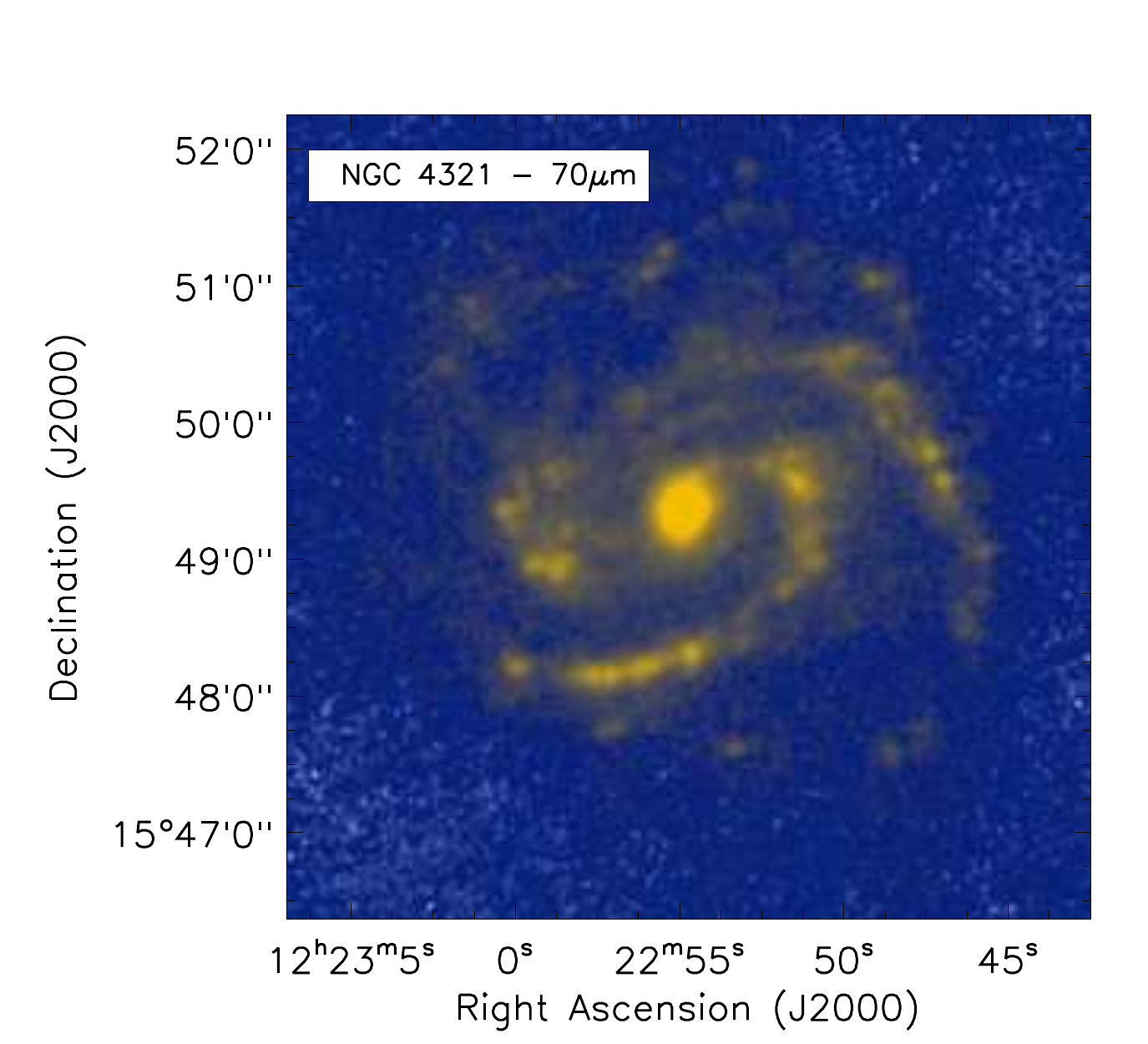}
		\includegraphics[scale=0.6]{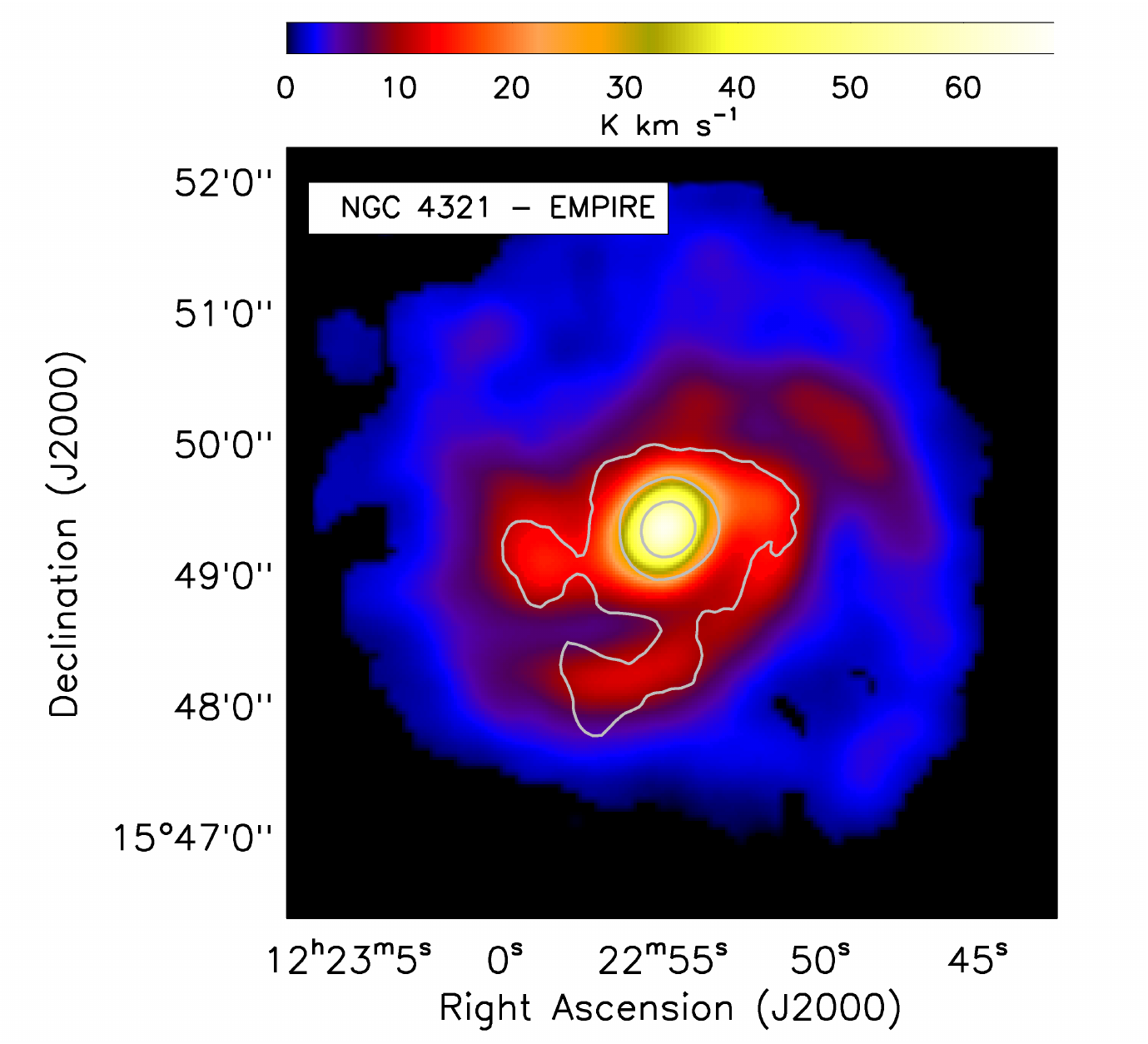}\\
		
		\includegraphics[scale=0.4]{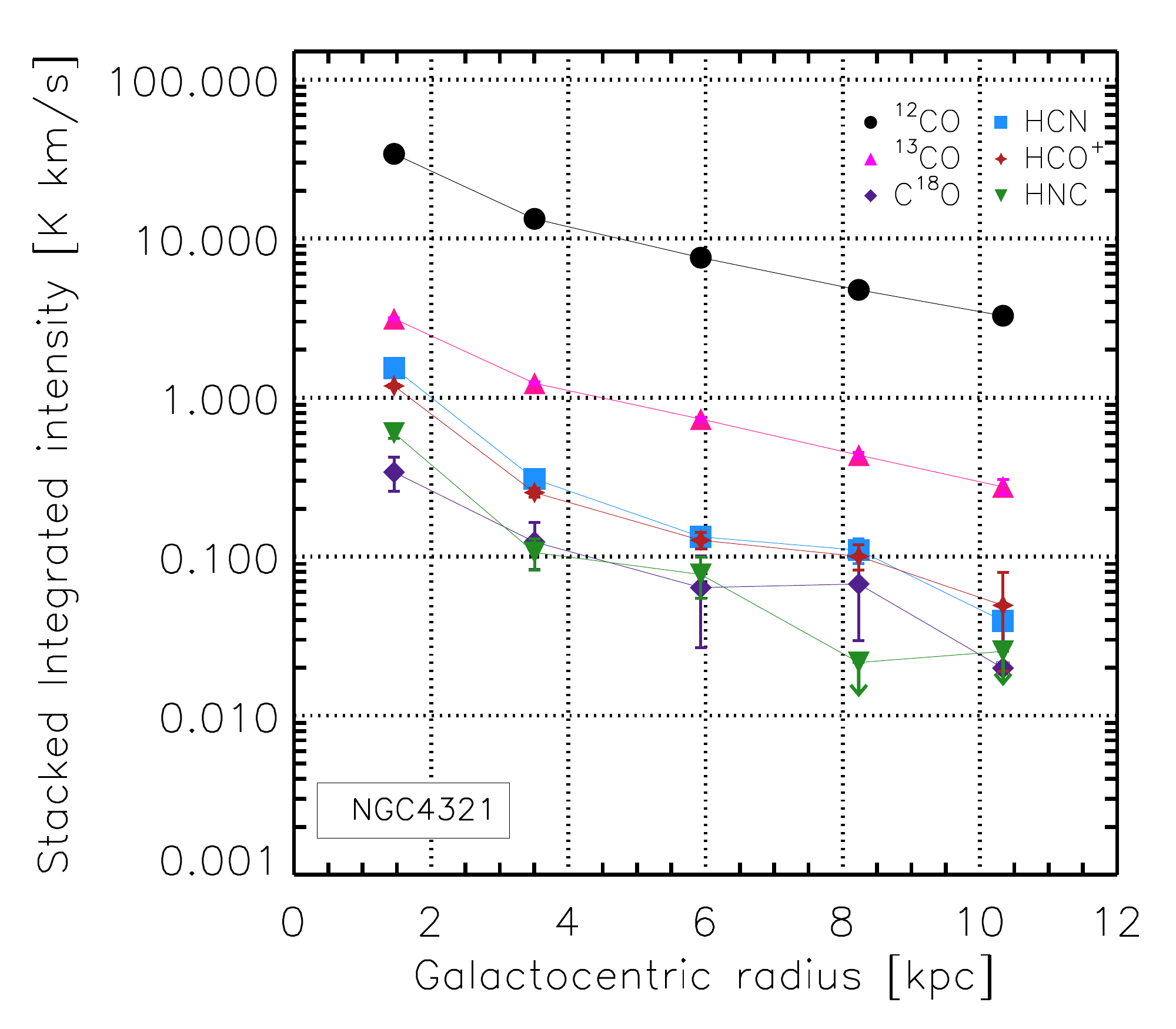}
		\includegraphics[scale=0.4]{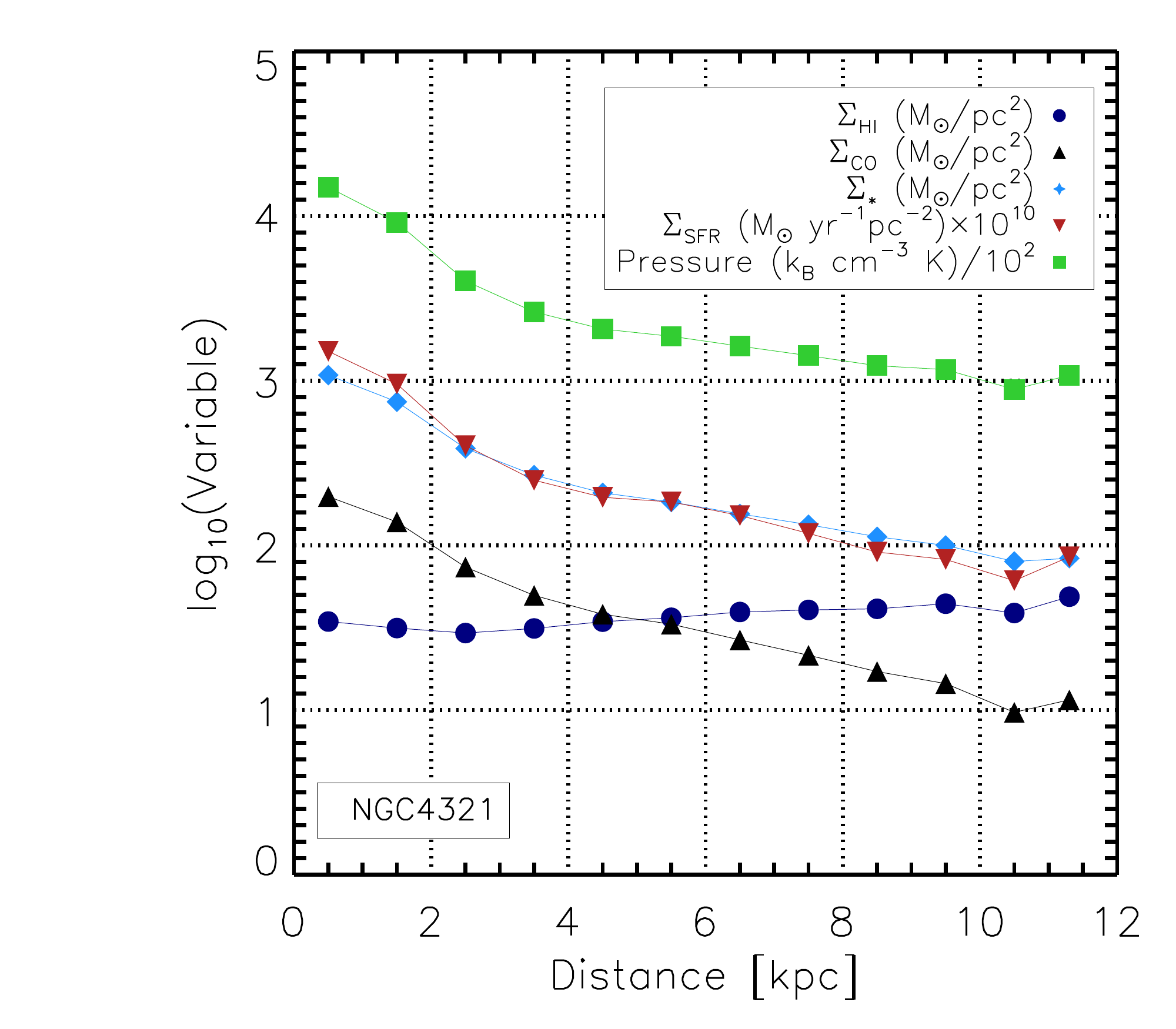}\\
		\includegraphics[scale=0.4]{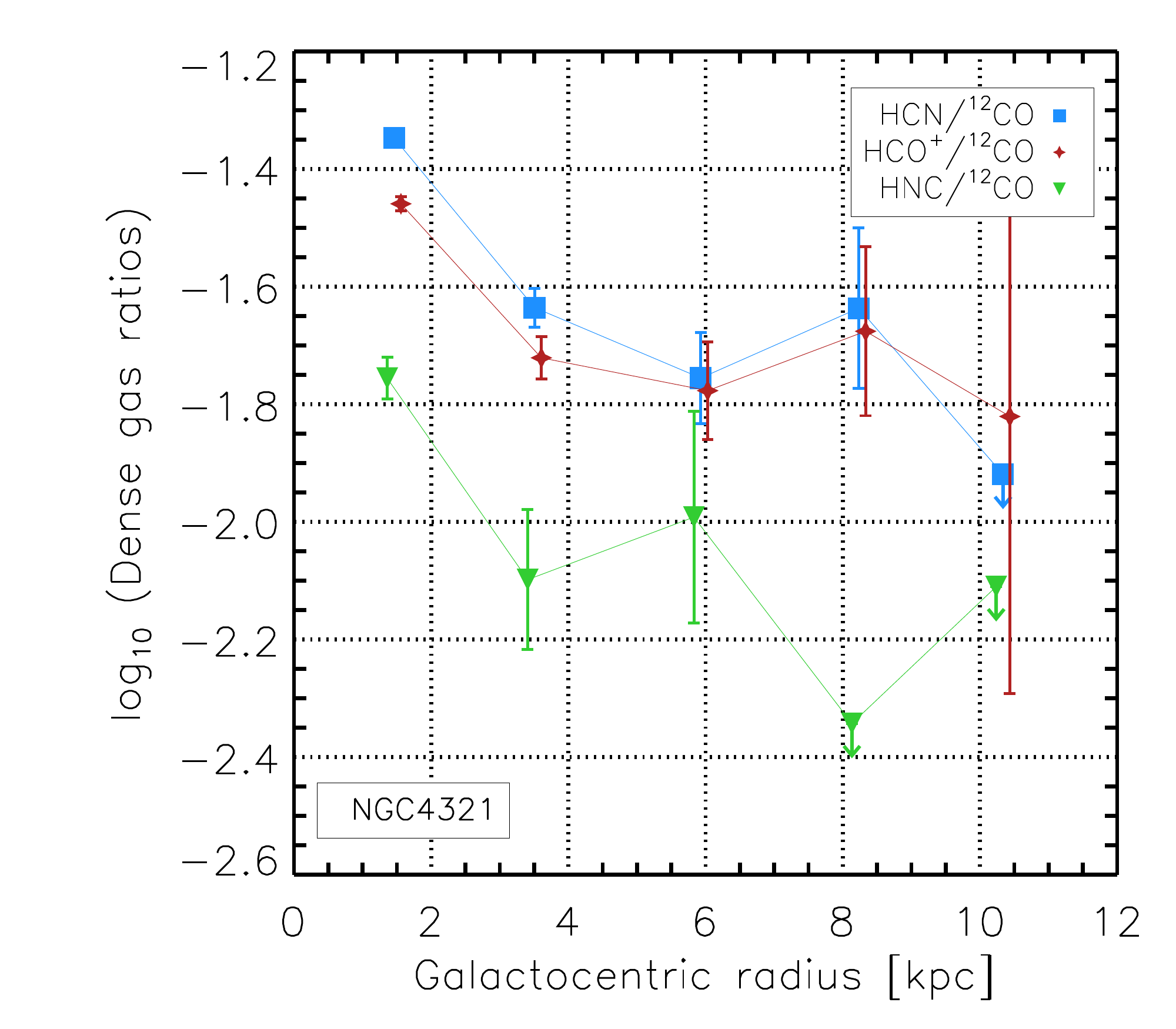}	
		\includegraphics[scale=0.4]{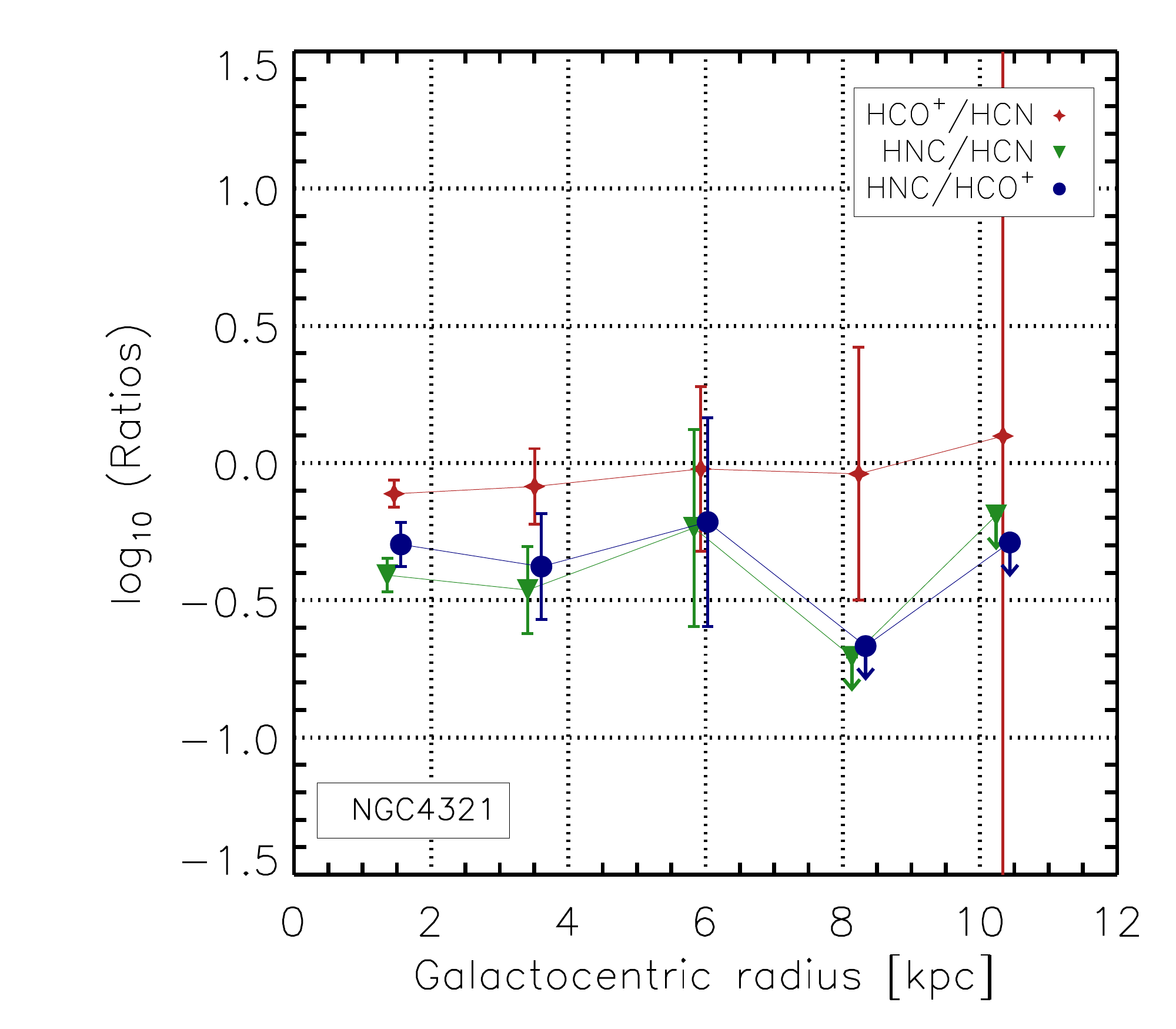}\\
	\end{center}
	\caption{Continued for NGC~4321. The HCN\,(1-0) contours employed are: 0.2, 1.0 and 2.0 K km s$^{-1}$.}
	\label{fig:maps6}
\end{figure*}

\begin{figure*}
	\begin{center}
		\includegraphics[scale=0.6]{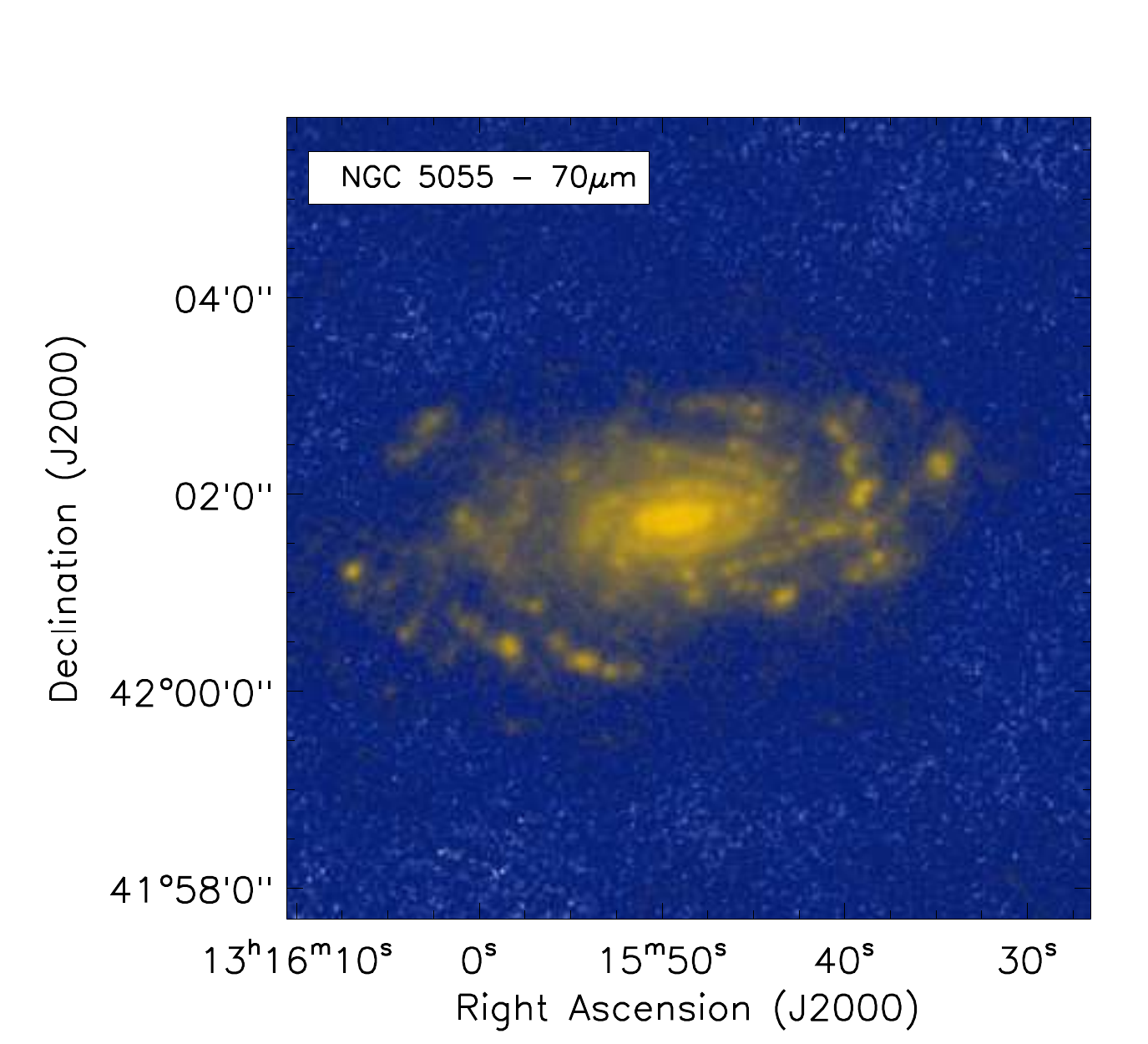}
		\includegraphics[scale=0.6]{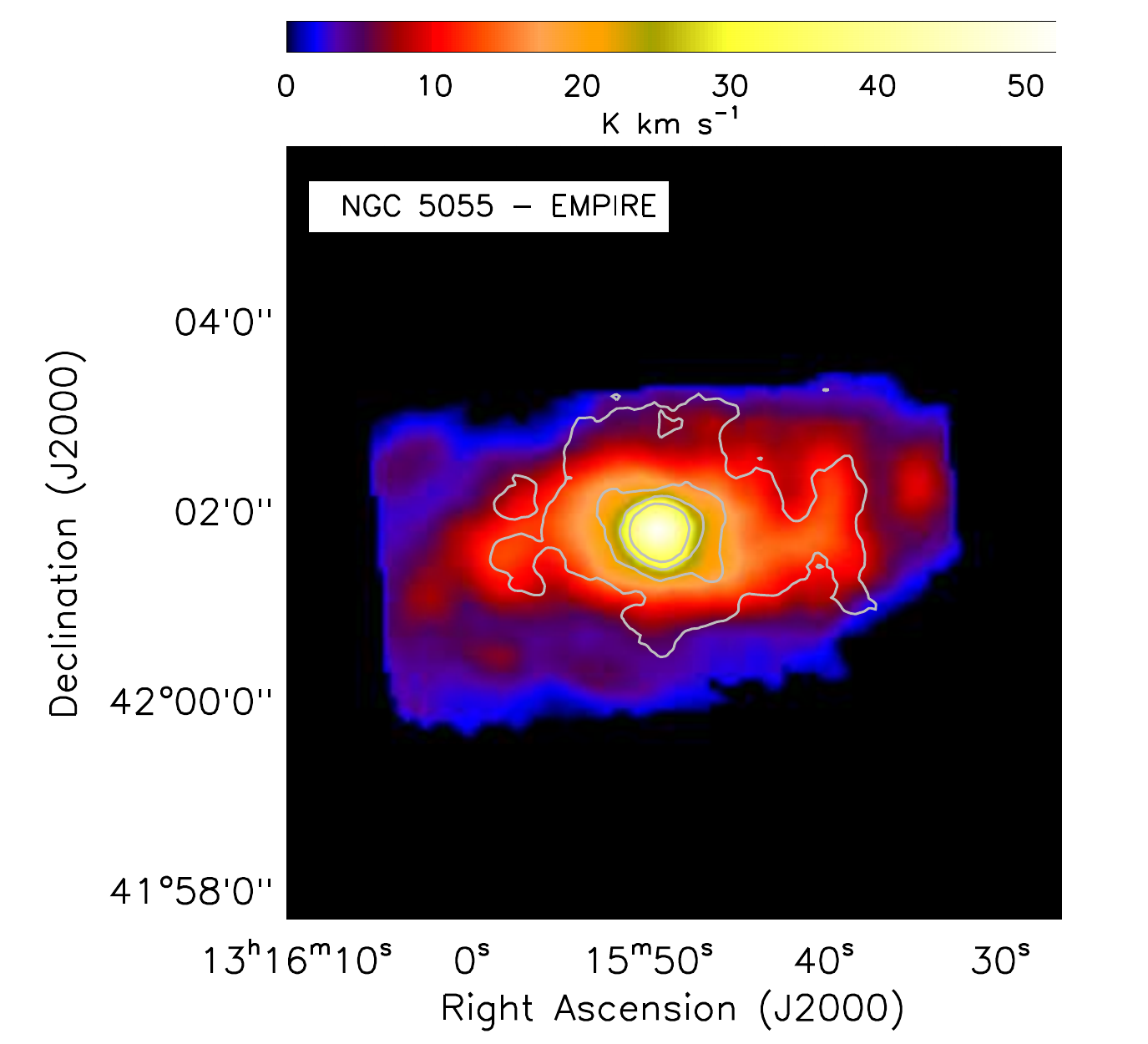}\\
		
		\includegraphics[scale=0.4]{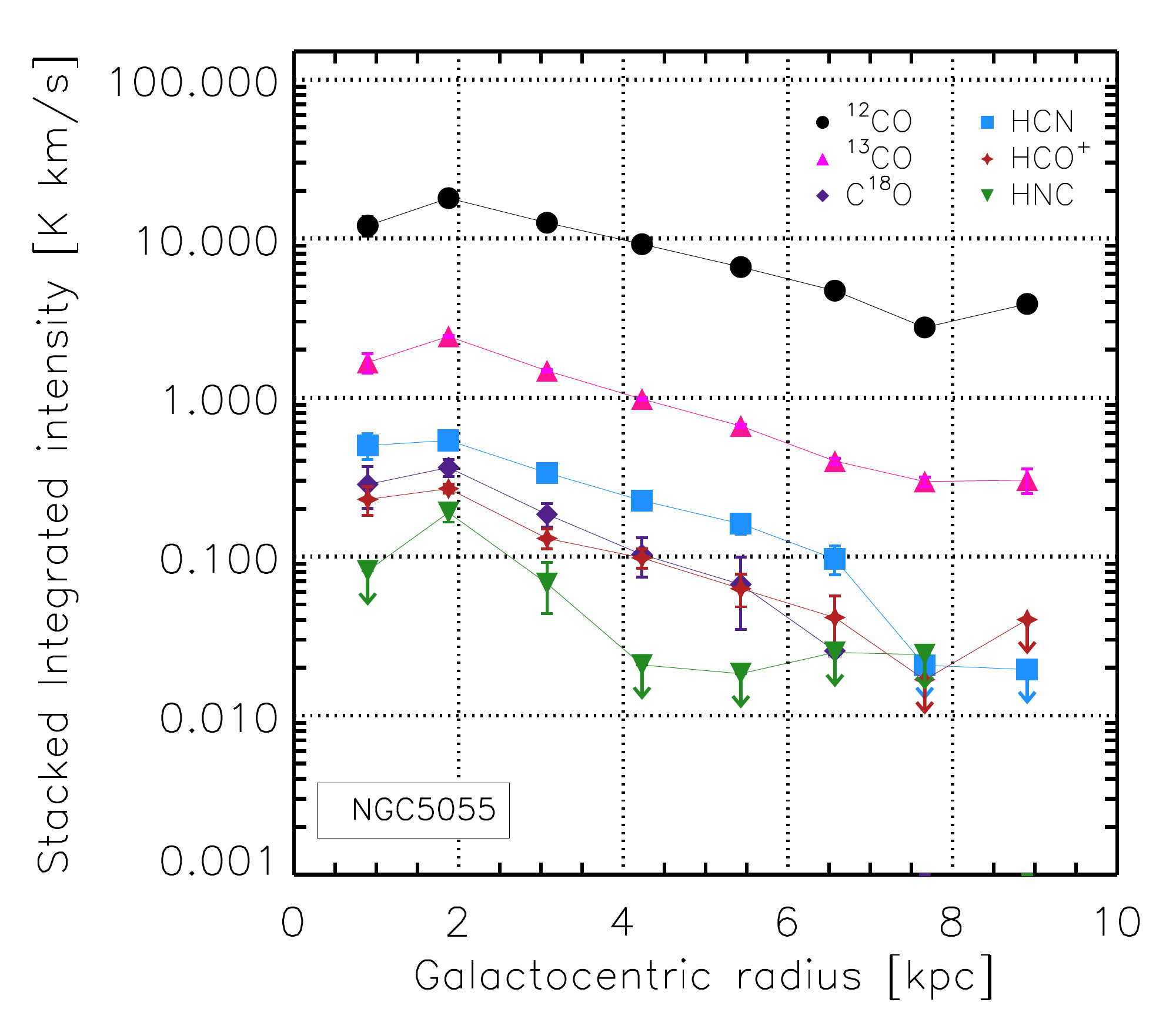}
		\includegraphics[scale=0.4]{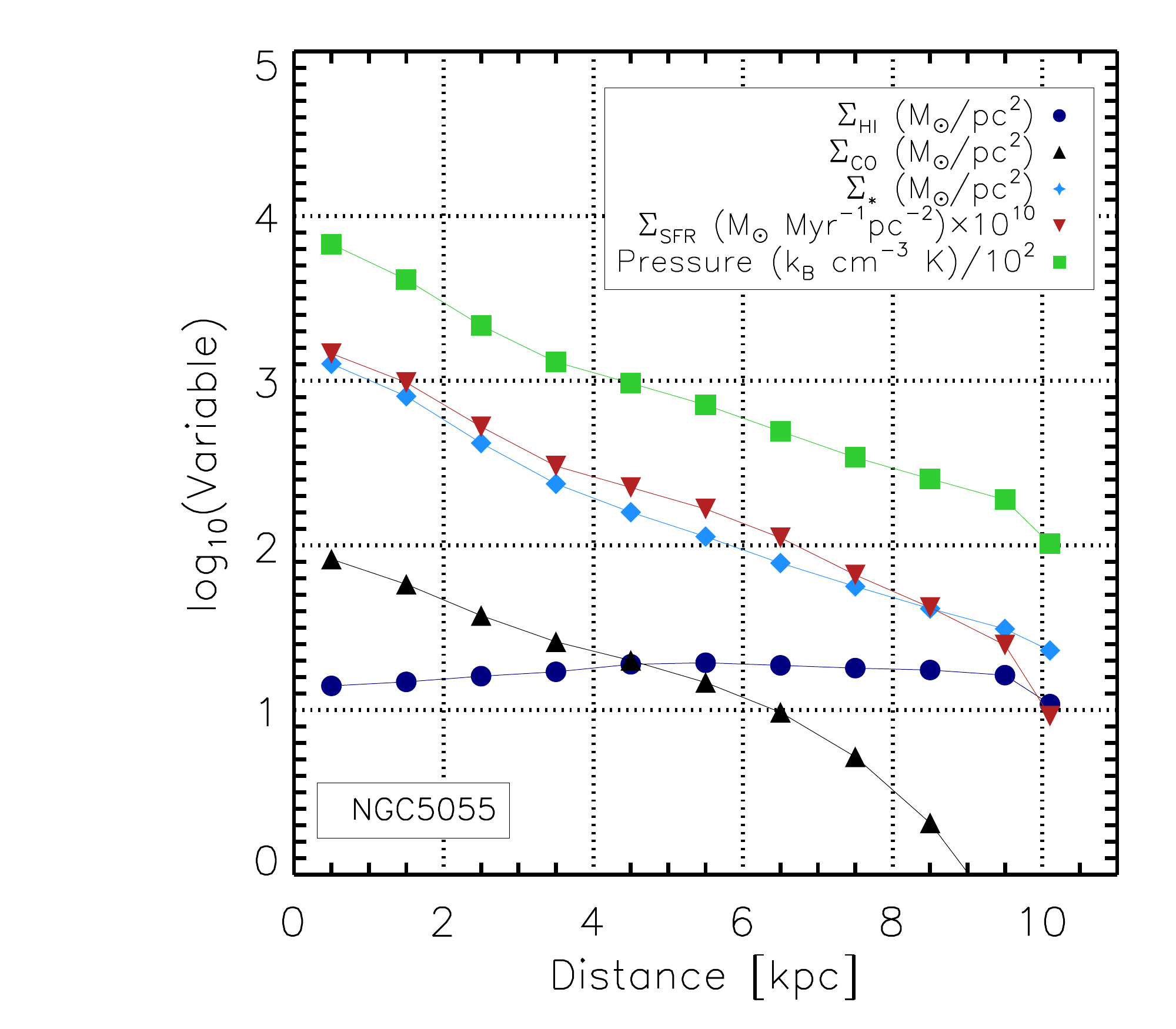}\\
		\includegraphics[scale=0.4]{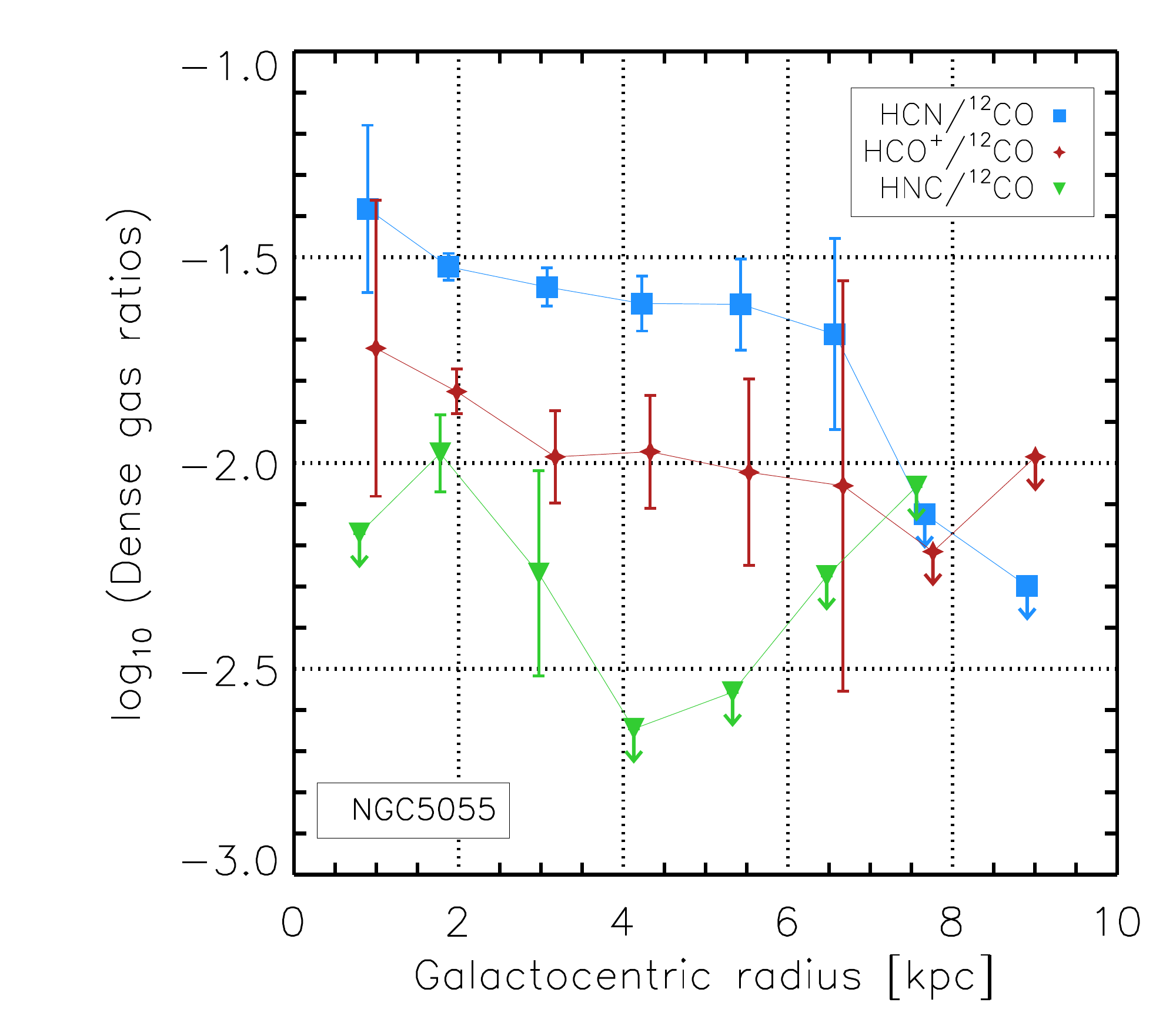}	
		\includegraphics[scale=0.4]{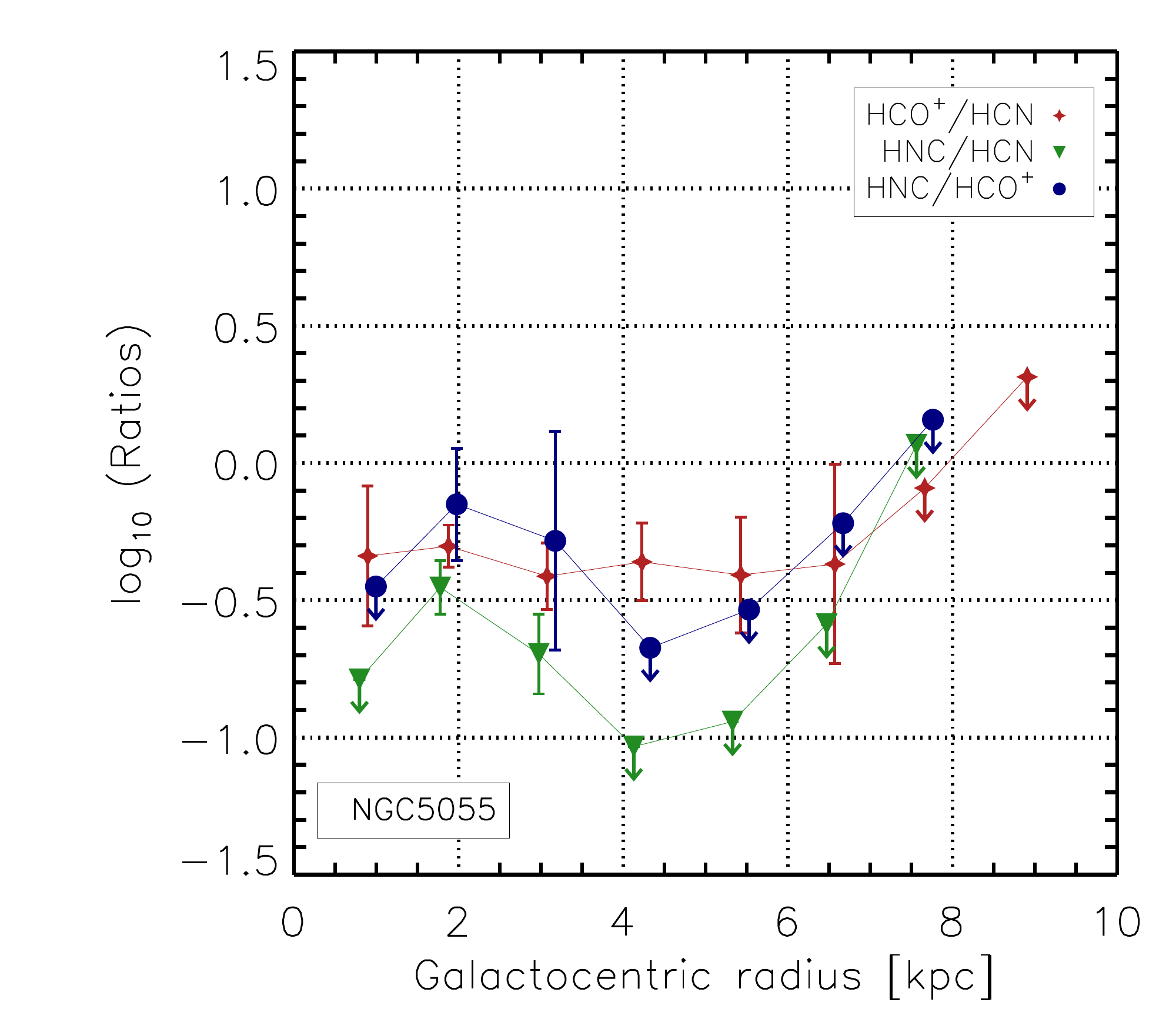}\\
	\end{center}
	\caption{Continued for NGC~5055. The HCN\,(1-0) contours employed are: 0.3, 0.9 and 1.6 K km s$^{-1}$.}
	\label{fig:maps7}
\end{figure*}

\begin{figure*}
	\begin{center}
		\includegraphics[scale=0.6]{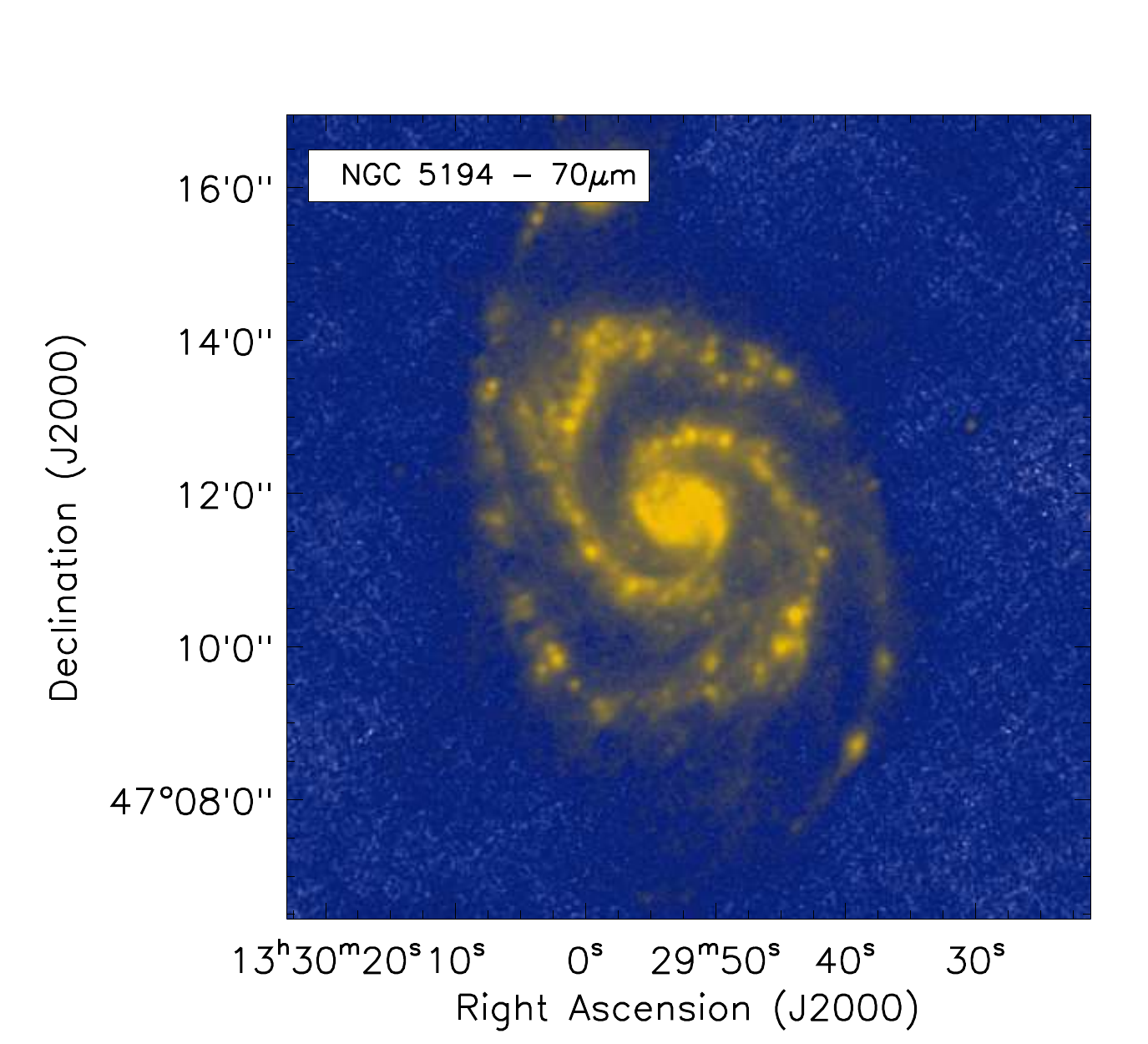}
		\includegraphics[scale=0.6]{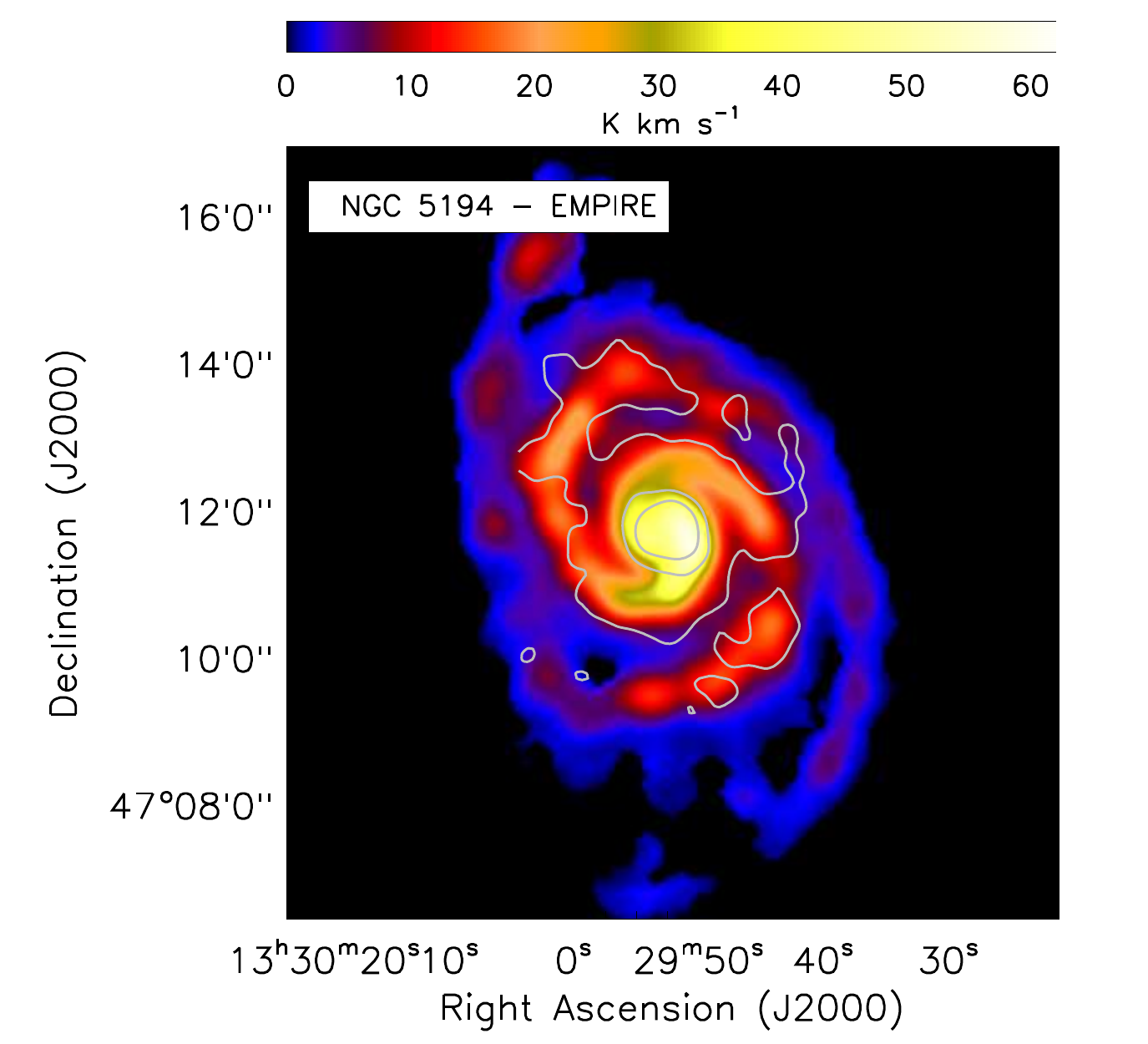}\\
		
		\includegraphics[scale=0.4]{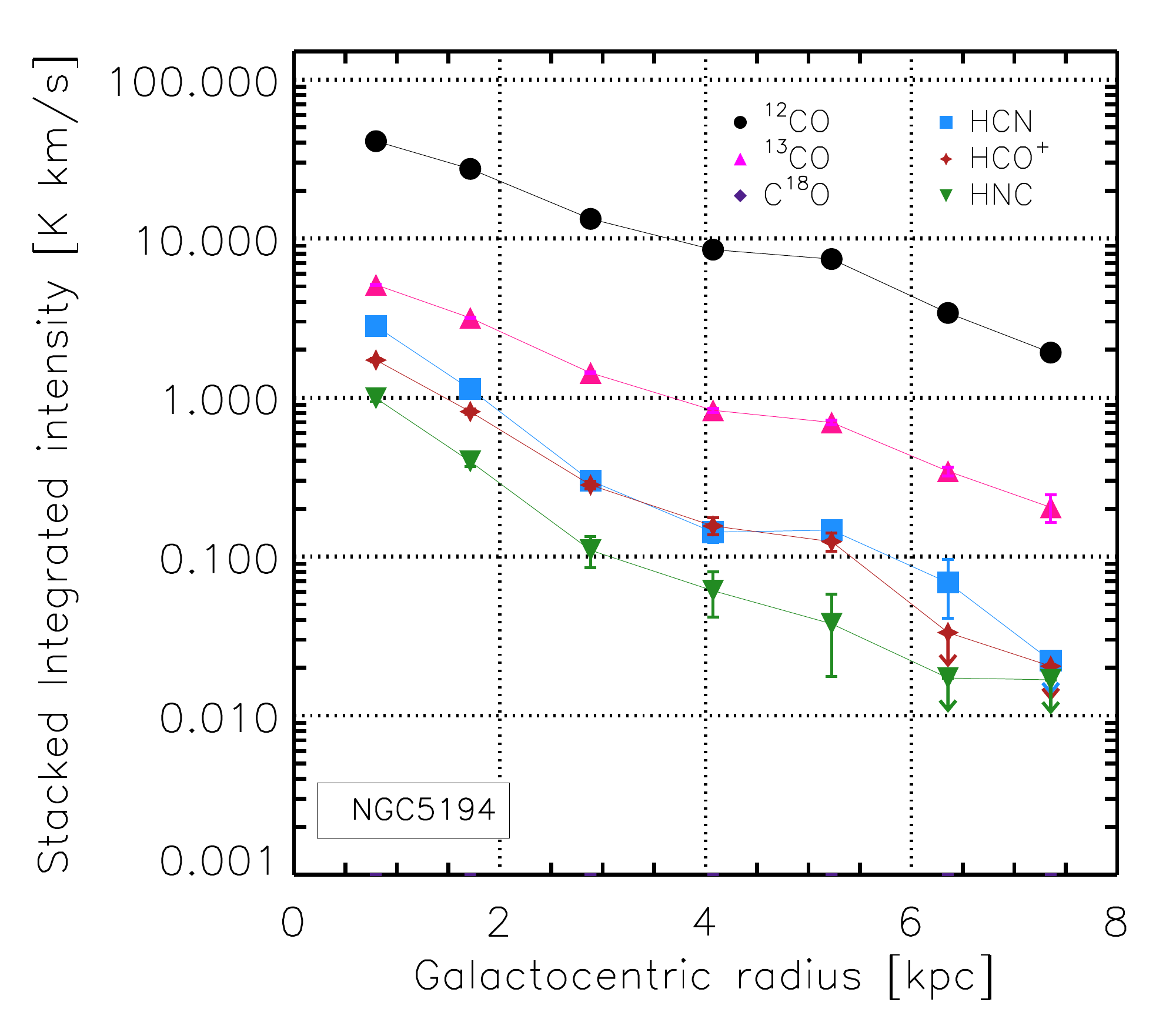}
		\includegraphics[scale=0.4]{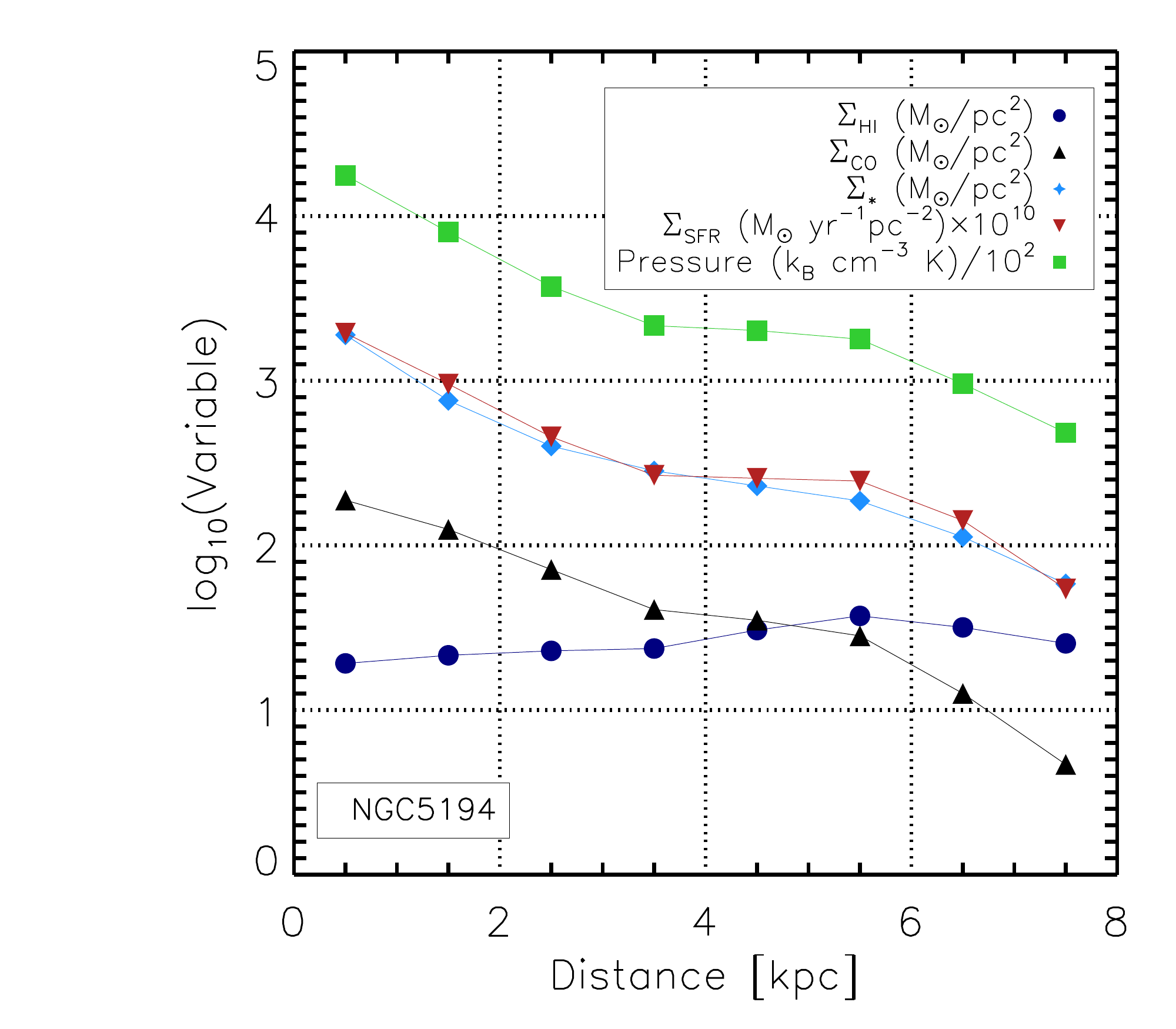}\\
		\includegraphics[scale=0.4]{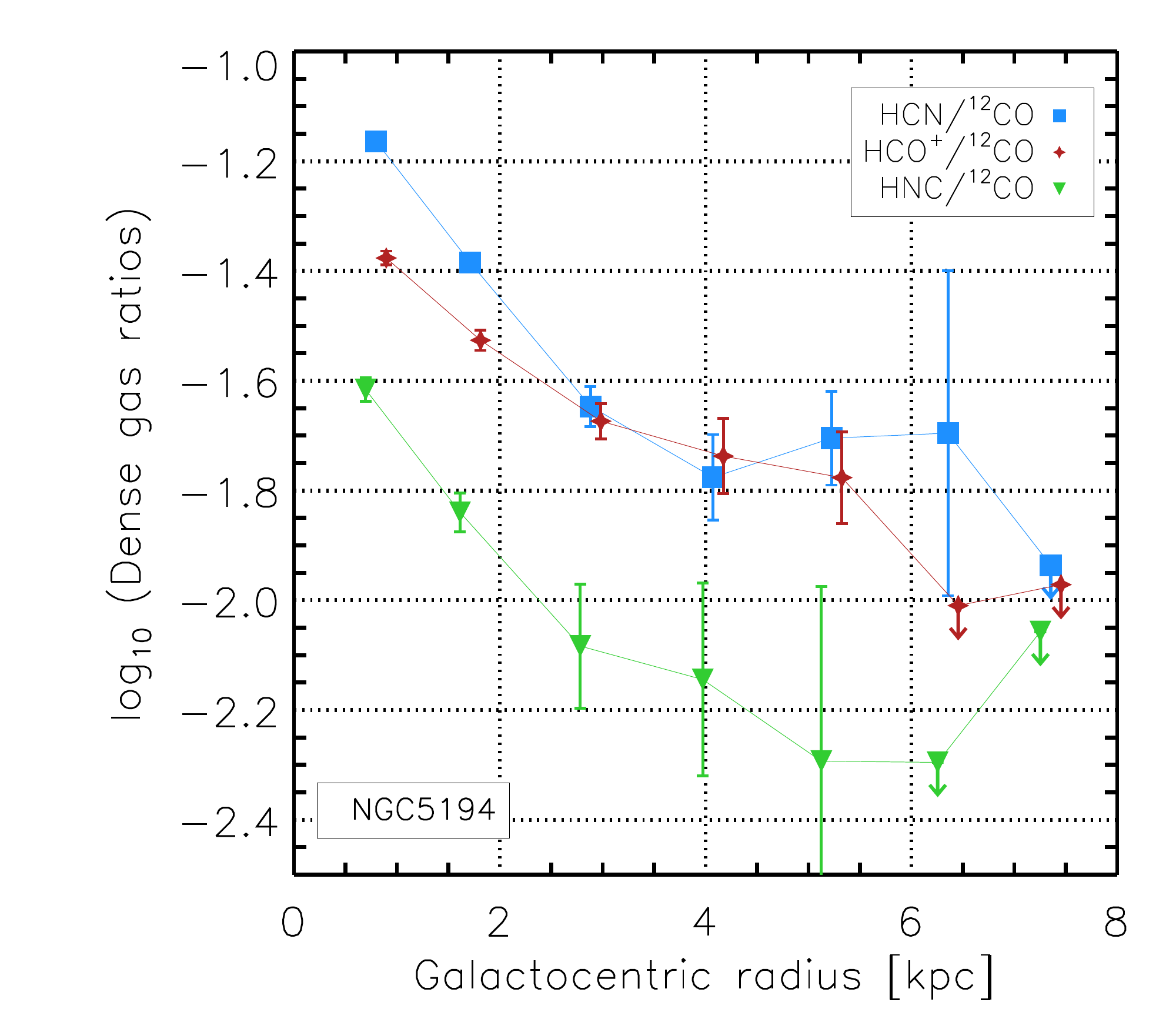}	
		\includegraphics[scale=0.4]{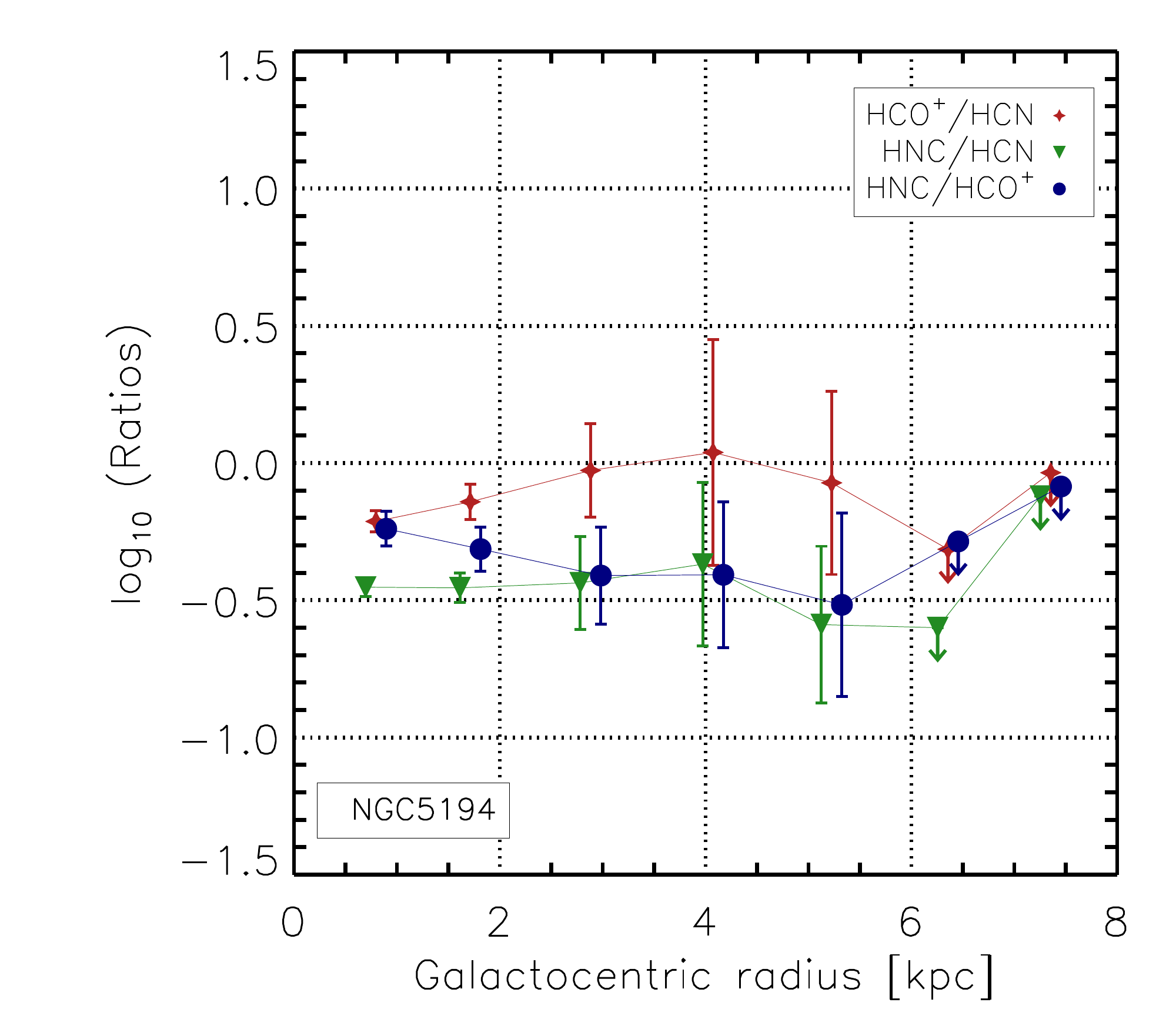}\\
	\end{center}
	\caption{Continued for NGC~5194. The HCN\,(1-0) contours employed are: 0.3, 0.9, 2.1 and 3.3 K km s$^{-1}$.}
	\label{fig:maps8}
\end{figure*}

\begin{figure*}
	\begin{center}
		\includegraphics[scale=0.6]{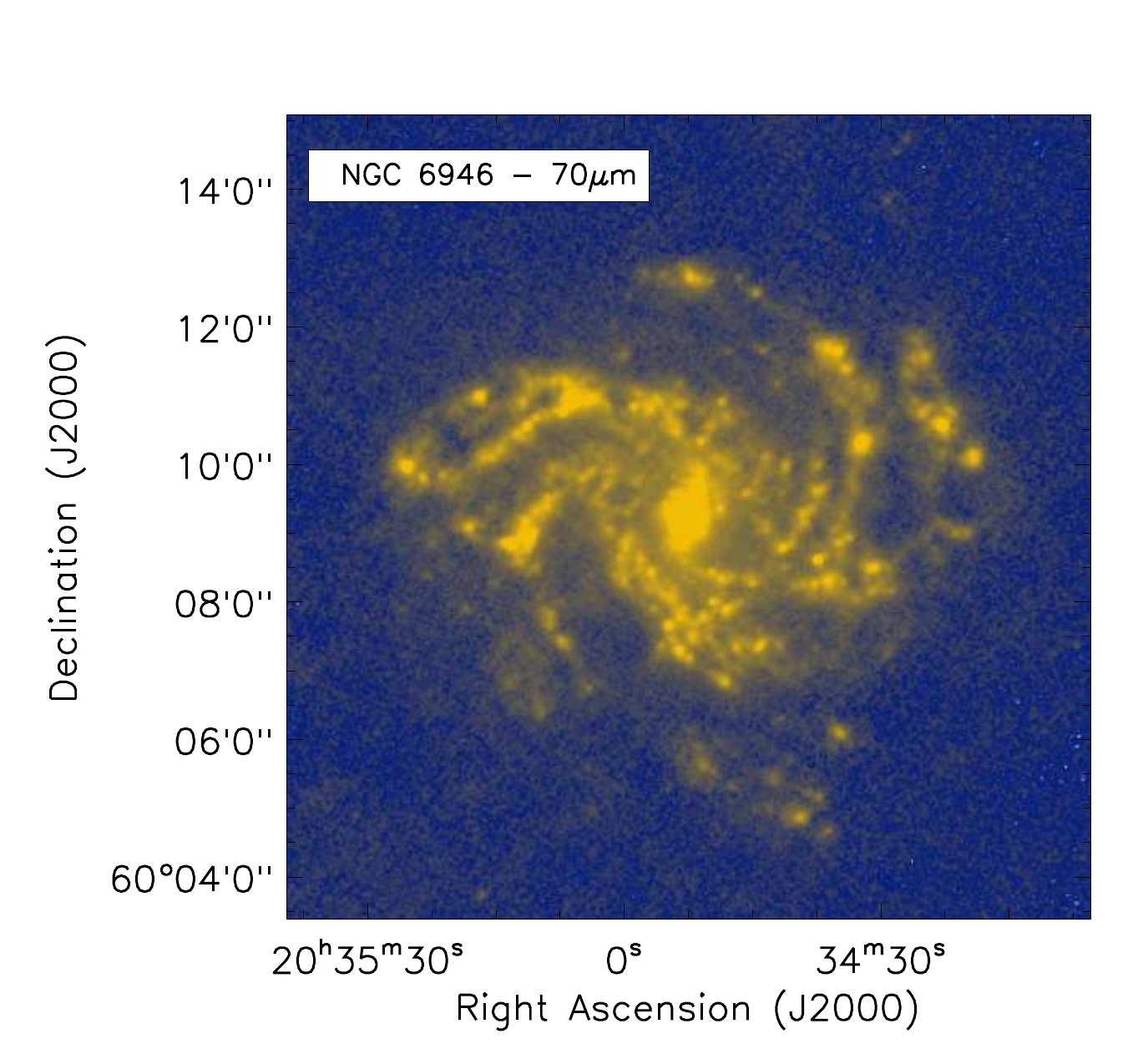}
		\includegraphics[scale=0.6]{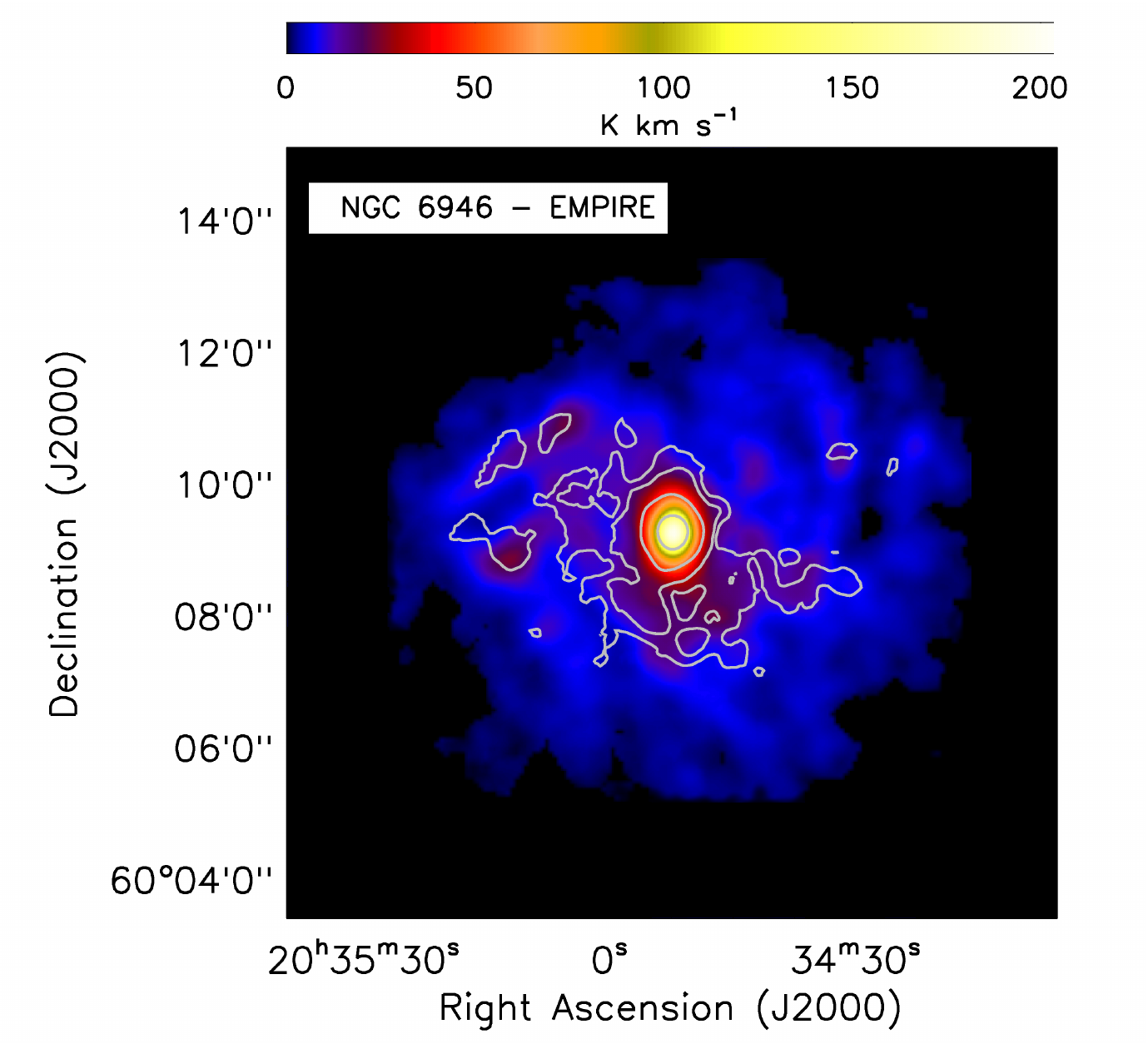}\\
		
		\includegraphics[scale=0.4]{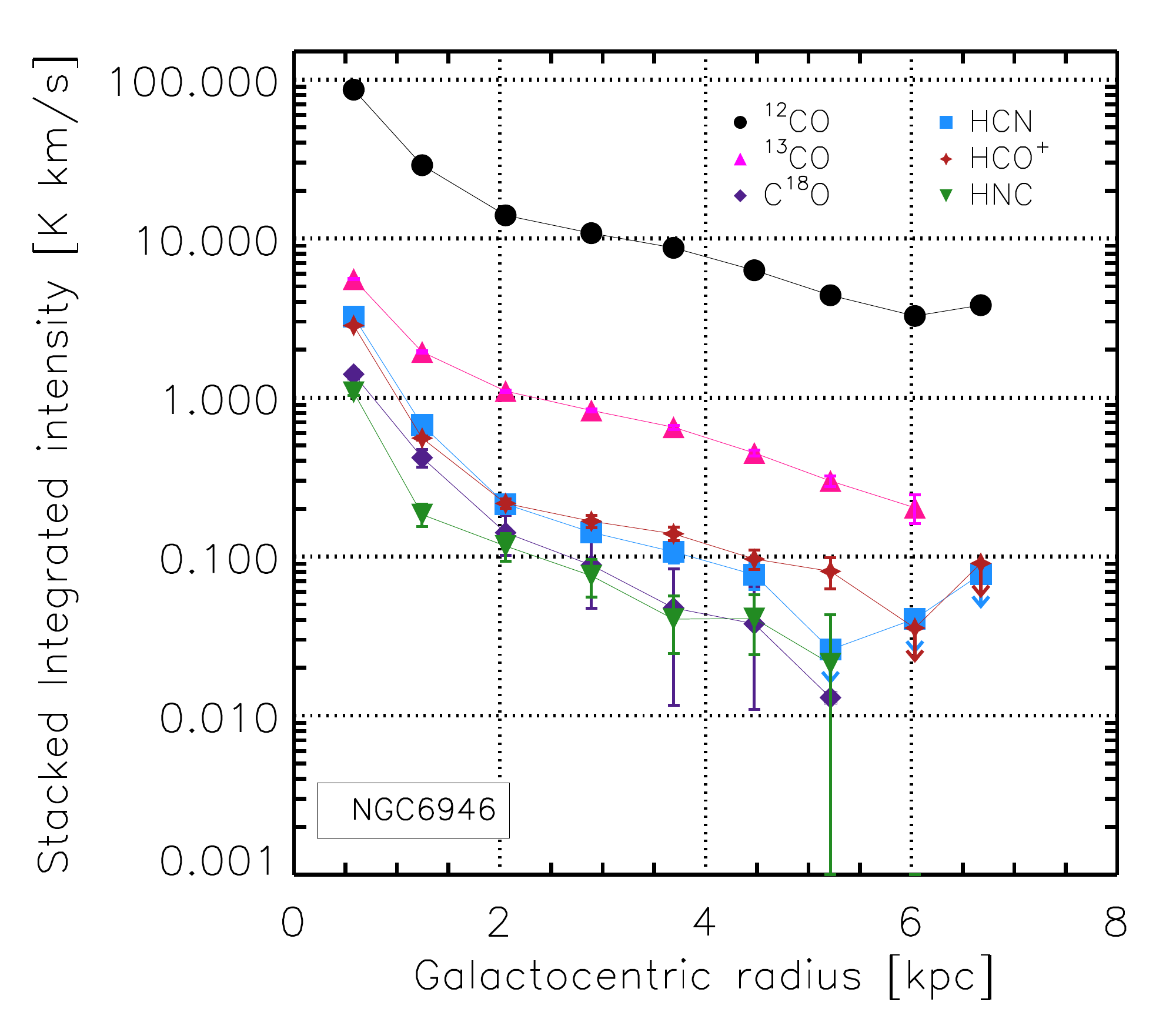}
		\includegraphics[scale=0.4]{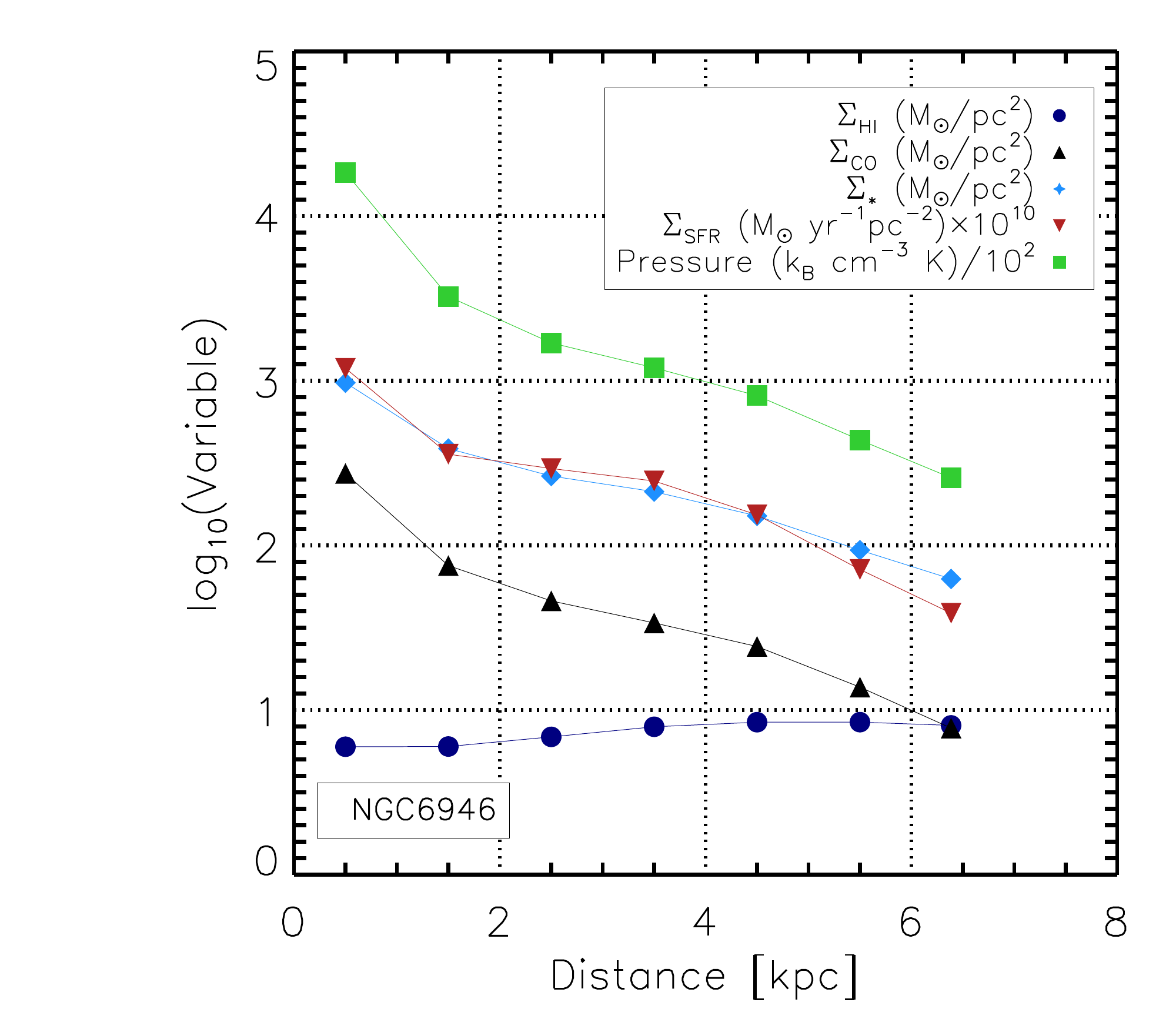}\\
		\includegraphics[scale=0.4]{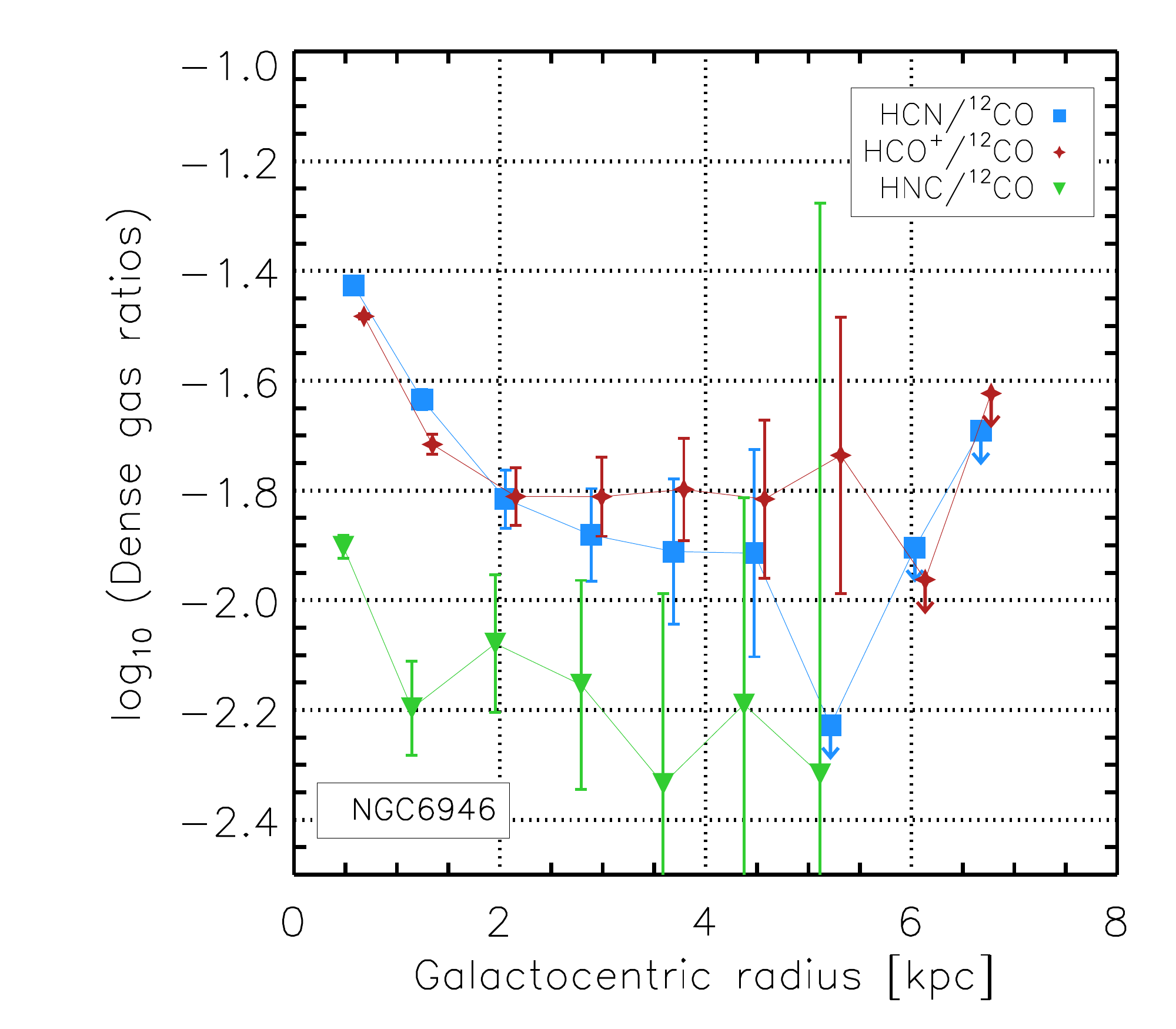}		
		\includegraphics[scale=0.4]{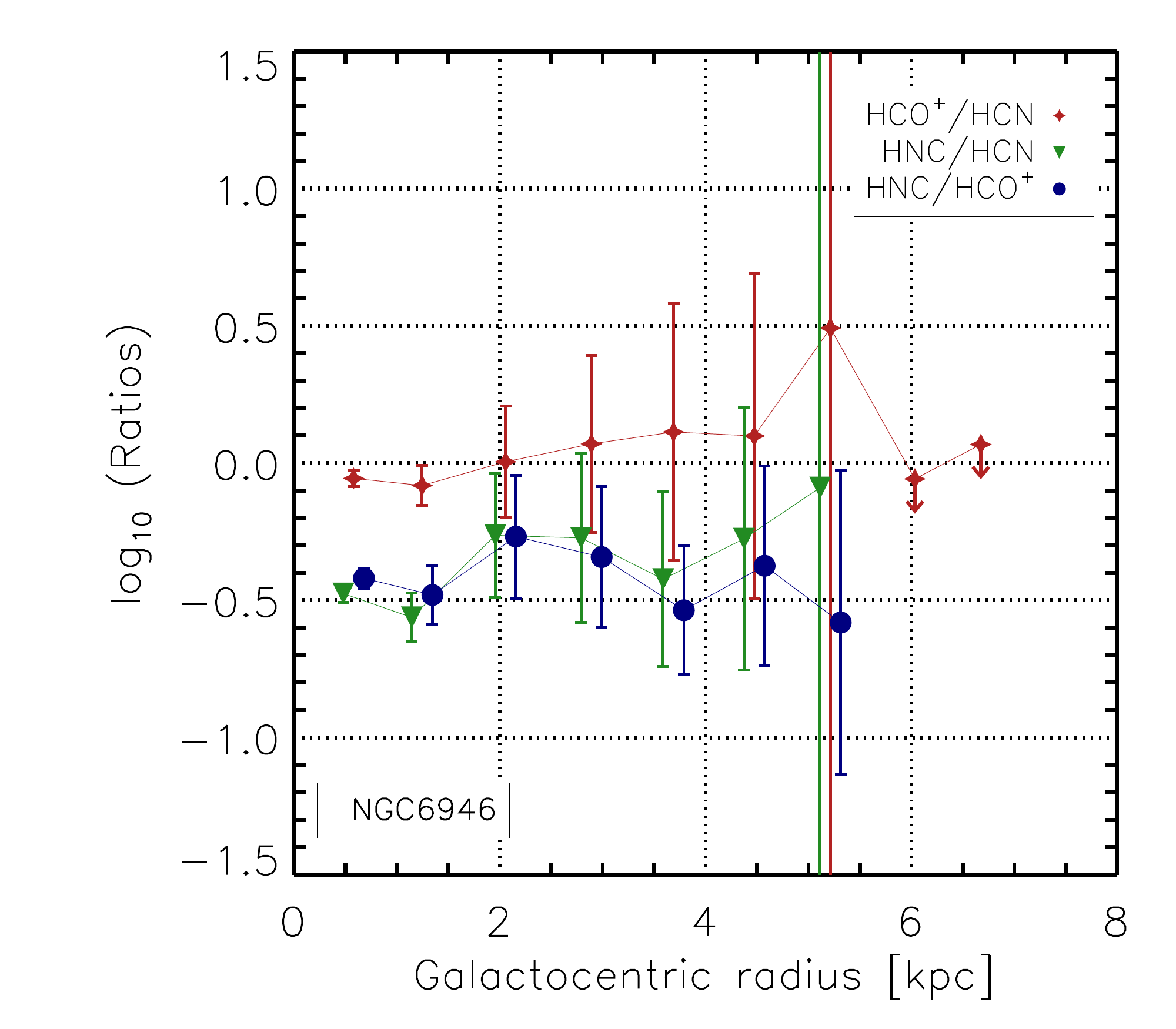}\\
	\end{center}
	\caption{Continued for NGC~6946. The HCN\,(1-0) contours employed are: 0.4, 1.4 and 4.7 K km s$^{-1}$.}
	\label{fig:maps9}
\end{figure*}

The radial profiles in Figures \ref{fig:maps1}-\ref{fig:maps9} illustrate the success of this stacking approach. We bin the data by galactocentric radius, using $30\arcsec$ wide bins ($\sim1-2$\,kpc at the distance of our sample). Within each bin, we create stacked spectra for each dense gas tracer and CO isotopologue (e.g., see Figure \ref{fig:dif_stacks}). Despite the faintness of HCN, we detect the average signal at high significance out to galactocentric radii $\sim 9{-}11$~kpc in HCN. This is similar to the radius of the Solar Circle. We also recover the average HCO$^+$ signal out to $\sim 7-10$\,kpc, and detect HNC out to $\sim 4-6$\,kpc. This represents the largest collection to date of extended, resolved profiles of dense gas in nearby galaxies.

\subsection{Molecular line ratios}
\label{sec:ratios}

\subsubsection{Dense gas tracers to CO}

In Figures \ref{fig:maps1}-\ref{fig:maps9} the stacked intensities of all lines ($^{12}$CO, $^{13}$CO, HCN, HCO$^+$ and HNC) decrease with increasing radius. On average, the emission of the dense gas tracers decreases more rapidly than that of lower density gas tracers $^{12}$CO and $^{13}$CO. In fact, in all galaxies except NGC~628, HCN/CO (blue in the bottom left panel) appears highest in the galaxy center and then decreases with increasing galactocentric radius.

On average, HCN/CO decreases by a factor of $\sim 2$ across the range of radii where we detect it. The decline in HCN/CO appears similar in our barred (NGC\,2903, NGC\,3184, NGC\,3627, NGC\,4321 and NGC\,6946) and unbarred (NGC\,628, NGC\,4254, NGC\,5055 and NGC\,5194) targets. We observe the largest HCN/CO declines in NGC~3627 ($\sim$0.60dex), NGC~4254 ($\sim$0.60dex), NGC~5194 ($\sim$0.60dex) and NGC~6946 ($\sim$0.55dex). Again, we see no strong morphological divide, NGC~3627 and NGC~6946 are strongly barred galaxies with prominent dense gas emission in their centers and bars, while NGC~4254 and NGC~5194 are unbarred galaxies rich in molecular gas.

\begin{figure*}
	\begin{center}
		\includegraphics[scale=0.14]{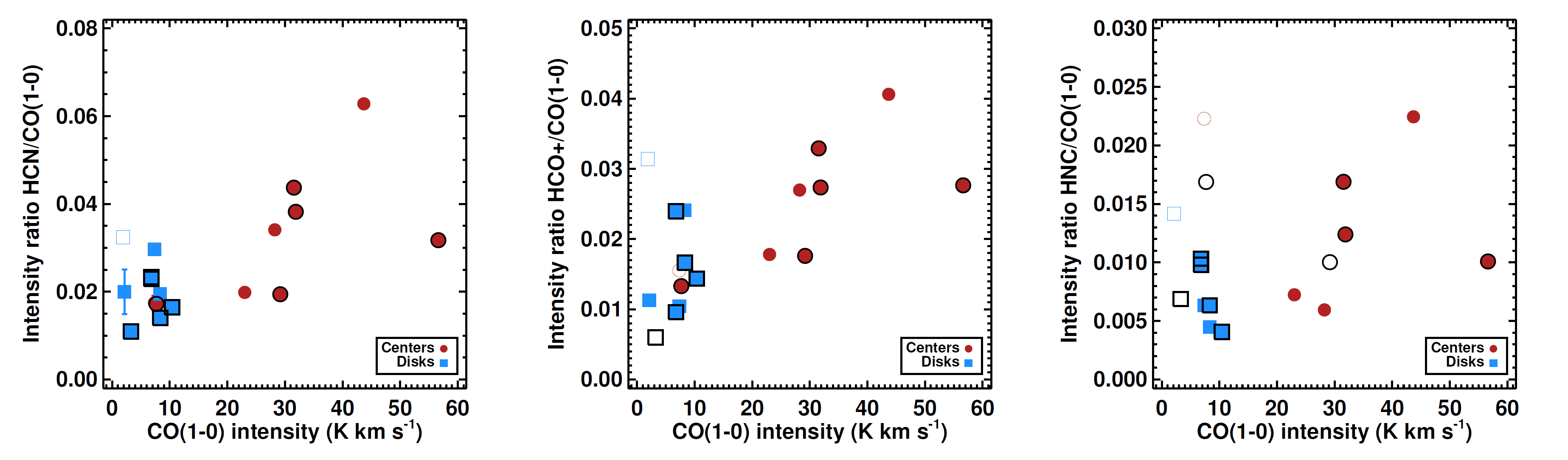}
	\end{center}
	\caption{Ratio of high critical density tracers to low critical density tracers for the central pointing (circles) and rest of the disk (squares) of the EMPIRE galaxies. The exact values and uncertainties are described in Table \ref{table:ratios}. Upper limits to the stacked line ratios are represented by open symbols. Symbols corresponding to barred galaxies are outlined with black contours.}
	\label{fig:ratios}
\end{figure*}

We quantify differences between the central pointing and the rest of each galaxy, which we refer to as the ``disk''. Following \citet{CORMIER18}, we take the ``center'' to have a radius of $16\arcsec \approx0.8\textrm{kpc}$ (i.e., one resolution element). The ``disk'' includes all other emission above a low intensity threshold ($\sim2\,\textrm{K km s}^{-1}$) and excludes ``center''.  For each galaxy center and each disk region, we create stacked spectra for each emission line. We use these to measure average intensities and line ratios. We calculate the mean ratios for barred galaxies and unbarred galaxies separately, as well as for the entire sample, and report these in Table \ref{table:ratios} and Figure \ref{fig:ratios}. We also note the implied dense gas fractions, adopting the fiducial conversion factors\footnote{We apply our adopted HCN conversion factor to HCO$^+$ and HNC. The dense gas fractions inferred from these lines should be taken as more approximate than that from HCN.}. 

From HCN/CO, we estimate dense gas fractions ($f_\textrm{dense}$) as described in Section \ref{sec:phys_parameters} of $6-10\,\%$ for the EMPIRE galaxy centers. They show twice as much dense gas than galaxy disks, where $f_\textrm{dense}\sim 5\,\%$. We also find similar HCN/CO, HCO$^+$/CO and HNC/CO ratios comparing barred and unbarred galaxies. 

While we do not find a clear link to bars, the concentration of gas at the galaxy center does appear to drive high HCN/CO ratios. Figure \ref{fig:ratios} shows that high HCN/CO, HCO$^+$/CO and HNC/CO values tend to appear in regions with high CO intensity. Modulo conversion factor effects, these high CO intensities indicate large concentrations of gas in the galaxy centers. Achieving a high dense gas fraction at the galaxy center appears to require concentrating a large amount of gas at the galaxy center. We return to this point in Section \ref{sec:fdense_sfe}.

Figure \ref{fig:ratios} shows similar trends in HCN/CO, HCO$^+$/CO and HNC/CO. The similarity among all three lines suggests that the results do reflect changing gas density. However, changing density may not be the only effect. Galaxy centers also host conditions that can lead to increased HCN excitation at fixed density. Increased gas temperatures by excitation by electrons, UV, X-rays, cosmic rays, and mechanical heating have all been suggested to increase HCN emission \citep[see e.g.,][]{KOHNO01,IZUMI13,BISBAS15,GOLDSMITH17}. We return to this in Sections \ref{sec:fdense_sfe} and \ref{sec:hcn_emission}.

Our dense gas fractions estimated from HCN/CO agree well with recent literature measurements in nearby galaxies. The recent higher resolution ($8\arcsec \sim 500$\,pc) study of nearby galaxies presented by \citet{GALLAGHER18}, shows that a median value of $10\%$ is characteristic of the inner kpc of four nearby galaxy disks (NGC\,3351, NGC\,3627, NGC\,4254 and NGC\,4321). Our EMPIRE central dense gas fractions are 9\%, 7\% and 10\% for NGC\,3627, NGC\,4254 and NGC\,4321, respectively. These are in good agreement with the results in \cite{GALLAGHER18}, the small differences are likely due to the larger beam size in EMPIRE, which will encompass more extended emission. We find very similar $f_\textrm{dense}$ values to those from \citet{USERO15} (median values of $8 \%$ in all disk pointings and $5 \%$ excluding the centers) and slightly lower values than those from \citet{GAO04} ($12 \%$). This is most likely attributed to the fact that EMPIRE median values are dominated by disk positions with overall lower $f_\textrm{dense}$, while \citet{GAO04} measured galaxy averages (with total luminosities dominated by the central enhancements) and focused on IR-bright and starburst galaxies. All of these studies also adopted our fiducial $\alpha_{\rm HCN}$.

\subsubsection{Ratios among dense gas tracers}

Table \ref{table:ratios} also provides the average line ratios among our high density tracers, HCO$^+$/HCN and HNC/HCN. We find average HCO$^+$/HCN values of $\sim$0.8 and HNC/HCN values of $\sim$0.5 across the disks of our targets. These measurements agree with observations of the Milky Way CMZ and Galactic GMCs \citep[e.g.,][$\sim 0.6$]{JONES12}, nearby galaxies \citep[such as M51, NGC\,253 and NGC\,6946][$\sim 0.6-1.1$]{MEIER14,MEIER15,CHEN15}, and a number of LIRGs \citep[e.g.,][$\sim 0.5-2.0$]{LOENEN08,PRIVON15}. 

We plot profiles of HCO$^+$/HCN, HCO$^+$/HNC and HNC/HCN in the bottom right panels of Figures \ref{fig:maps1}-\ref{fig:maps9}. The behavior of the HCO$^+$/HNC and HNC/HCN profiles is only weakly constrained due to our lower HNC detection rate. On the other hand, HCN and HCO$^+$ exhibit almost identical profiles in NGC\,3627, NGC\,4254, NGC\,4321, NGC\,5194 and NGC\,6946. In these targets we only observe significant differences in the galaxy centers and at large radii. This resembles the results seen at higher resolution over a smaller field of view by \citet{GALLAGHER18}. They find almost identical radial profiles for the two lines in the inner $\sim 4$\,kpc of their five targets.

The HCO$^+$/HCN line ratios measured across our galaxy sample are close to unity, and change little over the galaxy disks. This is suggestive of HCN and HCO$^+$ lines being slightly subthermal across the galaxy disks \citep{KNUDSEN07,MEIER15} if both are optically thick, as shown in e.g., \citet{JIMENEZDONAIRE17}.

Two targets, NGC\,4254 and NGC\,6946, show HCO$^+$-to-HCN profiles which are a function of increasing radius, where the typical ratio reaches values up to $\sim$1.5. These values are large in comparison to those found in their central regions ($\sim$0.8). Because HCO$^+$ has a lower critical density than HCN (see Table \ref{table:lines}), changing HCO$^+$/HCN values could simply be attributed to changing gas density across these disks. Larger HCO$^+$/HCN values are also found in lower metallicity systems such as IC\,10 \citep[1.1-2.8,][]{BRAINE17,Nishimura2016SpectralIC10,KEPLEY18}, M31 \citep[1.2,][]{BROUILLET05}, M33 \citep[1.1-2.5,][]{BUCHBENDER13,BRAINE17} or the Magellanic Clouds \citep[1.8-3,][]{CHIN97,CHIN98}, possibly due to less nitrogen produced by massive stars \citep{Vincenzo2016NitrogenUniverseb}. However the slightly rising HNC/HCN profiles in these two targets makes it difficult to conclude whether the larger HCO$^+$-to-HCN ratios can be associated to the reduction of nitrogen-bearing molecules like HCN or HNC. Alternatively, the HNC/HCN abundance ratio could increase at lower temperatures because the chemical balance between the two species is relatively more favourable to HNC. This would also cause the HCO$^+$/HCN profiles to increase in those regions.

\begin{table*}
	\caption{Average dense gas line ratios (observed integrated intensities, excluding upper limits), separated into central pointings and disks, across the EMPIRE spiral galaxies. The percentage numbers in parenthesis show the dense gas fractions ($f_\textrm{dense}$) computed using the fiducial conversion factors from Section \ref{sec:phys_parameters}. The quoted uncertainties are estimated as weighted means of the uncertainties derived as indicated in Section \ref{sec:stacking}.}
	\label{table:ratios}
	\centering
	\begin{tabular}{l c c c c r}
		\hline\hline
		  & \multicolumn{2}{c}{Center (inner $30\arcsec \sim 1-2\,$kpc)} & \multicolumn{2}{c}{Disk (excl. center)} & All \\
		\hline
        Ratio & Barred & Unbarred & Barred & Unbarred & \\
		\hline
		HCN/CO & $0.030(2)$ ($6.8 \pm 0.5$\%) & $0.034(2)$ ($7.7 \pm 0.5$\%) & $0.018(2)$ ($4.0 \pm 2.0$\%) & $0.024(5)$ ($5.4 \pm 1.0$\%) & 0.025 ($5.7 \pm 1.0$\%)\\
		HCO$^+$/CO & $0.024(2)$ ($5.4 \pm 0.5$\%) & $0.025(2)$ ($5.7 \pm 0.5$\%) & $0.014(5)$ ($3.2 \pm 1.0$\%) & $0.019(6)$ ($4.3 \pm 1.5$\%) & 0.018 ($4.0 \pm 2.0$\%)\\
		HNC/CO & $0.013(2)$ ($2.9 \pm 0.5$\%) & $0.014(2)$ ($3.2 \pm 0.5$\%) & $0.010(3)$ ($2.3 \pm 0.8$\%) & $0.014(2)$ ($3.2 \pm 1.0$\%) & 0.011 ($2.5 \pm 1.0$\%) \\
		\hline
		HCO$^+$/HCN & $0.8 \pm 0.1 $ & $0.7 \pm 0.1$ & $0.8 \pm 0.2 $ & $0.8 \pm 0.2 $ & $0.7 \pm 0.2 $\\
		HNC/HCN & $0.4 \pm 0.1 $ & $0.4 \pm 0.1$ & $0.6 \pm 0.2 $ & $0.6 \pm 0.2 $ & $0.4 \pm 0.2 $\\
		
		\hline
	\end{tabular}
\end{table*}

\subsection{IR-HCN Scaling relations}
\label{sec:scaling}

\begin{figure*}
	\begin{center}
		\includegraphics[scale=0.14]{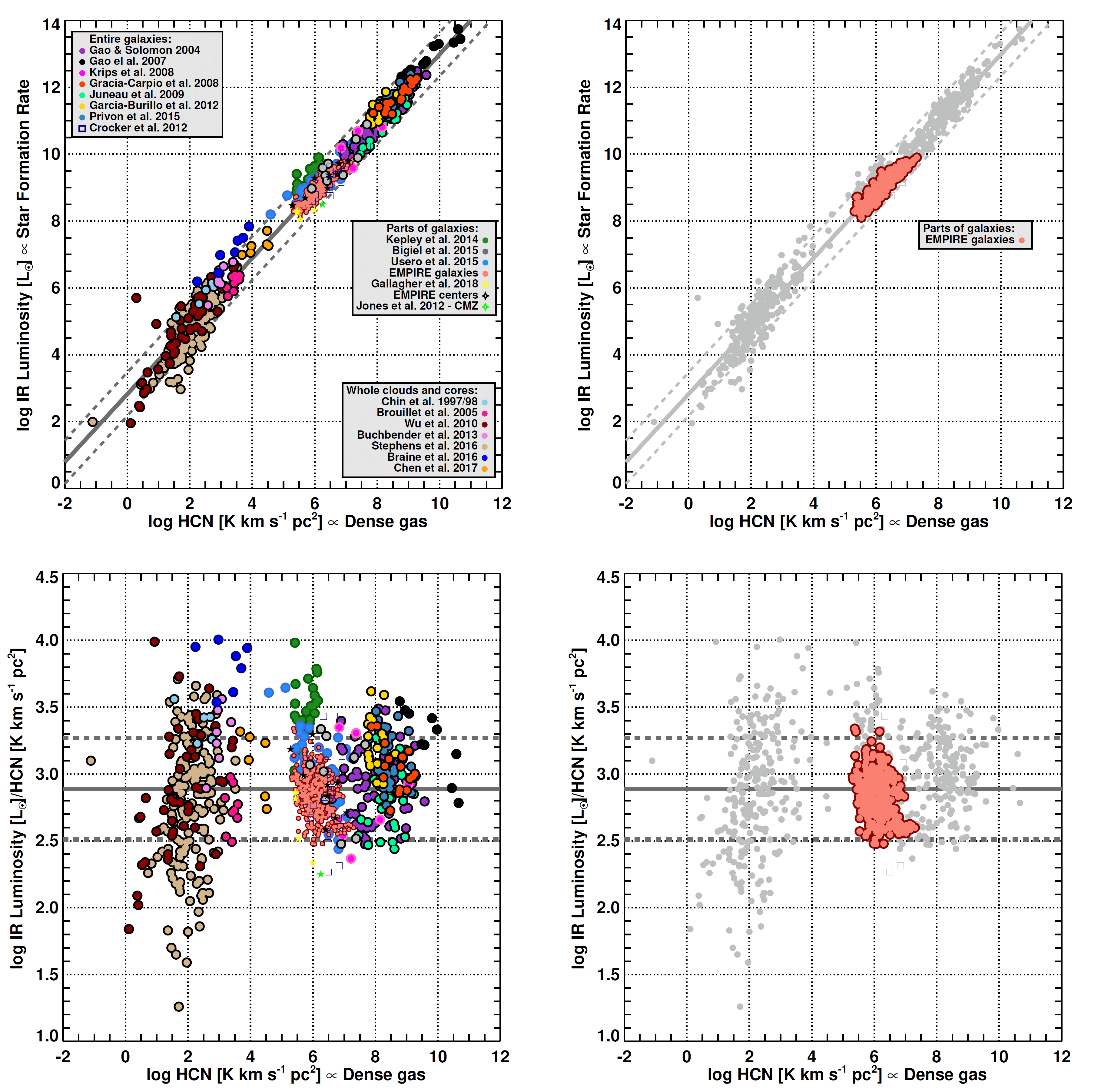}
	\end{center}
	\caption{{\it Top:} Luminosity-luminosity correlation between $L_\textrm{IR}$, as a tracer of the recent star formation rate (SFR) and $L_\textrm{HCN}$, tracing dense gas mass. {\it Bottom:} The star formation efficiency of dense gas (as traced by $L_\textrm{IR}/L_\textrm{HCN}$) plotted directly. Our literature compilation includes HCN observations ranging from Galactic clumps and cores \citep{WU10,STEPHENS16}, giant molecular clouds (GMCs) in the SMC, LMC and other low-metallicity galaxies \citep{CHIN97,CHIN98,BRAINE17}, giant molecular associations in nearby galaxies \citep{BROUILLET05,BUCHBENDER13,CHEN17}, resolved nearby galaxy disks \citep[][and this work]{KEPLEY14,USERO15,BIGIEL15,CHEN15,GALLAGHER18}, and whole galaxies and centers \citep{GAO04,GAO07,KRIPS08,GARCIACARPIO08,JUNEAU09,GARCIABURILLO12,PRIVON15}. The gray lines in both figures show the mean IR-to-HCN ratio derived from the combined dataset and quoted in Table \ref{table:totalfit}, and the dashed lines reflect the $1\sigma$ RMS scatter ($\pm$0.37dex) across all data in this plot. The right column highlights the EMPIRE datapoints overplotted on top of the literature compilation in gray.}
	\label{fig:all_hcn}
\end{figure*}

In Figure \ref{fig:all_hcn} we plot the IR luminosity, tracing the star-formation rate, as a function of the HCN luminosity, which traces the dense gas content. Light red dots show lines of sight from EMPIRE with signal-to-noise $>3$ HCN detections.  We also show an integrated measurement for each EMPIRE target as a filled gray circle.

We compare EMPIRE to an extensive compilation of literature measurements. This includes measurements of Galactic dense gas cores \citep{WU10,STEPHENS16}, individual giant molecular clouds (GMCs) in the SMC, LMC and other low-metallicity galaxies \citep{CHIN97,CHIN98,BRAINE17}, giant molecular associations in nearby galaxies \citep{BROUILLET05,BUCHBENDER13,CHEN17}, resolved nearby galaxy disks \citep[][]{KEPLEY14,USERO15,BIGIEL15,CHEN15,GALLAGHER18}, and whole galaxies and galaxy centers \citep{GAO04,GAO07,KRIPS08,GARCIACARPIO08,JUNEAU09,GARCIABURILLO12,CROCKER12,PRIVON15}. In total, we plot 225 data points for resolved cores and GMCs; 194 data points correspond to observations of entire galaxies or bright galaxy centers; and 415 data points (including the high signal-to-noise EMPIRE detections) for resolved ($\sim 0.3-2\,$kpc) galaxy disks. This literature collection is available\footnote{Studies employing measurements taken from the literature should cite the original works.} in Table \ref{table:literature}. The plots also include data for the Milky Way's central molecular zone (CMZ) \citep[i.e., the inner $\sim 500\,$pc][]{JONES12}.
The ensemble of data in Figure \ref{fig:all_hcn} follow the same relationship found by \citet{GAO04} relating IR and HCN emission in starbursts and IR-bright whole galaxies. As shown before \citep[e.g.,][]{WU05}, this scaling relation spans almost ten orders of magnitude in IR and HCN luminosity. More, the relationship appears approximately linear. Gray lines in Figure \ref{fig:all_hcn} show the mean IR-to-HCN ratio found across the entire data set (including EMPIRE). In Table \ref{table:totalfit}, we report the mean IR-to-HCN ratios for each type of data in the plot.

Our new EMPIRE measurements and the other resolved, kpc-scale data partly fill the gap between the resolved cores ($\sim 0.5\,$pc), individual clouds ($\sim 10{-}100$~pc), and the integrated emission from whole galaxies. We caution that while this represents an appealing way to visualize our data, the luminosity of a pointing in an EMPIRE disk is somewhat arbitrary. We could define larger or smaller regions and so shift the data in luminosity. As emphasized in the previous and next sections, the key physics in EMPIRE comes from resolved ratios among lines and tracers of recent star formation.

In that sense, the key point for Figure \ref{fig:all_hcn} is the good agreement between the IR-to-HCN ratio in EMPIRE and that from previous work. The bottom panels in Figure \ref{fig:all_hcn} plot this $L_\textrm{IR}$-to-$L_\textrm{HCN}$ ratio, which has been widely used as a tracer of the SFR per unit dense gas (SFE$_\textrm{dense}$). We find a mean ratio of $\sim776\,L_\odot/(\textrm{K km}^{-1}\textrm{pc}^2)$ across our whole compilation (Table \ref{table:totalfit}). 

Figure \ref{fig:all_hcn} also illustrates the significant scatter in IR-to-HCN across our data, which we also report in Table \ref{table:totalfit}. We find an RMS scatter of 0.37~dex across all objects. We find a smaller but still significant value of $\sim 0.3$~dex for whole galaxies and $\sim 0.25$~dex for  resolved regions in nearby galaxies (see Appendix \ref{appendix:montecarlo} for a detailed estimation of the physical scatter in EMPIRE measurements).

Some of this scatter reflects measurement uncertainty and on small scales stochasticity may play an important role. On the scale of resolved galaxy disks, much of this scatter has a physical origin. We see below that in EMPIRE the IR-to-HCN ratio shows systematic trends as a function of environment \citep[following][]{USERO15,BIGIEL16,GALLAGHER18}. As discussed in Section \ref{sec:background}, the Milky Way and other galaxy centers show low ratios of star formation to dense gas. This also points to a physical origin for much of the scatter in Figure \ref{fig:all_hcn}.

\subsection{Dense gas fraction and SFE$_\textrm{dense}$}
\label{sec:fdense_sfe}

We designed EMPIRE to measure how the dense gas fraction,  $f_\textrm{dense}$, and the star-formation efficiency of dense gas, SFE$_\textrm{dense}$, depend on location and local conditions inside a galaxy disk. Here, we address these questions using our brightest dense gas tracer, HCN\,(1-0). %We use HCN-to-CO as a proxy for $f_\textrm{dense}$ and IR-to-HCN as a proxy for the SFE$_\textrm{dense}$.

Figure \ref{fig:global_trends} plots $f_\textrm{dense}$ and SFE$_\textrm{dense}$, as likely functions of galactocentric radius and local conditions relevant to the formation and behavior of dense gas: stellar surface density ($\Sigma_*$), molecular gas surface density ($\Sigma_\textrm{mol}$), the ratio of the ISM in a molecular phase ($R_\textrm{mol}= \Sigma_\textrm{mol}/\Sigma_\textrm{HI}$), and the pressure ($P_{\rm DE}$). To construct these plots, we define bins in stellar surface density, molecular-to-atomic gas fraction and dynamical equilibrium pressure. We sort each galaxy by each quantity, identifying all lines of sight in each bin. Then, we stack all CO and HCN spectra and TIR intensities in each bin. We used these stacked CO, stacked HCN, and stacked TIR measurements to compute the average HCN-to-CO fraction and IR-to-HCN ratio in every bin. The error bars that we report in the plots include uncertainties from the statistical noise and from the spectral fitting.

In the main text, we focus on these stacked profiles to reveal the underlying physical trends in the data. This averaging technique is a core part of the EMPIRE experiment design and allows us to explore whole galaxy disks. In the Appendices (see Figures \ref{fig:fdense_rad}, \ref{fig:fdense_star}, \ref{fig:fdense_mol} and \ref{fig:fdense_pressure}), we present measurements of the same trends plotting each individual line of sight.

\subsubsection{Dense gas fraction}

The left panels of Figure \ref{fig:global_trends} show the variation of the HCN-to-CO ratio, tracing $f_{\rm dense}$, as a function of galactocentric radius, stellar surface density, molecular-to-atomic gas fraction and $P_{\rm DE}$.

In all targets $f_{\rm dense}$ increases towards galactic centers. We see similar stacked trends considering stellar surface density ($\Sigma_*$ up to $10^3-10^4\,M_\odot\,\textrm{pc}^{-2}$), molecular-to-atomic gas fractions ($R_\textrm{mol}\sim 10^2$) and equilibrium pressure ($P_\textrm{DE}/k_B \sim 10^6-10^7\,\textrm{K cm}^{-3}$). HCN-to-CO correlates positively with all of these quantities. Individual galaxies do show distinct relationships, so that the stacked trends appear offset among galaxies.

The positive correlations of HCN-to-CO with $\Sigma_*$, $f_\textrm{mol}$ and $P_\textrm{DE}$ agree with previous observations of $\sim$kpc-sized regions in nearby galaxies \citep[e.g.,][]{USERO15,CHEN15,BIGIEL16,GALLAGHER18}. The stacked trends in Figure \ref{fig:global_trends} cover the whole area of active star formation across a significant sample of whole galaxies. As a result, Figure \ref{fig:global_trends} represents the best systematic characterization to date of how $f_{\rm dense}$ depends on local conditions. Physically, all of the trends have the sense $f_{\rm dense}$ appears higher where there is higher stellar surface densities, higher molecular gas surface densities, higher molecular-to-atomic ratios and higher midplane pressures.

We measure the strength of the correlation between $f_{\rm dense}$ and our environmental measurements using the Spearman's rank correlation coefficient, $\rho$. Table \ref{table:rank_coef} reports $\rho$ for each environmental measure, each target, and all targets together. The $\rho$ coefficient quantifies the degree to which $f_{\rm dense}$ and the other quantity track one another monotonically in our binned measurements. To assess the uncertainty in our measured $\rho$, we repeatedly add noise to our measurements, with the magnitude reflecting the associated uncertainties. We take the scatter across 1,000 such Monte Carlo realizations to be the uncertainty in $\rho$. We do caution that because our bins have not been chosen for the purpose of rigorous statistical comparison, so $\rho$ should only be qualitatively compared between quantities. 

We also quantify the relationship between HCN-to-CO and local conditions using power law fits (Figure \ref{fig:global_trends}). We fit lines to the stacked trends in log-log space, neglecting upper limits (i.e., those bins with stacked SNR $< 3$ for HCN). We weight all bins equally. While these fits capture general trends, there remains large galaxy-to-galaxy scatter about each scaling relation. This implies that for each of these fits, significant additional physics beyond only the two variables considered affects the observed relation. As a complement, we provide weighted averages of fit parameters for each galaxy trend and discuss the scatter introduced by galaxy variations below.

\begin{table*}
	\caption{Mean $L_\textrm{IR}$-to-$L_\textrm{HCN}$ ratios representative of each sample (entire galaxies, resolved galaxy disks and MW clouds) used in Figure \ref{fig:all_hcn}. We include the $1\sigma$ RMS scatter found for each sample. The $L_\textrm{IR}$-to-$L_\textrm{HCN}$ Spearman's rank correlation coefficients and their $p-\textnormal{values}$ (in parenthesis), are also indicated in the table.}
	
	\label{table:totalfit}
	\centering
	\begin{tabular}{lccc}
		\hline\hline
		 Sample &  log$_{10}(L_\textrm{IR}/L_\textrm{HCN})$ & Scatter & Spearman's rank corr.\\
		  & $L_\odot\,/\,(\textrm{K km s}^{-1}\textrm{pc}^2)$ & & \\
		\hline
		Unresolved galaxies & 2.99 & $\pm$0.30dex & 0.91 ($<0.01$)\\
		Resolved galaxy disks & 2.85 & $\pm$0.24dex & 0.79 ($<0.01$)\\
		MW cores and nearby clouds & 2.85 & $\pm$0.47dex & 0.85 ($<0.01$)\\ 
		Combined & 2.89 & $\pm$0.37dex & 0.96 ($<0.01$)\\
		\hline
	\end{tabular}
\end{table*}

\bigskip

\textbf{Galactocentric radius:} HCN-to-CO anti-correlates with galactocentric radius. The following power-law describes our data:

\begin{equation}
\label{eq:fit_hcnco_rad}
\log_{10} \frac{\textrm{HCN}}{\textrm{CO}} = (-1.5\pm0.2) - (0.8\pm0.2) \, \log_{10} \frac{r}{r_\textrm{25}}~,
\end{equation}

\noindent where $r_{25}$ is the 25$^{\rm th}$ magnitude $B$-band isophotal radius from LEDA \citep{PATUREL03}. Our data mostly lie in the range $0{-}0.6~r_{25}$, and the fit should apply over this range. Individual binned measurements scatter by $\pm 0.20$~dex about this fit, with most scatter driven by galaxy-to-galaxy variations. Our central measurements out to $\sim 2\,$kpc in radius (confined in the inner two bins) show even more scatter, $\pm 0.35$~dex.

It is easy to understand an anti-correlation of galactocentric radius and $f_{\rm dense}$: Bars, interactions, and the inner parts of strong spiral arms can drive significant masses of gas towards the inner parts of disks \citep[e.g.,][]{ATHA92,KORMENDY04,SORMANI18}. As a result, galaxy centers tend to show higher gas surface densities than the rest of the disk. The central parts of galaxies also have the highest stellar surface densities in the disk. Thus there tends to be more gas and a deeper potential well in the inner parts of galaxies. Both factors should lead to higher gas densities at smaller radii. In turn, this leads us to expect both higher HCN intensities and higher HCN-to-CO ratios in the inner parts of galaxies. 

Individual galaxies show distinct trends in Figure \ref{fig:global_trends}.  Barred galaxies show stronger anti-correlations ($\rho \sim -0.7$) between $f_{\rm dense}$ and radius than unbarred galaxies ($\rho \sim -0.3$). Galaxies also appear offset from one another at fixed radius, reflecting that the same radius may correspond to different physical conditions in different galaxies. In the next few panels we plot $f_{\rm dense}$ as a function of gas or stellar surface density, we see greater similarity among all galaxies.

\medskip

\textbf{Stellar surface density:} We expect high $f_{\rm dense}$ where the gravitational potential is deeper \citep[e.g.,][]{HELFER97}. Stars represent the dominant mass component over the inner part of most galaxy disks. Therefore, following \citet{USERO15}, \citet{BIGIEL16}, and \citet{GALLAGHER18} the stellar surface density should be a good predictor of $f_{\rm dense}$, at least in regions with abundant gas.

Figure \ref{fig:global_trends} and Table \ref{table:rank_coef} indeed show a strong trend in each galaxy that also appears similar among galaxies. Our best-fit power-law relating $I_\textrm{HCN}/I_\textrm{CO}$ to $\Sigma_*$ is:

\begin{equation}
\label{eq:fit_hcnco_star}
\log_{10} \frac{\textrm{HCN}}{\textrm{CO}} = (-2.7\pm0.2) + (0.4\pm 0.1)\, \log_{10} \frac{\Sigma_*}{M_\odot\,\textrm{pc}^{-2}}.
\end{equation}

\noindent Which is valid at $\sim 1{-}2$~kpc resolution and mostly over the range $\Sigma_* \sim 100{-}1000$~M$_\odot$~pc$^{-2}$. Individual bins show a scatter of $\sim 0.2$~dex about this line, again driven mostly by galaxy-to-galaxy variations.

Our best-fit relation agrees well with that found by \citet{USERO15}. The slope is slightly shallower and offset by 0.6\,dex compared to the relation found in the resolved inner regions of nearby galaxies by \citet{GALLAGHER18}.

\medskip

\textbf{Molecular gas surface density:} In Section \ref{sec:ratios} we saw that high HCN-to-CO ratios correlate with the intensity of CO in galaxy centers. We can explain the larger concentrations of denser gas in regions with higher mean gas surface density, for the simple reason that these high surface densities indicate a large amount of gas concentrated in a small area. \citet{GALLAGHER18} found that HCN/CO correlates with $\Sigma_{\rm mol}$ at $\sim 500$~pc scales as well as with $\Sigma_{\rm mol}$ on cloud-scale \citep{GALLAGHER18b}.

We plot the observable ratio HCN/CO as a function of $\Sigma_{\rm mol}$ in Figure \ref{fig:global_trends} and fit the following scaling relation:

\begin{equation}
\label{eq:fit_hcnco_sigmol}
\log_{10} \frac{\textrm{HCN}}{\textrm{CO}} = (-2.4\pm0.2) + (0.5\pm0.1)\, \log_{10} \frac{\Sigma_\textrm{mol}}{M_\odot\,\textrm{pc}^{-2}}.
\end{equation}

\noindent This holds at $\sim 1{-}2$~kpc resolution over the range $\Sigma_{\rm mol} \sim 1{-}400$~M$_\odot$~pc$^{-2}$. We find $0.18$~dex scatter about the fit. Recall that we adopt a fixed $\alpha_{\rm CO} = 4.4~M_\odot$ pc$^{-2}$ (K km s$^{-1}$)$^{-1}$ and do not implement any environment-dependent conversion factor. Therefore, Equation \ref{eq:fit_hcnco_sigmol} formally captures the scaling between HCN-to-CO and $I_{\rm CO}$.

Thus HCN/CO, tracing the fraction of dense gas, correlates well with both stellar and gas surface densities in EMPIRE. More gas and a deeper stellar potential well imply higher gas densities. Our targets mostly show common behavior, with the main outlier being NGC~6946. This galaxy appears moderately displaced towards lower HCN-to-CO ratios at a fixed $\Sigma_{\rm mol}$ or $\Sigma_*$. NGC~6946 also shows evidence of a radius-dependent conversion factor \citep{SANDSTROM13}. Accounting for this effect should move the points from that galaxy into better agreement with the rest of our data, provided that the HCN conversion factor variations are milder.

\medskip

\textbf{Molecular-to-atomic gas ratio:} The local ratio of molecular to atomic gas reflects the interstellar density and pressure. Denser, higher pressure environments have a larger fraction of their gas in the molecular phase \citep[e.g.,][]{WONG02,BLITZ06,LEROY08}, though factors like the radiation field and dust abundance also play a role \citep[e.g.,][]{PELLEGRINI09,WOLFIRE10,STERNBERG14}. \citet{USERO15} showed that $f_{\rm dense}$ correlates with the molecular-to-atomic gas ratio. This implies that the same facts that cause gas to become molecular may also drive gas to higher densities.

We fit the following relation:

\begin{equation}
\label{eq:fit_hcnco_atom}
\log_{10} \frac{\textrm{HCN}}{\textrm{CO}} = (-1.8\pm0.1) + (0.42\pm0.04)\,\log_{10} R_\textrm{mol} ,
\end{equation}

\noindent The relationship holds at $1{-}2$~kpc resolution and over the range $R_{\rm mol} \sim 0.5{-}100$. We again find $\pm 0.2$~dex scatter from galaxy-to-galaxy at fixed $R_\textrm{mol}$. NGC\,6946 appears offset from the relations found for the rest of galaxy disks, likely due to a variable CO-to-molecular gas surface density conversion factor.

\medskip

\textbf{Dynamical Equilibrium Pressure:} The correlations with surface density and $R_{\rm mol}$ could be expected if the mean turbulent interstellar pressure couples closely to the gas density distribution \citep[see][]{HELFER97,USERO15,GALLAGHER18}. Assuming vertical hydrostatic equilibrium, the mean insterstellar pressure must balance the weight of the gas in the potential well (see references and discussion in Section \ref{sec:data_pressure}). We plot the $f_{\rm dense}$ as a function of pressure in Figure \ref{fig:global_trends}. There we do observe a clear correlation, though again with some notable outliers. 

We fit the following power law relating  $I_\textrm{HCN}/I_\textrm{CO}$ to $P_{\rm DE}$:

\begin{equation}
\label{eq:fit_hcnco_press}
\log_{10} \frac{\textrm{HCN}}{\textrm{CO}} = (-4.9\pm0.4) + (0.6\pm0.1)\,\log_{10}\,\left[\frac{P_{\rm DE}}{k_B\,\textrm{cm}^{-3}\,\textrm{K}}\right] .
\end{equation}

\noindent This holds at $1{-}2$~kpc resolution over the range $\log_{\rm 10} P_{\rm DE}/k_B$\,$[{\rm K~cm}^{-3}]$ $\sim 4.5{-}6.5$. Again, our individual binned stacks scatter by $\pm0.2$~dex RMS about the measurement. 

Similar to our results for $R_\textrm{mol}$, $I_\textrm{HCN}/I_\textrm{CO}$ correlates strongly with $P_{\rm DE}$ in each individual galaxy. This appears true for both barred ($\rho \sim 0.9$) and unbarred galaxies ($\rho \sim 0.8$). These correlations show similar slopes (within 10\%) for different galaxy disks. However, the overall correlation appears weaker because the stacked relations show considerable offset from one another. Again NGC~6946 appears as a significant outlier, possibly due to conversion factor effects. In this plot, NGC~4254 also appears as a significant outlier.

In theory, $P_{\rm DE}$ represents the most direct physical driver of density that we test. If $I_\textrm{HCN}/I_\textrm{CO}$ traces $f_{\rm dense}$ and we estimate $P_{\rm DE}$ correctly, then our observations imply that, while the $\sim 1{-}2$~kpc mean pressure scales with $f_{\rm dense}$ other physics also play an important role. In addition to the conversion factor effects discussed above, we might also expect the structure of the gas within our large beam to play a role. Our observations do not distinguish between gas concentrated into a few massive, dense clouds and gas spread through a diffuse layer. Comparisons to higher resolution CO mapping of our targets \citep[e.g.,][]{SUN18,GALLAGHER18b} will help to clarify the role of detailed ISM structure in producing this galaxy-to-galaxy scatter.
\medskip

\textbf{Dense gas fraction and environment:} The measurements in this section represent the most thorough view to date of how the HCN-to-CO ratio, tracing $f_{\rm dense}$, depends on environment in nearby galaxies. Our results agree well with previous work by \citet{USERO15}, \citet{BIGIEL16}, and \citet{GALLAGHER18}, but extend these studies to wider area and more complete coverage of a sample of galaxies. EMPIRE recovers HCN emission out to radii similar to the Solar Circle and spanning a wide range of local conditions: out to $\sim 8-10$\,kpc in galactic radius, $\Sigma_*$ ranging from $\sim 30-3200\,M_\odot\,\textrm{pc}^{-2}$, molecular gas surface densities up to $\Sigma_\textrm{mol} \sim 300\,M_\odot\,\textrm{pc}^{-2}$, three orders of magnitude in $R_\textrm{mol}$ (typical ratios range from $\sim 0.1-2$), and more than two orders of magnitude in $P_{\rm DE}$. The correlations, fits, and stacked profiles that we present should provide a basic reference for how the HCN-to-CO ratio behaves across galaxies.

We find $f_{\rm dense}$ traced by HCN-to-CO to vary significantly as a function of local environment. $f_{\rm dense}$ appears higher in regions with high stellar and gas surface densities, high interstellar pressures, and high molecular gas fractions. These conditions tend to occur more in the inner parts of galaxies, and we also observe that $f_{\rm dense}$ appears to anti-correlate with galactocentric radius. These correlations are often very strong for individual galaxies, appearing almost monotonic in our binned data. However the relationship between $f_{\rm dense}$ and each of these quantities still shows significant galaxy-to-galaxy scatter, with typical RMS scatter $\pm0.2$~dex. This exceeds our measurement errors and highlights additional physics still at play. In addition to uncertainties in physical parameter estimation, we highlight an important possible role for ISM structure beneath the $1{-}2$~kpc resolution of our data (i.e. beam filling factor variations). We also emphasize that the role of galactic dynamics (other than vertical force balance) remains relatively unexplored so far \citep[but see,][]{MEIDT16,MEIDT18}.

\subsubsection{Star formation efficiency of dense gas} 

At face value, the ratio of TIR-to-HCN emission traces the star formation efficiency of dense gas, SFE$_\textrm{dense} \equiv {\rm SFR}/M_{\rm dense}$. In the right panels of Figure \ref{fig:global_trends}, the lower part of Table \ref{table:rank_coef}, and the fits in this section we measure how SFE$_\textrm{dense}$ depends on environment in EMPIRE. In the main text, we again focus on stacked trends. In the Appendix, we show every individual kpc-measurement (Figures \ref{fig:sfe_rad}, \ref{fig:sfe_star}, \ref{fig:sfe_mol} and \ref{fig:sfe_pressure}).

Figure \ref{fig:global_trends} shows that SFE$_\textrm{dense} \propto\,$TIR/HCN generally increases towards large galactocentric radii, and systematically decreases towards regions of high stellar surface density, high molecular gas surface density, high molecular fraction, and high pressure. These trends all contrast with what we observed for $f_\textrm{dense}$, where the systematic behavior has the opposite sense. The clear correlation of SFE$_\textrm{dense}$ with environment shows that the observed scatter about the $L_{\rm IR}$-$L_{\rm HCN}$ scaling relation (Section \ref{sec:scaling}) reflects real correlations of SFE$_{\rm dense}$ with local environment.

\medskip

\textbf{Galactocentric radius:} SFE$_\textrm{dense}$ tends to rise with increasing galactocentric radius, but with large galaxy-to-galaxy scatter. Our best fit between $I_\textrm{TIR}/I_\textrm{HCN}$ and galactocentric radius is:

\begin{equation}
\label{eq:fit_irhcn_rad}
\log_{10} \frac{\textrm{TIR}}{\textrm{HCN}} = (2.8\pm0.3) + (0.6\pm0.2) \, \frac{r}{r_\textrm{25}},
\end{equation}

with an overall scatter of $\pm0.30$dex from galaxy to galaxy at fixed radius. We quote the fit here in terms of the TIR-to-HCN ratio, which has units of $L_\odot/(\textnormal{K km s}^{-1}\,\textnormal{pc}^2)$.

SFE$_\textrm{dense}$ increases with increasing galactocentric radius in most of our targets. But the large galaxy-to-galaxy scatter means that radius alone does not do a good job predicting the TIR-to-HCN ratio. As with $f_{\rm dense}$, the same radius in different galaxies corresponds to different physical conditions in a way that affects SFE$_\textrm{dense}$.

The increase in SFE$_\textrm{dense}$ with radius appears in both barred ($\rho \sim 0.7$) and unbarred ($\rho \sim 0.6$) galaxies. We do observe some difference in the shape of the profile between these two groups, however. For unbarred galaxies the $I_\textrm{TIR}/I_\textrm{HCN}$ profile often appears quite flat across most of the galaxy disk, with a lower value in the inner $\sim 2$\,kpc of the galaxy. In barred galaxies, the profiles appear smoother, with TIR-to-HCN steadily increasing with increasing radius.

We note that, with the resolution of the EMPIRE data, we cannot rule out the effects of galaxy dynamics in these radial trends. More specifically, barred galaxies in which bars are smaller than their corotation regions, often show pile-ups of gas in the leading edges of the bar \citep[e.g.,][]{DOWNES96,SHETH05,BEUTHER16}. This particular orbit structure creates shear motions in the molecular gas and little star formation occurs, thus lower SFE, which is typically restricted to the resonances of the bar. In addition to that, nuclear bars such as those present in NGC\,4321 \citep{SAKAMOTO95,GARCIABURILLO98} may also be responsible for this orbit structure and contribute to the lower SFE observed in barred galaxy centers.

\medskip

\textbf{Stellar surface density:} The middle right panel of Figure \ref{fig:global_trends} demonstrates an overall anti-correlation between $I_\textrm{IR}/I_\textrm{HCN}$ and stellar surface density. In fact, all individual galaxies show an anticorrelation between IR/HCN and $\Sigma_*$, and all but three galaxies show a strong anti-correlation (see Table \ref{table:rank_coef}). A good fit to our data is given by:

\begin{equation}
\label{eq:fit_irhcn_star}
\log_{10} \frac{\textrm{TIR}}{\textrm{HCN}} = (4.0\pm0.3) - (0.40\pm0.1)\, \log_{10}\, \left[\frac{\Sigma_*}{M_\odot\,\textrm{pc}^{-2}}\right].
\end{equation}

\noindent This fit holds for disk galaxies at $\sim 1{-}2$~kpc resolution over the range $\Sigma_* \sim 100{-}1000$~M$_\odot$~pc$^{-2}$. Individual binned measurements show RMS scatter of $\pm0.26$~dex about the fit. 

These trends with stellar surface density resemble those seen by \citet{USERO15}, \citet{BIGIEL16}, and \citet{GALLAGHER18}. The same conditions that make the gas denser on average also appear to drive SFE$_{\rm dense}$ to lower values.

\medskip

\textbf{Molecular gas surface density:} Above, we find higher $f_{\rm dense}$ in regions with higher $\Sigma_{\rm mol}$. Considering SFE$_{\rm dense}$, the trend reverses. We find lower $I_\textrm{IR}/I_\textrm{HCN}$ in regions of high $\Sigma_\textrm{mol}$. Similar to the case for $\Sigma_*$, the entire sample shows an overall strong anti-correlation ($\rho = -0.64$) between $\Sigma_\textrm{mol}$ and $I_\textrm{IR}/I_\textrm{HCN}$. This anti-correlation appears even stronger in many individual galaxies, again reflecting offset trends among galaxies.

Our EMPIRE data are well-described by:

\begin{equation}
\label{eq:fit_irhcn_sigco}
\log_{10} \frac{\textrm{TIR}}{\textrm{HCN}} = (3.5\pm0.7) - (0.4\pm0.1)\, \log_{10}\, \frac{\Sigma_\textrm{mol}}{M_\odot\,\textrm{pc}^{-2}}~.
\end{equation}

\noindent The fit holds over $\Sigma_{\rm mol} \sim 10{-}100$~M$_\odot$~pc$^{-2}$ at $1{-}2$~kpc resolution. The individual measurements scatter by $\pm 0.22$~dex scatter about the global fit. As in the $f_{\rm dense}$-$\Sigma_{\rm mol}$ correlation, NGC~6946 appears moderately displaced towards higher TIR-to-HCN ratios at fixed $\Sigma_{\rm mol}$. This could be related to the radius-dependent conversion factor seen in \citet{SANDSTROM13}.
%\medskip

To explain similar trends, \citet{USERO15}, \citet{BIGIEL16}, and \citet{GALLAGHER18} suggested a context-dependent role for the ``dense'' gas traced by HCN. In this scenario, which we discuss more below, the anti-correlations observed between SFE$_{\rm dense}$ and $\Sigma_*$ or $\Sigma_{\rm mol}$ may occur because the mean density of the ISM rises in high $\Sigma_*$, high $\Sigma_{\rm mol}$ regions \citep[this appears to be the case in EMPIRE and][]{GALLAGHER18b}. In this case, the HCN may trace gas at lower density than the local density needed for gas to collapse and subsequently form stars. 

\medskip

\textbf{Molecular-to-atomic gas ratio:} Given that SFE$_{\rm dense}$ anti-correlates with $\Sigma_*$ and $\Sigma_{\rm mol}$, we also expect an anti-correlation with the ratio of molecular to atomic gas,  $R_\textrm{mol}$. We observe an anti-correlation in most targets, but the scatter among galaxies is large compared to the dynamic range of the observations. 

We find a best fit relation

\begin{equation}
\label{eq:fit_irhcn_atom}
\log_{10} \frac{\textrm{TIR}}{\textrm{HCN}} = (3.1\pm0.1) - (0.30\pm0.06)\, R_\textrm{mol} ,
\end{equation}

\noindent which holds over the range $R_{\rm mol} \sim 0.3{-}10$ at $\sim 1{-}2$~kpc resolution. Compared to the trends with $\Sigma_*$, $\Sigma_{\rm mol}$, the correlation of SFE$_{\rm dense}$ shows a significantly weaker correlation coefficient of -0.38, and the data scatter about the fit with RMS $\pm 0.28$~dex. Again, most of this scatter is due to offsets among galaxies. The typical $R_{\rm mol}$ varies by more than an order of magnitude across our sample, and binning by $R_{\rm mol}$ does not appear to reveal a strong common underlying relation. $R_\textrm{mol}$ alone seems insufficient to predict the SFE$_{\rm dense}$ with high precision.

\medskip

\textbf{Dynamical equilibrium pressure:} If SFE$_{\rm dense}$, traced by TIR-to-HCN, anti-correlates with the mean ISM density, then it should anti-correlate with $P_{\rm DE}$, our environmental measure directly related to the mean midplane density. We do observe a clear anti-correlation between SFE$_{\rm dense}$ and $P_{\rm DE}$, though again we find significant galaxy-to-galaxy scatter.

A least-square minimization of our data yields:

\begin{equation}
\label{eq:fit_irhcn_press}
\log_{10} \frac{\textrm{TIR}}{\textrm{HCN}} = (4.7\pm0.4) - (0.3\pm0.1)\,\frac{P_{\rm DE}}{k_B\,\textrm{cm}^{-3}\,\textrm{K}},
\end{equation}

\noindent valid at $1{-}2$~kpc resolution over the range \\ $\log_{\rm 10} P_{\rm DE}/k_B [{\rm K~cm}^{-3}]$ $\sim 4.5{-}6.5$. Individual data show $\pm0.24$~dex scatter about the fit at fixed $P_{\rm DE}$. 

SFE$_{\rm dense}$ anti-correlates with $P_{\rm DE}$ in the expected sense, but does not offer a better predictor of SFE$_{\rm dense}$ than $\Sigma_{\rm mol}$ or $\Sigma_*$ \citep[similar to the finding by][]{GALLAGHER18}, though it shows a clearer relation than $R_{\rm mol}$ or $r_{\rm gal}$.

We note that this spread in pressures is correlated with the total IR emission: for a fixed TIR-to-HCN ratio, galaxies with higher SFR on average (see middle right panel in Figures \ref{fig:maps1}-\ref{fig:maps9}) show much larger characteristic $P_{\rm DE}$ values in their centers.

\medskip

\textbf{Star formation efficiency of dense gas:} Taking the TIR-to-HCN ratio to trace SFE$_\textrm{dense}$, we observe a systematic dependence of SFE$_{\rm dense}$ on environment across the EMPIRE sample. These have the sense that the inner, high pressure, high gas surface density regions of galaxy disks appear to be the most inefficient at forming stars out of dense molecular gas. These results also demonstrates that the scatter in the $L_{\rm IR}$-$L_{\rm HCN}$ scaling relation (Section \ref{sec:scaling}) has a physical origin. Using the scaling relations in this section, one could predict whether an EMPIRE data point would fall above or below the scaling relation, on average.

Our results agree well with previous observations of dense gas in nearby galaxy disks \citep[see][]{USERO15,CHEN15,BIGIEL16,GALLAGHER18}. As we emphasize above, EMPIRE represents the best systematic measurement to date. The combination of whole-galaxy mapping and a significant sample mean that our results can serve as a reference for the behavior of nearby disk galaxies at $\sim 1{-}2$~kpc resolution. In this sense, our observations help establish that the SFE$_{\rm dense}$ variations observed in previous studies are not restricted to the nucleus of galaxies, nor are they the result of biased sampling.

\citet{GALLAGHER18} speculate that the behavior that we see could be expected if environment affects the mean density of molecular clouds \citep[which does appear to be the case, e.g.,][]{SUN18} and star formation occurs in regions of local overdensity. In this case as one moves to regions with high mean cloud densities, e.g., high pressure regions like  galaxy centers, the gas traced by HCN represents less and less of an overdensity relative to the mean. This scenario \citep[see also][]{KRUMHOLZ07,NARAYANAN08,USERO15} would qualitatively explain our results, but raises some other issues in turn. We return to this in Section \ref{sec:discussion}).

We show that while SFE$_{\rm dense}$ anti-correlates with $\Sigma_*$, $\Sigma_{\rm mol}$, $R_\textrm{mol}$ and $P_{\rm DE}$, none of these quantities places all of the EMPIRE targets on a single scaling relation. This suggests that several of these variables need to be taken into account, or that there must be additional physics at play regulating SFE$_{\rm dense}$. We highlight the likely roles of sub-beam structure, i.e., different gas structure within our $1{-}2$~kpc beams and dynamics. \citet{QUEREJETA19} show a strong relationship between SFE$_\textrm{dense}$ and velocity dispersion in M51, and kinematics in M51 also strongly correlate with the star formation efficiency of the total molecular gas \citep[see][]{MEIDT13,LEROY17B}. Comparing EMPIRE-based SFE$_{\rm dense}$ to kinematic information will be an important next step.

Finally, we emphasize that {\it our fitted scaling relations should not be extrapolated far outside the regime where we measure them.} \citet{GAO04}, \citet{GARCIABURILLO12}, and \citet{USERO15} have all shown that the TIR-to-HCN ratio in (U)LIRGs is {\it not} heavily suppressed relative to that in disks (see Section \ref{sec:scaling}. Extrapolating our relationships to arbitrarily high $P_{\rm DE}$, $\Sigma_{\rm mol}$, or $\Sigma_*$ would thus yield incorrect results.

\begin{figure*}
	\centering
		\includegraphics[scale=0.43]{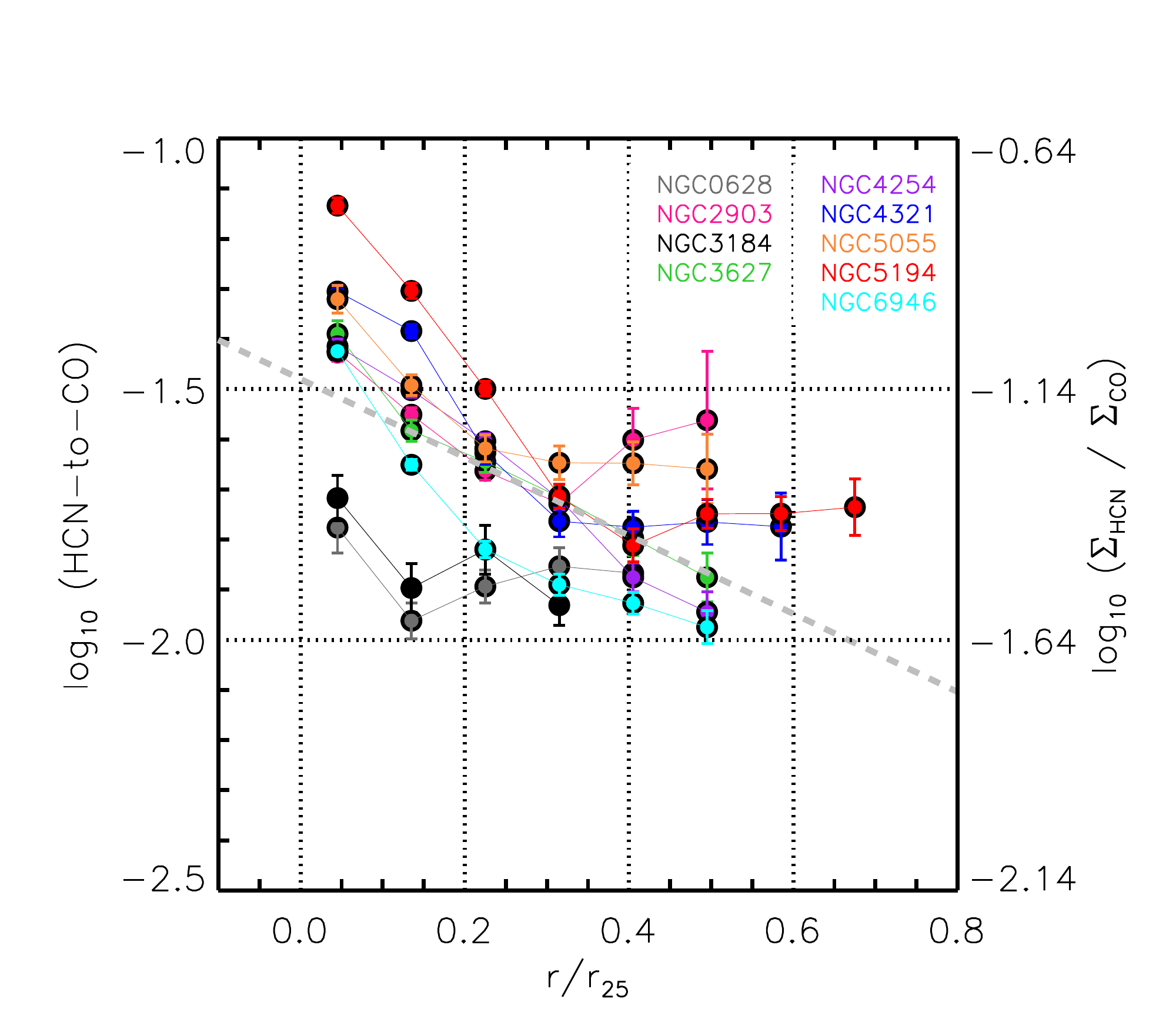}\,
		\includegraphics[scale=0.43]{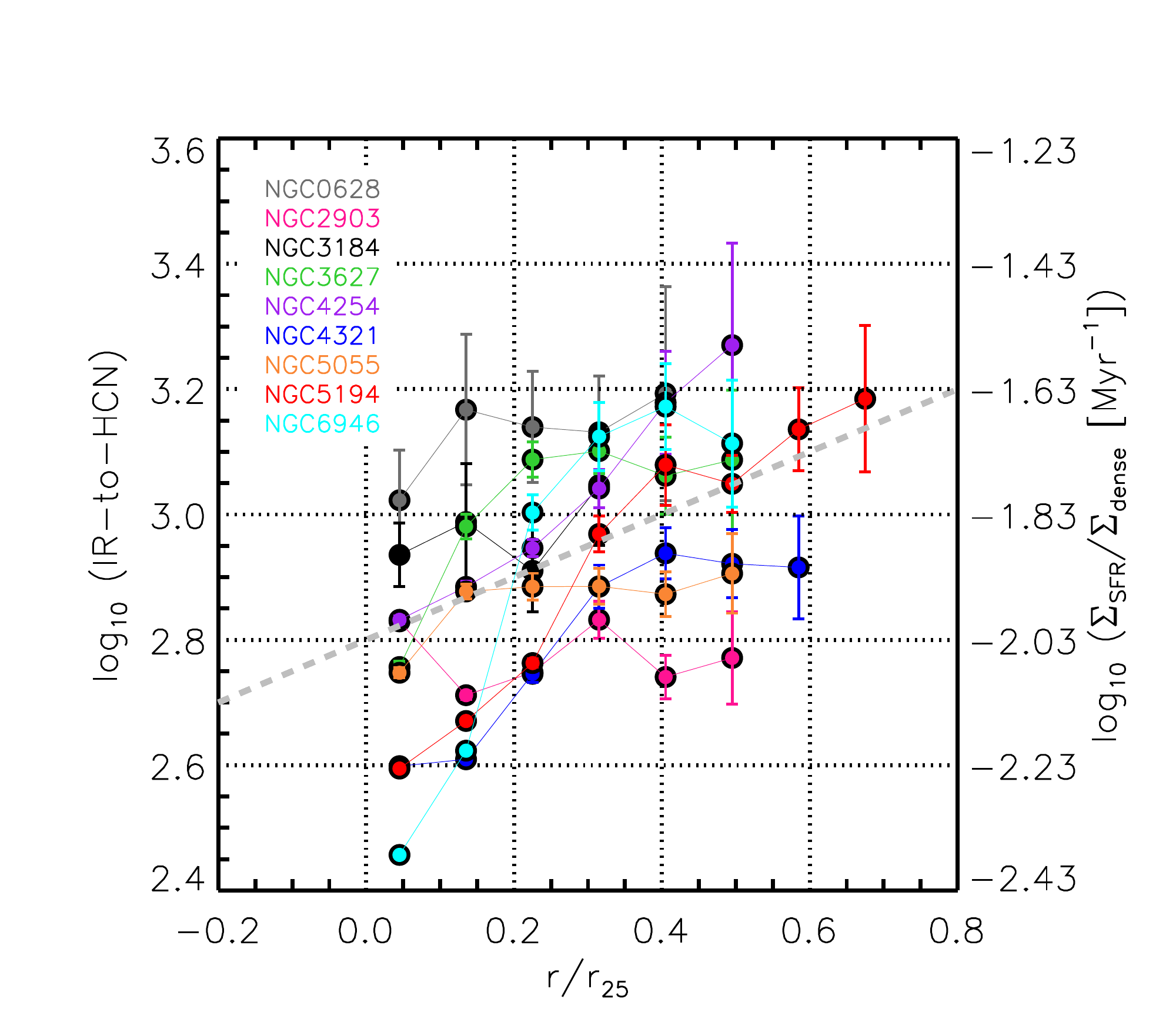}\\
		\includegraphics[scale=0.43]{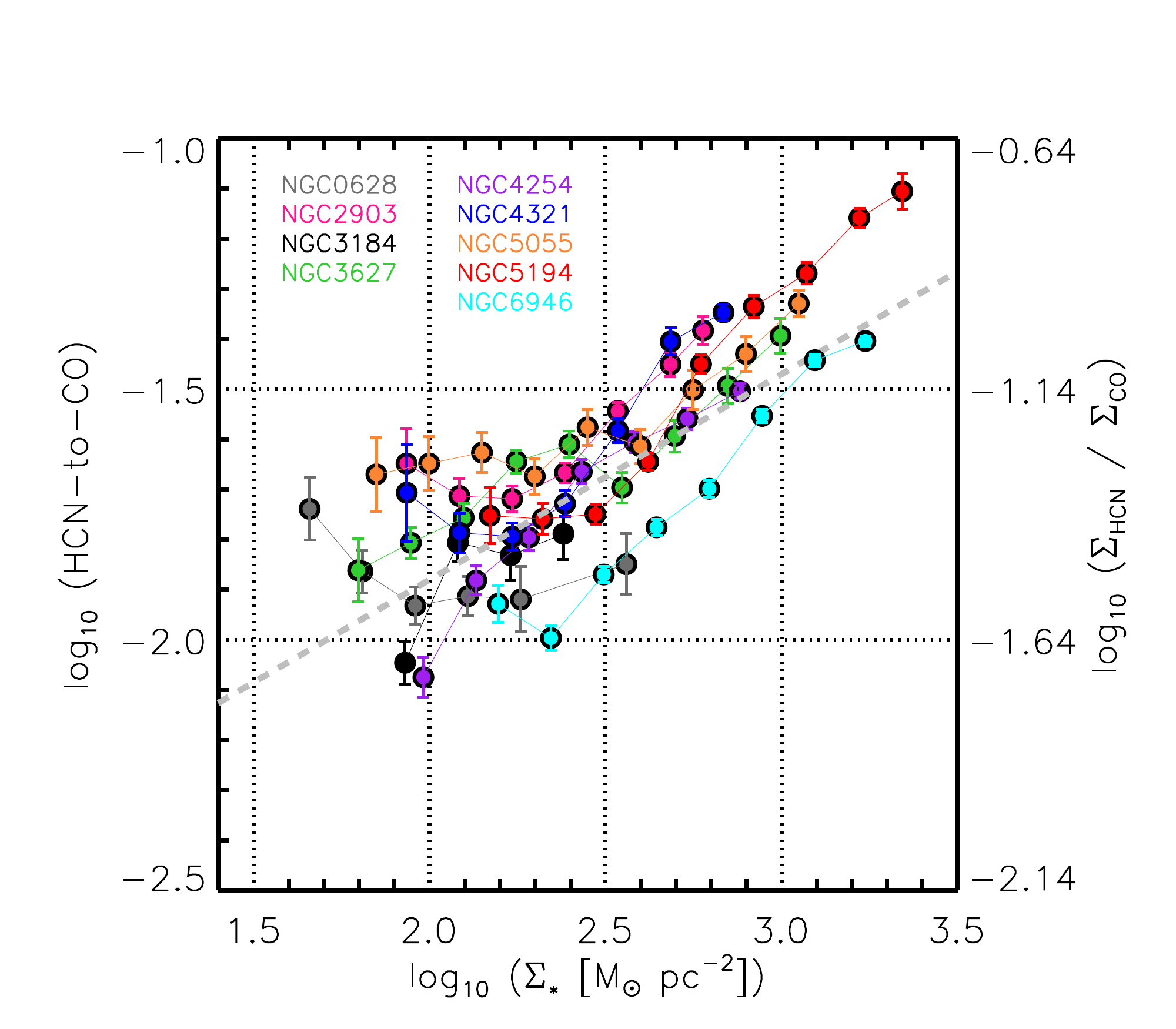}\, \includegraphics[scale=0.43]{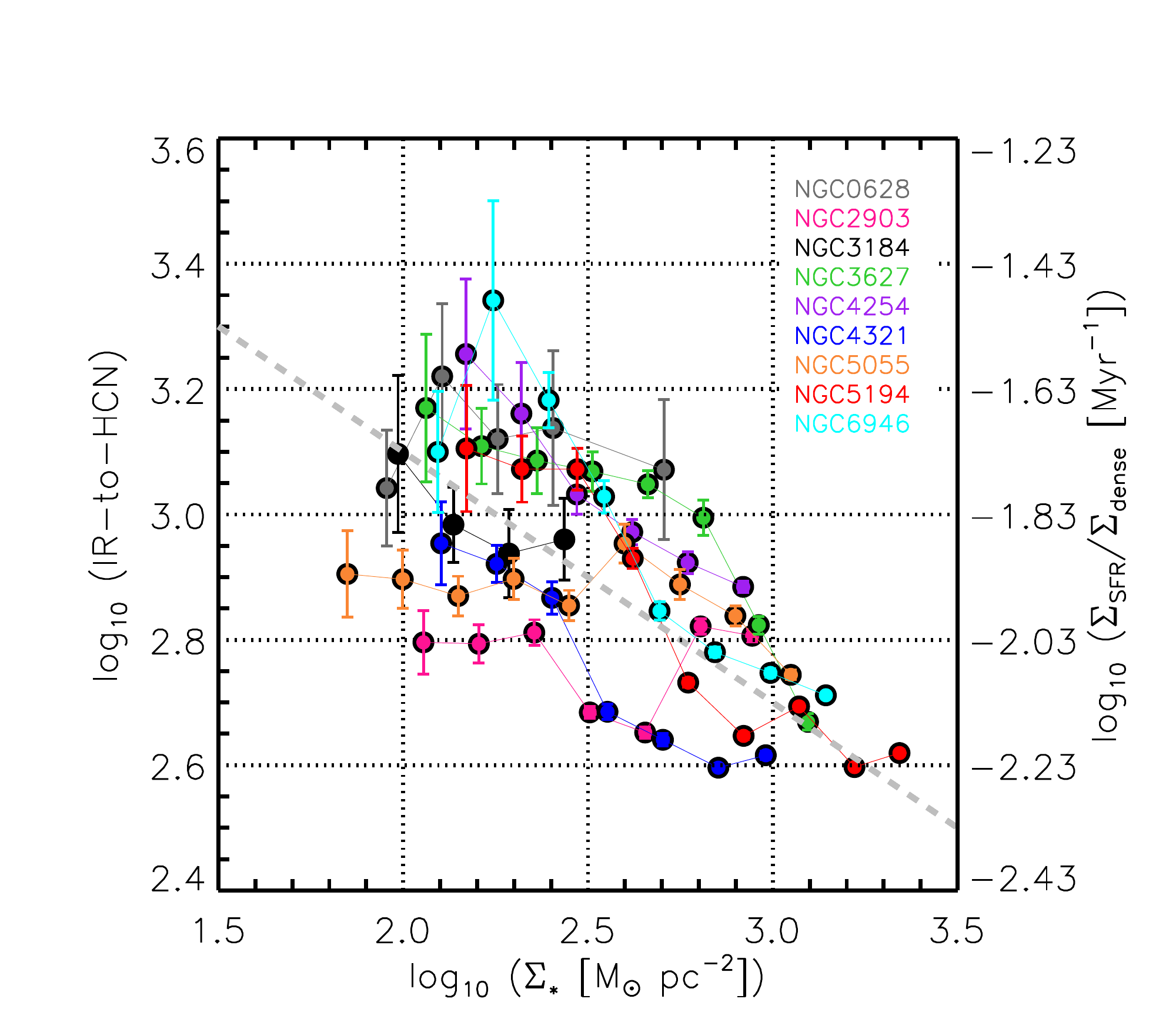}\\
		\includegraphics[scale=0.43]{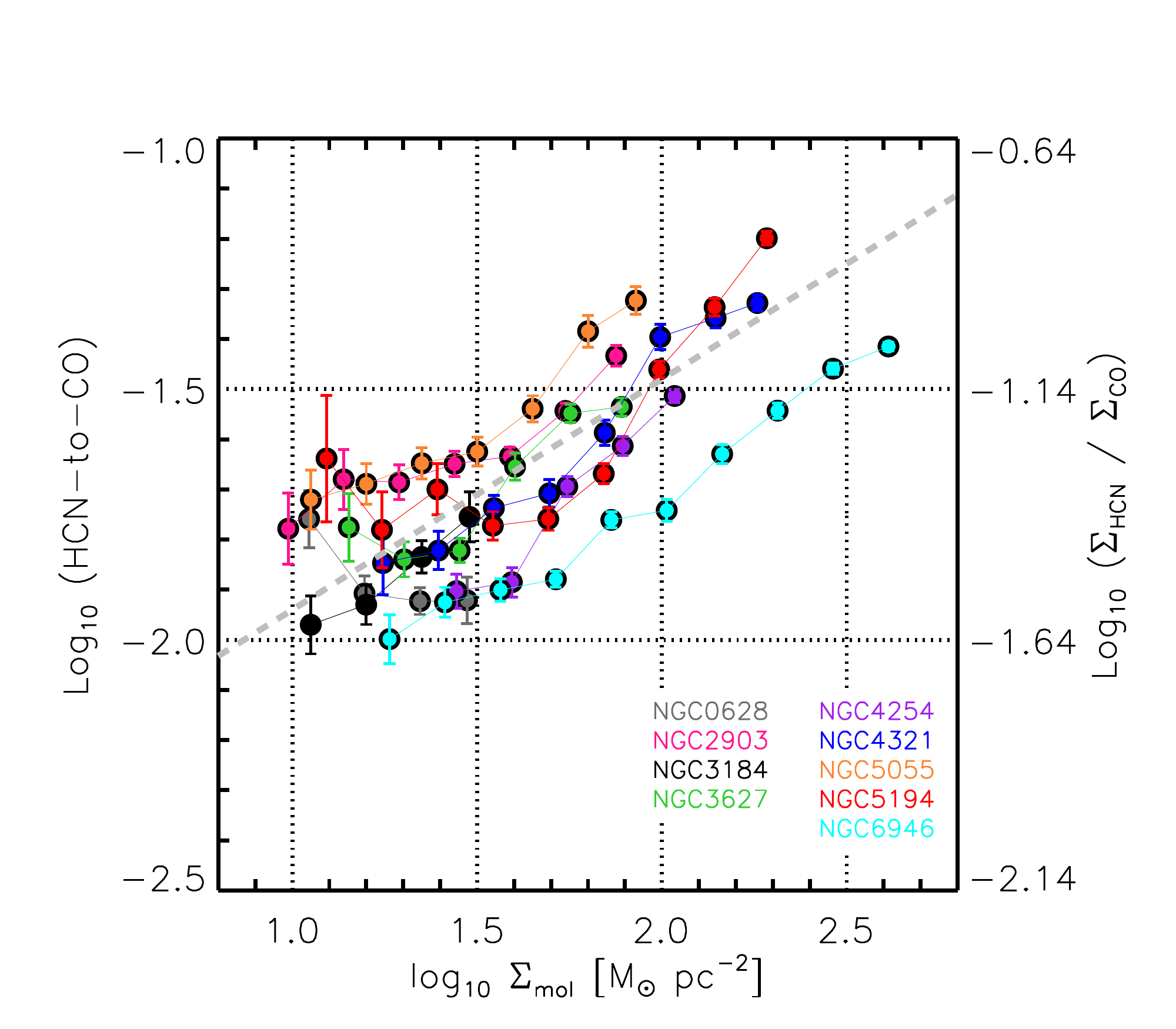}\, \includegraphics[scale=0.43]{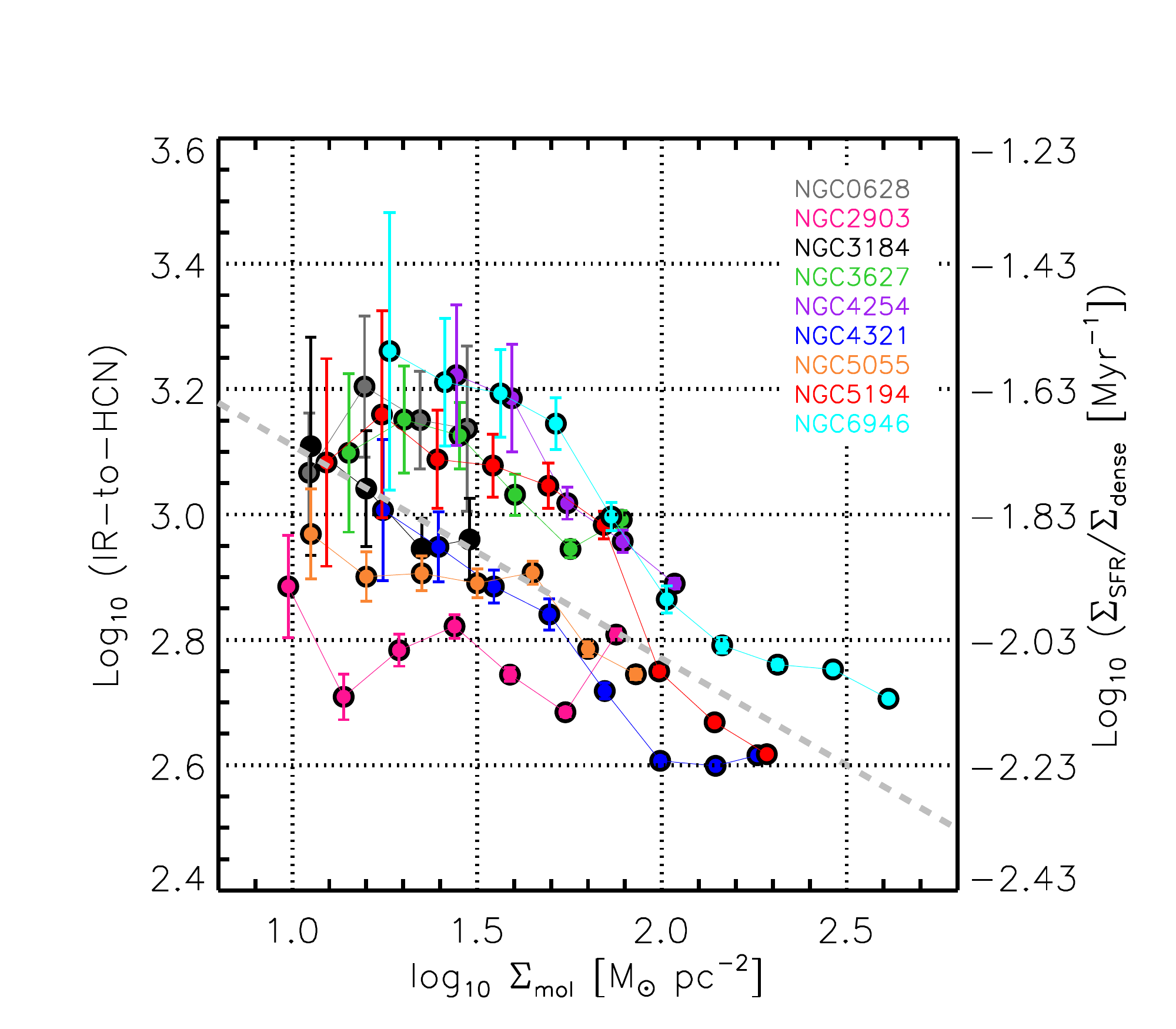}\\
		\caption{{\it Left}: HCN-to-CO ratio (left y-axis) as a proxy for $f_\textrm{dense}$ (right y-axis). {\it Right}: TIR-to-HCN ratio (left y-axis) as a proxy for SFE$_\textrm{dense}$ (right y-axis). Both quantities are plotted versus the normalized galactocentric radius ($r/r_{25}$, top), stellar surface densities (middle) and molecular gas surface densities (bottom). Individual circles show the stacked measurements with respect to each environmental parameter. The grey dashed line indicates the fits to all EMPIRE data.}
		\label{fig:global_trends}
\end{figure*}

\setcounter{figure}{13}

\begin{figure*}
	\centering
		\includegraphics[scale=0.43]{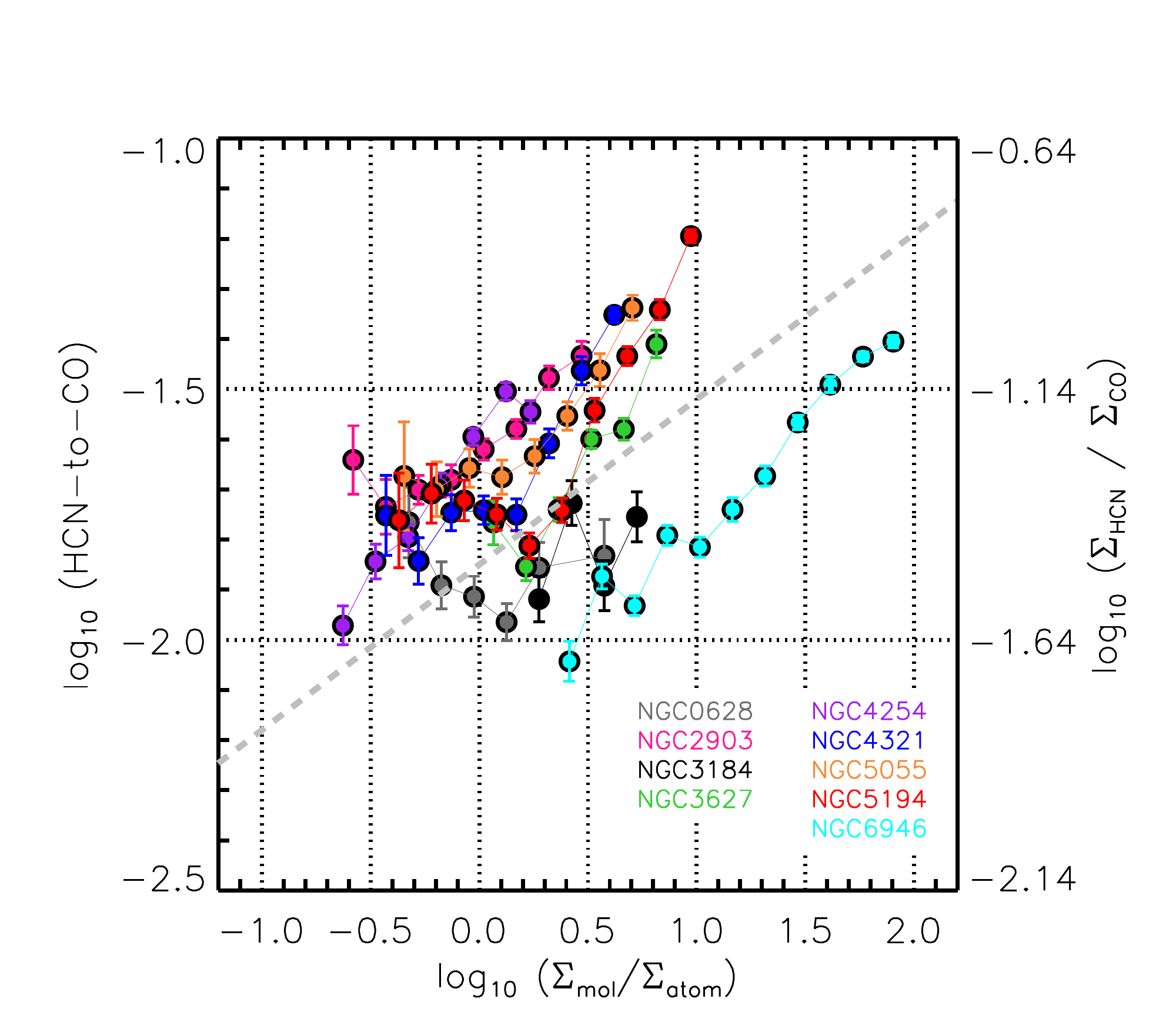}\, \includegraphics[scale=0.43]{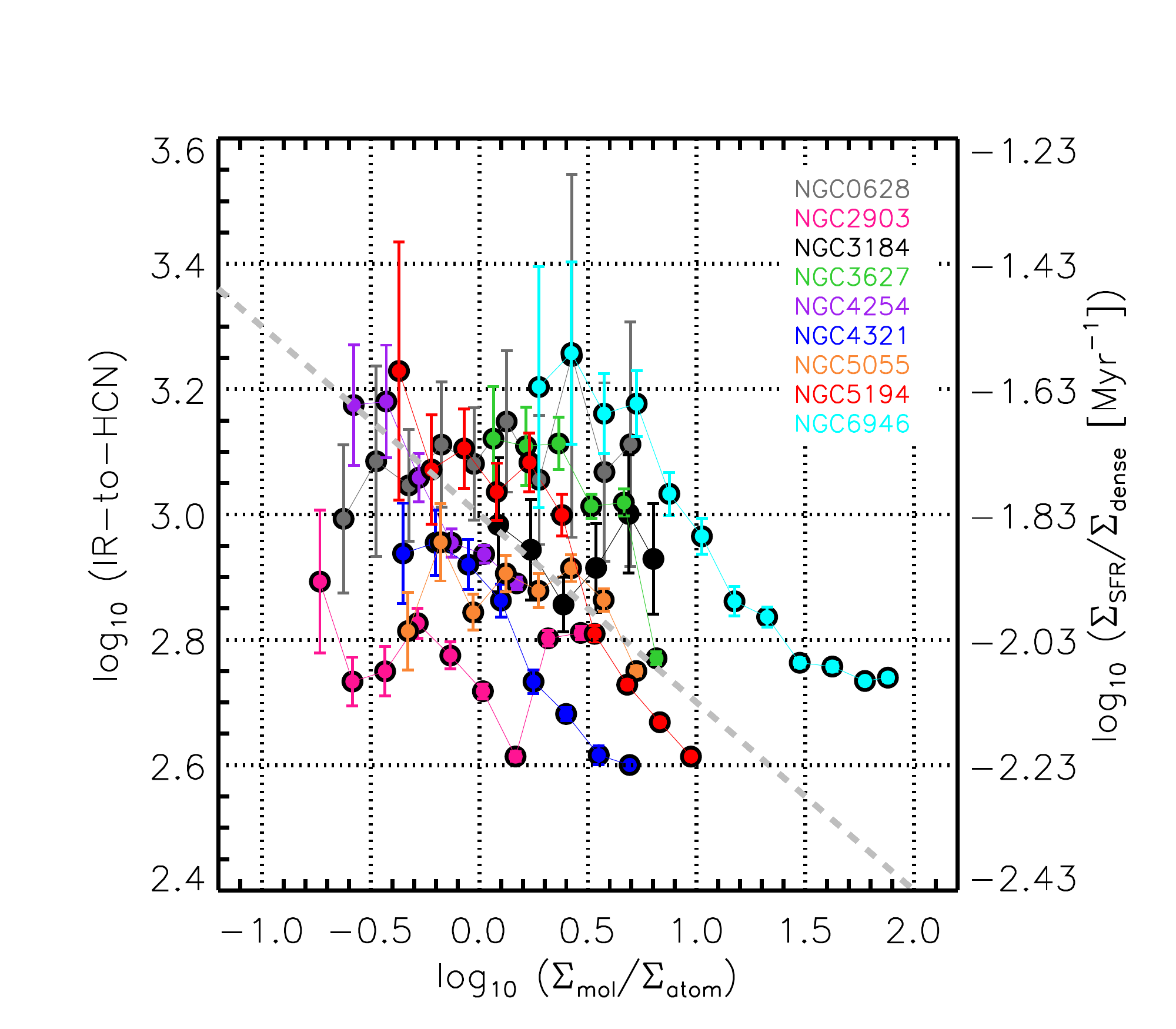}\\
		\includegraphics[scale=0.43]{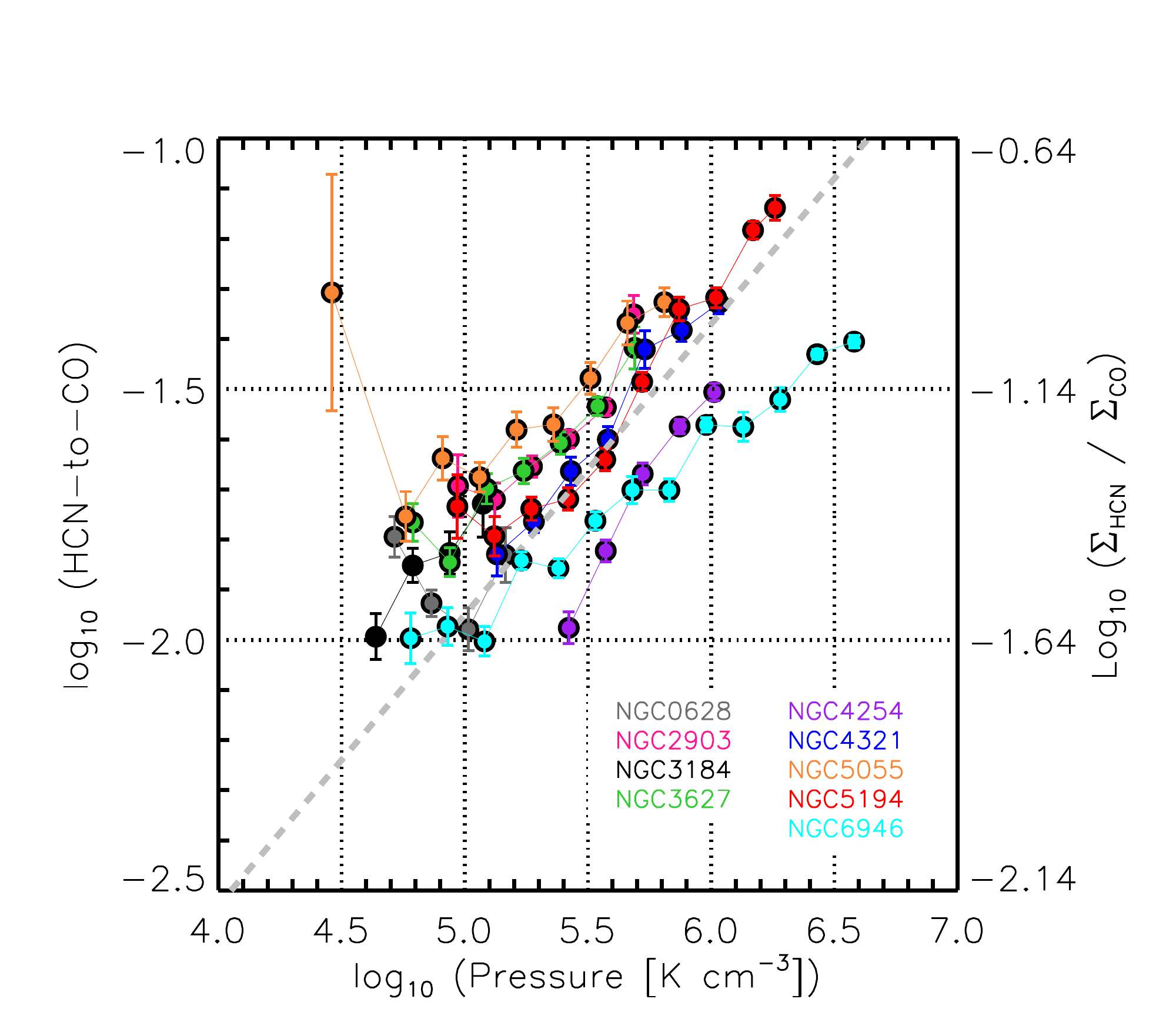}\, \includegraphics[scale=0.43]{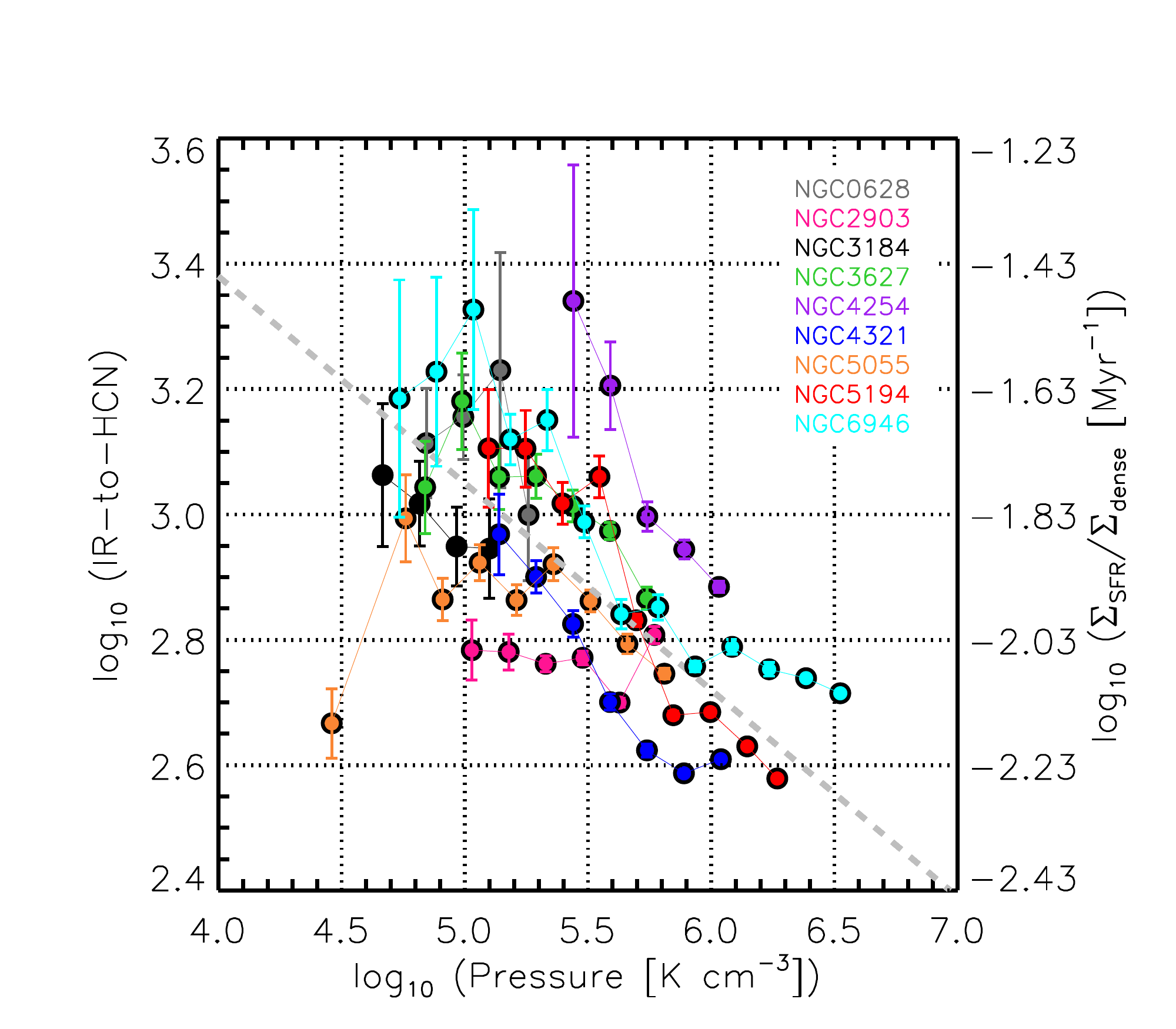}\\
		\caption{Continued. Dense gas fraction (left) and star formation efficiency of dense gas (right) versus molecular gas fraction (top) and dynamical equilibrium pressure (bottom).}
\end{figure*}

\begin{table*}
	\caption{Rank correlation coefficients of HCN/CO (proxy for dense gas fraction) and TIR/HCN (proxy for star formation efficiency of dense gas) as a function of galactoncetric radius, stellar surface density, molecular gas surface density, ratio of molecular-to-atomic gas, and local dynamical equilibrium pressure.}
	
	\label{table:rank_coef}
	\centering
	\begin{tabular}{lccccc}
		\hline\hline
		$I_\textrm{HCN}/I_\textrm{CO}$ vs & $r/r_{25}$ & $\Sigma_*$ & $\Sigma_\textrm{mol}$ & $R_\textrm{mol}$ & $P_{\rm DE}$\\
		\hline
		Galaxy & & & &\\
		\hline
		NGC~628$^a$ & 0.20 (0.06) & 0.25 (0.09) & -0.80 (0.02) & -0.10 (0.04) & 0.30 (0.05)\\
		NGC~2903$^b$ & -0.61 (0.02) & 0.75 (0.04) & 0.96 (0.03) & 0.90 (0.02) & 0.94 (0.03)\\
		NGC~3184$^b$ & -0.75 (0.02) & 0.80 (0.10) & 1.0 (0.04) & 0.70 (0.04) & 1.00 (0.04)\\ 
		NGC~3627$^b$ & -1.00 (0.01) & 0.95 (0.03) & 0.82 (0.01) & 0.96 (0.02)& 0.97 (0.01)\\
		NGC~4254$^a$ & -1.00 (0.02) & 1.00 (0.01) & 1.00 (0.01) & 0.96 (0.01)& 1.00 (0.01)\\ 
		NGC~4321$^b$ & -0.97 (0.03) & 0.83 (0.04) & 1.00 (0.01) & 0.93 (0.04)& 1.00 (0.01)\\
		NGC~5055$^a$ & -0.90 (0.04) & 0.92 (0.03) & 1.00 (0.01) & 0.91 (0.02) & 0.55 (0.03)\\
		NGC~5194$^a$ & -0.70 (0.02) & 0.98 (0.02) & 1.00 (0.02) & 0.75 (0.02)& 0.96 (0.01)\\
		NGC~6946$^b$ & -0.93 (0.02) & 0.98 (0.01) & 1.00 (0.01) & 0.98 (0.01)& 0.97 (0.01)\\
		\hline
		All data & -0.60 (0.04) & 0.80 (0.03) & 0.80 (0.02) & 0.45 (0.03) & 0.60 (0.03)\\
		\hline
		\hline
		$I_\textrm{IR}/I_\textrm{HCN}$ vs & $r/r_{25}$ & $\Sigma_*$ & $\Sigma_\textrm{mol}$ & $R_\textrm{mol}$ & $P_{\rm DE}$\\
		\hline
		NGC~628$^a$ & 0.10 (0.05) & -0.37 (0.08) & -0.40 (0.06) & 0.50 (0.09) & -0.20 (0.08)\\
		NGC~2903$^b$ & 0.04 (0.05) & -0.35 (0.10) & -0.75 (0.05) & -0.55 (0.05) & -0.94 (0.09)\\
		NGC~3184$^b$ & -0.05 (0.08) & -0.90 (0.08) & -1.00 (0.08) & -0.61 (0.08) & -1.00 (0.08)\\ 
		NGC~3627$^b$ & 0.25 (0.06) & -1.00 (0.04) & -0.82 (0.04) & -0.93 (0.04) & -0.83 (0.03)\\
		NGC~4254$^a$ & 1.00 (0.06) & -1.00 (0.04) & -1.00 (0.04) & -0.96 (0.03) & -1.00 (0.05)\\ 
		NGC~4321$^b$ & 0.42 (0.05) & -1.00 (0.02) & -0.92 (0.03) & -0.98 (0.05) & -0.31 (0.01)\\
		NGC~5055$^a$ & 0.55 (0.07) & -0.17 (0.05) & -0.75 (0.03) & -0.18 (0.02) & -0.06 (0.04)\\
		NGC~5194$^a$ & 0.98 (0.05) & -0.98 (0.03) & -0.95 (0.06) & -0.94 (0.05) & -0.97 (0.02)\\
		NGC~6946$^b$ & 0.90 (0.06) & -0.92 (0.03) & -1.00 (0.06) & -0.99 (0.03) & -0.96 (0.03)\\
		\hline
		All data & 0.35 (0.05) & -0.70 (0.04) & -0.64 (0.06) & -0.38 (0.05) & -0.50 (0.04)\\
		\hline
	\end{tabular}
	\\ \flushleft{{\bf Notes:} $(a)$ Unbarred galaxies. $(b)$ Barred galaxies. The numbers in parenthesis indicate the corresponding $p$-values.}
\end{table*}

\subsection{The relation between $f_\textrm{dense}$ and SFE$_\textrm{mol}$}
\label{sec:centers}

\begin{figure*}
	\centering
    \includegraphics[scale=0.57]{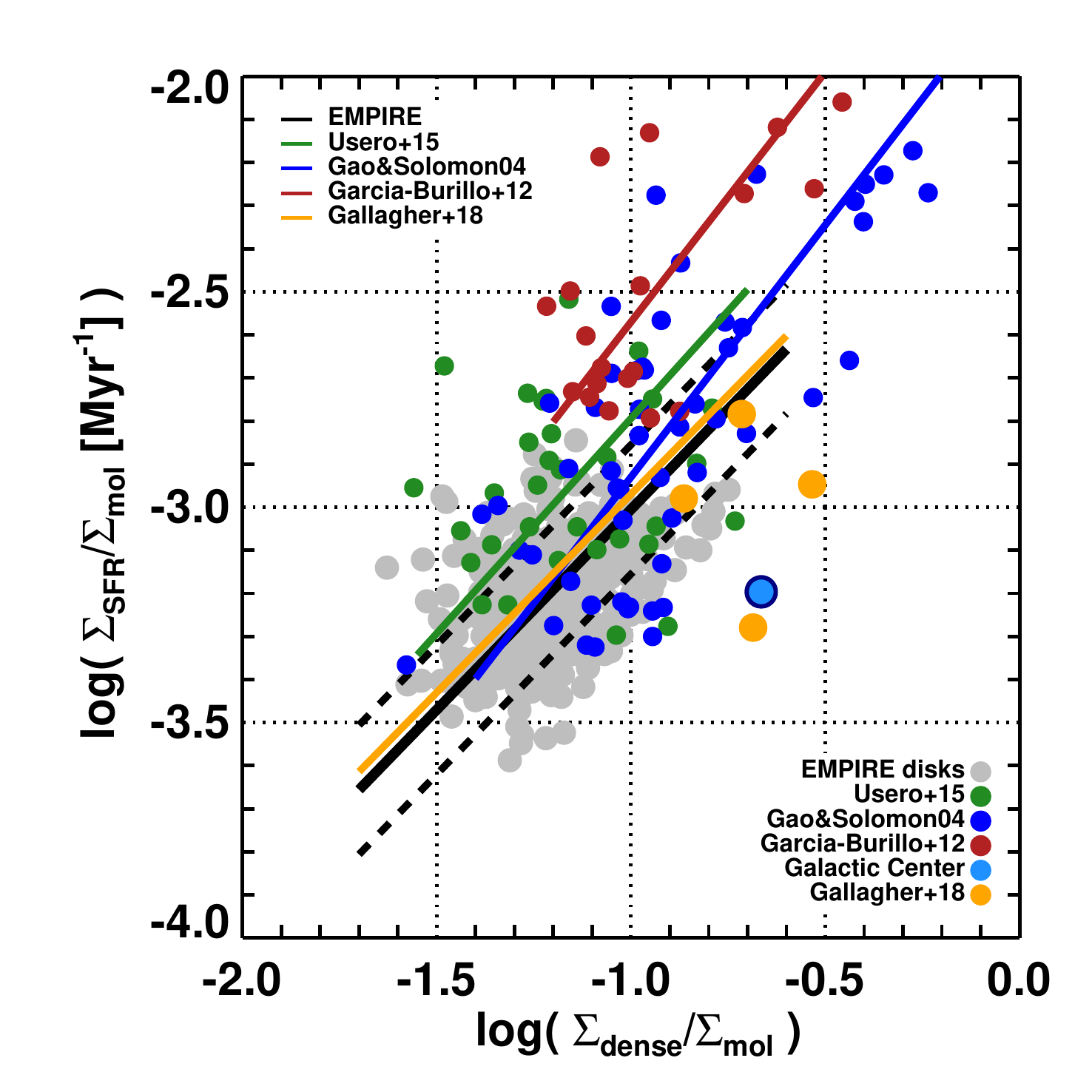}
    \includegraphics[scale=0.57]{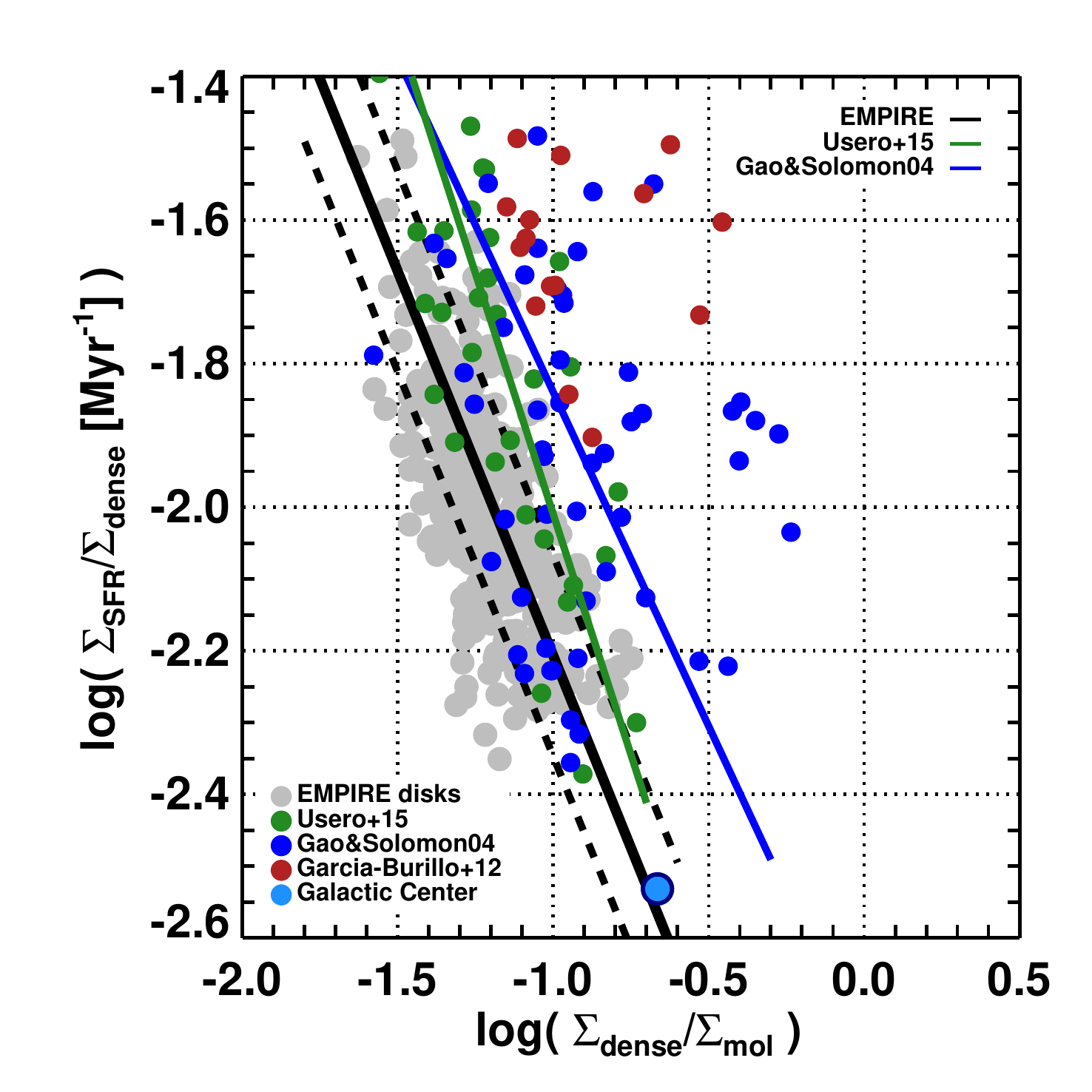}
	\caption{$\Sigma_\textrm{SFR}$-to-$\Sigma_\textrm{mol}$ ratio ({\it left}), as a proxy for the star formation efficiency of the bulk molecular gas, and the $\Sigma_\textrm{SFR}$-to-$\Sigma_\textrm{dense}$ ratio ({\it right}), as a proxy for the star formation efficiency of the dense gas, versus the $\Sigma_\textrm{dense}$-to-$\Sigma_\textrm{mol}$ ratio, a proxy for the dense gas fraction. In gray circles we display the EMPIRE disk measurements and we compare them with the samples from \citet{USERO15}, \citet{GARCIABURILLO12} and \citet{GAO04} shown as green, red and dark blue circles, respectively. We include observations for the CMZ in CO \citep{DAME01}, HCN \citep{JONES12} and TIR \citep{BARNES17} for a comparison with a well-studied extreme environment. The orange points show the galaxy centers (inner $\sim500$\,pc) from the nearby galaxies in \citet{GALLAGHER18}. The black line indicates the best fit power law to the EMPIRE measurements, while the dashed lines show the $1\sigma$ scatter about the mean. The red, blue, green and orange solid lines indicate the trends found by \cite{GARCIABURILLO12}, \cite{GAO04} (significantly driven by (U)LIRGs in their samples), \cite{USERO15} and \cite{GALLAGHER18}.}
	\label{fig:centers}
\end{figure*}

At face value, our EMPIRE results show a variable SFE$_\textrm{dense}$ as traced by the observable TIR-to-HCN ratio across and within galaxies. These results are difficult to explain within the framework of density threshold models. As noted by \citet{USERO15} and \citet{GALLAGHER18}, among others, in a density threshold model one expects variations in the star formation efficiency of the {\it total} molecular gas, SFE$_{\rm mol}$ to track $f_{\rm dense}$ with no change in SFE$_{\rm dense}$  \citep[e.g., see][]{GAO04,LADA12}.

Following \citet{GALLAGHER18}, \citet{USERO15}, and \citet{GAO04}, we test the density threshold hypothesis by measuring the strength of the  correlation between SFE$_\textrm{mol}=\Sigma_\textrm{SFR}/\Sigma_\textrm{mol}$, as traced by the TIR-to-CO ratio, and the dense gas fraction ($f_{\rm dense} = \Sigma_\textrm{dense}/\Sigma_\textrm{mol}$) indicated by HCN/CO. EMPIRE allows us to test this hypothesis across the whole area of nearby galaxies, in the process extending to lower $\Sigma_\textrm{dense}/\Sigma_\textrm{mol}$ and $\Sigma_\textrm{SFR}/\Sigma_\textrm{mol}$ than previous tests. 

The left panel in Figure \ref{fig:centers} displays $\Sigma_\textrm{SFR}/\Sigma_\textrm{mol}$ as a function of $\Sigma_\textrm{dense}/\Sigma_\textrm{mol}$. EMPIRE $>3\sigma$ disk measurements appear as gray points. For a comparison, we show the \citet{USERO15} pointed observations (green points), integrated galaxy measurements from \citet[][red points]{GARCIABURILLO12} and \citet[][dark blue points]{GAO04}, data from the Milky Way's Central Molecular Zone (CMZ)\footnote{We define the CMZ region  as a rectangle centered on $l = 0.545^{\circ}$, $b = −0.035$ with width$ = 151.0' $ and height$ = 29.8' $. This width corresponds to a linear size of $\sim500\,$pc.} \citep[light blue, data from][]{JONES12,BARNES17}, and data from other galaxy centers from \citet[][orange points]{GALLAGHER18}.

The plot clearly indicates a relationship between SFE$_{\rm mol}$ and $f_{\rm dense}$. The thick black line indicates an ordinary least squares bisector fit to the significant EMPIRE measurements (gray points). This fit has the form:

\begin{equation}
\label{eq:fit_sfemol_fdense}
    \textrm{log}_{10}\,\frac{\Sigma_\textrm{SFR}}{\Sigma_\textrm{mol}} = -2.07 + 0.93\,\textrm{log}_{10}\,\frac{\Sigma_\textrm{dense}}{\Sigma_\textrm{mol}}~,    
\end{equation}

\noindent where $\Sigma_\textrm{SFR}/\Sigma_\textrm{mol}$ has units of Myr$^{-1}$. The accompanying black dashed lines show the $\pm 1\sigma$ scatter about the relation. This relation applies at $1{-}2$~kpc resolution mainly over the range log$_{10}\,f_{\rm dense} \sim -1.5{-}-1.0$ and $\log_{10} \textrm{SFE}_{\rm mol} [{\rm Myr}^{-1}] \sim -3.5{-}-3.0$ (i.e., molecular gas depletion times of $\sim 1{-}3$~Gyr). We caution that our fit does not extend to the starburst regime included in the samples of \citet{GARCIABURILLO12} and \citet{GAO04}.

For comparison, we plot fits using the same methodology applied to samples of \citet[][all data, not only the plotted galaxy centers]{GALLAGHER18}, \citet{USERO15}, \citet{GARCIABURILLO12}, and \citet{GAO04}, shown as orange, green, red and blue lines, respectively. In Table \ref{table:sfedense} we report the fits, rank correlation coefficients and its significance relating $\Sigma_\textrm{SFR}/\Sigma_\textrm{mol}$ and $\Sigma_\textrm{dense}/\Sigma_\textrm{mol}$ in each sample. We also note the scatter in SFE$_{\rm dense}$ for each sample in parentheses.

All the datasets in Figure \ref{fig:centers} show some correlation between SFE$_{\rm mol}$ and $\Sigma_\textrm{dense}/\Sigma_\textrm{mol}$. But the choice of data sets can significantly alter the best fit scaling relation \citep[a conclusion also emphasized by][]{GALLAGHER18}. Our best fit to EMPIRE appears almost identical to our fit to the \citet{GALLAGHER18} measurements. Both appear similar, but modestly offset from the fit to the \citet{USERO15} points. The \citet{USERO15} points show systematically higher SFE$_{\rm mol}$, perhaps consistent with the selection of bright disk regions for that sample. Similarly, our fit to the \citet{GAO04} data intersects the EMPIRE data but shows a steeper slope, reflecting the high SFE$_{\rm mol}$, high $f_{\rm dense}$ points that make up most of their sample. Meanwhile, the \citet{GARCIABURILLO12} relation appears displaced from ours. Their sample covers starbursts with higher $f_{\rm dense}$ and higher SFE$_{\rm dense}$ than EMPIRE, so a direct comparison requires extrapolations. 
 
These discrepancies are also reflected in the median SFE$_\textrm{dense}$ for each sample (see Table \ref{table:sfedense}). The EMPIRE median ratio, log$_{10} \Sigma_\textrm{SFR}/\Sigma_\textrm{dense}\,[{\rm Myr}^{-1}]=1.05\times10^{-2}$ is similar to those found by \citet{GALLAGHER18} ($1.10\times10^{-2}$) and slightly lower than that found by \citet{USERO15} ($1.80\times10^{-2}$). EMPIRE shows notably lower scatter than both the \citet{GALLAGHER18} and \citet{USERO15} data ($\sim 0.3$~dex). This likely reflects that those two studies emphasized a range of environments by focusing on the disk-center contrast or selecting a few bright pointings per galaxy. We suggest to take the EMPIRE $0.2$~dex as indicative of the true scatter of SFE$_{\rm dense}$ treating all (detected) $\sim 1{-}2$~kpc points equally in disk galaxies. The median EMPIRE SFE$_{\rm dense}$ also resembles the value found by \citet{GAO04} ($1.31\times10^{-2}$). Their scatter $\sim 0.25$~dex reflects galaxy-to-galaxy variations. Our SFE$_{\rm dense}$ appears significantly lower than that found by \citet{GARCIABURILLO12} $2.45\times10^{-2}$. 

This scatter, $\sim 0.2{-}0.3$~dex depending on the sample, is significant when compared to the dynamic range in either $f_{\rm dense}$ or SFE$_{\rm mol}$. More, we show above that the scatter has a physical origin. Figure \ref{fig:centers} thus offers at best qualified support for a threshold model: overall, higher $f_{\rm dense}$ does correspond to higher SFE$_{\rm mol}$, but only with an RMS accuracy of $0.2{-}0.3$~dex, and the scatter about the relation is physical. This suggests that the dense gas fraction as calculated from the observable $I_\textrm{HCN}$-to-$I_\textrm{CO}$ ratio may not be an accurate predictor of $\Sigma_\textrm{SFR}/\Sigma_\textrm{mol}$ ($\propto$ SFE$_\textrm{mol}$) across all systems.

The left panel of Figure \ref{fig:centers} and the right panel of Figure \ref{fig:VULTI} clearly display the suppression of SFE$_\textrm{dense}$ in galaxy centers (see Section \ref{sec:background}), also seen in Figure \ref{fig:all_hcn}. To see this, one can contrast the central datapoints from the \citet{GALLAGHER18} ALMA sample (orange points) and the Milky Way CMZ (light blue point) with the EMPIRE scaling relation. The Milky Way's CMZ and the \citet{GALLAGHER18} central points appear inefficient at forming stars relative to their dense gas content \citep[e.g.,][]{JONES12,LONGMORE13,BARNES17,MILLS17}. The reason for the low SFE$_{\rm dense}$ in the CMZ remains under debate, with dynamical explanations among the most popular (e.g., large velocity fields, turbulence or large scale ``breathing'' modes, see \citet{BENINCASA16,BATTERSBY17,KAUFFMANN17}). Similar processes seem likely to be at play in the central regions of these other nearby galaxy disks, and perhaps to operate at a lower level to create the scatter in the EMPIRE data.

\subsubsection{Scatter in SFE$_{\rm dense}$ and SFE$_{\rm mol}$}

\citet{VUTI16} compared the SFR per mass of molecular gas (SFE$_\textrm{mol}$), and the SFR per mass of dense gas (SFE$_\textrm{dense}$) to the molecular gas mass and dense gas mass (which they call ``aggregate mass,'' in the case of external galaxies). They studied a sample of molecular clouds from high-mass star forming regions in the Galactic Plane and compared their results to other nearby clouds from \citet{EVANS14}, M51 \citep{CHEN15}, and a sample of unresolved starburst galaxies \citep[][]{LIU15}. \citet{VUTI16} concluded that the mass of dense gas appears to be a better predictor of SFE than the total molecular gas mass across all environments. Treating the scatter in the SFE as a figure of merit, they found that dense gas predicted the SFE with roughly one third of the scatter found when using all molecular gas mass. 

In Figure \ref{fig:VULTI} we replicate the calculation of \citet{VUTI16}. We plot SFE$_\textrm{mol}$ as a function of molecular gas mass in the left panel and SFE$_\textrm{dense}$ as a function of dense gas mass in the right panel. In addition to the measurements by \citet{VUTI16}, we show the EMPIRE disk measurements (gray), the samples of resolved and unresolved extragalactic systems used in Figure \ref{fig:centers}, and measurements from the Milky Way's CMZ.

If we follow \citet{VUTI16} and treat the scatter in SFE$_\textrm{mol}$ or SFE$_\textrm{dense}$ as the figure of merit, then we reach similar conclusions to that paper. We find a mean SFE$_\textrm{dense}$ of $-1.97\pm0.22$, which is very similar to the average value found in \citet{VUTI16}. The scatter we find in SFE$_\textrm{dense}$ ($\pm$0.22dex) is smaller than what we find for SFE$_\textrm{mol}$ ($\pm$0.31dex). This also closely resembles the original arguments made by \citet{GAO04} regarding HCN and CO. In the most basic terms, HCN emission does appear to represent a more basic predictor of the star formation rate than CO. However, note that the SFR-CO relation is non-linear for starbursts galaxies and (U)LIRGs at $\sim$kpc scales, as observed by 
\citet{GAO04,GARCIABURILLO12,USERO15}. This non-linearity contributes to the higher spread of points observed in the SFE-$M_\textrm{mol}$ 
relation in the left panel of Figure \ref{fig:VULTI}. If a linear relation is required, and the absolute scatter in SFE$_{\rm mol}$ or SFE$_{\rm dense}$ is treated as the figure of merit, then the arguments of \citet{VUTI16} hold. But if a non-linear relation is adopted allowed for CO, then the situation becomes more nuanced.

We do caution that the absolute molecular and dense gas masses for parts of galaxies plotted in Figure \ref{fig:VULTI} have limited physical meaning. The integrated mass in one of our EMPIRE measurements depends on a number of quantities (e.g., inclination, distance) in addition to the physical properties of that part of the galaxy.

\subsubsection{Scatter Within Individual EMPIRE Galaxies}

\begin{figure*}
	\centering
    \includegraphics[scale=0.57]{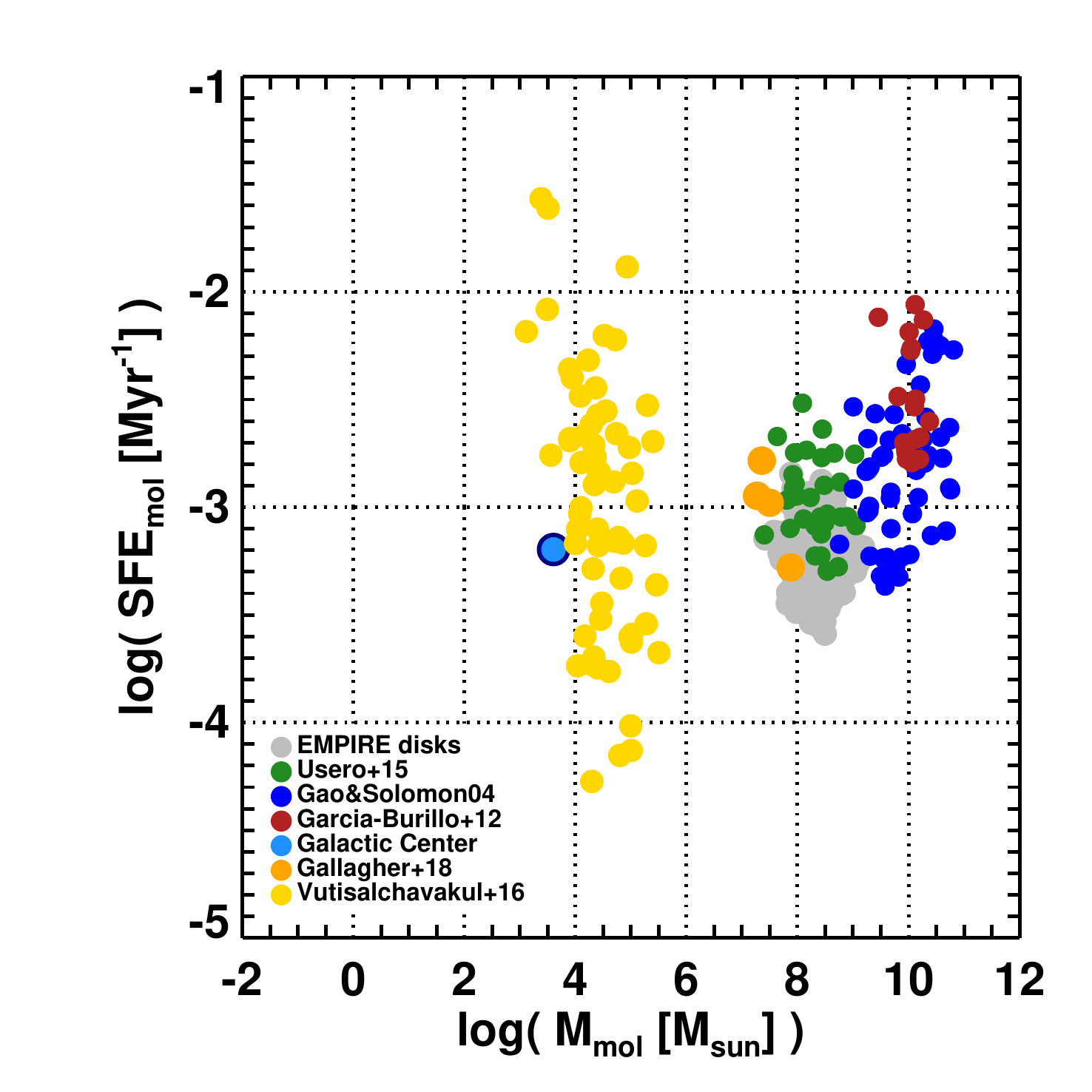}
    \includegraphics[scale=0.57]{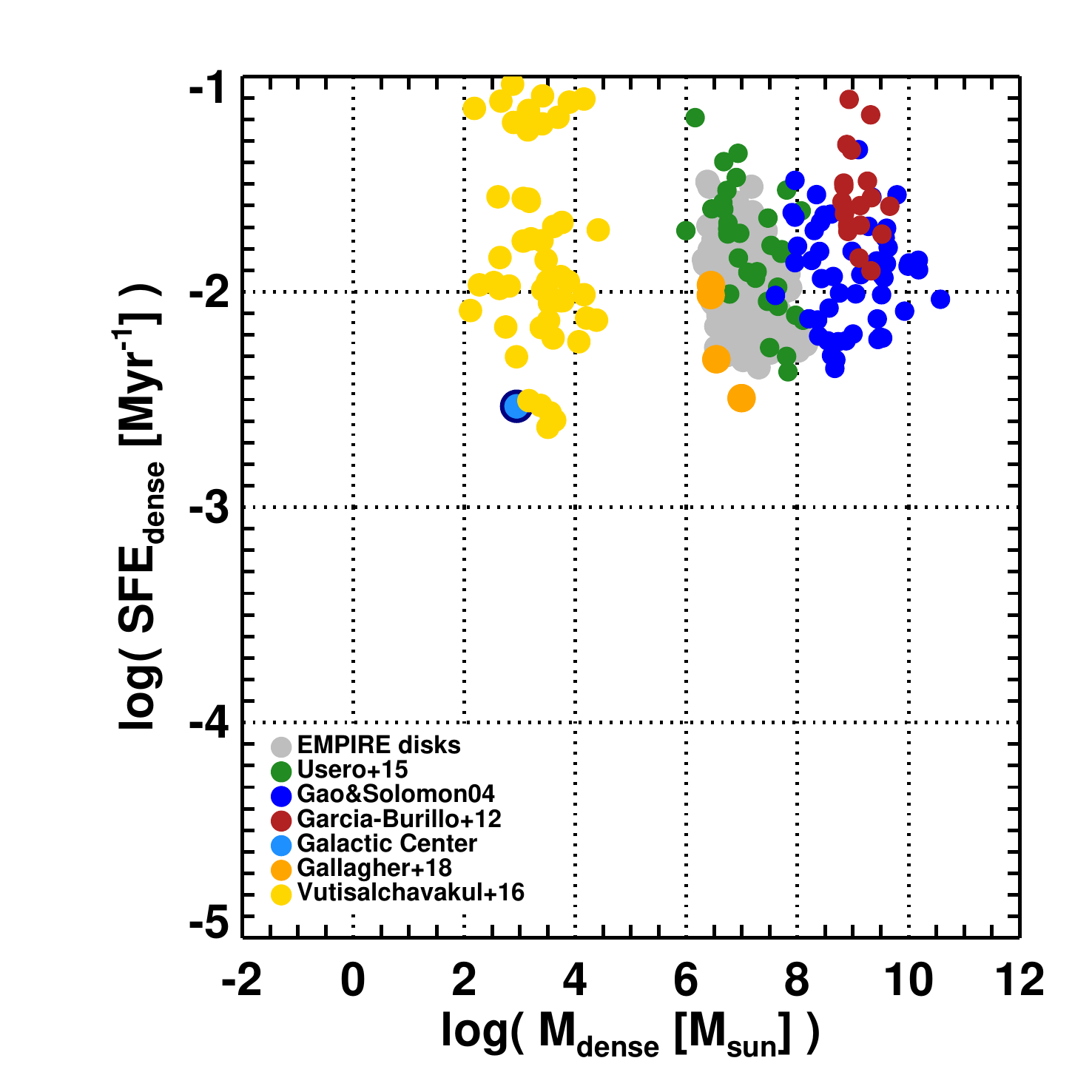}
	\caption{SFR per mass of molecular gas ({\it left}), as a function of the mass of molecular gas, and SFR per mass of dense gas ({\it right}), as a function of the mass of dense gas. In gray circles we display the EMPIRE disk measurements and we compare them with the extragalactic samples from \citet{USERO15}, \citet{GARCIABURILLO12}, \citet{GAO04} and \citet{GALLAGHER18} (same as Figure \ref{fig:centers}), and included the data from \citet{VUTI16} for a comparison with Galactic molecular clouds.}
	\label{fig:VULTI}
\end{figure*}

As discussed and seen in the Figures in Section \ref{sec:fdense_sfe}, much of the dispersion in SFE$_{\rm dense}$ and $f_{\rm dense}$ that we find EMPIRE appears as offsets among galaxies in the scaling relations. In Figure \ref{fig:VULTI} the scatter in the gray points mixes both galaxy-to-galaxy offsets and the intrinsic scatter within individual galaxies. 

To better quantify the relative importance of these contributions, we calculated the 1-$\sigma$ dispersion in SFE$_\textrm{dense}$ for each individual galaxy ($\sigma-\textrm{SFE}_\textrm{dense}$). We find an average of $\sigma-\textrm{SFE}_\textrm{dense}=0.12\pm0.02$ dex in our sample. We caution that this value includes only regions with significant detections and so suffers from some bias. Taken at face value, this scatter is comparable to the lowest values found for SFE$_{\rm mol}$ in individual galaxies \citep[e.g., see Figure 12 in][]{LEROY13}. This again highlights a tighter local correlation between HCN and SFR than CO and SFR, though a rigorous statistical analysis will require either careful statistical modeling or data with individually higher S/N than EMPIRE. Again, treating the scatter in the simplest terms, we can subtract this local scatter in quadrature from the global $\pm$0.22dex scatter found for the entirety of the EMPIRE sample (Table \ref{table:vuti}). The result is that $\sim 0.18$~dex of the point-by-point scatter in the well-detected EMPIRE regions is due to galaxy-to-galaxy variations, while $\sim 0.12$~dex comes from intra-galaxy variations.

\begin{table}
	\caption{SFE$_\textrm{mol}$ vs. $\Sigma_\textrm{dense}/\Sigma_\textrm{mol}$}
	
	\label{table:sfedense}
	\centering
	\begin{tabular}{lcc}
		\hline\hline
		Dataset & $\rho$ & log$_{10}\,\Sigma_\textrm{SFR}/\Sigma_\textrm{dense}$\\
		\hline
		EMPIRE (this work) & 0.33 (0.02) & -1.98 ($\pm$0.20)\\
		\citet{GALLAGHER18} & 0.13 (0.003) & -1.96 ($\pm$0.33)\\
		\citet{USERO15} & 0.42 (0.005) & -1.73 ($\pm$0.29)\\
		\citet{GARCIABURILLO12} & 0.35 (0.07) & -1.62 ($\pm$0.20)\\
		\citet{GAO04} & 0.65 (0.001) & -1.88 ($\pm$0.25)\\
		\hline
	\end{tabular}
	\\ \flushleft{{\bf Notes:} Rank correlation coeffients, $\rho$, and $p$-value in parentheses. We quote the median of the logarithm of the ratio $\Sigma_\textrm{SFR}/\Sigma_\textrm{dense}$, as well as its $1\sigma$ RMS scatter in units of log$_{10}$\,(Myr$^{-1}$).}
\end{table}

\begin{table}
	\caption{Mean SFE$_\textrm{mol}$ and mean SFE$_\textrm{dense}$.}
	
	\label{table:vuti}
	\centering
	\begin{tabular}{lcc}
		\hline\hline
		Average & \citet{VUTI16} & This paper\\
		\hline
		SFE$_\textrm{mol}$ & -2.83 ($\pm$0.42) & -3.09 ($\pm$0.31) \\
		SFE$_\textrm{dense}$ & -1.82 ($\pm$0.19) & -1.97 ($\pm$0.22) \\
		\hline
	\end{tabular}
	\\ \flushleft{{\bf Notes:} We quote the mean of the logarithm of SFE$_\textrm{mol}$ and SFE$_\textrm{dense}$, as well as its $1\sigma$ RMS scatter (in parenthesis) in units of log$_{10}$\,(Myr$^{-1}$).}
\end{table}

\section{Discussion}
\label{sec:discussion}

EMPIRE reveals a systematic dependence of $I_\textrm{HCN}/I_\textrm{CO}$ and $I_\textrm{TIR}/I_\textrm{HCN}$ on local conditions. These variations appear in all EMPIRE targets, with magnitude $\sim 0.2$~dex up to one order of magnitude, depending on the trend in question. At face value, $I_\textrm{HCN}/I_\textrm{CO}$ traces the fraction of dense gas, while $I_\textrm{TIR}/I_\textrm{HCN}$ traces the star formation rate per unit dense gas mass. Both interpretations have important caveats, however. In this section we discuss the implications of our observations in the context of galactic star formation and then lay out some key caveats regarding the translation of observed to physical quantities.

\subsection{The gas density distribution depends on environment}
\label{sec:disk_density}

Our observations of $I_\textrm{HCN}/I_\textrm{CO}$ indicate that the density distribution in molecular gas depends on local environment and changes across galaxy disks. Given the large difference in effective critical density between the two species, the ratio HCN-to-CO will certainly be sensitive to density variations, although there are important subtleties (Section \ref{sec:hcn_emission}).

Two important pieces of evidence support the idea that HCN-to-CO traces density variations. First, we see qualitative agreement among different dense gas tracers (J. Puschnig et al. in preparation will present a quantitative comparison). In both this paper and \citet{GALLAGHER18}, the radial profiles of HCO$^+$ agree well with those of HCN. To a lesser degree, the profiles of HNC (here) and CS \citep{GALLAGHER18} also show the same shape, though these are limited by signal-to-noise and lack of short spacing correction for  \citet{GALLAGHER18}. Qualitative agreement among the high critical density lines of species not chemically coupled suggests that variations of the HCN abundance are not the main driver for our results. Second-order variations in excitation and chemical abundances certainly remain important topics, however (Section \ref{sec:hcn_emission}).

Second, \citet{GALLAGHER18b} showed, using EMPIRE and ALMA data, that the HCN-to-CO ratio measured at $\sim$\,kpc scales correlates, on average, with the mass-weighted average of the 120~pc resolution molecular gas surface density, traced by CO~(2-1) emission, inside the beam. That is, changes in the HCN-to-CO ratio correlate with changes in the apparent surface density of molecular clouds. Regions with high HCN-to-CO also show high surface brightness CO emission and apparently dense clouds at $120$~pc (FWHM) resolution.

These arguments give us good reason to expect that HCN-to-CO traces the gas density distribution to first order. We have phrased the associated physical quantity as $f_{\rm dense}$, but we also expect that HCN-to-CO traces the mass-weighted mean density. For any somewhat universal gas density distribution, e.g., the lognormal density distribution expected for isothermal turbulence \citep[e.g.,][]{VAZQUEZSEMADENI94,PADOAN02} or a power law, the two properties will correlate. The \citet{GALLAGHER18b} results suggest a close association.

Our results indicate that the gas density PDF changes across galaxy disks. The sense of the variations is that regions with more gas and deeper potential wells (as traced by $\Sigma_*$) on average, also show denser gas on small scales.

Following \citet{HELFER97} and \citet{GALLAGHER18}, one useful way to express this dependence is that $f_{\rm dense}$ appears to correlate with the dynamical equilibrium pressure, $P_{\rm DE}$, estimated from hydrostatic equilibrium (see \ref{sec:data_pressure}). The $f_{\rm dense}$-$P_{\rm DE}$ correlation is not the tightest one that we observe, but it has a solid physical underpinning: when we look at parts of a disk with higher mean pressure, we find higher density gas. We do still observe significant galaxy-to-galaxy scatter in each scaling relation. We suggest that galactic dynamics and ISM structure below the scale of our beam (e.g., flows along bars, spiral arms, etc.) represent the natural next environmental factors to consider.

Our results relate directly to the evolving literature relating molecular cloud properties to local environment. If the sub-beam gas density distribution, traced by the HCN-to-CO ratio, reflects environment, then the gas density at intermediate scales, traced by the properties of giant molecular clouds, likely does as well. Molecular clouds in high pressure environments should have higher mean densities, and this should relate to their internal gas density distribution, traced by our spectroscopic measurements.

Recent work on this topic does suggest that molecular cloud masses and surface densities correlate with environment  \citep[e.g.][A. Schruba et al. submitted, J. Sun et al. in preparation]{HUGHES13,COLOMBO14B,LEROY16,SUN18}. The sense of the observed correlations agrees with what we find here: higher mass galaxies and the centers of galaxies host more massive, higher surface density clouds. The internal pressure of molecular clouds appears to correlate with the mean pressure in the environment \citep{HUGHES13}. The observed CO line widths within molecular clouds also increase in higher pressure environments \cite{SUN18}. These observed larger line widths are directly related to larger Mach numbers, a measure of the intracloud turbulence. Thus, in turbulent models for star formation, this would also lead to more dense gas because the Mach number drives the width of the density distribution \citep[e.g., see][]{PADOAN02}.

One next major step on this topic will involve detailed comparison of molecular cloud structure to spectroscopic observations like EMPIRE. \citet{GALLAGHER18b} take an important first step here, showing that cloud-scale gas properties do correlate with our EMPIRE HCN-to-CO ratios.

Another key next step will be to constrain the shape of the density distribution using the full suite of molecular line data available from EMPIRE and other surveys. We mainly focus on HCN-to-CO. The combination of all EMPIRE lines allows the prospect to measure the relative amounts of low, intermediate, and high density gas, though abundance variations remain a key concern \citep[e.g., see][]{LEROY17}. This work is ongoing in EMPIRE and will be presented in J. Puschnig et al. (in preparation).

Finally, these results have the prospect to inform and test turbulent models of star formation \citep[e.g.,][]{PADOAN02,KRUMHOLZ07,FEDERRATH12}. In these models, the mean density, virial parameter, Mach number, and other properties of clouds affect the gas density PDF and star formation in the cloud. Coupling these models, our measurements of density and star formation to cloud-scale molecular gas properties can test these models. Comparing all three types of measurements to key environmental factors, e.g., $\Sigma_*$, $\Sigma_{\rm mol}$, $P_{\rm DE}$, etc., allows the prospect of a holistic disk-to-core model of star formation. 

\subsection{An environment-dependent role for gas density in star formation?}
\label{sec:disk_pressure}

In agreement with \citet[][]{USERO15}, \citet{BIGIEL16}, and \citet{GALLAGHER18} we find that $I_\textrm{TIR}/I_\textrm{HCN}$, tracing the efficiency of dense gas to form stars, anit-correlates with the surface density of stars and molecular gas, $P_{\rm DE}$, and $R_{\rm mol}$. These same quantities correlate with $f_{\rm dense}$, so that as gas becomes denser, the dense gas traced by HCN also appears less efficient at forming stars. This observation agrees with recent work targeting the Milky Way's Central Molecular Zone \citep{LONGMORE13,BARNES17,MILLS17}.

As above, the use of HCN to trace dense gas is pivotal to this interpretation. We review caveats on this below (Section \ref{sec:hcn_emission}). The arguments above hold here, too. Other dense gas tracers yield a qualitatively similar picture and comparison to cloud-scale gas properties does suggest that the HCN-to-CO ratio traces density. More, the Galactic center work has employed a variety of gas tracers, not only HCN, to reach qualitatively similar conclusions in the Milky Way \citep[e.g.,][]{BATTERSBY17,WALKER18}. \citet{USERO15} present a detailed discussion of plausible scenarios for $\alpha_{\rm HCN}$ variations and conclude that the observed trends are unlikely to be exclusively driven by conversion factor effects.

Previous studies of dense gas in nearby galaxies \citep{USERO15,BIGIEL16,GALLAGHER18} interpreted similar observations as evidence that density plays a context-dependent role for star formation. The simplest interpretation would be that star formation occurs in the densest parts of clouds, so that contrast with the mean density, not absolute density, represents the key quantity. This might be expected if the mean dynamical state of clouds is approximately universal (e.g., clouds are all virialized, on average), but the mean density of clouds varies due to changes, reflecting the mean density and pressure in the disk. To first order, HCN traces only a fixed density where the free-fall time should not change significantly. In low pressure regions, this density may capture star-forming overdensities. In higher pressures regions, like galactic centers, HCN may trace a larger fraction of the emission, extending into the ``bulk'' molecular material. In practice, this would translate into a lower apparent star formation efficiency of dense gas, in line with the increasing HCN-to-CO and decreasing TIR-to-HCN ratios observed in individual EMPIRE galaxy disks. Aspects of this argument have been made by \citet{KRUMHOLZ07}, \citet{NARAYANAN08}, \citet{USERO15}, \citet{BIGIEL16}, and \citet{GALLAGHER18}.

\begin{figure}
	\centering
    \includegraphics[scale=0.6]{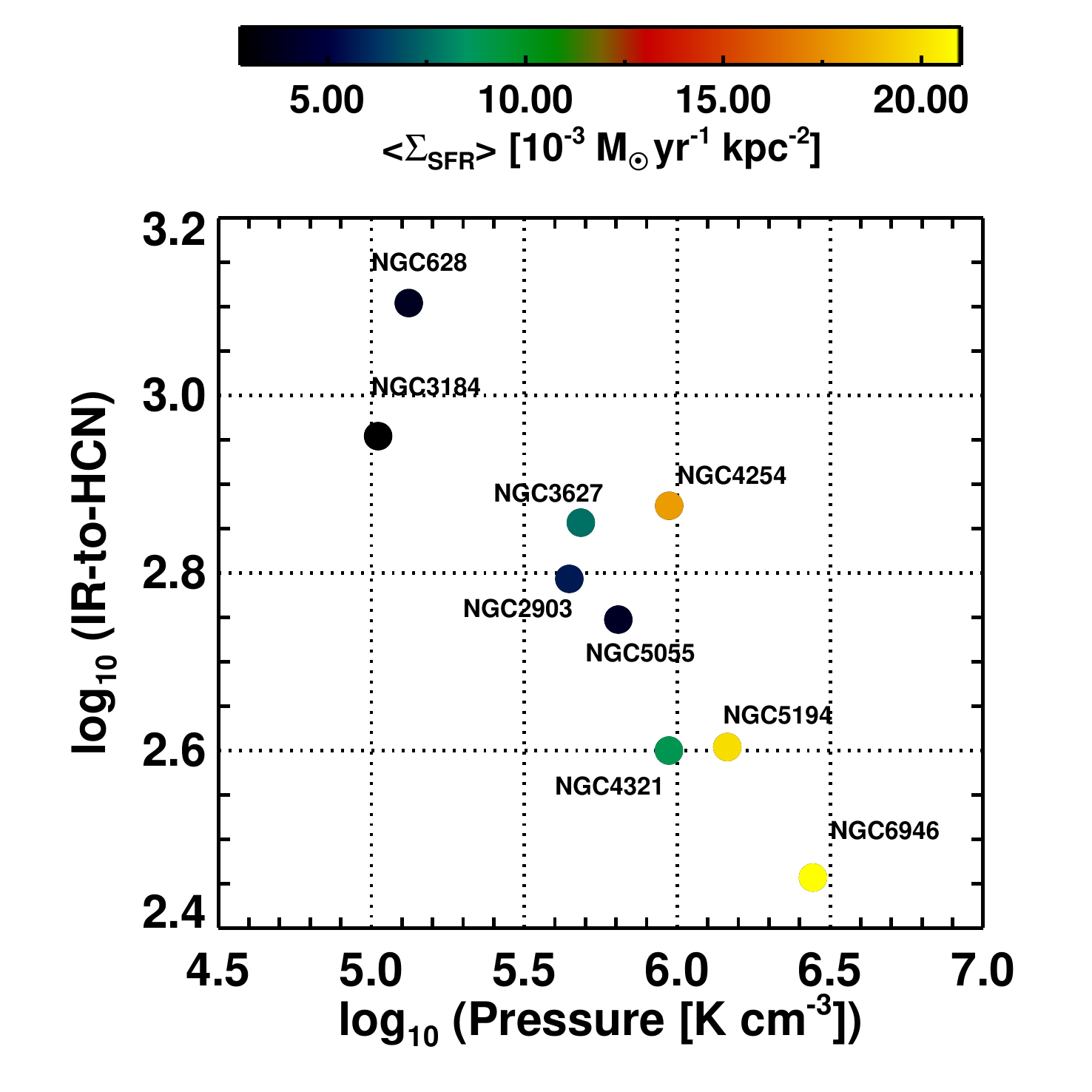}
	\caption{IR-to-HCN ratio, tracing the star formation efficiency of the dense gas, as a function of the dynamical equilibrium pressure for the EMPIRE galaxy centers (inner $30\arcsec \sim1-2$\,kpc). The data points are color coded by the average $\Sigma_{\textrm{SFR}}$ from Table \ref{table:sample}. Galaxy centers from high star-forming galaxies appear at higher pressures and, on average, they seem less efficient at forming stars out of dense gas.}
	\label{fig:test_centers}
\end{figure}

Figure \ref{fig:test_centers} illustrates that the suppression of SFE$_{\rm dense}$ in galaxy centers in EMPIRE does appear more related to the mean pressure. We
show the average observed TIR-to-HCN ratio as a function of the mean $P_{\rm DE}$ in the same region. The centers of our EMPIRE galaxies are characterized by their high pressures, but the exact central $P_{\rm DE}$ varies from galaxy to galaxy by more than one order of magnitude. Figure \ref{fig:test_centers} shows that the centers with the highest $P_{\rm DE}$ appear, on average, to form less stars per unit dense gas mass. 

This simple view clashes with the popular claim that the star formation efficiency per free fall time is approximately fixed across scale and density \citep[e.g.,][]{KRUMHOLZTAN07,UTOMO18}. If this were true, then HCN-emitting gas would show approximately the same SFE$_{\rm dense}$ everywhere, regardless of whether HCN traced ``bulk'' or ``star-forming'' gas. 

In practice, most turbulent models of star formation contain additional physical parameters related to dynamics, e.g., the Mach number and virial parameter, which can affect the density distribution and star formation efficiency. Considering the models of \citet{KRUMHOLZ05} and \citet{KRUMHOLZ07}, \citet{USERO15} showed that the observed IR, CO, and HCN data for nearby galaxies and starburst galaxies could all be explained by allowing Mach number and density to both vary. Our observations agree well with those of \citet{USERO15}, and a similar case should hold for these data too. A key next test will be to infer the Mach number and mean density from high resolution observations \citep[e.g.,][]{SUN18,GALLAGHER18b,QUEREJETA19} and test for consistency with these models when the physical parameters are constrained.

As with $f_{\rm dense}$, significant galaxy-to-galaxy scatter remains in all of our observed SFE$_{\rm dense}$ scaling relations. The strong correlations between the HCN-to-CO ratio and the local $P_\textrm{DE}$ seen in every individual galaxy disk suggest that the ambient pressure (set by the hydrostatic midplane pressure of a galaxy disk) does play a key role by setting the natal density distribution of molecular clouds, initially in hydrostatic equilibrium. Recent semi-analytic modelling by \citet{RAHNER17} and \citet{RAHNER18} have shown exactly this effect: ensembles of identical clouds can evolve differntly when they are initially set in different pressure environments. In addition to dynamics and ISM structure, timescale effects should also play an important role. Recent modelling by \citet{RAHNER17} and \citet{GRUDRIC18} shows that larger and more massive star-forming clouds evolve and expand more slowly with high internal pressures. High SFR regions, formed out of larger and more massive clouds, would typically show much higher internal cloud pressures. This could contribute to the horizontal shift seen in the global EMPIRE trends with respect to pressure in Figures \ref{fig:global_trends} and \ref{fig:test_centers}. If this evolutionary sequence is slow and individual galaxies are dominated by only a handful of clouds, then the scatter among galaxies might capture evolutionary effects. Alternatively, if large scale dynamics synchronizes star formation in some way, these timescale effects might play a key role. Such ``breathing modes'' have been suggested based on simulations by \citet{BENINCASA16,ORR18}, though it is possible that short dynamical timescale associated with dense gas might wash these effects out.

\subsection{Relation to the $L_{\rm HCN}$-$L_{\rm TIR}$ scaling relation and SFE$_{\rm mol}$}

The scaling relation between IR and HCN luminosity, most influentially shown by \citet{GAO04}, has been interpreted to indicate a universal role in star formation for the gas traced by HCN \citep[e.g.,][]{LADA10,LADA12}. Our EMPIRE data do fall on this scaling relation, intermediate between individual cores and clouds and whole galaxies. 

Thus, our observation of an environment-dependent TIR-to-HCN ratio should not be taken to invalidate the scaling relation. Rather, our observations show that the scatter about the relation is physical in nature. For detected regions of resolved galaxy disks, and treating each unit area the same, the RMS scatter is $0.2{-}0.3$~dex and correlates with environment as described above. In practice, the quantitative scatter about the relation depends on the adopted sampling scheme. For example, weighting equally by area tends to de-emphasize galaxy centers. Weighting by luminosity de-emphasizes outer disks with little star formation. In any case, we find significant physical scatter about the IR-HCN scaling relation and have quantified the dependence of the TIR-to-HCN ratio on environment.

An important corollary, already emphasized above, is that the trends that we observe relating TIR-to-HCN to environment cannot be extrapolated indefinitely. \citet{GAO04} and \citet{GARCIABURILLO12}, among others, show that the TIR-to-HCN ratio in starburst galaxies with high $\Sigma_{\rm mol}$ and high $P_{\rm DE}$ is ``normal.'' Galaxy centers do not extrapolate into the (U)LIRG population correctly, perhaps due to the different dynamics at play in the different environments.

\citet{GAO04} and several following papers also highlighted that the HCN-to-CO ratio, $f_{\rm dense}$, could predict the star formation efficiency of the total molecular gas, SFE$_{\rm mol}$ or TIR-to-CO. We do find that SFE$_{\rm mol}$ correlates with $f_{\rm dense}$. That is, variations in SFE$_{\rm dense}$ and $f_{\rm dense}$ do not totally offset. But the exact scaling inferred depends sensitively on the data sets considered because the TIR-to-HCN ratio varies. In that sense, our work agrees with \citet{GALLAGHER18} and \citet{USERO15} in finding that a density threshold model above which SFE$_{\rm dense}$ remains constant appears too simple to explain the observations of IR, HCN, and CO in nearby galaxies.

\subsection{Caveats}
\label{sec:hcn_emission}

In this work, we focus on the content of dense gas in nearby galaxies by analyzing the emission of lines with high critical densities such as HCN\,(1-0). However, the interpretation of the HCN emission and thus the {\it dense gas mass} remains an open issue (see Section \ref{sec:factors}) especially in the context of clouds with varying density probability distribution functions (PDFs).

\subsubsection{The mass of dense gas from HCN observations}
\label{sec:factors}

If we assume that HCN is a good tracer of dense gas, the second major limiting factor needed for a well-calibrated relationship between HCN emission, dense gas mass and star formation is the conversion factor $\alpha_\textrm{HCN}$. Thus, the observational constraints we can place on any star formation theory are sensitive to the conversion factors that translate line luminosities into masses of dense molecular gas.

The first estimation of the HCN conversion factor is detailed in the seminal work by \citet{GAO04}. The authors derived a dense gas conversion factor assuming virialized (self-gravitating), optically thick dense gas cores with $n\sim3\times10^4\textrm{cm}^{-3}$ and constant brightness temperatures of 35\,K \citep[e.g.,][]{RADFORD91}. In this way, they obtained a simple relation:
\begin{equation}
\alpha_\textrm{HCN} = 2.1\,\sqrt{n(\textrm{H}_2)}/T_b = 10\,M_\odot\,(\textrm{K\,km\,s}^{-1}\textrm{pc}^2)^{-1}.
\end{equation}

Any conversion factor calculated under these assumptions would then depend on the gas density in molecular clouds and its temperature, which will in turn depend on the gas excitation and the beam filling fraction. Later on, \citet{WU10} also derived a dense gas conversion factor by comparing the HCN luminosity in massive Galactic clumps, to their virial mass and found $\alpha_\textrm{HCN} = 20 \pm 1\,M_\odot\,(\textrm{K\,km\,s}^{-1}\textrm{pc}^2)^{-1}$. However, the physical conditions observed in individual Galactic clumps likely differ from those of the bulk dense gas in galaxies.

Generally $\alpha_\textrm{CO}$ has been observed to increase with decreasing metallicity and to drop where the gas is more turbulent \citep[galaxy centers and starbursts, e.g.,][]{GARCIACARPIO08,GARCIABURILLO12}. Moreover, excitation effects can also drive changes in the molecular gas conversion factor $\alpha_\textrm{CO}$. While the dense gas conversion factor is harder to constrain due to the scarce data situation and challenging observations, one can expect a similar dependence on turbulence and excitation.

In that regard, \citet{SHIMAJIRI17} estimate the mass of dense gas in Galactic clouds using dust column densities from {\it Herschel} and compare them with HCN luminosities to obtain empirical conversion factors. The authors claim that variations in $\alpha_\textrm{HCN}$ in Galactic clouds could be related to variations in the FUV field (which is significantly stronger towards galaxy centers) and so to gas excitation and/or chemistry variations. Could the observed variations in SFE$_\textrm{dense}$ (Figure \ref{fig:global_trends}) be explained by gas excitation variations? Their dust temperature maps from {\it Herschel}, however, blend different ISM phases when used at extragalactic scales ($>100\,\textrm{pc}$), which could introduce additional uncertainty in $G_0$ and the strength of the FUV field. Thus, assessing dense gas excitation through direct observations of several rotational transitions (e.g., $J=3-2$ to $J=1-0$) is of crucial importance.

In terms of modelling, most estimates of dense gas conversion factors are performed using idealized clouds and density distributions \citep{KRUMHOLZ07,LEROY17}, or simulate very small regions withing molecular clouds \citep{ONUS18}, which are likely not representative of the varying conditions across galaxy disks. Recent work by \citet{VOLLMER17} on analytic modeling also considers large-scale properties of entire galaxies (e.g., surface density, turbulent velocity, disk height) and also models the line emission from individual self-gravitating clouds using detailed chemical networks and an escape probability formalism. By comparing their calculated dense gas masses and line emission in star-forming galaxies, they predict $\alpha_\textrm{HCN} = 21 \pm 6$, $33 \pm 17$, and  $59 \pm 21 \,M_\odot\,(\textrm{K\,km\,s}^{-1}\textrm{pc}^2)^{-1}$ for local spiral galaxies and ULIRGs, submillimeter galaxies, and high-$z$ galaxies, respectively.

A more accurate determination of the dense gas conversion factor (e.g., between HCN and the dense molecular gas mass) is needed for a complete understanding of the relation between HCN emission, gas density and star formation efficiency of dense gas. This requires detailed knowledge of HCN emission in different systems (e.g., ULIRGs, starbursts, low-metallicity galaxies) and further constraints on HCN excitation conditions. Assessing dense gas excitation using higher-$J$ lines of these molecules is possible with facilities like ALMA, NOEMA or the Submillimeter Array, and will be an important step forward to resolve the tension between competing star formation theories. High-resolution observations of dust continuum emission in nearby galaxies and its comparison to GMC-scale observations of CO and HCN will additionally provide insight into empirical molecular and dense gas conversion factors, as well as its variation across different galactic environments.

\subsubsection{HCN emissivity and critical density}

The line ratios we employ in our study (e.g., HCN/CO) are sensitive to density changes but also could reflect opacity, chemical, or excitation effects. Thus, analyzing and interpreting line ratios arising from sub-beam density distributions requires additional knowledge to probe gas densities. In this regard more extragalactic observations of optically thin isotopologues remain crucial to probe the optical depth, effective critical density ($n_\textrm{eff}$) and isotopic abundance of high-density tracers \citep{JIMENEZDONAIRE17,JIMENEZDONAIRE17B}. As detailed in \citet{LEROY17}, differential excitation also plays an important role in determining true optical depths and characterizing the emissivity properties of gas tracers. Multi-$J$ observations of high-density tracers like HCN, HCO$^+$ and HNC in external galaxies have the prospect to constrain $T_\textrm{kin}$ and $n$ of the gas.

Ideally, we are interested in knowing how much gas mass is emitting at a given density, and how this emissivity changes as a function of density. The HCN\,(1-0) transition has a high effective critical density compared to the low-$J$ CO lines. But the mean density of HCN emitting gas remains uncertain because even gas below the effective critical density can emit (the emission is {\it subthermal}), albeit with lower emissivity. In real molecular clouds, there is much less mass at high density than at low densities. If this imbalance is large enough, then despite the lower emissivity, the high abundance of low density gas can lead to a case where almost all emission comes from sub-thermally excited gas. There is observational evidence that this sub-thermal emission constitutes a significant contribution to the dense gas luminosity of entire galaxies \citep[e.g.,][]{PAPADOPOULOS07,ARAVENA14}. Thus, knowing how much of the emission we detect comes from high-density gas is crucial to interpret the observed HCN-to-CO and TIR-to-HCN variations.

Current Galactic surveys focusing on resolved sub-parts of star-forming regions have investigated whether commonly used dense gas tracers, including HCN\,(1-0), are good tracers of dense gas. One of the key results from the ORION-B survey \citep{PETY17} and first conclusions from the LEGO survey \citep{KAUFFMANN17} is that most of the HCN emission comes from gas densities $n \leq 10^4\,\textrm{cm}^{-3}$. \citet{PETY17} and \citet{KAUFFMANN17} find that N$_2$H$^+$\,(1-0) is the only tracer sensitive to high column densities ($>10^{22}\,\textrm{cm}^{-2}$). It is important to note that their direct observables are, however, column densities ($N$) instead of volume densities, $n$. Additionally, they are focused on very specific physical conditions (strong interstellar radiation fields by young stars and almost no embedded stars) that are inherent to small ($< 10\,$pc in diameter) sub-regions within Orion. 
In particular, observations of pre-stellar cores and cold filaments in Orion A and B have shown that freeze-out of molecules onto dust grains reduces the gas phase abundance of CO and other molecules, with the notable exception of N$_2$H$^+$, which stays in the gas phase for a long time \citep[e.g.,][]{HACAR18}. Along those lines, high-resolution simulations analyzed by \citet{ONUS18} also show that a significant portion of the HCN\,(1-0) emission comes from gas with mean densities a factor of 10 lower than the HCN critical density. However, most of the HCN emission originates in gas at densities $\sim 2.5-5$ times greater than the mean density of the gas \citep{ONUS18}.

Recent efforts modelling line emission from sub-beam density distributions \citep[e.g.,][]{LISZT16,LEROY17} show that, while gas can indeed emit effectively below its effective critical density, transitions with critical densities higher than the average density of the gas show emissivities that vary strongly as the density distribution changes. Therefore the line ratios we employ as proxies (HCN-to-CO) should be good probes of the fraction of dense gas. These models however are subject to uncertain factors such as fixed $T_\textrm{kin}$ and fixed abundances, and they only account for one main collider.

Additionally, other physical mechanisms at play in the ISM of galaxies could increase the emissivity of HCN\,(1-0) at lower densities. This is mainly motivated by the fact that \citet{PETY17} find that the spatial extent of the emission of high-density tracers like HCN\,(1-0) does not correlate with the $\textrm{H}_2$ density that is required for collisional excitation. Two possible causes of low-density HCN\,(1-0) excitation are cosmic ray heating and electron collisions. Recent work by \citet{VOLLMER17} presents an analytic model of galactic clumpy gas disks where, given physical properties of galaxies (e.g., size, rotation curve, stellar mas profile), they are able to simultaneously calculate quantities such as the total gas mass, gas velocity dispersion, TIR luminosity, CO SLED and HCN\,(1-0) luminosity. They show that, while cosmic ray heating does not significantly alter the CO emission, it can increase the HCN\,(1-0) emission by at most a factor of two. They also show that this factor is indeed necessary to reproduce the observed HCN emission in ULIRGs. \citet{GOLDSMITH17} and \citet{KAUFFMANN17} additionally suggest that HCN can also be excited by collisions with electrons. In  \citet{GOLDSMITH17}, the authors compute the collisional excitation of the rotational levels of HCN, HCO$^+$, CN, and CS by electrons and $\textrm{H}_2$ molecules. They conclude that electron excitation of HCN\,(1-0) is important at densities $n<n_\textrm{crit}$ if the electron abundance is $X(\textrm{e}^-)>10^{-5}$, that is electron collisions dominate the excitation of HCN molecules in regions where most carbon is ionized but hydrogen remains molecular.

Thus, there are a number of factors responsible for increasing the HCN emissivity (e.g., UV, X-rays, cosmic rays, mechanical heating) that will always depend on the details of the chemistry models. While taking all these factors into account is extremely complex, a key path forward would have to involve contrasting large scale Galactic and extragalactic observations with predictions from simulations of ensembles of molecular clouds, equipped with detailed chemistry models \citep{RAHNER18,SEIFRIED19,BISBAS19}.

\section{Summary and conclusions}
\label{sec:summary}

We present EMPIRE, a spectral line mapping survey that targeted $\lambda = 3-4$~mm tracers of dense molecular gas (HCN, HCO$^+$, HNC) and the bulk-gas-tracing CO isotopologues ($^{12}$CO, $^{13}$CO, C$^{18}$O). EMPIRE covered the whole star-forming disk (typically out to $\sim 8-10$\,kpc) of nine nearby, massive galaxies and so provides the first sample of whole-galaxy resolved (1-2~kpc resolution) dense gas maps. 

Here we describe the survey products, which will be publicly available from the IRAM repository and the EMPIRE website. We use these data to investigate how the dense gas fraction $f_{\rm dense}$, as traced by the HCN-to-CO line ratio, and the efficiency with which this gas forms stars, SFE$_{\rm dense}$, as traced by the TIR-to-HCN line ratio, depend on environment and host galaxy. Our main results are:

\begin{enumerate}
	\item We detect dense gas as traced by HCN\,(1-0), HCO$^+$\,(1-0) and HNC\,(1-0) emission across the entire galaxy sample. We employ stacking techniques to recover the emission from low signal-to-noise regions. This allows us to detect HCN out to radii of $\sim 9-11\,$kpc, i.e., beyond the radius of the Solar Circle in the Milky Way. We detect HCO$^+$ out to $\sim 7-10\,$kpc, and HNC out to $\sim 4-6\,$kpc. To first order, the HCN integrated intensity maps show similar large-scale structure to the CO and 70 $\mu$m emission.
	
	\item Emission from the three dense gas tracers appears faint. On average across all EMPIRE galaxies, the HCN-to-CO line ratio is $0.025$ and the HCO$^+$-to-CO ratio is $0.018$. HNC appears fainter, with a typical HNC-to-CO ratio of $0.011$. Following this, the average HCO$^+$-to-HCN is $0.7$, while the average HNC-to-HCN ratio is $0.4$. HCO$^+$ shows, on average, a similar radial profile to HCN but we identify a few cases where the HCO$^+$-to-HCN ratio shows a systematic increase with radius. Adopting a (highly uncertain) standard conversion from HCN integrated intensity to dense gas suggests that on average $\sim 6\%$ of the molecular gas across the EMPIRE targets is dense HCN-emitting material.
	
	\item EMPIRE reveals a clear dependence of the dense gas fraction, $f_{\rm dense}$, on local conditions in a galaxy disk. $f_{\rm dense}$ appears highest in galaxy centers and decreases with increasing galactocentric radius in all targets. At our 1-2~kpc resolution $f_{\rm dense}$ correlates with the local stellar mass surface density, the local molecular gas mass surface density, the molecular-to-atomic gas ratio, and the local dynamical equilibrium pressure $P_{\rm DE}$ estimated from hydrostatic equilibrium. All of these trends have the sense that concentrating more gas in a deeper potential well leads to a larger fraction of dense gas. Our measurements agree well with those seen in previous work \citep[e.g.,][]{USERO15,CHEN15,BIGIEL16,GALLAGHER18}. With EMPIRE, we quantity the relations across the whole area of a sample of galaxies, providing the best systematic measurement to date.
	
	\item Well-detected individual regions from EMPIRE follow the same global infrared-HCN luminosity scaling as a large compilation of literature observations targeting Galactic cores, individual clouds, and whole galaxies \citep[i.e., our data agree with][among many others]{GAO04,WU05,GARCIABURILLO12}. That is, on average, EMPIRE shows the same TIR-to-HCN ratio as starburst galaxies and Galactic cores. In detail, there is significant scatter about this global scaling relation. Our observations show that there are physical, systematic variations causing this large scatter, and that it is not the result of random statistics.
	
	\item The TIR-to-HCN ratio also shows a systematic dependence on local environment. SFE$_{\rm dense}$ anti-correlates with the stellar mass surface density, molecular gas mass surface density, molecular-to-atomic gas ratio, and the dynamical equilibrium pressure. As a result, the inner regions of our targets, especially the inner 1-2\,kpc, appear inefficient at forming stars relative to their (high) dense gas content. Our results agree with other recent studies of nearby galaxies \citep{USERO15,CHEN15,BIGIEL16,GALLAGHER18} and resemble findings for the Milky Way's Central Molecular Zone, which also shows low SFE$_{\rm dense}$ \citep[e.g.,][]{LONGMORE13,BARNES17,MILLS17}. These results reinforce that the role of gas density in star formation is at least somewhat context-dependent. As with $f_{\rm dense}$, the wide field of view and complete coverage of EMPIRE should render our measured relationships more general than previous work.
	
    \item We find a correlation between dense gas fraction, $f_{\rm dense}$, and the overall star formation efficiency of the total molecular gas, SFE$_{\rm mol}$, as expected by density threshold models. However, there is considerable scatter, $\sim 0.2{-}0.3$~dex in the relationship between SFE$_{\rm mol}$ and $f_{\rm dense}$ due to the systematic, physical variations in SFE$_{\rm dense}$ described above. Thus EMPIRE shows that in normal star-forming galaxies, dense gas threshold models can only hold with an accuracy of $\sim 0.2{-}0.3$~dex, which is large compared to the dynamic range in $f_{\rm dense}$.
    
    \item We observe significant, $\sim 0.2$~dex, galaxy-to-galaxy scatter in the scaling relations between $f_{\rm dense}$ and SFE$_{\rm dense}$ to environment. Much of this scatter appears as offsets among individual galaxies. We suggest that galactic dynamics and sub-beam gas structure may be important additional factors at play. We also highlight the importance of HCN excitation studies and further investigations into how our adopted line ratios trace the underlying gas density distribution.
	
\end{enumerate}

\acknowledgments {\bf Acknowledgements.} The IRAM 30m large program EMPIRE was carried out under project number 206-14 (PI Bigiel), the $^{12}$CO(1-0) observations under projects 061-15 and 059-16 (PI Jim\'enez-Donaire) and D15-12 (PI Cormier). IRAM is supported by INSU/CNRS (France), MPG (Germany) and IGN (Spain). The authors would like to thank S. Hony and E. Pellegrini for useful discussions, and the referee, Neal Evans, for a constructive and helpful report. MJJD would like to thank the International Max Planck Research School for
Astronomy and Cosmic Physics at the University of Heidelberg (IMPRS-HD) for the support of this work. MJJD acknowledges support from the Smithsonian Institution as a Submillimeter Array (SMA) Fellow. FB and JP acknowledge funding from the European Union's Horizon 2020 research and innovation programme (grant agreement No 726384). AU acknowledges support from Spanish MINECO grants ESP2015-68964 and AYA2016-79006. The work of AKL and MJG is partially supported by the National Science Foundation under Grants No.~1615105, 1615109,and 1653300. AKL also acknowledges partial support from NASA ADAP grants NNX16AF48G and NNX17AF39G. The work of MJG is partially supported by a National Radio Observatory (NRAO) student observing support award (\# 359067). AH was supported by the Programme National Cosmology et Galaxies (PNCG) of CNRS/INSU with INP and IN2P3, co-funded by CEA and CNES, and by the Programme National 'Physique et Chimie du Milieu Interstellaire' (PCMI) of CNRS/INSU with INC/INP, co-funded by CEA and CNES. ADB acknowledges partial support from NSF-AST1615960. The National Radio Astronomy Observatory is a facility of the National Science Foundation operated under cooperative agreement by Associated Universities, Inc.

%\software{GILDAS/CLASS \citep[][GILDAS team 2013]{PETY05}}.

\bigskip

\setcounter{figure}{0}
\setcounter{table}{0}
\renewcommand{\thefigure}{A\arabic{figure}}
\renewcommand{\thetable}{A\arabic{table}}

\section*{APPENDIX A. Line calibrators}

\begin{figure*}
	\begin{center}
		\includegraphics[scale=0.6]{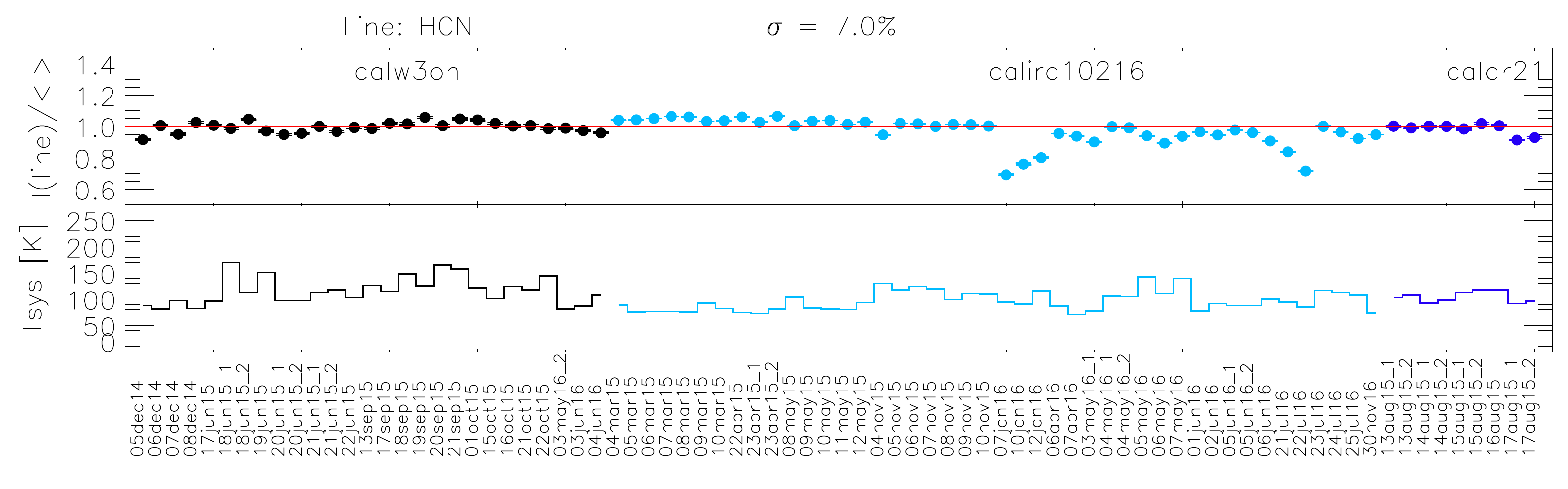}
	\end{center}
	\caption{HCN\,(1-0) integrated intensity for each day and line calibrator, divided by the mean of all measured intensities (top panel). During the EMPIRE observing runs, three different line calibrators were used: W3(OH), IRC 10216 and DR21(OH).}
	\label{calibrators}
\end{figure*}

Figure \ref{calibrators} show the different line calibrators observed for the EMPIRE survey: W3(OH), IRC 10216 and DR21(OH). The bottom panel shows the typical system temperatures during the observations. The variations seen over the course of the observations are of the order of only 7\%, which implies a very stable relative calibration of our observed lines.

\setcounter{figure}{0}
\setcounter{table}{0}
\renewcommand{\thefigure}{B\arabic{figure}}
\renewcommand{\thetable}{B\arabic{table}}

\section*{APPENDIX B. Individual measurements and radial stacks}
\label{appendix:radialprof}

We provide every individual line-of-sight measurements performed for the EMPIRE galaxy sample. Table \ref{table:rad_measurements}, which appears as electronic material only, includes the integrated intensities and respective uncertainties for each molecular line mapped (HCN~(1-0), HCO$^+$~(1-0), HNC~(1-0), $^{12}$CO~(1-0), $^{13}$CO~(1-0) and C$^{18}$O~(1-0)) at each galactocentric radii.

\begin{figure*}
	\centering
	\includegraphics[scale=0.13]{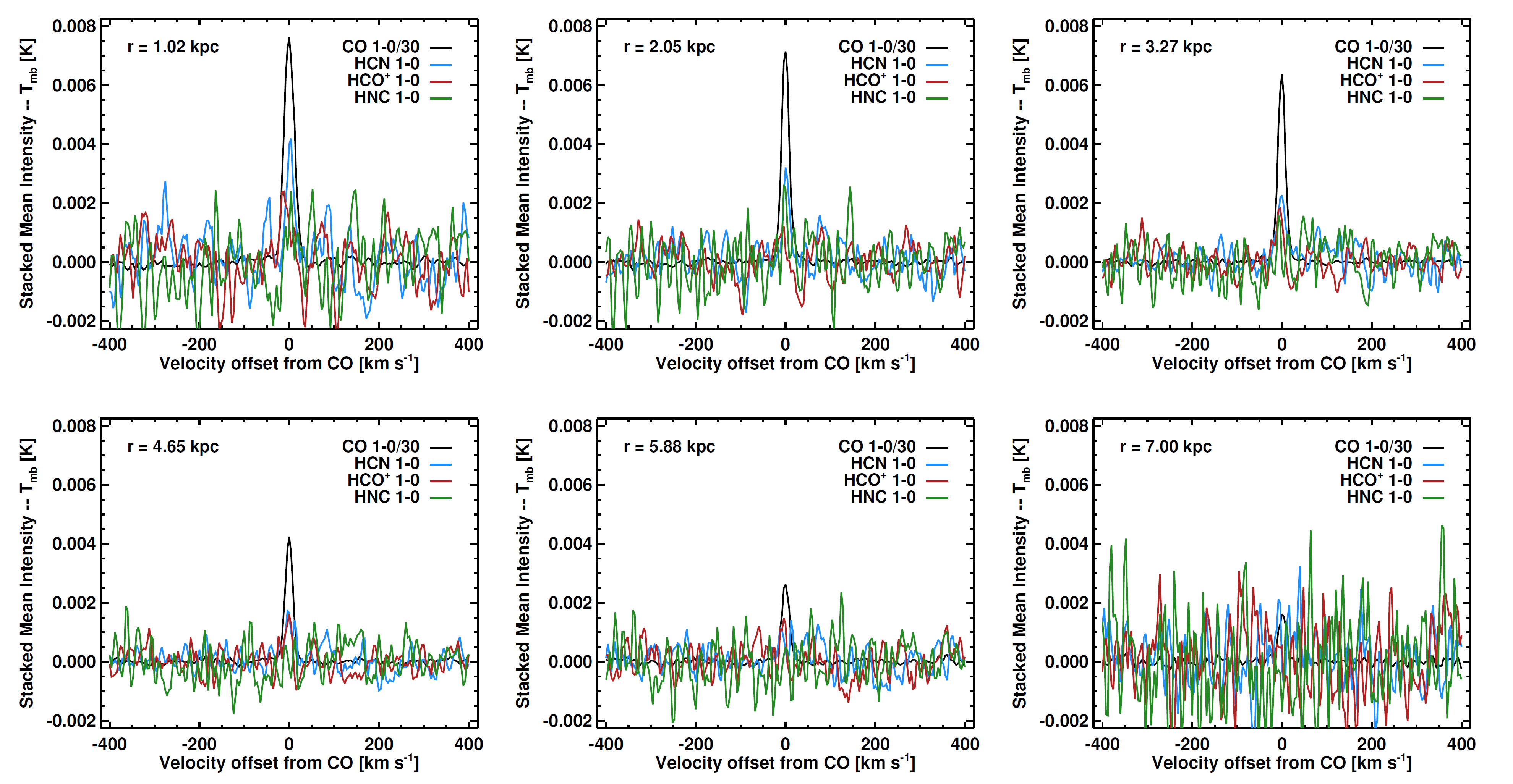}
	\caption{Same as Figure \ref{fig:dif_stacks} but for NGC\,0628. Stacked CO\,(1-0), HCN(1-0), HCO$^+$\,(1-0) and HNC(1-0) in $30\arcsec$ ($\sim 1.5\,\textrm{kpc}$) radial bins. Galactocentric radii are shown in units of kpc.}
	\label{fig:dif_stacks2}
\end{figure*}

\begin{figure*}
	\centering
	\includegraphics[scale=0.13]{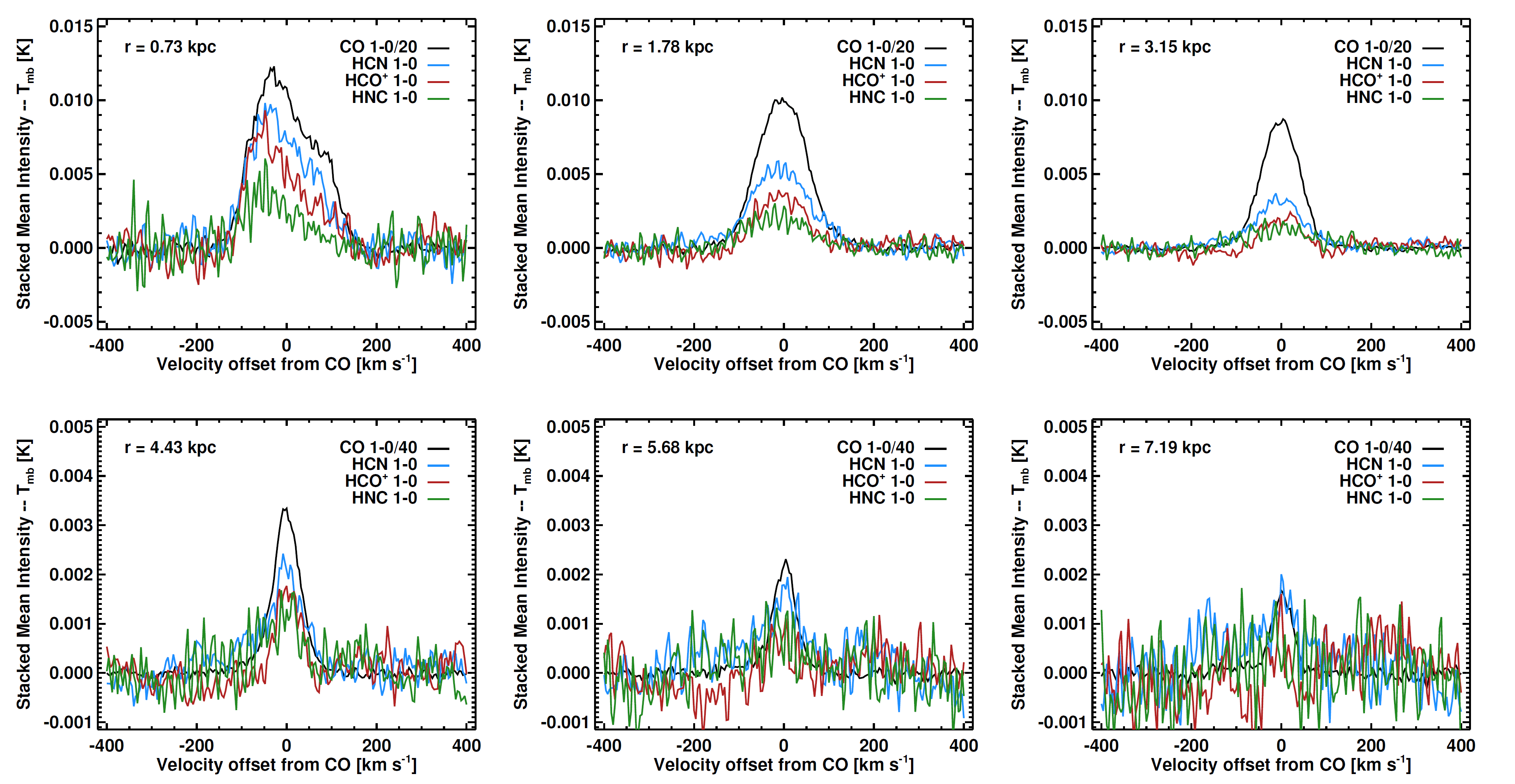}
	\caption{Same as Figure \ref{fig:dif_stacks} but for NGC\,2903. Stacked CO\,(1-0), HCN(1-0), HCO$^+$\,(1-0) and HNC(1-0) in $30\arcsec$ ($\sim 1.5\,\textrm{kpc}$) radial bins. Galactocentric radii are shown in units of kpc.}
	\label{fig:dif_stacks3}
\end{figure*}

\begin{figure*}
	\centering
	\includegraphics[scale=0.13]{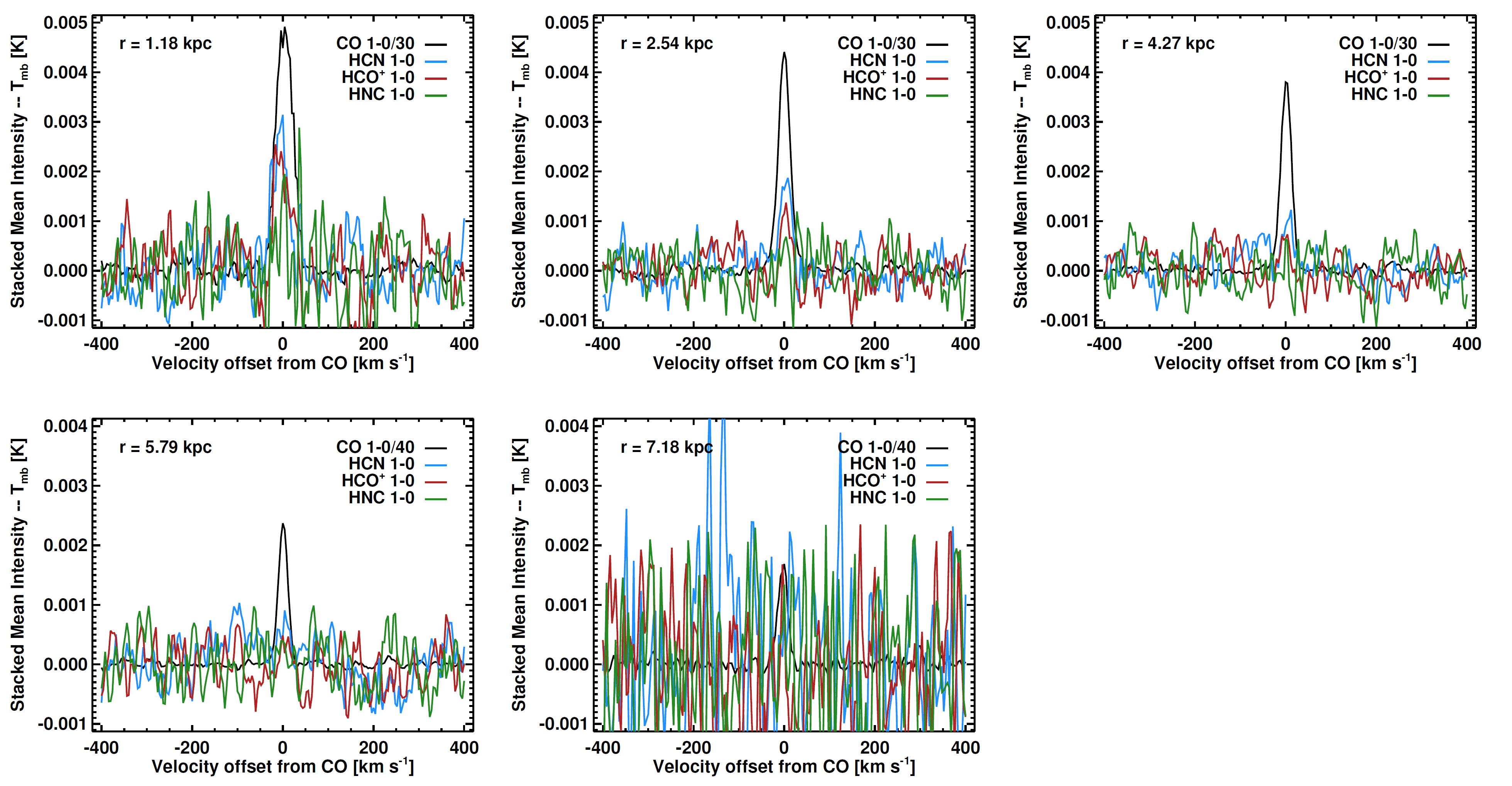}
	\caption{Same as Figure \ref{fig:dif_stacks} but for NGC\,3184. Stacked CO\,(1-0), HCN(1-0), HCO$^+$\,(1-0) and HNC(1-0) in $30\arcsec$ ($\sim 1.5\,\textrm{kpc}$) radial bins. Galactocentric radii are shown in units of kpc.}
	\label{fig:dif_stacks4}
\end{figure*}

\begin{figure*}
	\centering
	\includegraphics[scale=0.13]{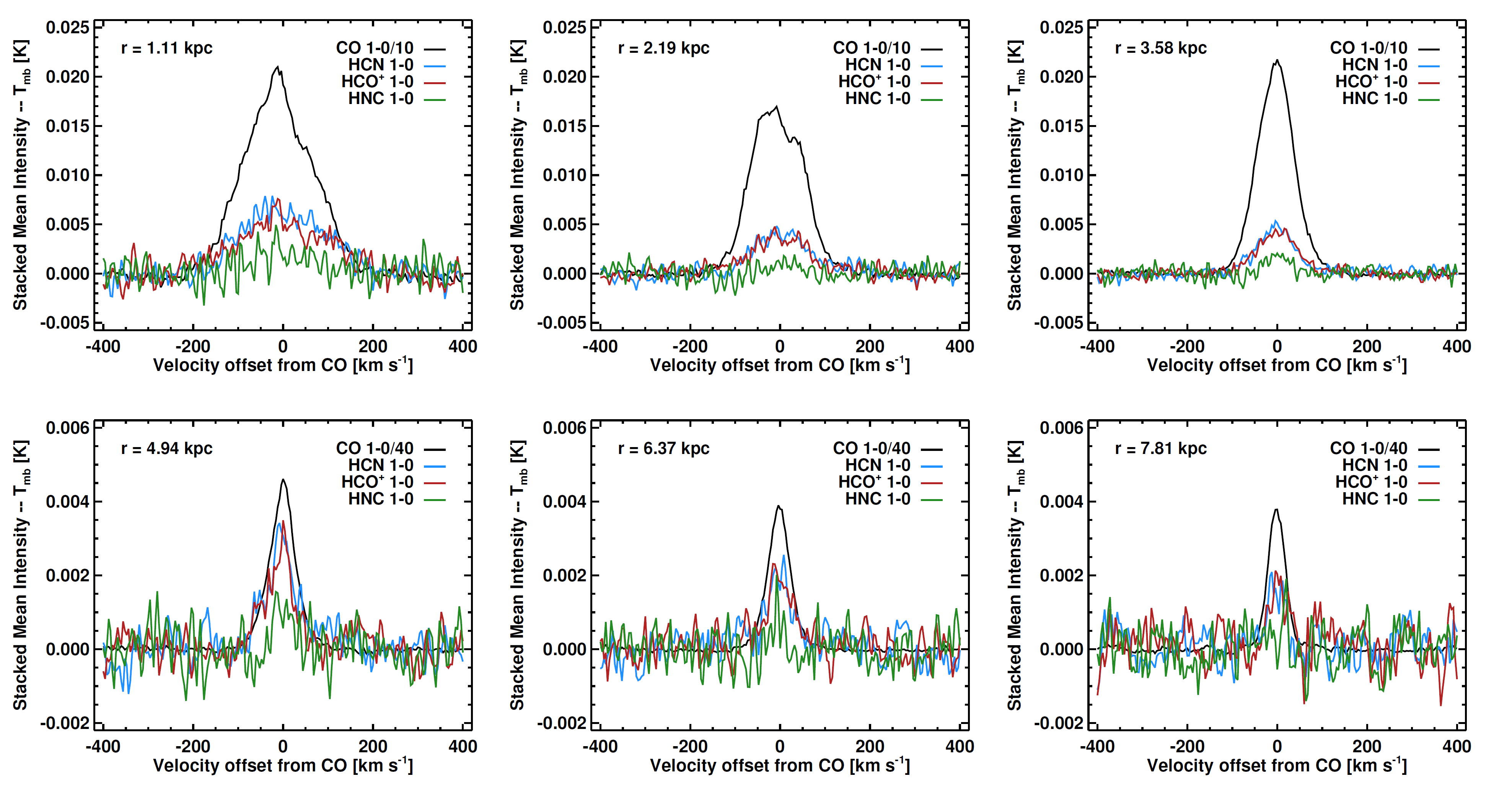}
	\caption{Same as Figure \ref{fig:dif_stacks} but for NGC\,3627. Stacked CO\,(1-0), HCN(1-0), HCO$^+$\,(1-0) and HNC(1-0) in $30\arcsec$ ($\sim 1.5\,\textrm{kpc}$) radial bins. Galactocentric radii are shown in units of kpc.}
	\label{fig:dif_stacks5}
\end{figure*}

\begin{figure*}
	\centering
	\includegraphics[scale=0.13]{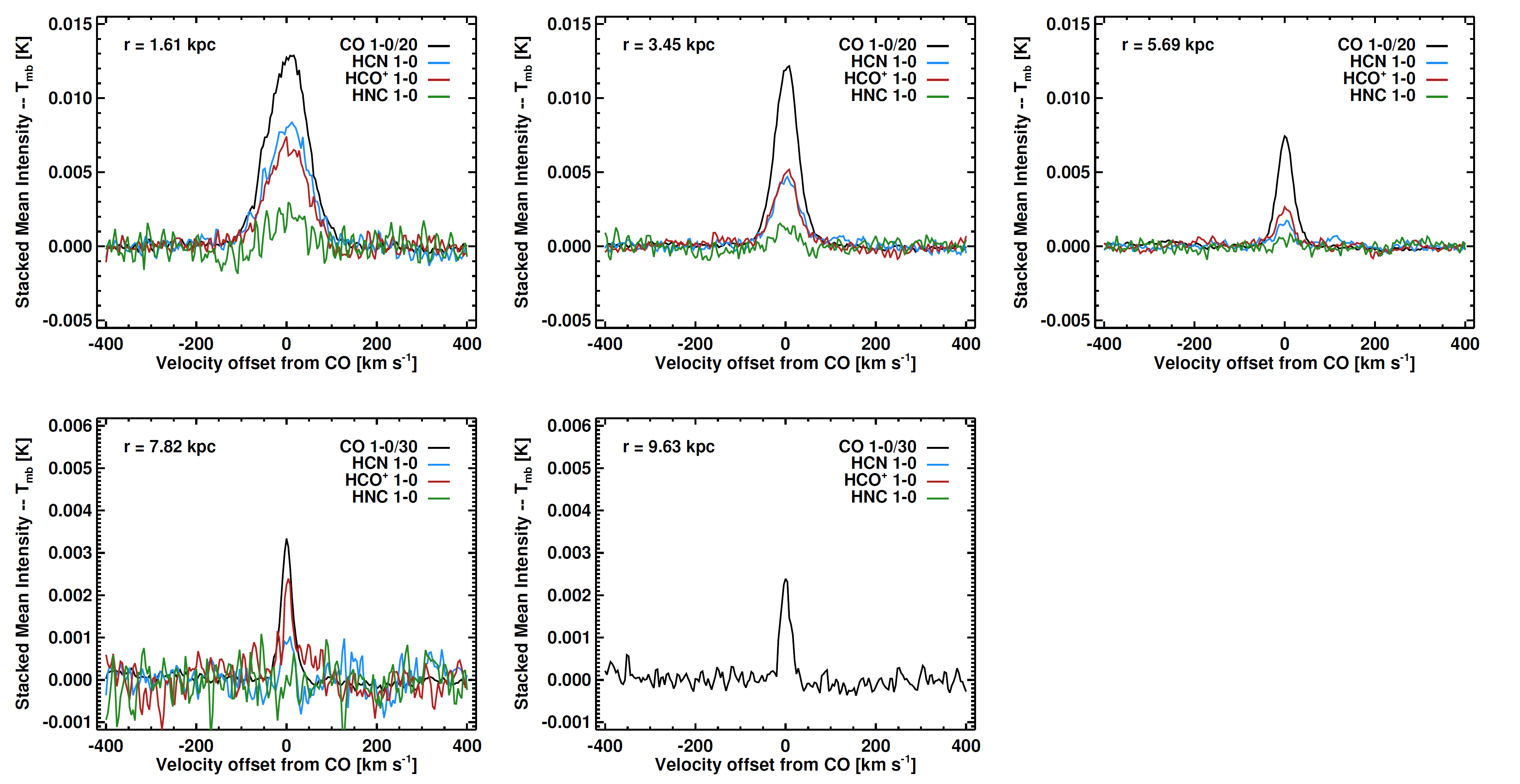}
	\caption{Same as Figure \ref{fig:dif_stacks} but for NGC\,4254. Stacked CO\,(1-0), HCN(1-0), HCO$^+$\,(1-0) and HNC(1-0) in $30\arcsec$ ($\sim 1.5\,\textrm{kpc}$) radial bins. Galactocentric radii are shown in units of kpc.}
	\label{fig:dif_stacks6}
\end{figure*}

\begin{figure*}
	\centering
	\includegraphics[scale=0.13]{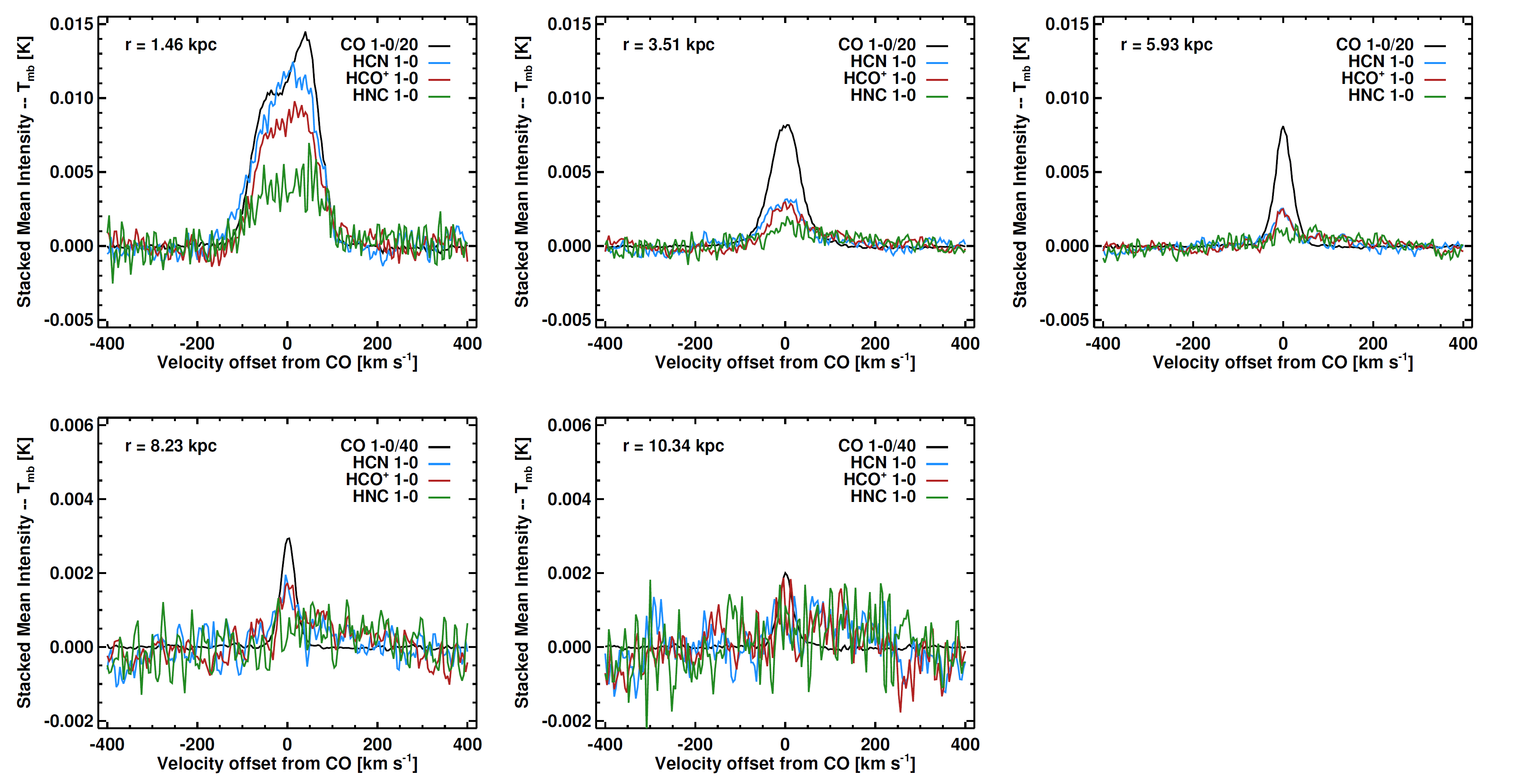}
	\caption{Same as Figure \ref{fig:dif_stacks} but for NGC\,4321. Stacked CO\,(1-0), HCN(1-0), HCO$^+$\,(1-0) and HNC(1-0) in $30\arcsec$ ($\sim 1.5\,\textrm{kpc}$) radial bins. Galactocentric radii are shown in units of kpc.}
	\label{fig:dif_stacks7}
\end{figure*}

\begin{figure*}
	\centering
	\includegraphics[scale=0.13]{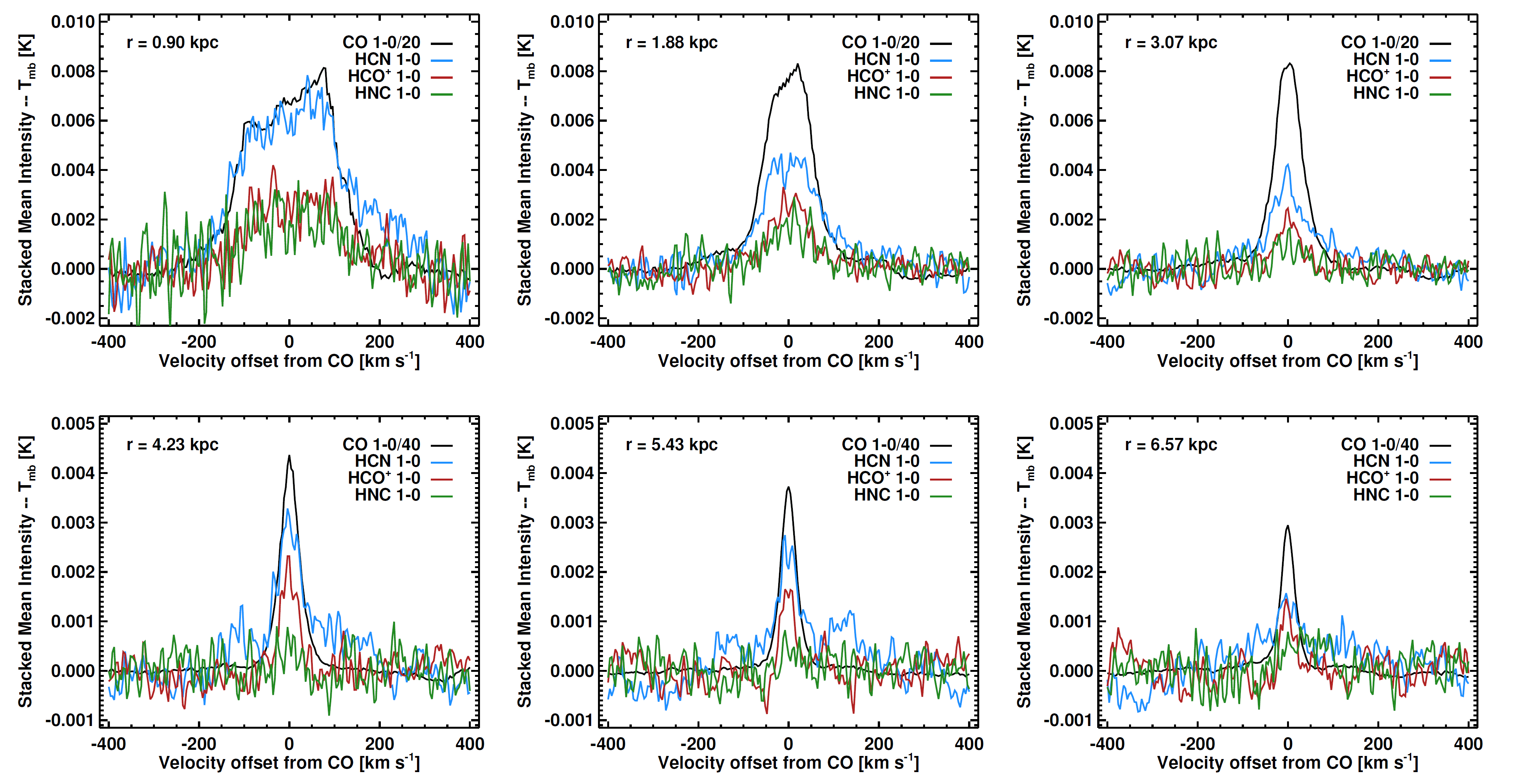}
	\caption{Same as Figure \ref{fig:dif_stacks} but for NGC\,5055. Stacked CO\,(1-0), HCN(1-0), HCO$^+$\,(1-0) and HNC(1-0) in $30\arcsec$ ($\sim 1.5\,\textrm{kpc}$) radial bins. Galactocentric radii are shown in units of kpc.}
	\label{fig:dif_stacks8}
\end{figure*}

\begin{figure*}
	\centering
	\includegraphics[scale=0.13]{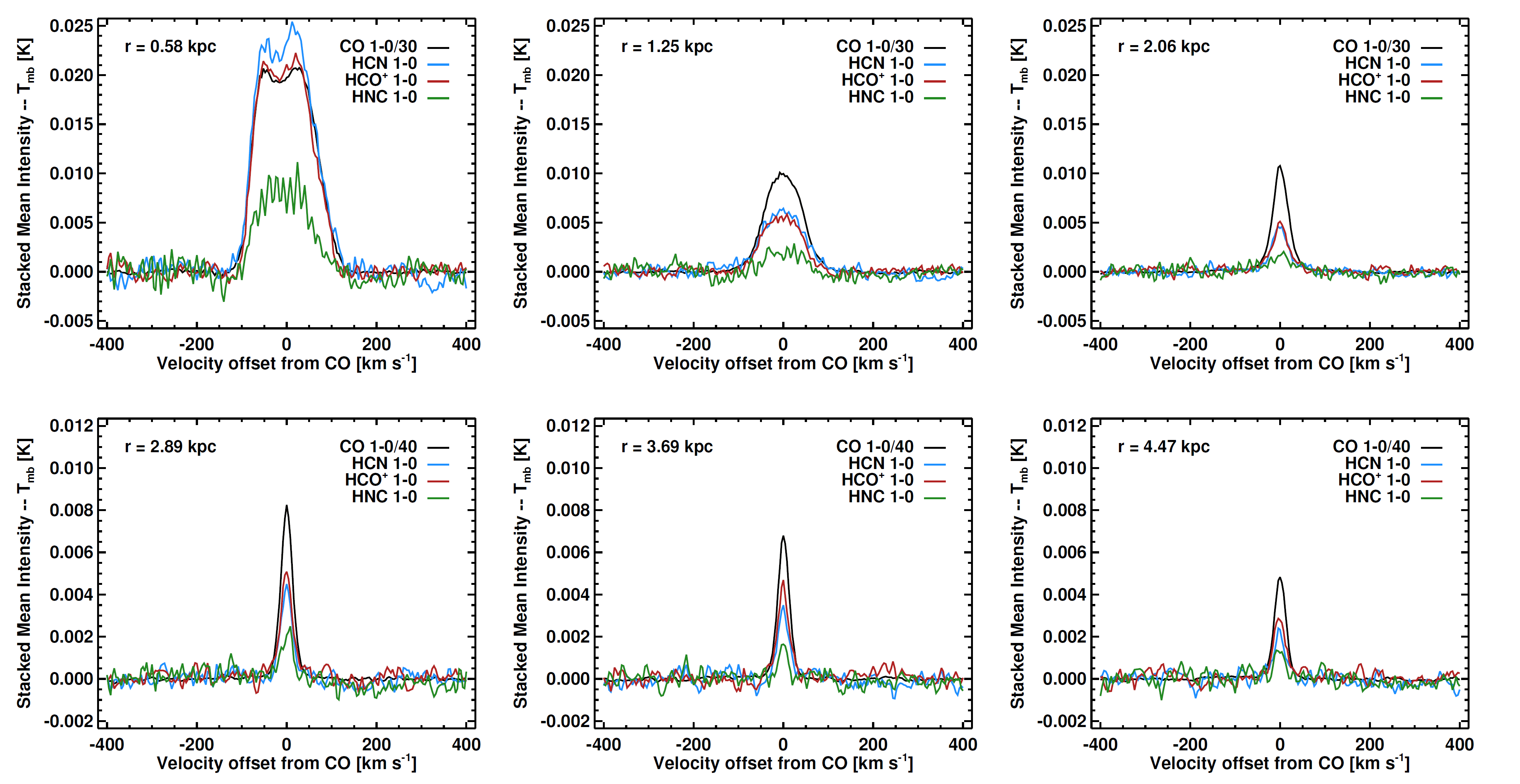}
	\caption{Same as Figure \ref{fig:dif_stacks} but for NGC\,6946. Stacked CO\,(1-0), HCN(1-0), HCO$^+$\,(1-0) and HNC(1-0) in $30\arcsec$ ($\sim 1.5\,\textrm{kpc}$) radial bins. Galactocentric radii are shown in units of kpc.}
	\label{fig:dif_stacks9}
\end{figure*}

Figures \ref{fig:dif_stacks2}-\ref{fig:dif_stacks9} show the result from our spectral stacking technique, applied to regions of increasing radii in the EMPIRE galaxy sample. As detailed in Section \ref{sec:stacking}, we employed our well detected CO\,(1-0) data as a prior to average independent spectra from the weaker, high critical density lines (HCN, HCO$^+$ and HNC). We perform this averaging over extended radial regions of 30\arcsec in angular size ($\sim 1-2$\,kpc), which roughly corresponds to the angular resolution of our observations. 

Table \ref{table:rad_profiles} provides the measured integrated intensities in each of the radial bins displayed in Figures \ref{fig:dif_stacks2}-\ref{fig:dif_stacks9} as a result of our stacking procedure. The full version of this table appears as an electronic table only.

\begin{table*}
	\caption{Table of individual line-of-sight measurements}
	
	\label{table:rad_measurements}
	\centering
	\begin{tabular}{lccccccccc}
		\hline\hline
		Galaxy & Radius & $I_\textrm{HCN}$ & $\Delta_\textrm{HCN}$ & $I_\textrm{HCO+}$ & $\Delta_\textrm{HCO+}$ & $I_\textrm{HNC}$ & $\Delta_\textrm{HNC}$ & $I_{^{12}\textrm{CO}}$ & \nodata\\
		\hline
		 & $r_{25}$ & (K km s$^{-1}$) & (K km s$^{-1}$) & (K km s$^{-1}$) & (K km s$^{-1}$) & (K km s$^{-1}$) & (K km s$^{-1}$) & (K km s$^{-1}$) & \nodata\\
		\hline
NGC~628 &  0.00 &  0.21 &  0.05 &   NaN &  0.14 &   NaN &  0.22 &  7.67 &  \nodata\\
NGC~628 &  0.06 &  0.14 &  0.04 &   NaN &  0.13 &   NaN &  0.22 &  7.12 &  \nodata\\
NGC~628 &  0.06 &   NaN &  0.14 &   NaN &  0.14 &   NaN &  0.23 &  6.86 &  \nodata\\
NGC~628 &  0.06 &   NaN &  0.14 &   NaN &  0.15 &   NaN &  0.22 &  7.04 &  \nodata\\
NGC~628 &  0.06 &   NaN &  0.16 &   NaN &  0.14 &   NaN &  0.21 &  6.88 &  \nodata\\
NGC~628 &  0.06 &   NaN &  0.15 &   NaN &  0.14 &   NaN &  0.19 &  7.05 &  \nodata\\
NGC~628 &  0.06 &   NaN &  0.16 &   NaN &  0.16 &   NaN &  0.24 &  6.86 &  \nodata\\
		\nodata & \nodata & \nodata & \nodata & \nodata & \nodata & \nodata & \nodata & \nodata & \nodata\\
		\hline
	\end{tabular}
	\\ \flushleft{{\bf Notes:} Uncertainties:
1) Where intensity measurements are below the significance threshold ($3\sigma$ RMS), columns 3-14 contain NaN for the integrated intensities, and upper limits to the emission in the respective uncertainty column.
2) The data were sampled at a common angular resolution of $33\arcsec$.
3) The full version of this table appears as online only material.}
\end{table*}

\begin{table*}
	\caption{Table of individual radial profiles}
	
	\label{table:rad_profiles}
	\centering
	\begin{tabular}{lccccccccc}
		\hline\hline
		Galaxy & Radius & $I_\textrm{HCN}$ & $\Delta_\textrm{HCN}$ & $I_\textrm{HCO+}$ & $\Delta_\textrm{HCO+}$ & $I_\textrm{HNC}$ & $\Delta_\textrm{HNC}$ & $I_{^{12}\textrm{CO}}$ & \nodata\\
		\hline
		 & (kpc) & (K km s$^{-1}$) & (K km s$^{-1}$) & (K km s$^{-1}$) & (K km s$^{-1}$) & (K km s$^{-1}$) & (K km s$^{-1}$) & (K km s$^{-1}$) & \nodata\\
		\hline
NGC~628 &  1.02 &  0.08 &  0.04 &  0.06 &  0.00 &  0.04 &  0.00 &  6.54 &  \nodata\\
NGC~628 &  2.05 &  0.05 &  0.03 &  0.02 &  0.00 &  0.04 &  0.00 &  4.62 &  \nodata\\
NGC~628 &  3.27 &  0.05 &  0.02 &  0.03 &  0.02 &  0.02 &  0.00 &  4.01 &  \nodata\\
NGC~628 &  4.65 &  0.04 &  0.02 &  0.03 &  0.02 &   NaN &   NaN &  2.69 &  \nodata\\
NGC~628 &  7.00 &  0.04 &   NaN &  0.01 &   NaN &   NaN &   NaN &  1.09 &  \nodata\\
\hline
NGC~2903 & 0.73 &  1.16 &  0.04 &  0.81 &  0.05 &  0.31 &  0.06 & 31.38 &  \nodata\\
NGC~2903 & 1.78 &  0.68 &  0.02 &  0.42 &  0.03 &  0.29 &  0.02 & 23.96 &  \nodata\\
		\nodata & \nodata & \nodata & \nodata & \nodata & \nodata & \nodata & \nodata & \nodata & \nodata\\
		\hline
	\end{tabular}
	\\ \flushleft{{\bf Notes:} Uncertainties:
1) Where intensity measurements are below the significance threshold ($3\sigma$ RMS), columns 3-14 contain NaN for the integrated intensities, and upper limits to the emission in the respective uncertainty column.
2) The full version of this table appears as online only material. The radius provided corresponds to the outer edge of each selected ring.}
\end{table*}

\section*{APPENDIX C. Literature data}
\label{appendix:literature}

\setcounter{figure}{0}
\setcounter{table}{0}
\renewcommand{\thefigure}{C\arabic{figure}}
\renewcommand{\thetable}{C\arabic{table}}

Table \ref{table:literature} provides a subset of the most up-to-date dense gas observations in the literature, as traced by the HCN\,(1-0) emission line. Its full version appears as online material only. This compilation includes the data used for constructing Figure \ref{fig:all_hcn}. When using this compilation table, please refer to the original studies of the various datasets included.

\begin{table*}
	\caption{Dense gas literature compilation}
	
	\label{table:literature}
	\centering
	\begin{tabular}{lcc}
		\hline\hline
		Reference & $\textrm{log}_{10}(L_{\textrm{IR}})$ & $\textrm{log}_{10}(L_{\textrm{HCN}})$ \\
		\hline
		 & ($L_\odot$) & (K km s$^{-1}\,\textrm{pc}^{2}$)\\
		\hline
		\citet{GARCIACARPIO08} &  12.26 &       9.26 \\
        &       12.24 &       9.06 \\
        &       12.18 &       9.19 \\
        &       12.07 &       9.10 \\
        &       11.99 &       8.86 \\
        &       11.98 &       8.85 \\
        &       11.88 &       9.00 \\
        &       11.86 &       8.81 \\
        &       11.66 &       8.77 \\
        &       11.61 &       8.75 \\
        &       11.54 &       8.43 \\
        &       11.53 &       8.30 \\
        &       11.41 &       8.05 \\
        &       11.23 &       7.87 \\
        &       11.36 &       8.30 \\
		\nodata & \nodata & \nodata \\
		\hline
	\end{tabular}
	\\ \flushleft{{\bf Notes:} This table is a literature compilation, and its full version appears as online only material. Please ensure to cite each individual study when making use of the contents of this table.}
\end{table*}

\section*{APPENDIX D. Individual galaxy trends}
\label{appendix:trends}

\setcounter{figure}{0}
\setcounter{table}{0}
\renewcommand{\thefigure}{D\arabic{figure}}
\renewcommand{\thetable}{D\arabic{table}}

\begin{figure*}
	\includegraphics[scale=0.35]{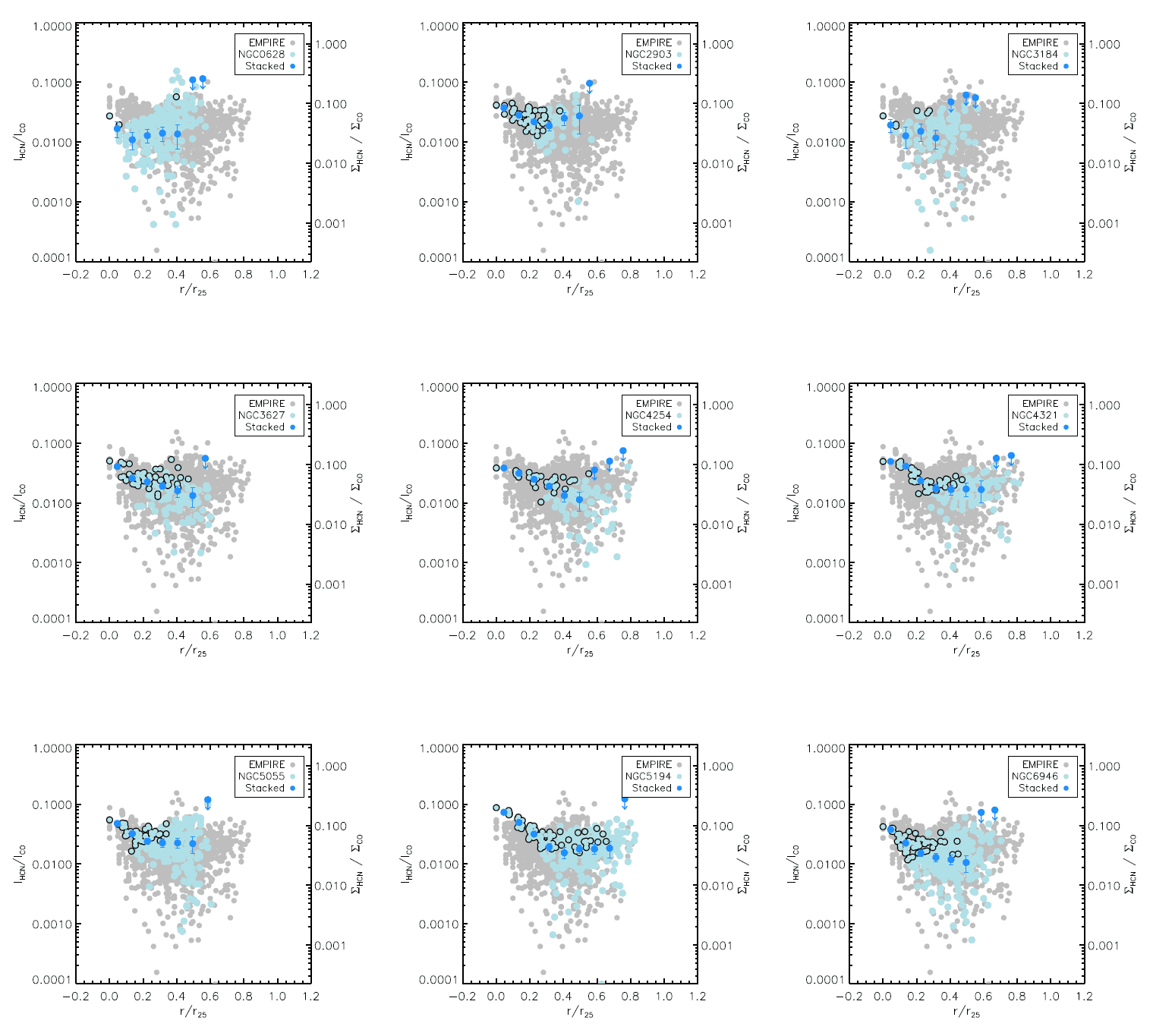}
	\caption{HCN-to-CO ratio (left axis) and $\Sigma_{\textrm{HCN}}/\Sigma_{\textrm{CO}}$ (right axis), tracing the dense gas fraction, as a function of the normalized galactocentric radius $r/r_{25}$. Each data point represents a $\sim$kpc-size measurement per line of sight. Grey points show all EMPIRE lines of sight. Light blue points indicate meaurements for each selected galaxy. Points with black outlines show regions where HCN is detected at S/N$>$3. Dark blue points show the stacked HCN data, which recovers signal in low S/N regions; downward arrows give a lower limit to the ratio in those regions where HCN is not detected. The dense gas fraction in all galaxy disks appears to decrease at larger galactocentric radii. We note that the plots above are in logarithmic scale, therefore non-detections with negative values cannot be represented. These are, however, taken into account in the stacked intensities.}
	\label{fig:fdense_rad}
\end{figure*}

\begin{figure*}
	\includegraphics[scale=0.35]{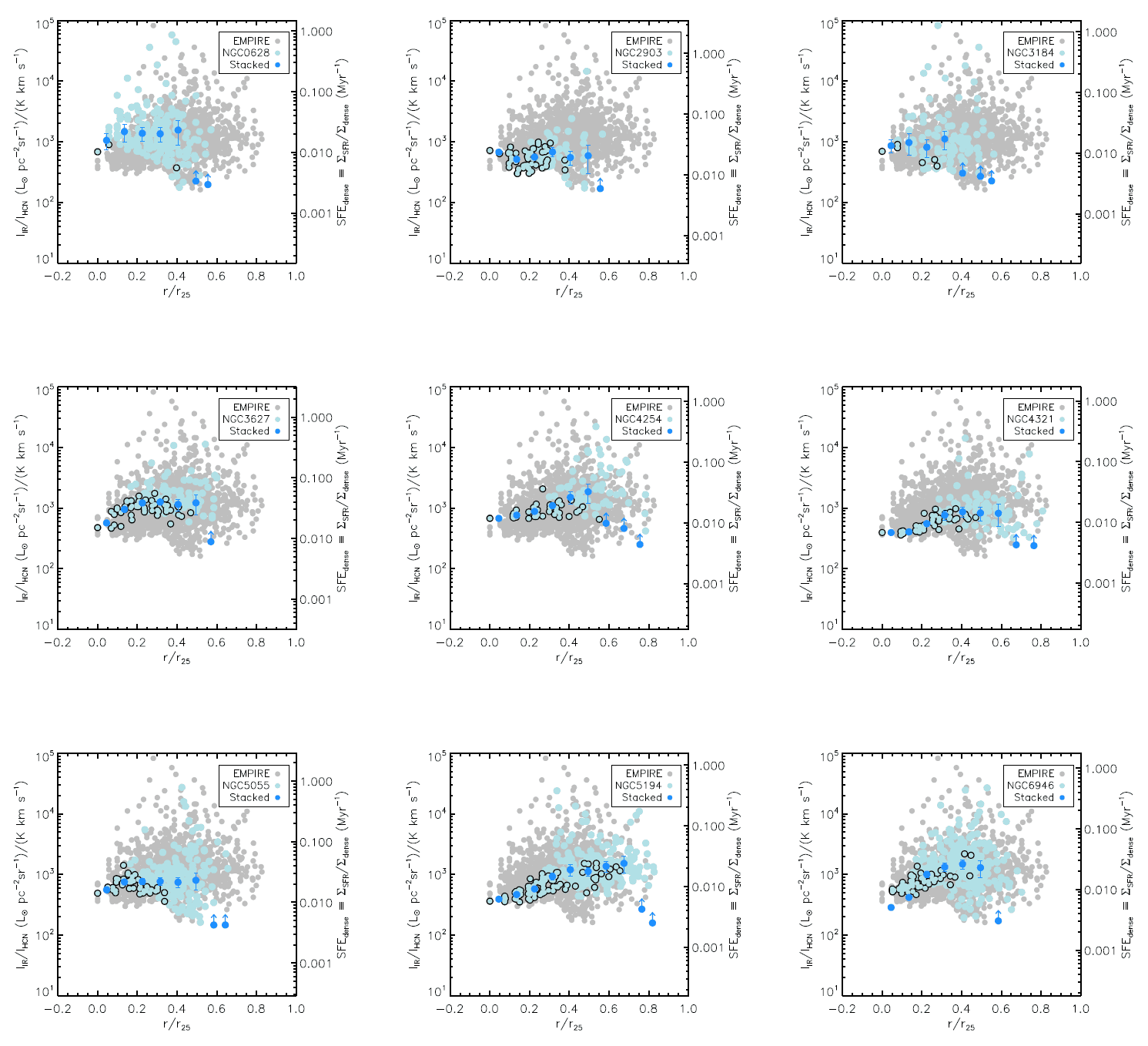}
	\caption{IR-to-HCN ratio (left axis) and $\Sigma_{\textrm{SFR}}/\Sigma_{\textrm{HCN}}$ (right axis), tracing the efficiency of dense gas, as a function of the normalized galactocentric radius $r/r_{25}$. The description of each data point is the same as in Figure \ref{fig:fdense_rad}. The dense gas efficiency appears to decrease for higher values of $\Sigma_*$. We emphasize that NGC\,2903 lacks Herschel data, and its SFR is calculated only using the available 24$\mu$m, which is a less accurate procedure.}
	\label{fig:sfe_rad}
\end{figure*}

\begin{figure*}
	\includegraphics[scale=0.35]{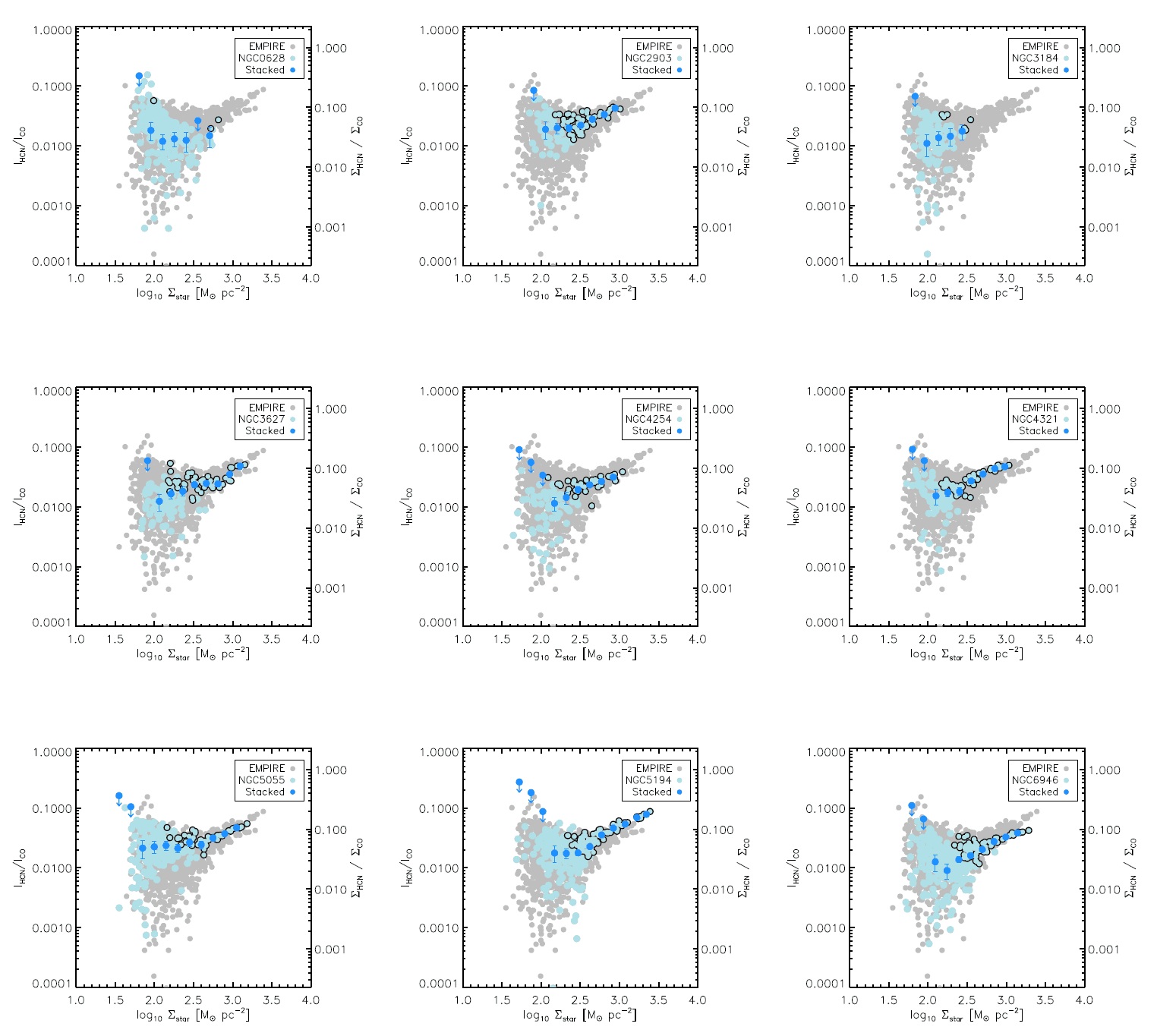}
	\caption{HCN-to-CO ratio (left axis) and $\Sigma_{\textrm{HCN}}/\Sigma_{\textrm{CO}}$ (right axis), tracing the dense gas fraction, as a function of the surface density of stars, $\Sigma_*$. The description of each data point is the same as in Figure \ref{fig:fdense_rad}. The dense gas fraction in all galaxy disks appears to decrease for higher values of $\Sigma_*$. }
	\label{fig:fdense_star}
\end{figure*}

\begin{figure*}
	\includegraphics[scale=0.35]{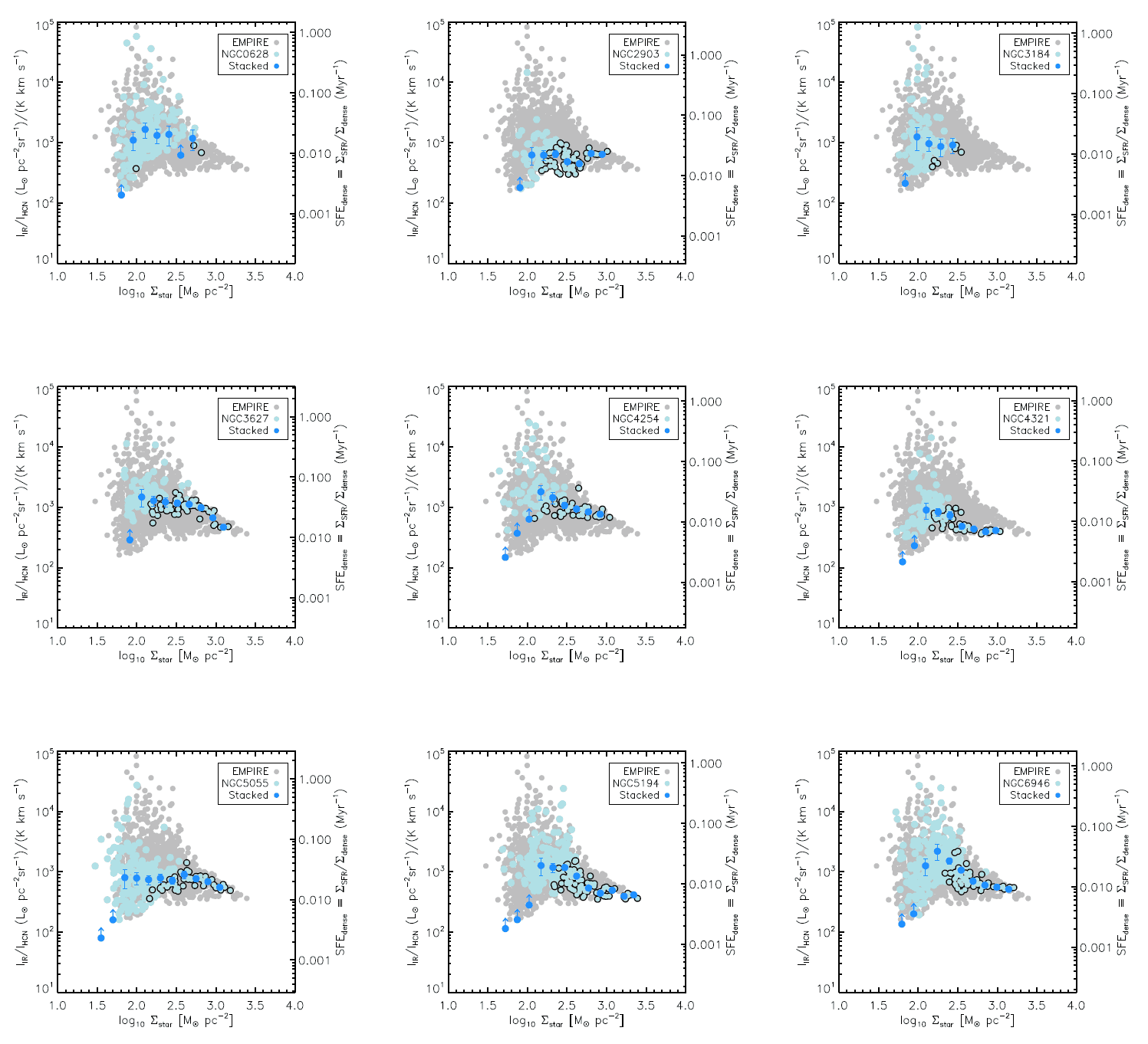}
	\caption{IR-to-HCN ratio (left axis) and $\Sigma_{\textrm{SFR}}/\Sigma_{\textrm{HCN}}$ (right axis), tracing the efficiency of dense gas, as a function of the surface density of stars, $\Sigma_*$. The description of each data point is the same as in Figure \ref{fig:fdense_rad}. The dense gas efficiency appears to decrease for higher values of $\Sigma_*$. We emphasize that NGC\,2903 lacks Herschel data, and its SFR is calculated only using the available 24$\mu$m, which is a less accurate procedure.}
	\label{fig:sfe_star}
\end{figure*}

In this Section we present the individual line-of-sight measurements of the observed HCN-to-CO and TIR-to-HCN line ratios in every galaxy disk as a function of galactocentric radius (Figures \ref{fig:fdense_rad} and \ref{fig:sfe_rad}), stellar surface density $\Sigma_*$ (Figures \ref{fig:fdense_star} and \ref{fig:sfe_star}), molecular-to-atomic gas ratio $R_\textrm{mol}$ (Figures \ref{fig:fdense_mol} and \ref{fig:sfe_star}) and the local dynamical equilibrium pressure $P_{\rm DE}$ (Figures \ref{fig:fdense_pressure} and \ref{fig:sfe_pressure}). In all figures the light gray datapoints represent the entire EMPIRE survey, while the light blue datapoints represent the line-of-sight measurements for each individual galaxy. Light blue points with black outlines show points in the galaxy where HCN is detected at S/N$>3$. Dark blue points show the stacked trends shown in Figure \ref{fig:global_trends}, which indicate systematic variations of the HCN-to-CO (as a proxy for the dense gas fraction) and IR-to-HCN (as a proxy for the star formation efficiency of the dense gas) line ratios as a function of galactic environment.

\begin{figure*}
	\includegraphics[scale=0.14]{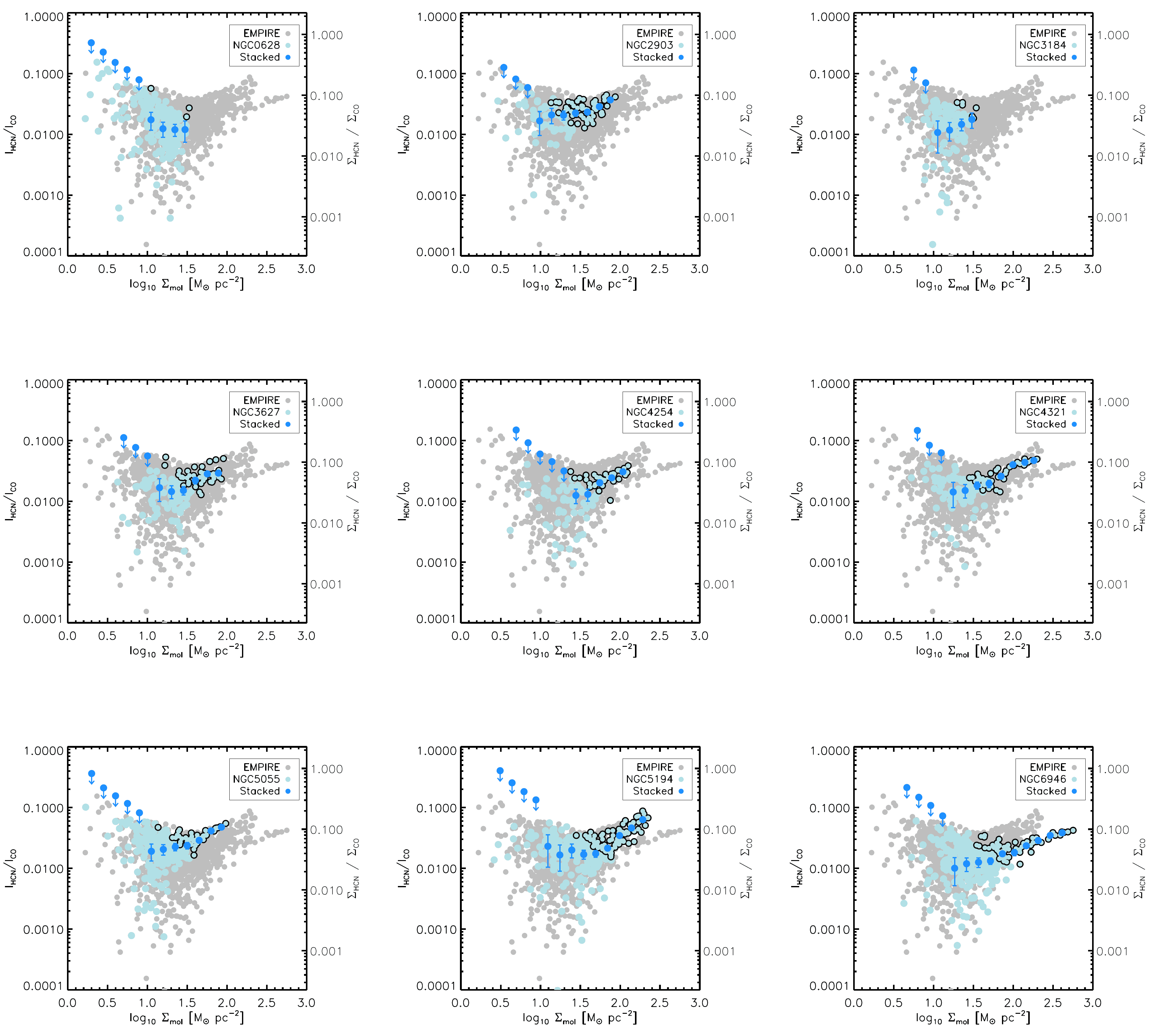}
	\caption{HCN-to-CO ratio (left axis) and $\Sigma_{\textrm{HCN}}/\Sigma_{\textrm{CO}}$ (right axis), tracing the dense gas fraction, as a function of the surface density of molecular gas as traced by CO. The description of each data point is the same as in Figure \ref{fig:fdense_rad}. The dense gas fraction in all galaxy disks appears to decrease for higher values of $\Sigma_*$. }
	\label{fig:fdense_gas}
\end{figure*}

\begin{figure*}
	\includegraphics[scale=0.14]{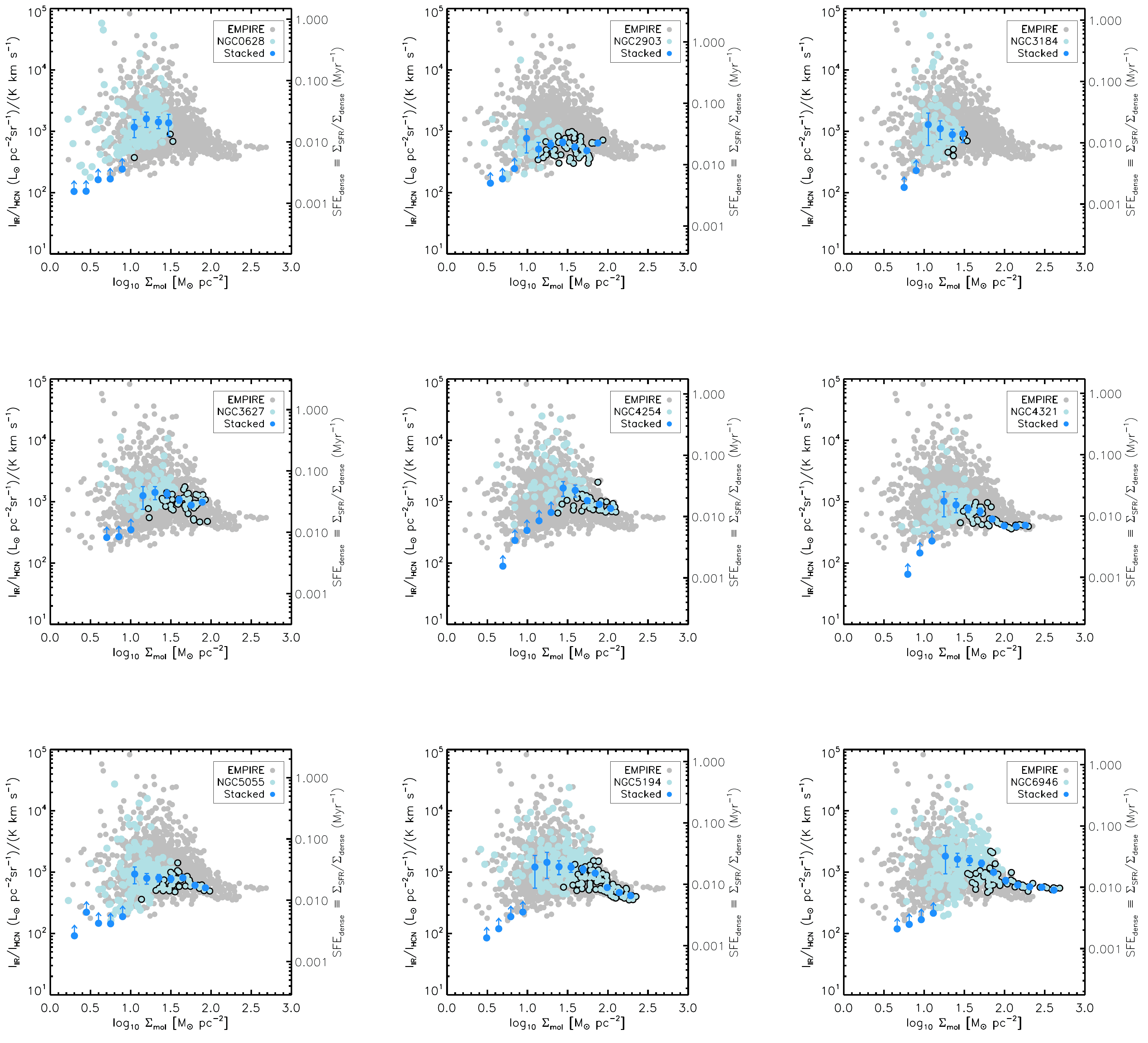}
	\caption{IR-to-HCN ratio (left axis) and $\Sigma_{\textrm{SFR}}/\Sigma_{\textrm{HCN}}$ (right axis), tracing the efficiency of dense gas, as a function of the surface density of molecular gas as traced by CO. The description of each data point is the same as in Figure \ref{fig:fdense_rad}. The dense gas efficiency appears to decrease for higher values of $I_\textrm{CO}$. We emphasize that NGC\,2903 lacks Herschel data, and its SFR is calculated only using the available 24$\mu$m, which is a less accurate procedure.}
	\label{fig:sfe_gas}
\end{figure*}

\begin{figure*}
	\includegraphics[scale=0.35]{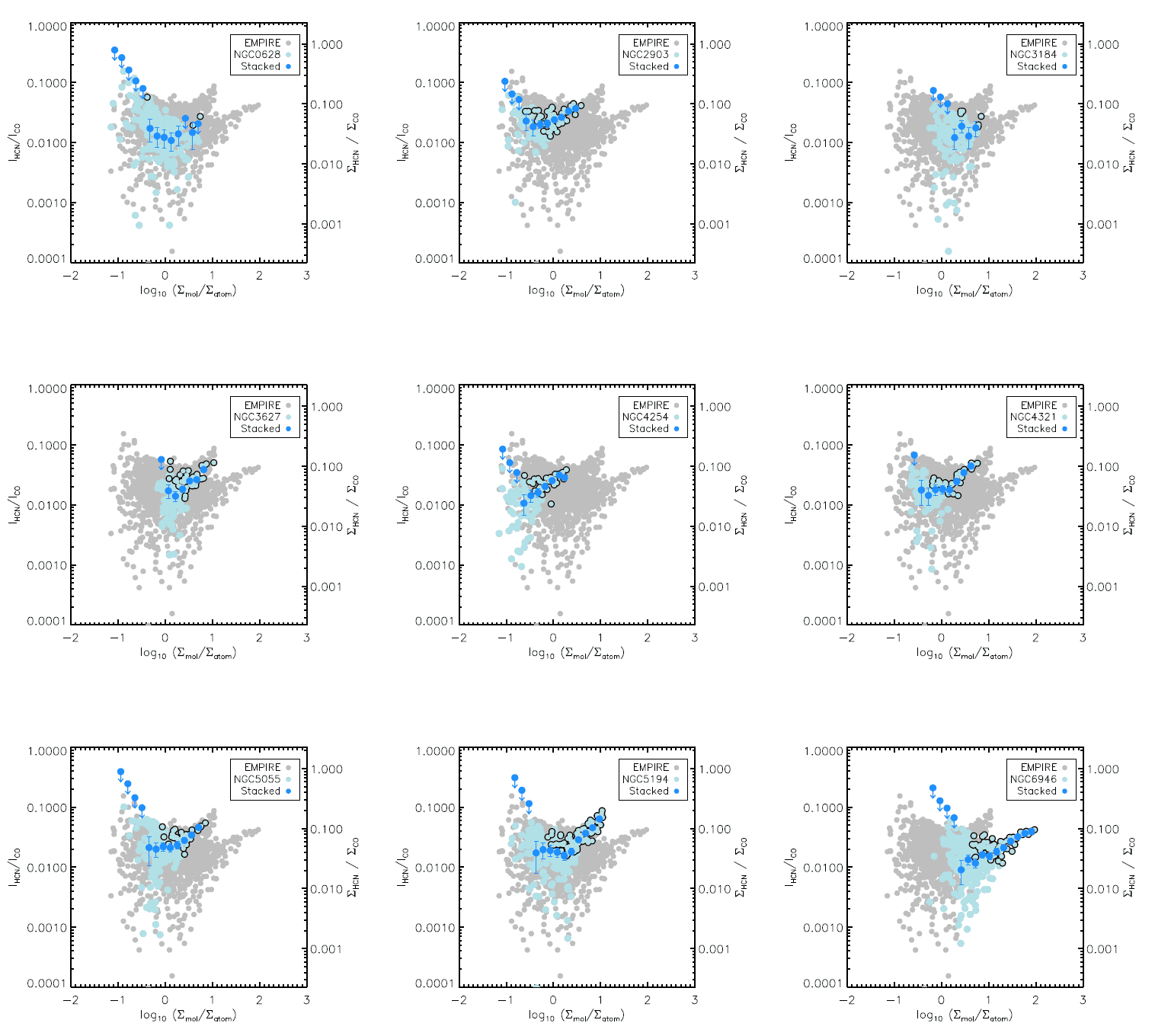}
	\caption{HCN-to-CO ratio (left axis) and $\Sigma_{\textrm{HCN}}/\Sigma_{\textrm{CO}}$ (right axis), tracing the dense gas fraction, as a function of the molecular-to-atomic gas ratio, $R_\textrm{mol}=\Sigma_{\textrm{CO}}/\Sigma_{\textrm{H\,I}}$. The description of each data point is the same as in Figure \ref{fig:fdense_star}. The dense gas fraction in all galaxy disks appears to correlate with the molecular gas fraction, as traced by $f_\textrm{mol}$.}
	\label{fig:fdense_mol}
\end{figure*}

\begin{figure*}
	\includegraphics[scale=0.14]{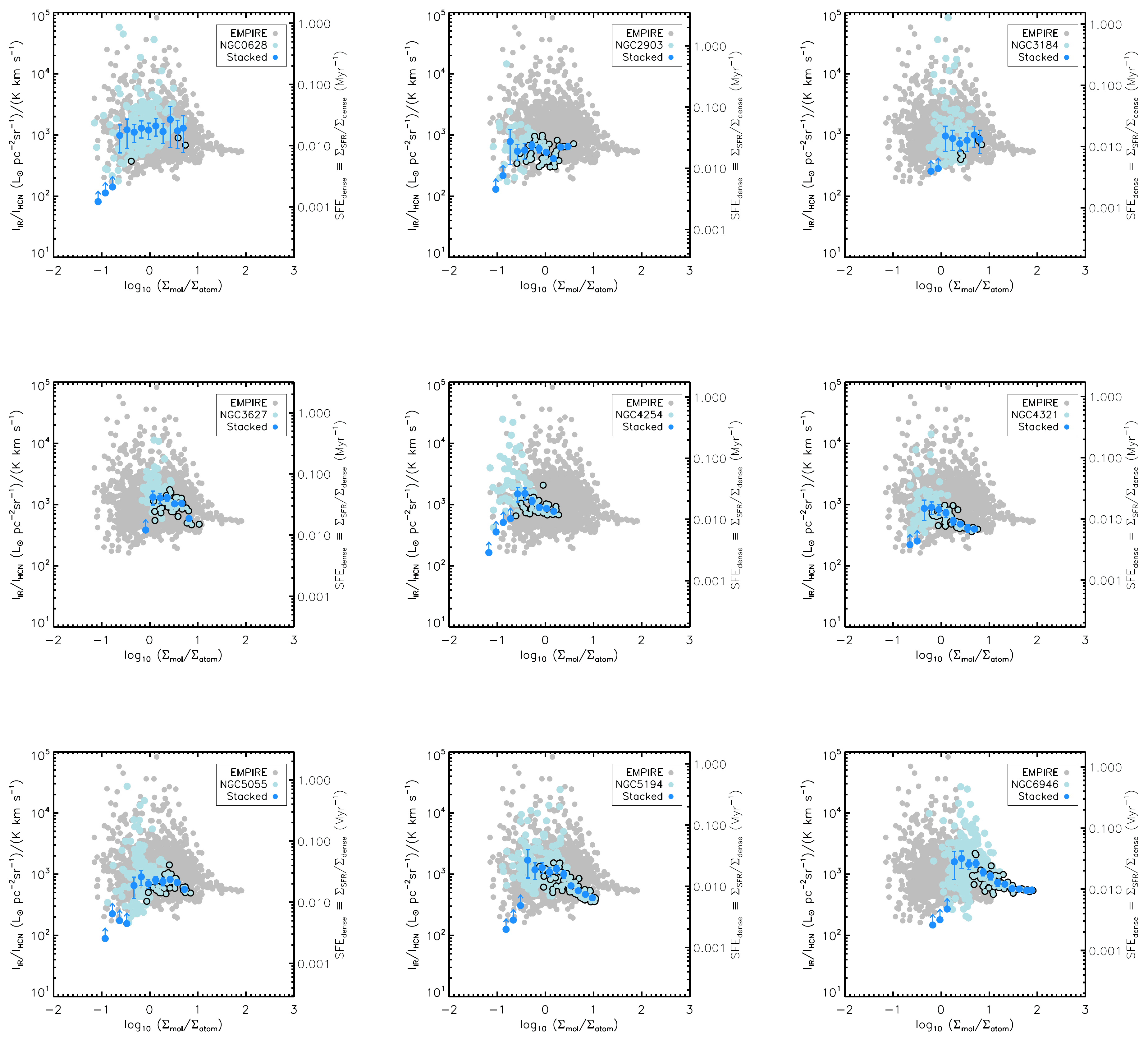}
	\caption{IR-to-HCN ratio (left axis) and $\Sigma_{\textrm{SFR}}/\Sigma_{\textrm{HCN}}$ (right axis), tracing the efficiency of dense gas, as a function of the molecular-to-atomic gas ratio, $R_\textrm{mol}=\Sigma_{\textrm{CO}}/\Sigma_{\textrm{H\,I}}$. The description of each data point is the same as in Figure \ref{fig:fdense_star}. The dense gas efficiency appears to decrease for higher values of $f_\textrm{mol}$.}
	\label{fig:sfe_mol}
\end{figure*}

\begin{figure*}
	\includegraphics[scale=0.35]{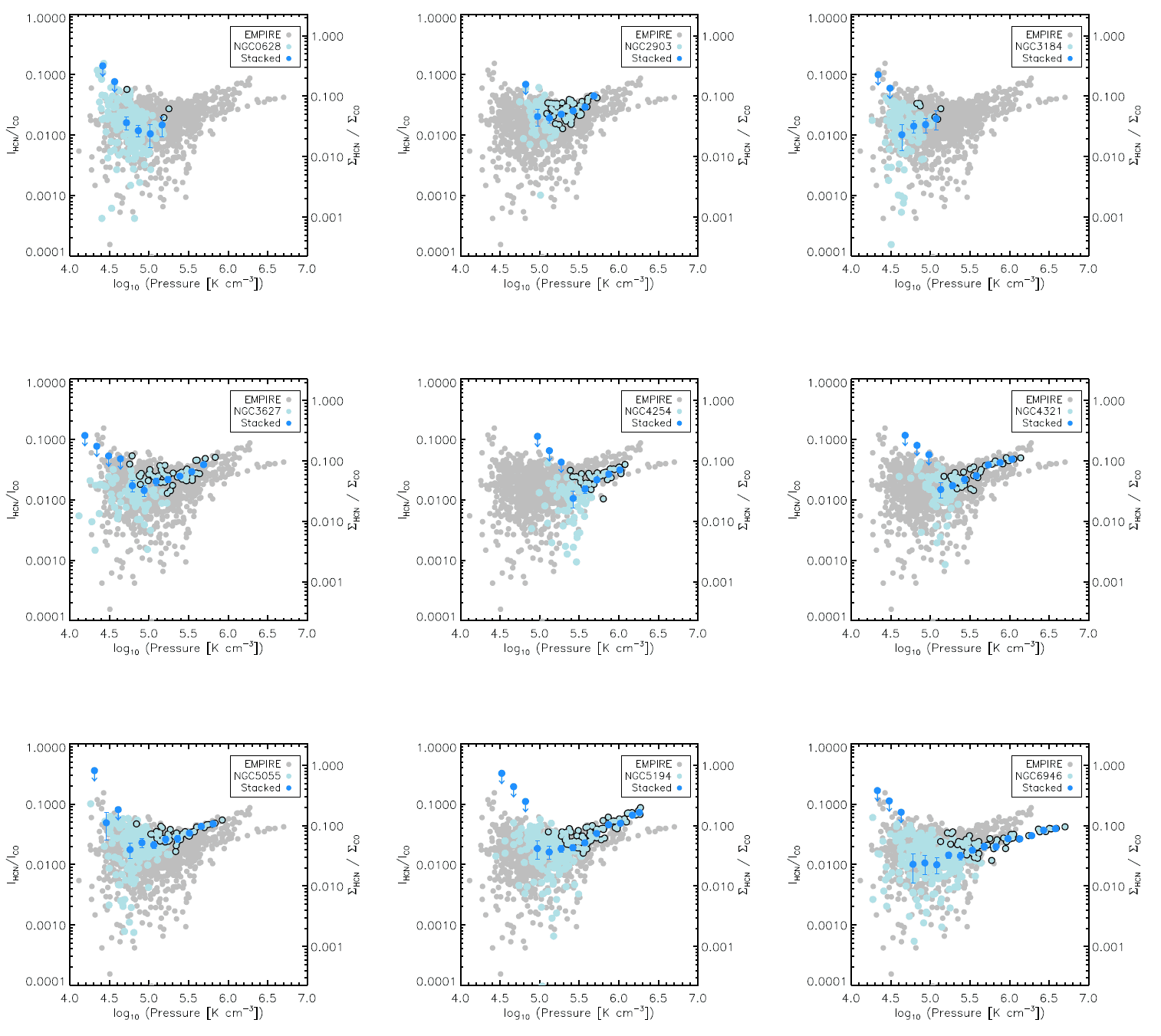}
	\caption{HCN-to-CO ratio (left axis) and $\Sigma_{\textrm{HCN}}/\Sigma_{\textrm{CO}}$ (right axis), tracing the dense gas fraction, as a function of the local dynamical equilibrium pressure, $P_{\rm DE}$. The description of each data point is the same as in Figure \ref{fig:fdense_rad}. The dense gas fraction in all galaxy disks appears to increase for higher ISM pressures.}
	\label{fig:fdense_pressure}
\end{figure*}

\begin{figure*}
    \includegraphics[scale=0.35]{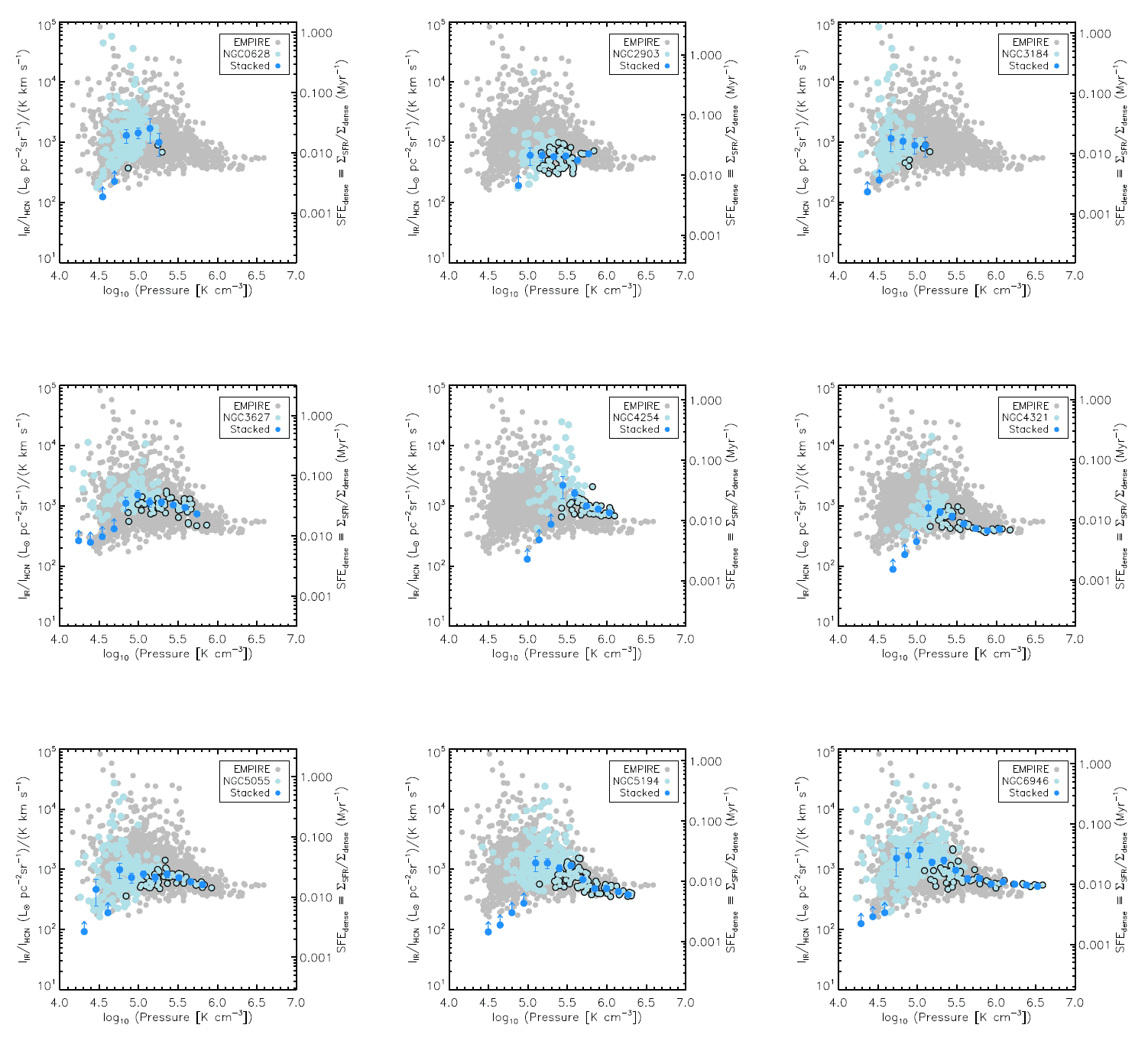}
	\caption{IR-to-HCN ratio (left axis) and $\Sigma_{\textrm{SFR}}/\Sigma_{\textrm{HCN}}$ (right axis), tracing the efficiency of dense gas, as a function of the local dynamical equilibrium pressure, $P_{\rm DE}$. The description of each data point is the same as in Figure \ref{fig:fdense_pressure}. The dense gas efficiency appears to decrease for higher values of the ISM pressure.}
	\label{fig:sfe_pressure}
\end{figure*}

\section*{APPENDIX E. Alternative SFR tracers}
\label{appendix:sfr-tracers}

\setcounter{figure}{0}
\setcounter{table}{0}
\renewcommand{\thefigure}{E\arabic{figure}}
\renewcommand{\thetable}{E\arabic{table}}

In this paper we make use of TIR emission, calculated from a combination of $\lambda=70$, 160 and 250\,$\mu$m {\it Herschel} bands following the prescription from \citet{GALAMETZ13}, as our main SFR tracer. TIR emission is a common SFR tracer in other galaxies \citep[e.g.,][]{GAO04,GARCIABURILLO12,USERO15,BIGIEL16}, which makes it our preferred tracer for a comparison to prior work. The method rests on probing IR emission over the full IR range from dust heated by UV emission from recent star formation. However, one of the main caveats to its usage is that TIR emission is sensitive to stellar populations up to $\sim100$\,Myr \citep{KENNICUTT12}.

An alternative approach is to use a tracer sensitive to more recent massive star formation, like H$\alpha$ emission ($\sim10$\,Myr). These need to be carefully corrected for extinction, however, which is commonly done in other galaxies by combining them with mid-IR measurements accounting for reprocessed starlight at shorter wavelengths. Here we use two of these ``hybrid'' SFR tracers to test for systematic effects in our results by our specific choice of SFR. The first calibration is a linear combination of H$\alpha$ and $24\,\mu$m emission following \citet{CALZETTI07}: 
\begin{equation}
    \label{eq:sfr_calzetti}
\frac{\Sigma_\textrm{SFR}}{M_\odot\,\textrm{yr}^{-1}\,\textrm{kpc}^{-2}}=634\,I_{\textrm{H}\alpha}+0.0025\,I_{24\mu m}, 
\end{equation}

\noindent where $I_{\textrm{H}\alpha}$ is in units of erg s$^{-1}$ cm$^{-2}$ sr$^{-1}$ and $I_{24\mu m}$ in MJy sr$^{-1}$. We also employ a linear combination of FUV intensity and $24\,\mu$m emission, as proposed by \citep{LEROY12},
\begin{equation}
    \label{eq:sfr_leroy12}
\frac{\Sigma_\textrm{SFR}}{M_\odot\,\textrm{yr}^{-1}\,\textrm{kpc}^{-2}}=0.081\,I_{\textrm{FUV}}+0.0032\,I_{24\mu m}, 
\end{equation}

\noindent where $I_{\textrm{FUV}}$ is in units MJy sr$^{-1}$. 

\begin{figure*}
    \centering
    \includegraphics[scale=0.28]{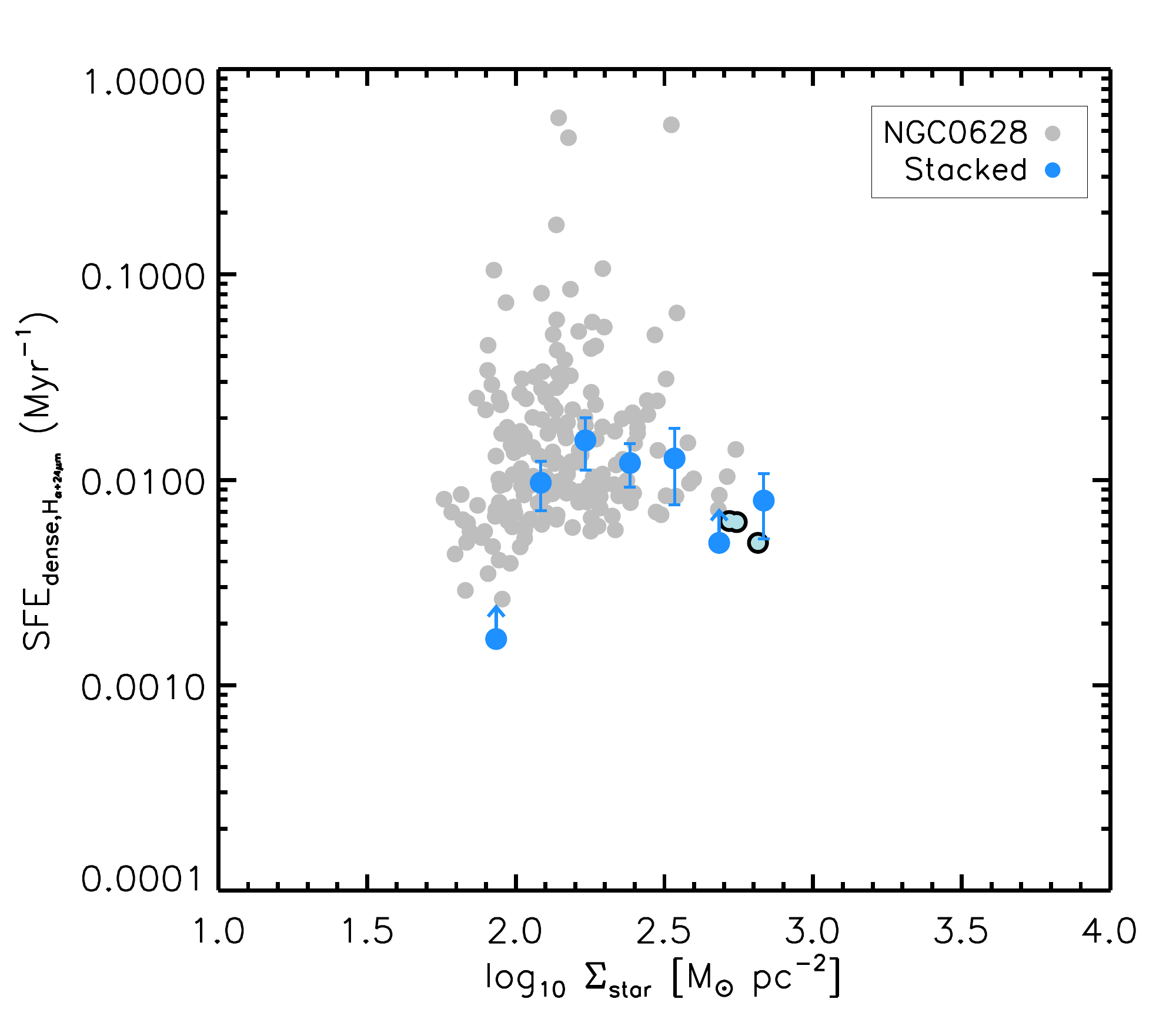}\,\includegraphics[scale=0.28]{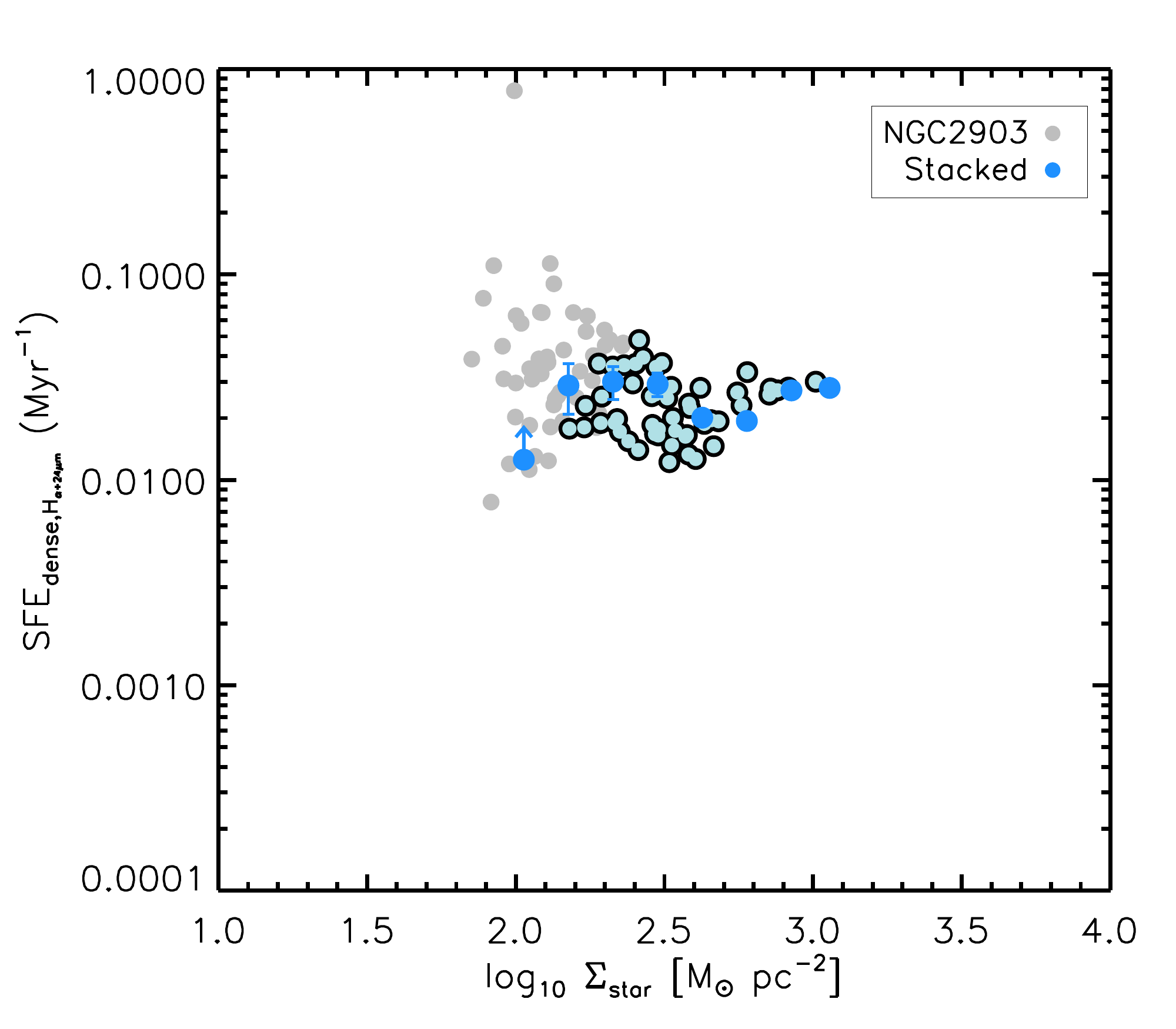}\,\includegraphics[scale=0.28]{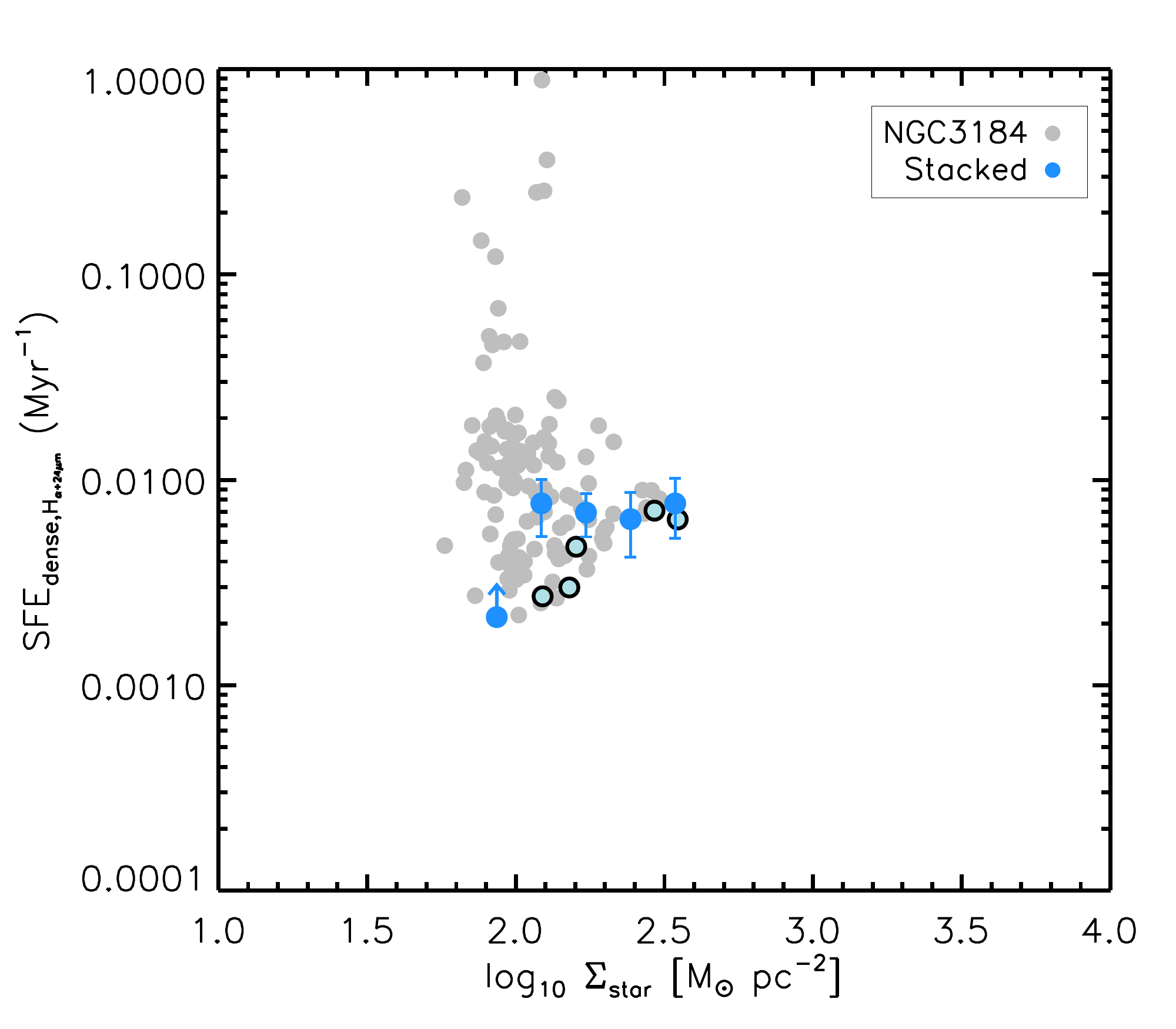} \\
    \includegraphics[scale=0.28]{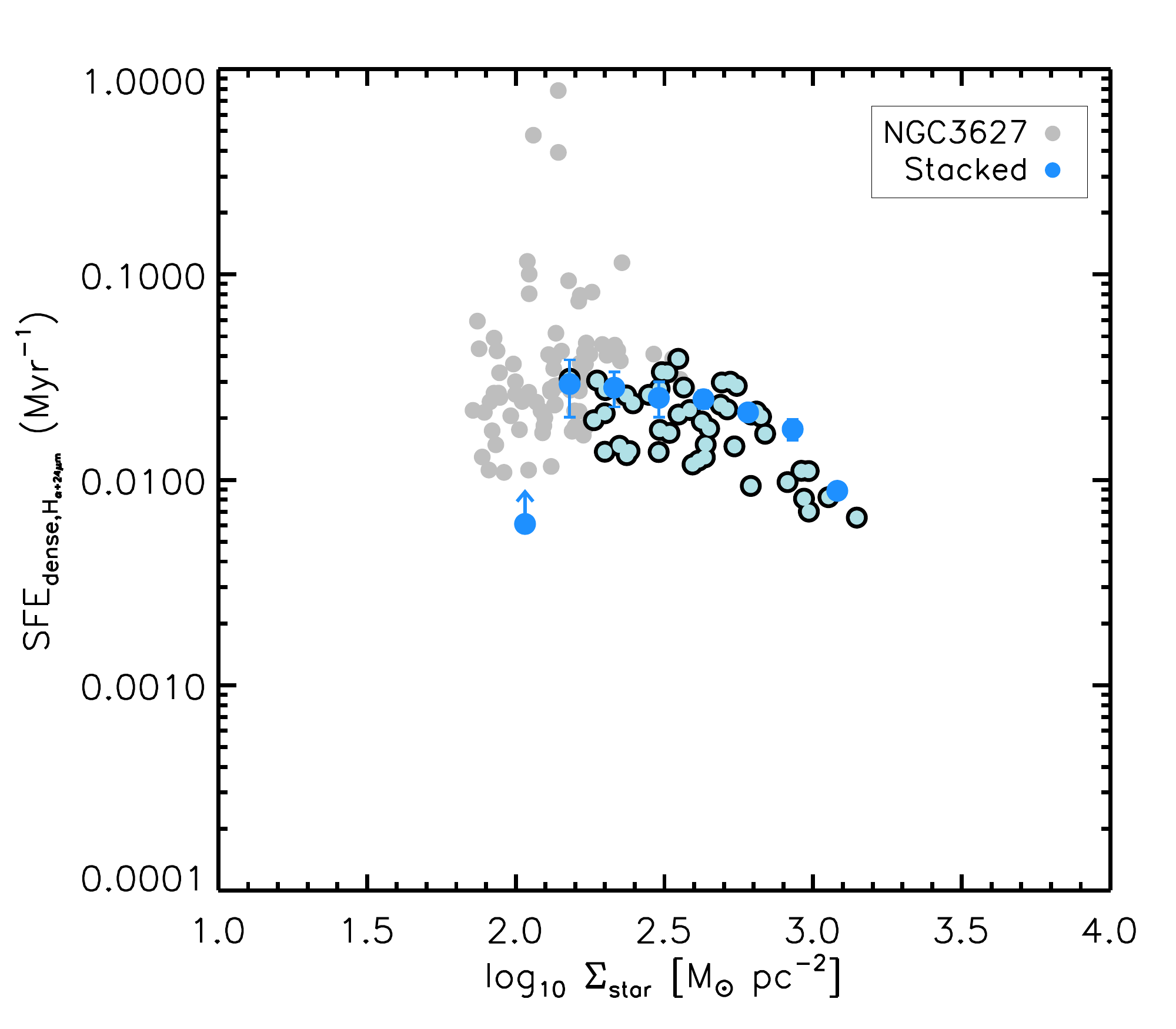}\,\includegraphics[scale=0.28]{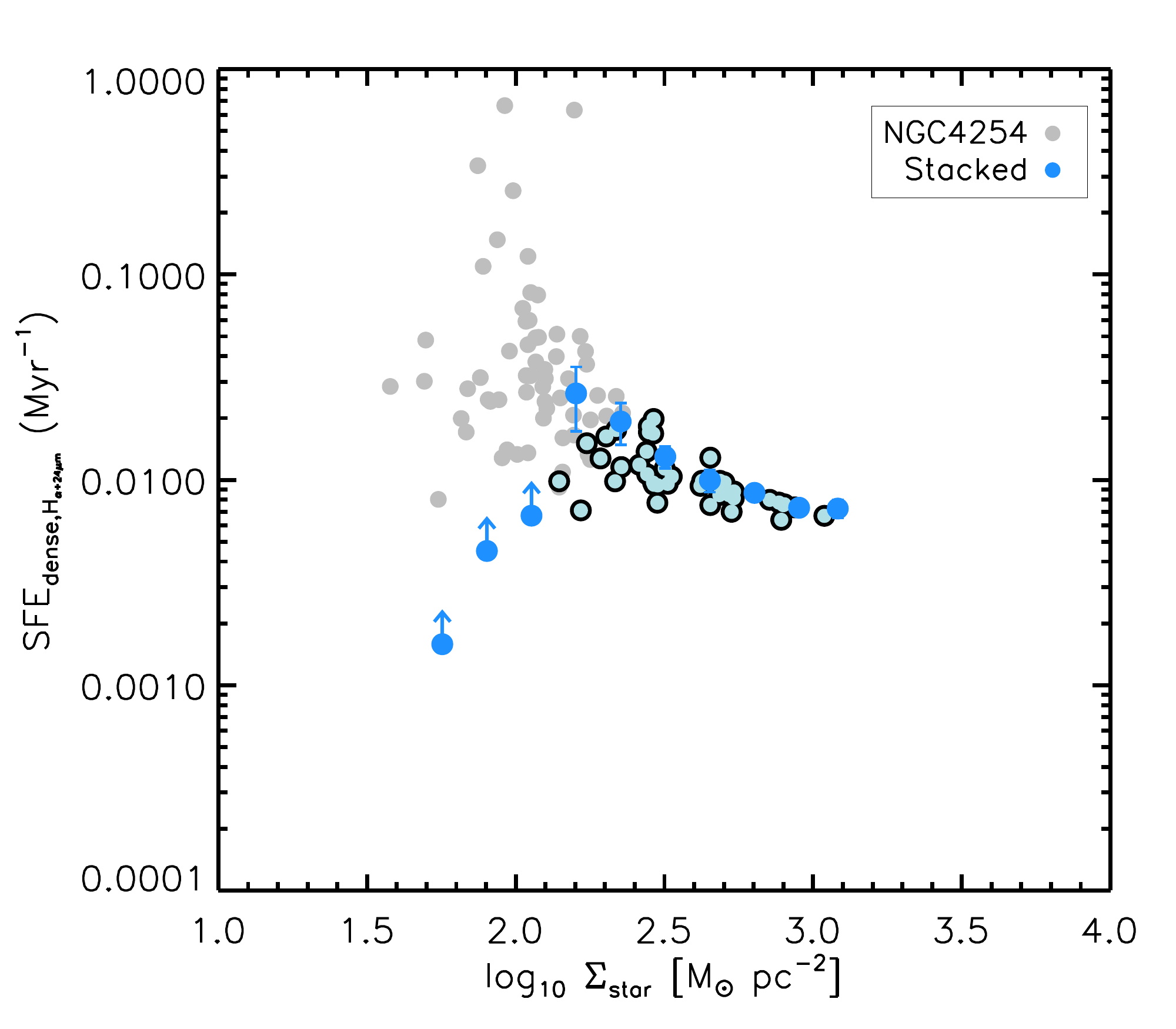}\,\includegraphics[scale=0.28]{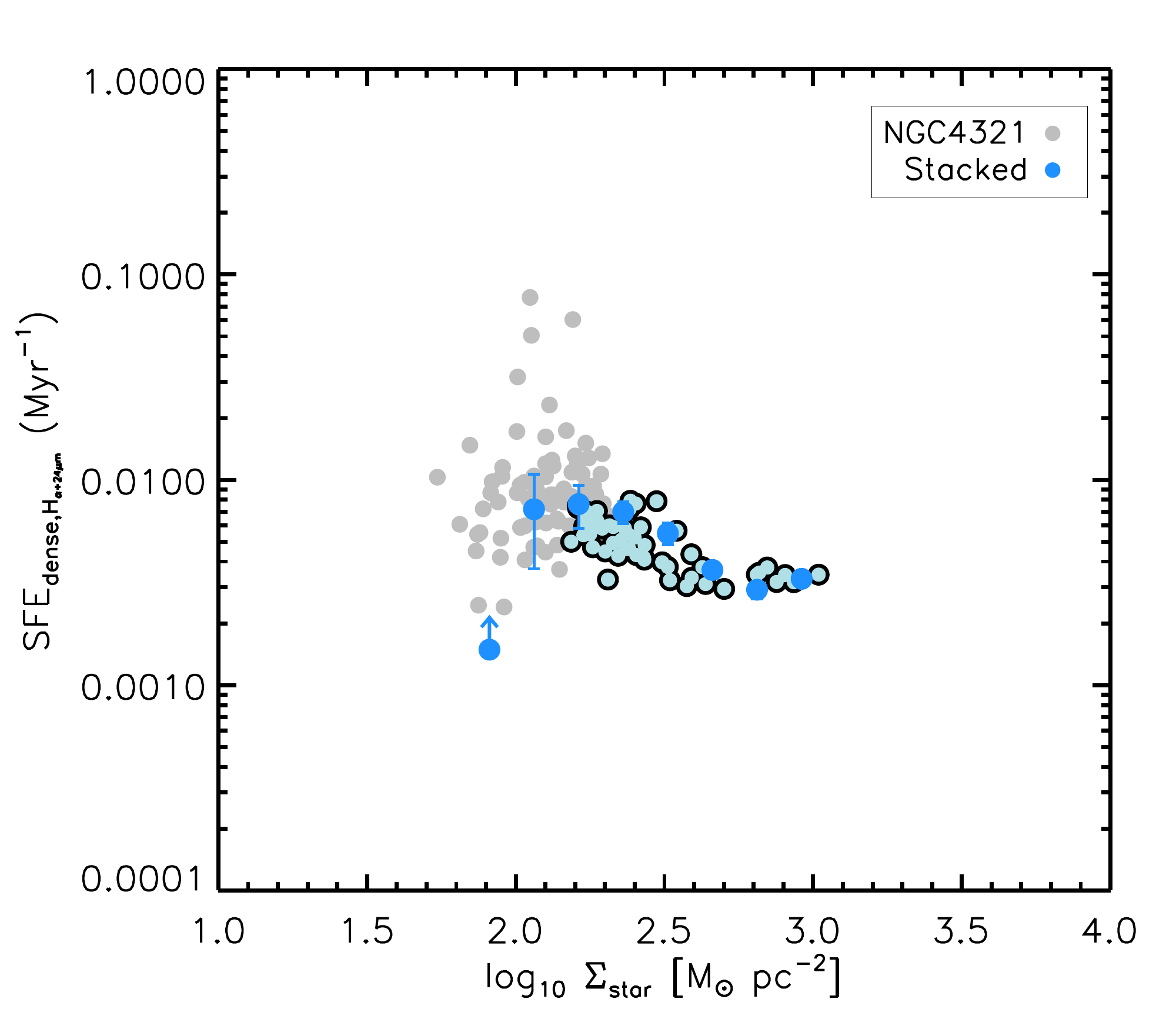} \\
	\includegraphics[scale=0.28]{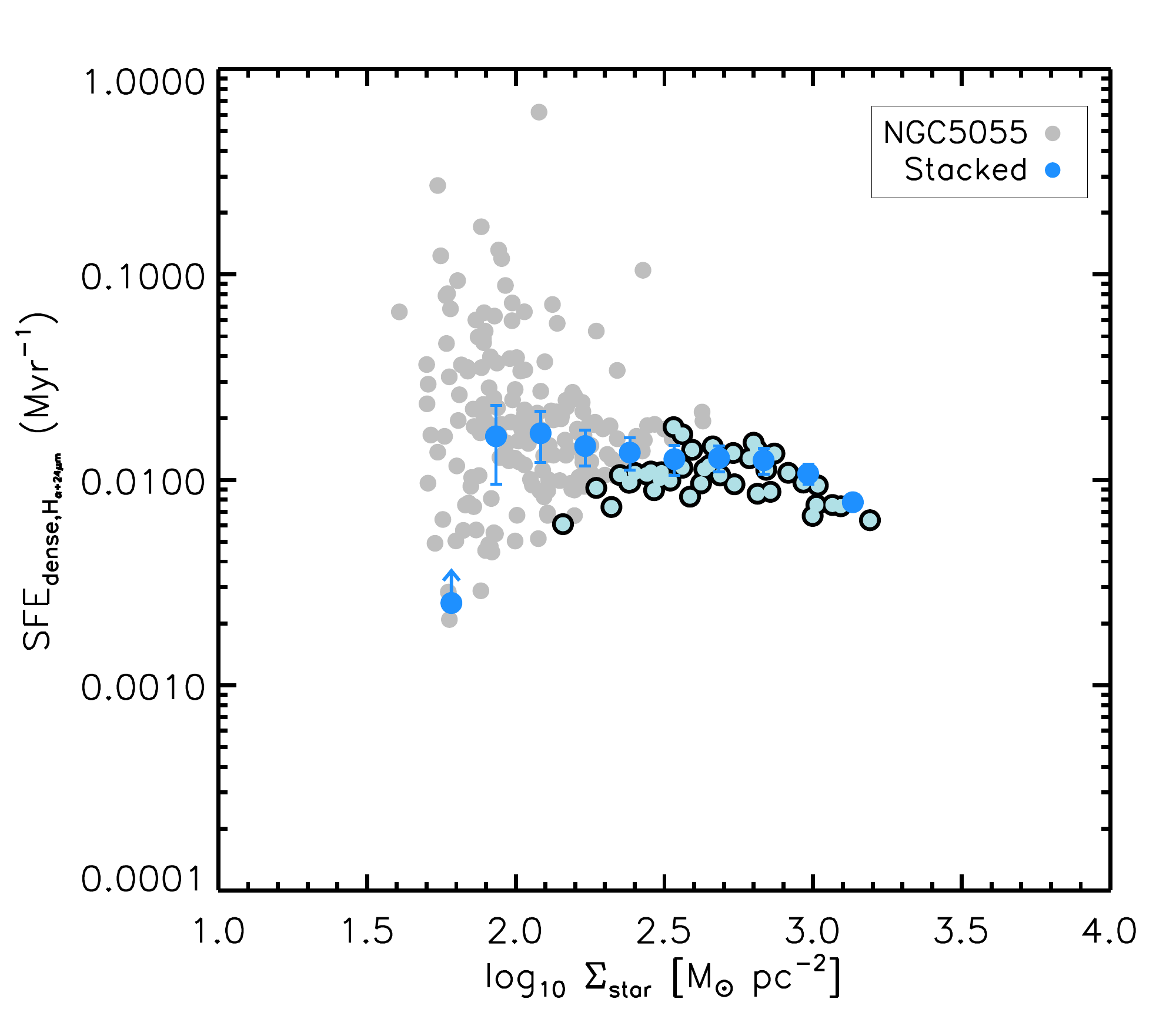}\,\includegraphics[scale=0.28]{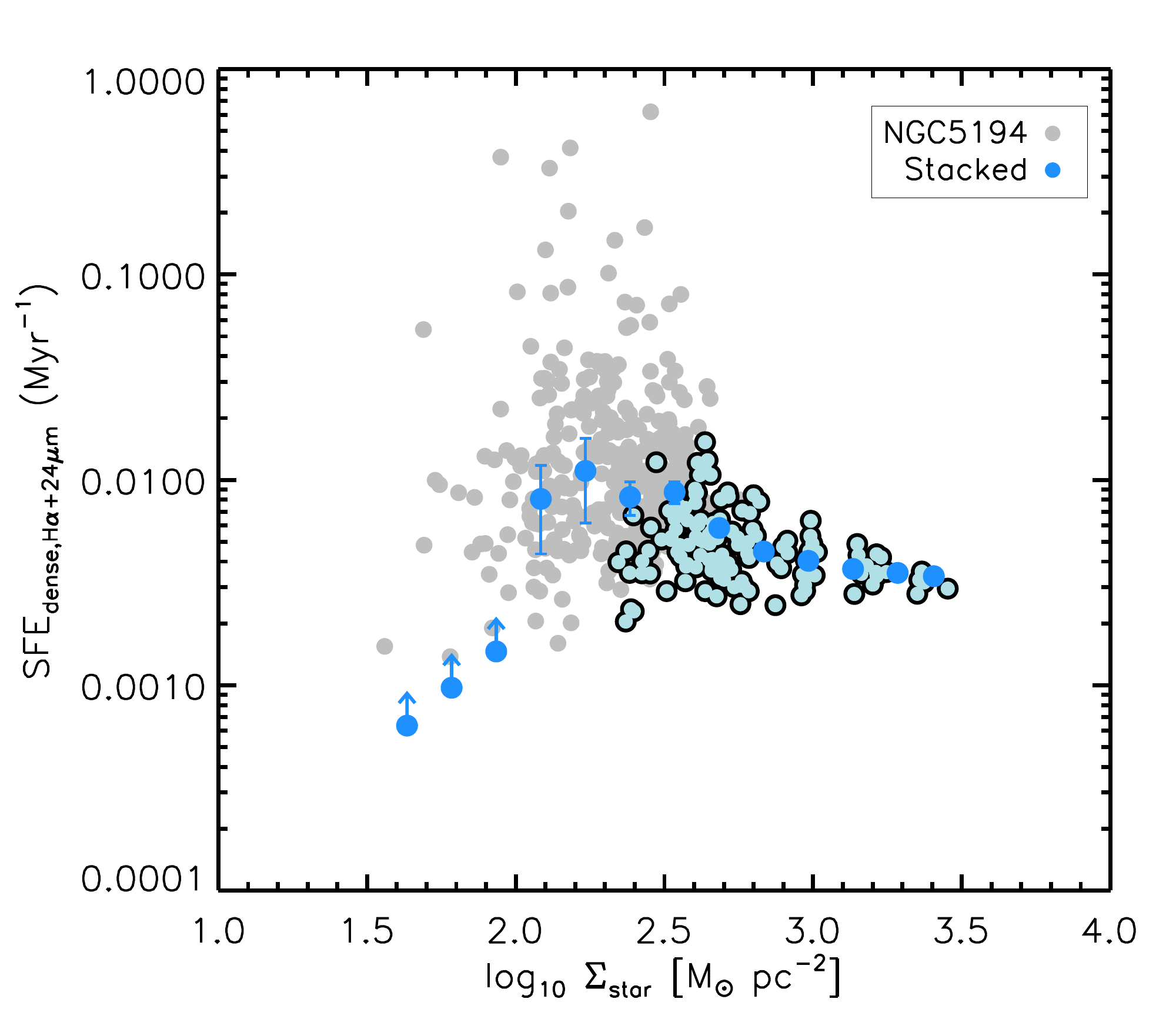}\,\includegraphics[scale=0.28]{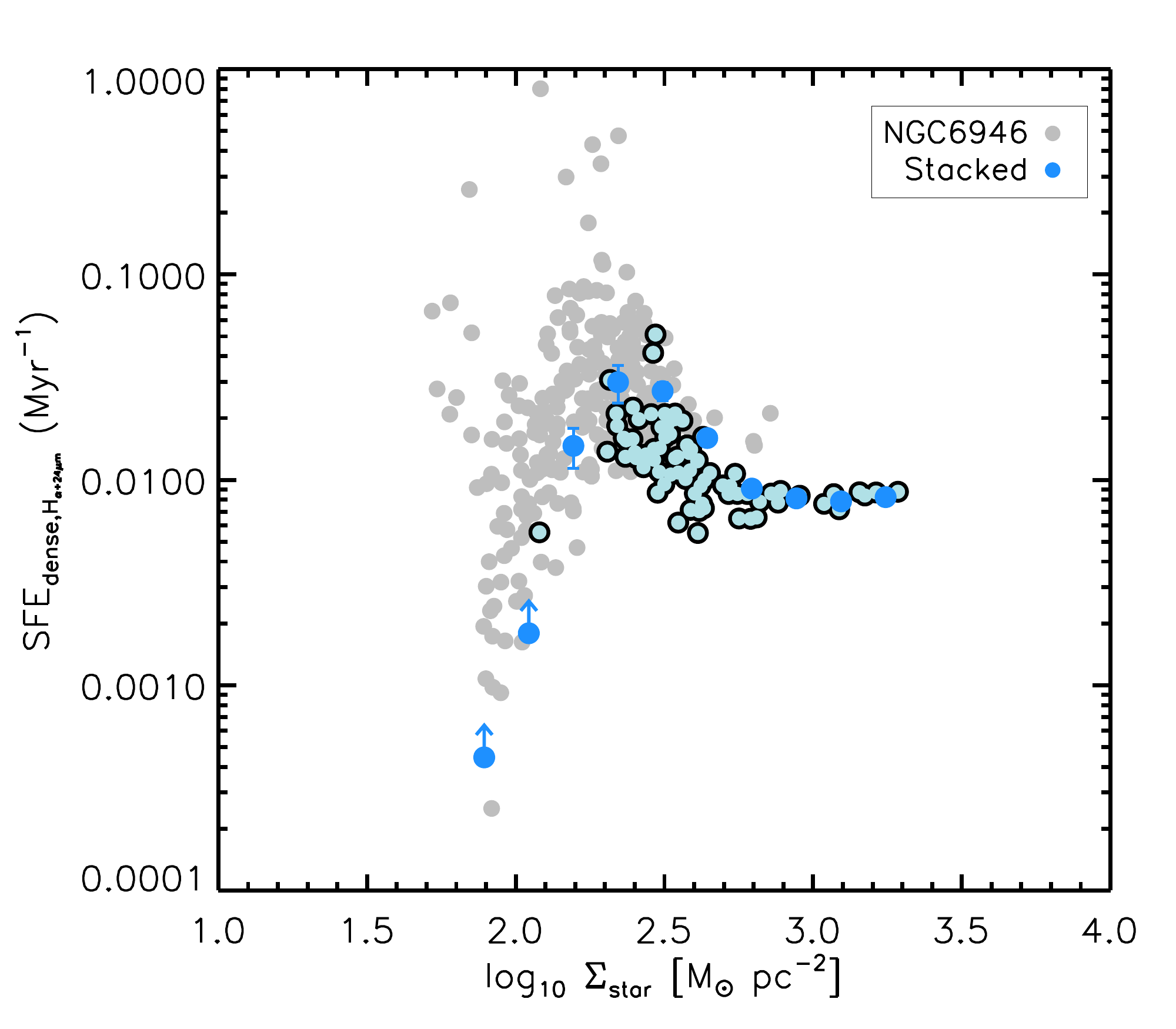}
	\caption{SFE$_\textrm{dense}$ as a function of the stellar surface density of stars for the EMPIRE galaxies. The SFE$_\textrm{dense}$ has been calculated using a linear combination of 24\,$\mu$m and H$\alpha$ intensities as a SFR tracer. The grey points correspond to every line-of-sight measurement, light blue points surrounded by black circles represent measurements above the $3\sigma$ detection limit, and dark blue points represent our stacked results.}
	\label{fig:sfe-tracersha24}
\end{figure*}

\begin{figure*}
    \centering
    \includegraphics[scale=0.28]{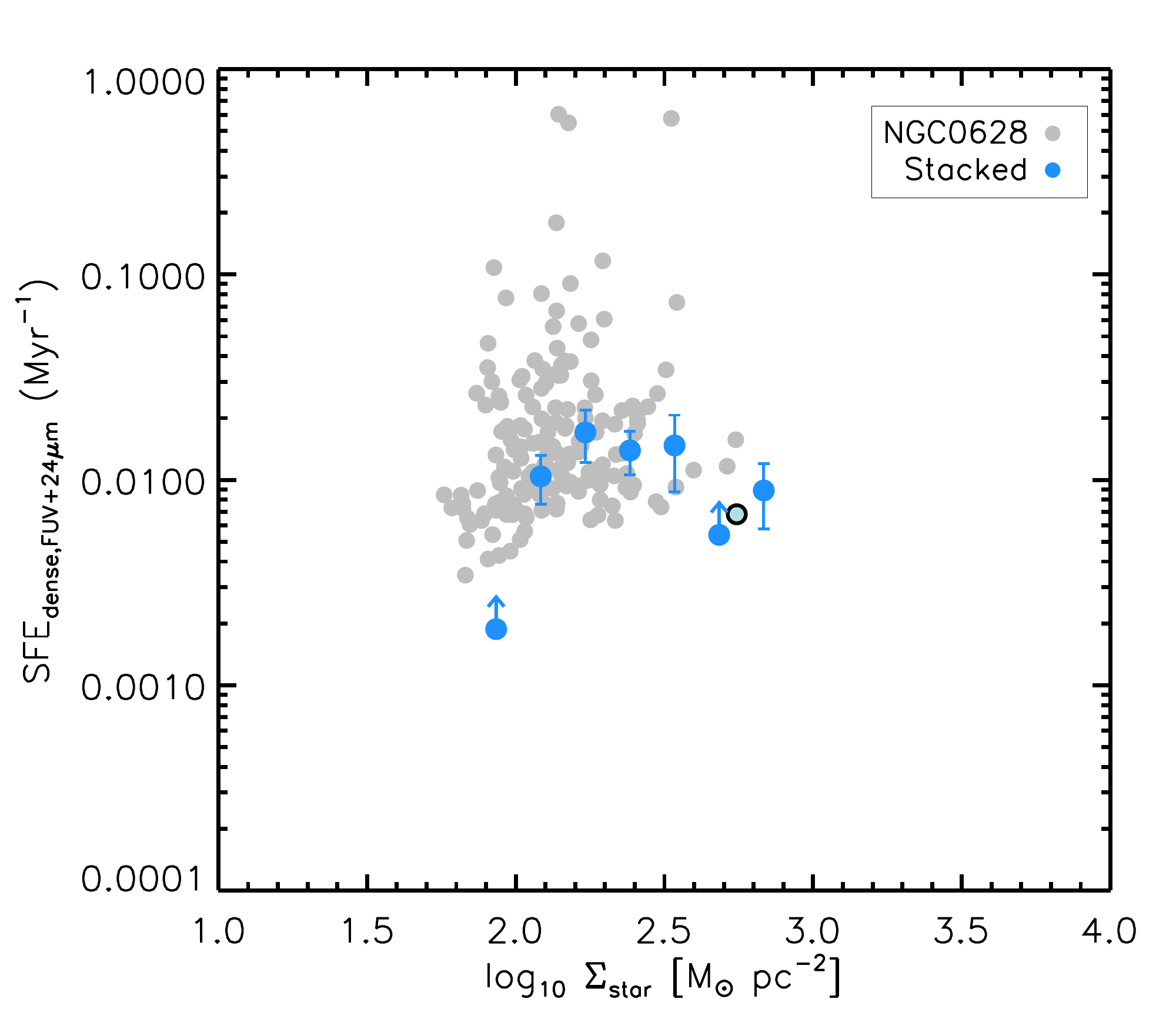}\,\includegraphics[scale=0.28]{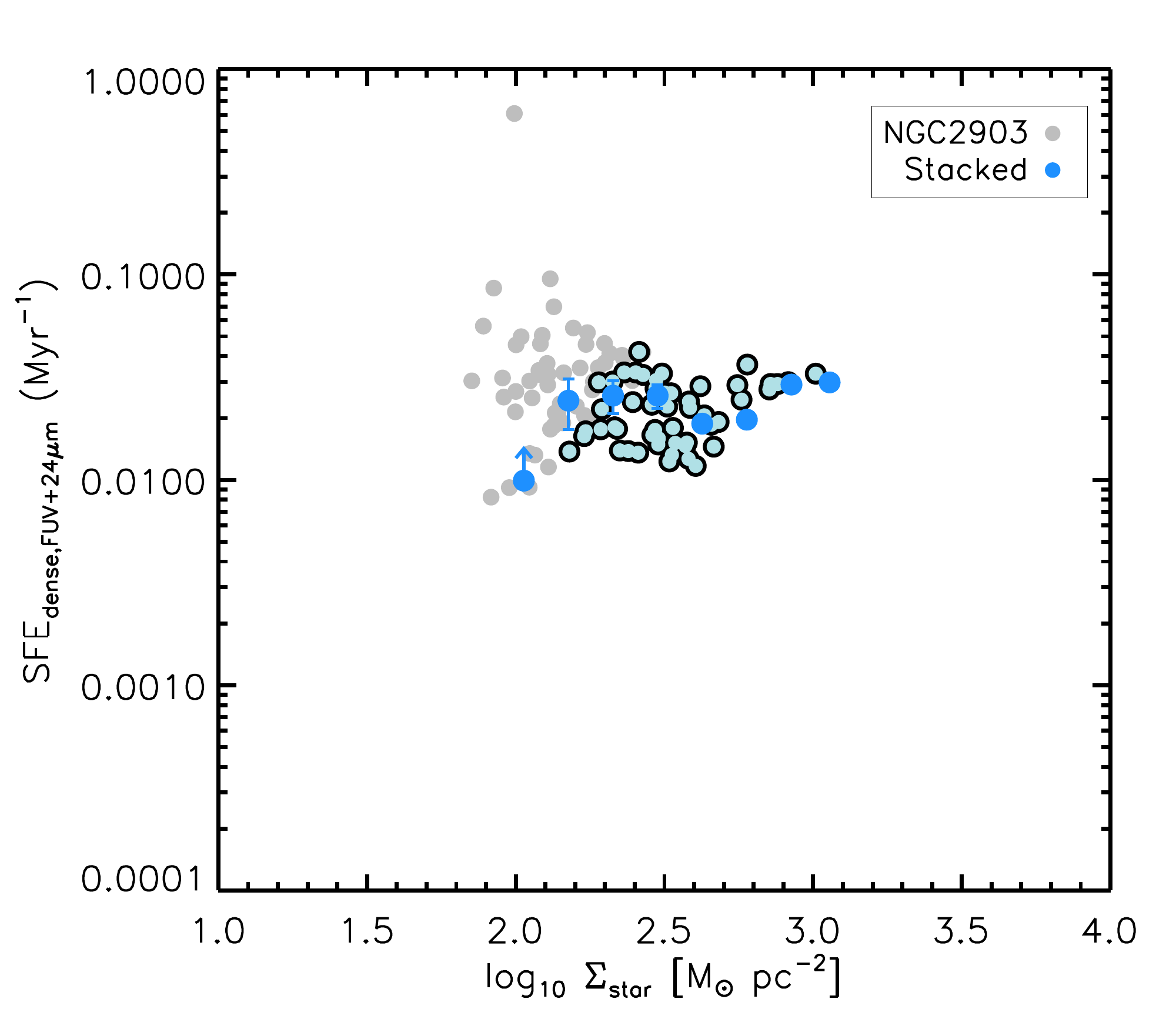}\,\includegraphics[scale=0.28]{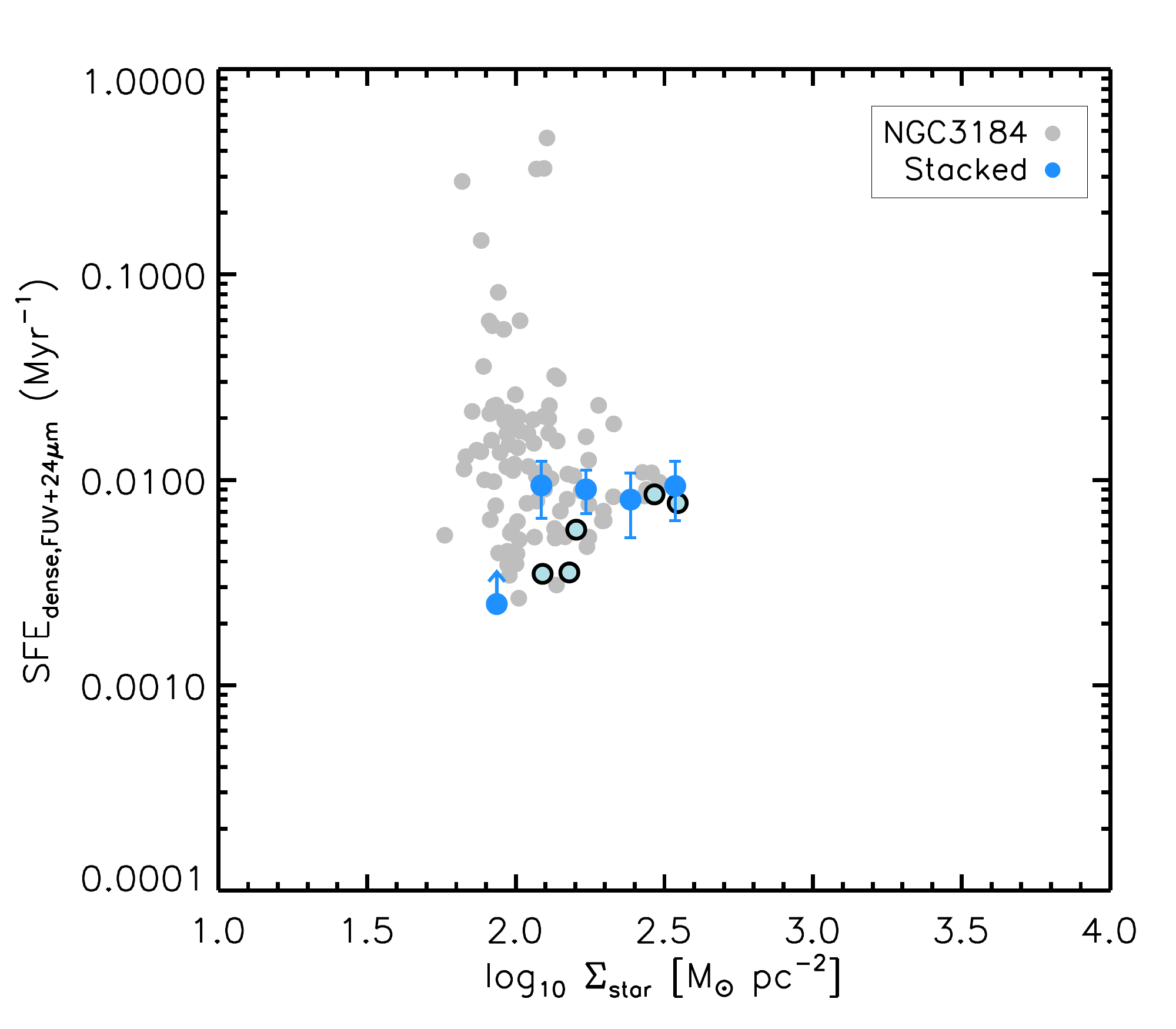} \\
    \includegraphics[scale=0.28]{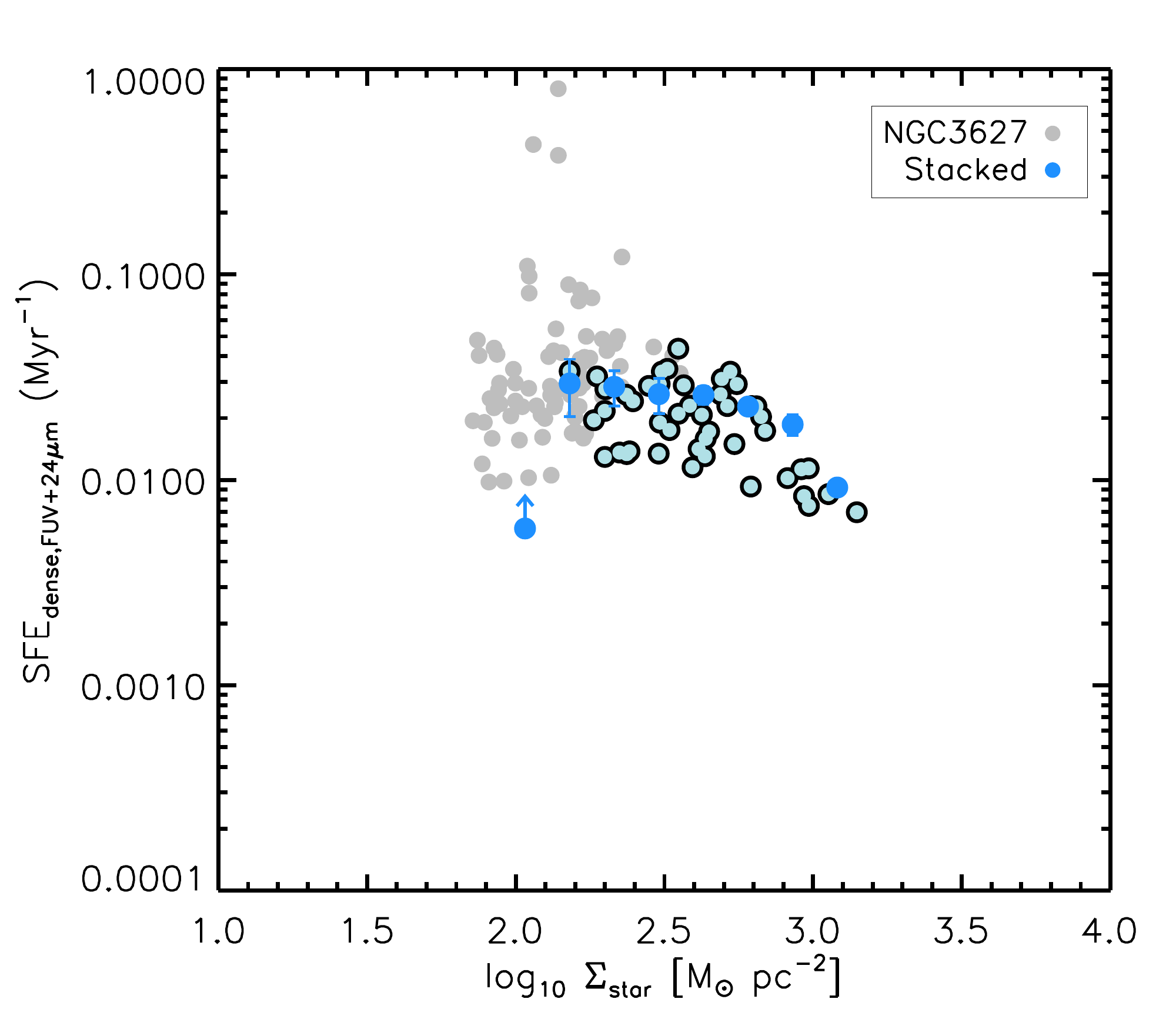}\,\includegraphics[scale=0.28]{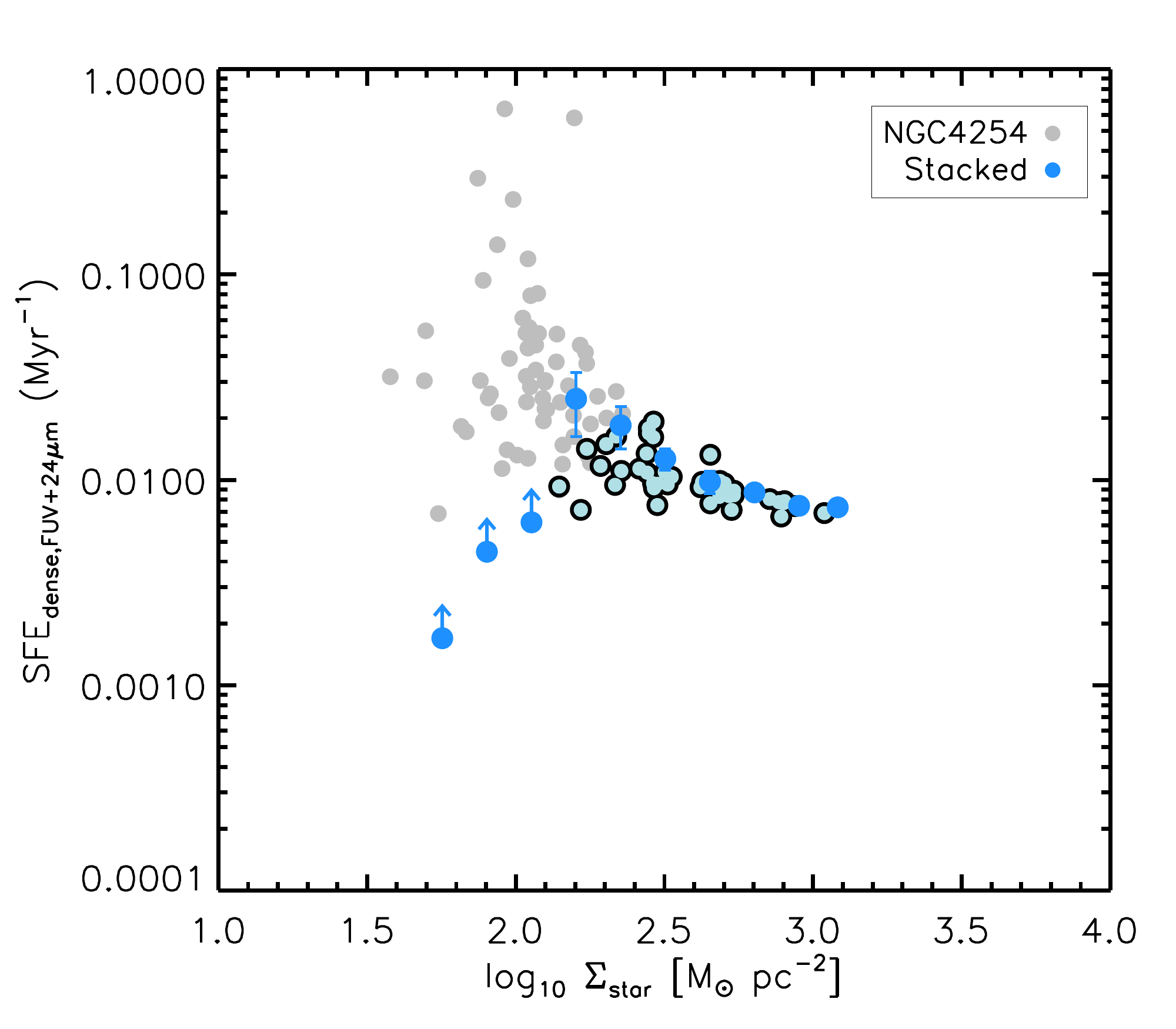}\,\includegraphics[scale=0.28]{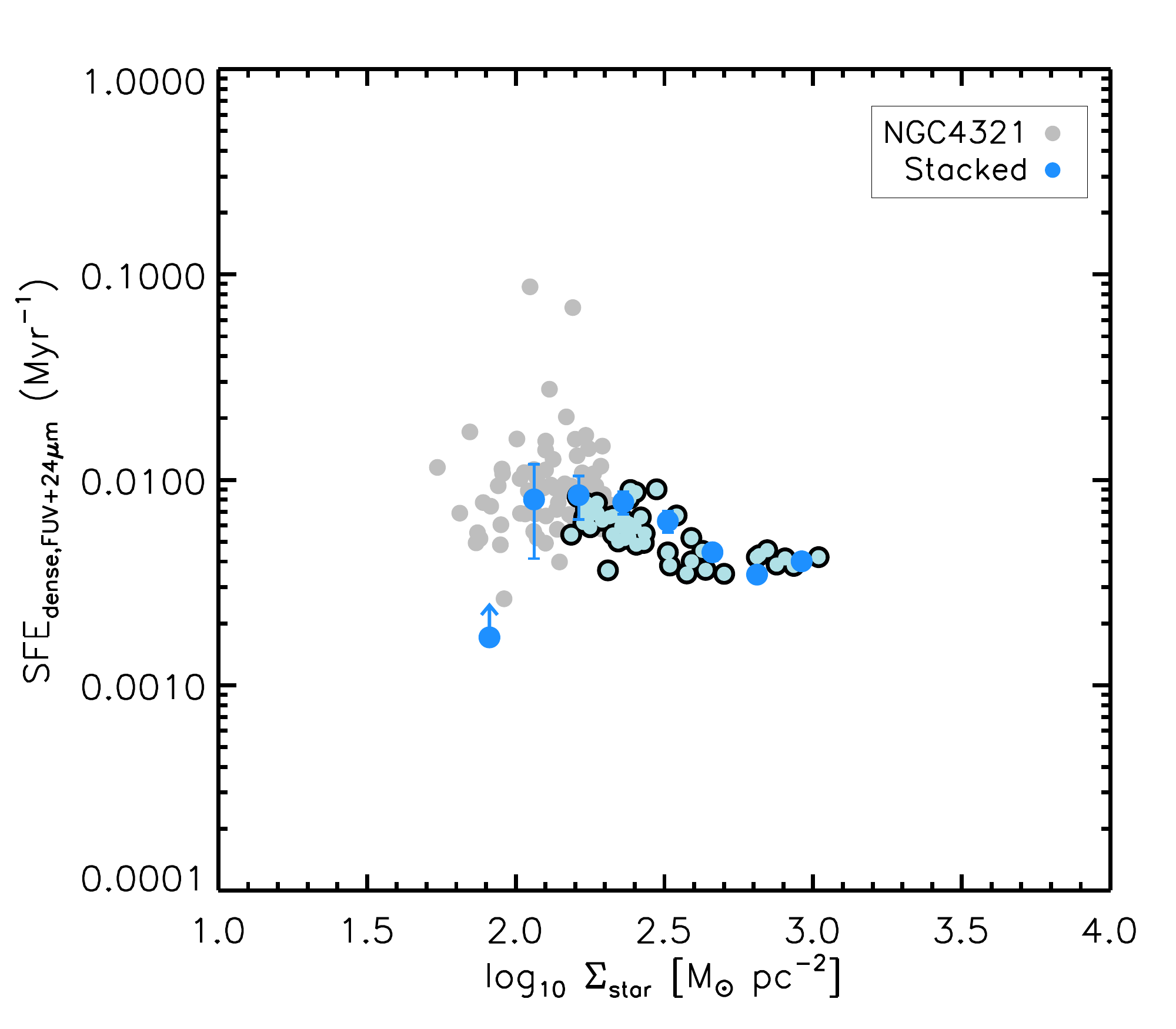} \\
	\includegraphics[scale=0.28]{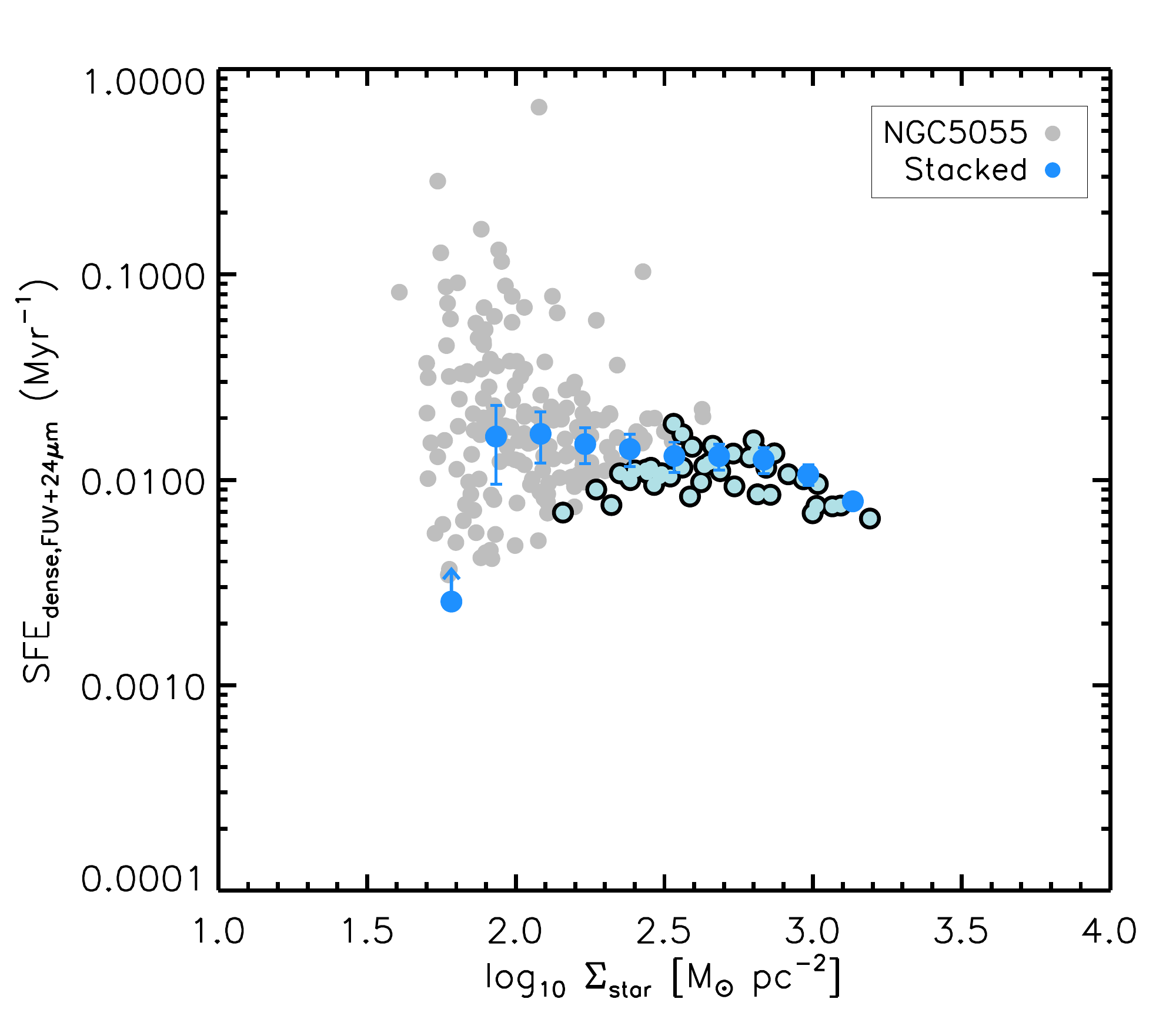}\,\includegraphics[scale=0.28]{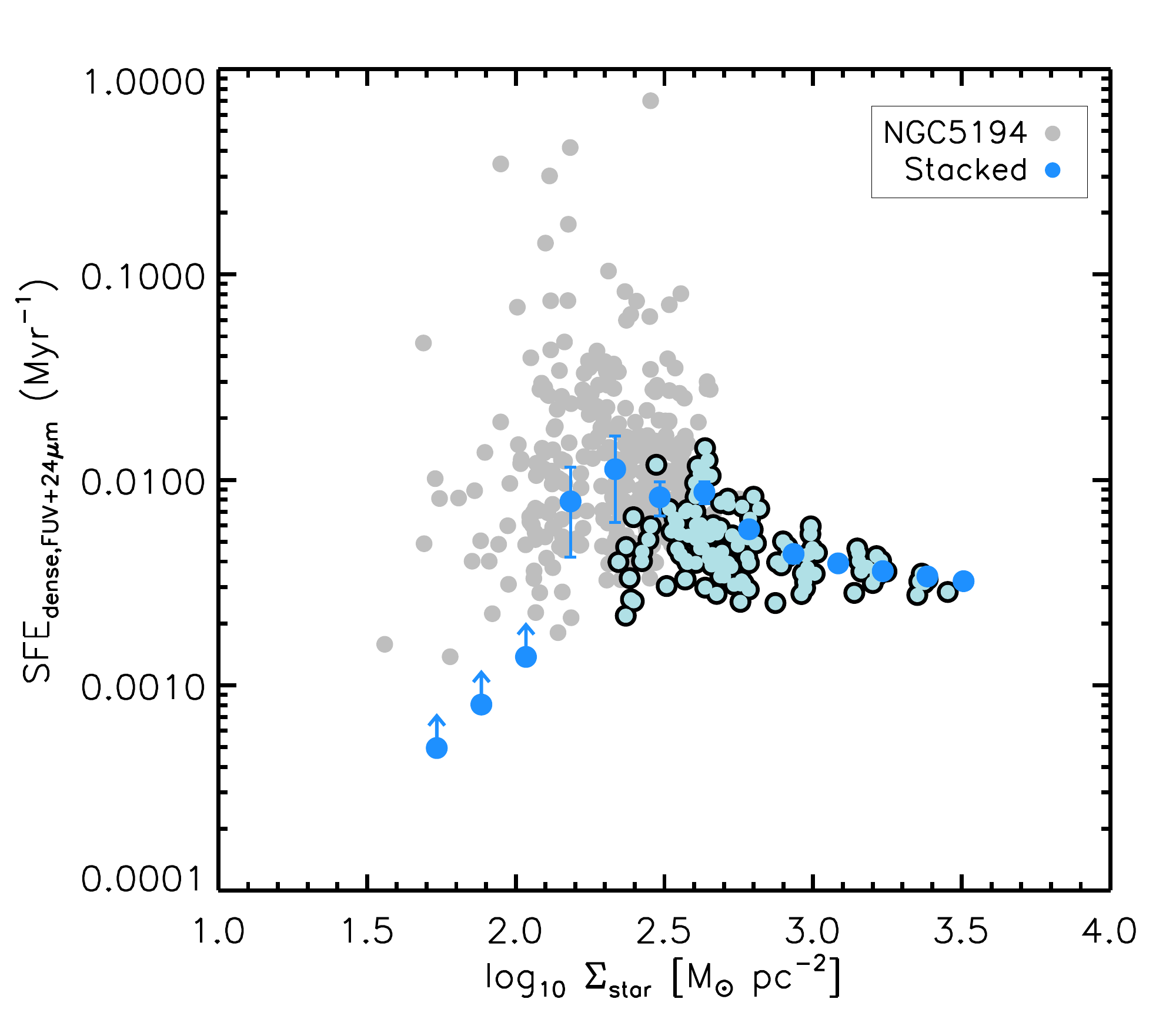}\,\includegraphics[scale=0.28]{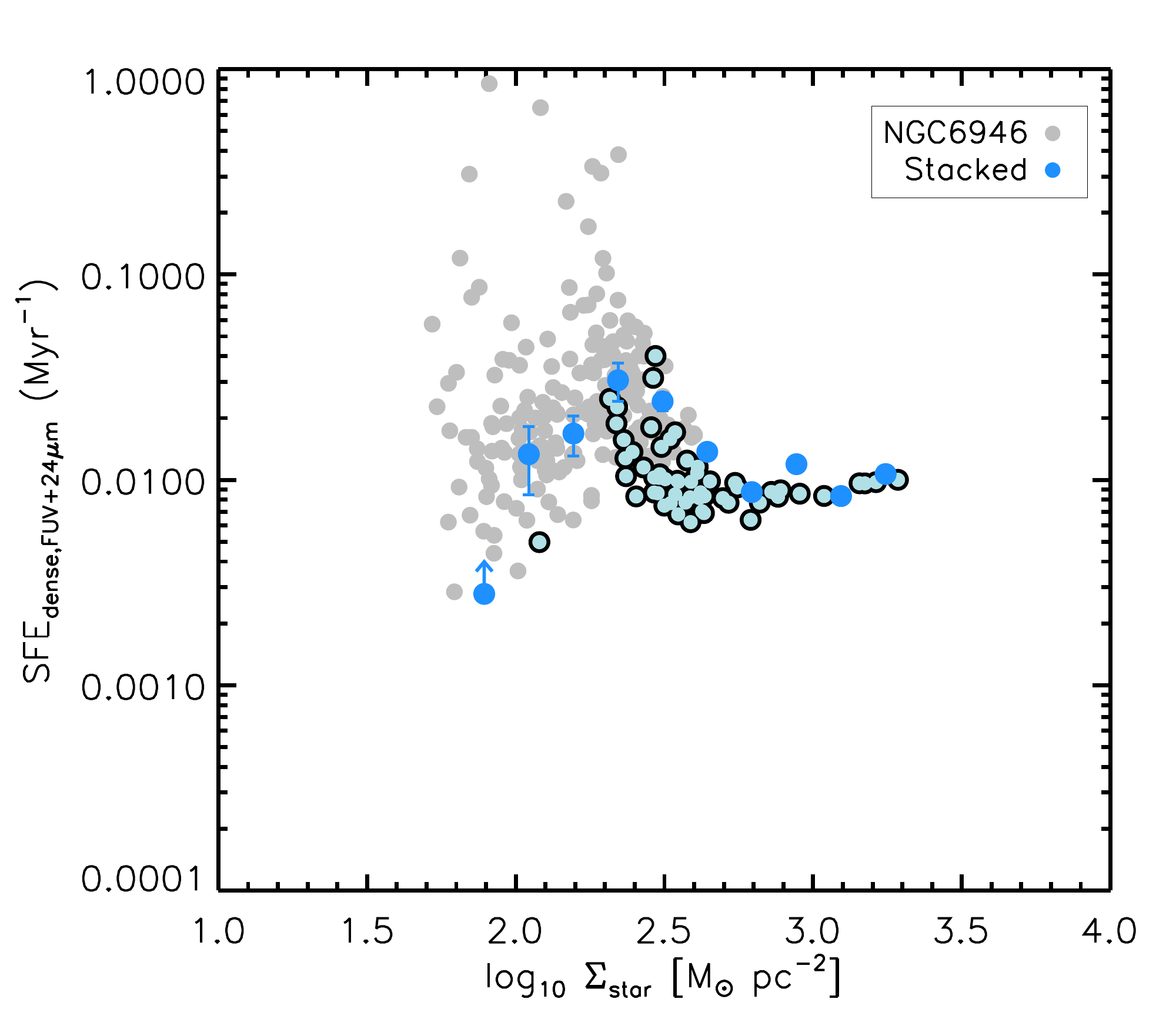}
	\caption{Same as Figure \ref{fig:sfe-tracersha24}, but using a combination of 24\,$\mu$m and FUV intensities to calculate the SFR and SFE$_\textrm{dense}$.}
	\label{fig:sfe-tracersfuv}
\end{figure*}

In Figures \ref{fig:sfe-tracersha24} and \ref{fig:sfe-tracersfuv} we plot the SFE$_\textrm{dense}$ calculated using these two SFR tracers, instead of the TIR emission, as a function of one of our environmental parameters, $\Sigma_*$. We do this for all the galaxies observed in EMPIRE. When compared to SFE$_\textrm{dense}$ from TIR emission in Figure \ref{fig:sfe_star}, Figures \ref{fig:sfe-tracersha24} and \ref{fig:sfe-tracersfuv} show that there are minimal differences in our trends (up to a $\sim 20\%$ level), suggesting that most of the radiation associated with recent star formation is reprocessed by dust. The results presented in this paper appear, therefore, robust with respect to the choice of SFR tracer.

For a more detailed study of the choice of SFR tracers, we refer the reader to the previous analyses in \citet{USERO15} and \citet{GALLAGHER18}. They find the same trends exist when using a selection of SFR tracers: TIR, H$\alpha$, 24$\mu$m and FUV data, as well as the hybrid combinations of 24$\mu$m and H$\alpha$, and 24$\mu$m and FUV. In \citet{GALLAGHER18} they additionally find that, for the majority of regions of interest (inner $\sim 3-5\,$kpc) in their galaxy sample, the contribution from the unobscured FUV and H$\alpha$ emission is significantly much lower than any estimate that involves IR emission.

\section*{APPENDIX F. Physical variation in the IR-to-HCN ratio}
\label{appendix:montecarlo}

\setcounter{figure}{0}
\setcounter{table}{0}
\renewcommand{\thefigure}{F\arabic{figure}}
\renewcommand{\thetable}{F\arabic{table}}

Our main results for star formation efficiencies in the different galaxy disks we analyzed suggest a variation in the SFE with respect to several environmental parameters somehow interconnected (radius, stellar surface densities, molecular-to-atomic gas ratios and local dynamical equilibrium pressure, Figures \ref{fig:fdense_star} to \ref{fig:sfe_pressure}). Before we continue to interpret the physical mechanism behind the line ratio variations, we assess whether they could be driven by the noise in the HCN data. We build a simple Montecarlo test that includes the uncertainties on the data which can be significant. The null hypothesis of this model is that the underlying true ratios:
\begin{equation}
	\frac{\textrm{HCN}}{\textrm{TIR}}\propto C,
\end{equation}
\noindent are constant, $C$. We compute random values that will be added to our HCN emission measurements as randomly generated values within the range of actually observed HCN uncertainties in our data. After that, we compute new, {\it synthetic} ratios as:
\begin{equation}
	\frac{\textrm{HCN}}{\textrm{TIR}}=\frac{C\,\times \textrm{TIR}+\Delta(\textrm{HCN})}{\textrm{IR}}.
\end{equation}

As seen above, we perturb the HCN observations with different, random values per data point, realistic within our observations to check if the observed scatter in Figures \ref{fig:fdense_pressure} to \ref{fig:sfe_pressure} can be explained. We perform $10^5$ realizations of the experiment and we compute the mean IR-to-HCN ratio obtained among them, as well as the standard deviation from the mean. We also repeat the experiment for a range of possible values of $C$, which is initially determined from the positions in the galaxy disks where we have high SNR measurements of HCN.

\begin{figure*}
	\centering
	\includegraphics[scale=0.8]{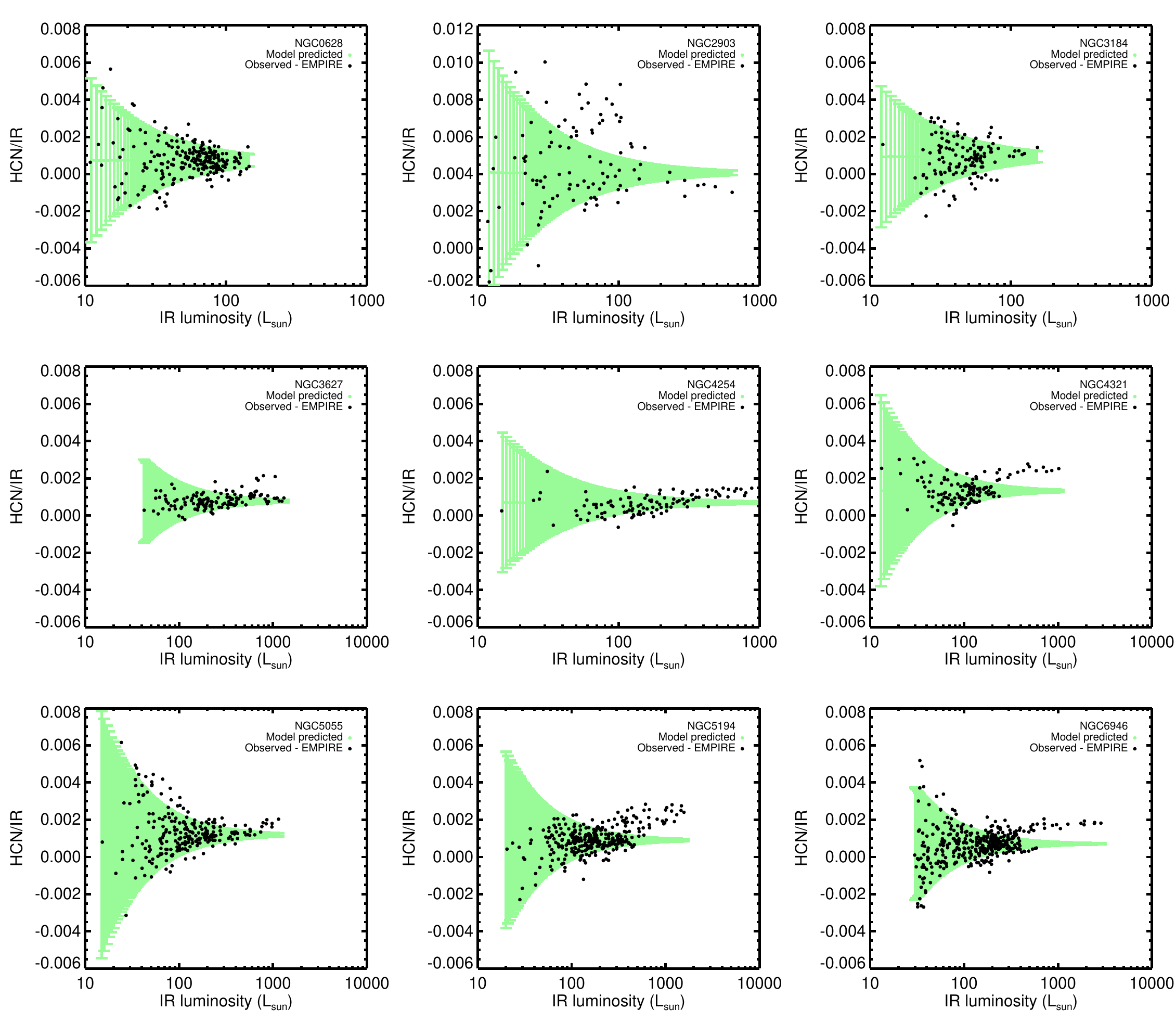}
	\caption{Montecarlo realizations for the observed galaxies. The EMPIRE original data are shown as black points, whereas the perturbed points for a null hypothesis of $\textrm{SFE}=C$ are shown in light green areas. The different panels show the case for each galaxy, for a typical $C$ value equal to the median $\textrm{HCN}/\textrm{IR}$ value in each galaxy. The error bars correspond to the $1\sigma$ standard deviations from the modeled points in a sample of $10^5$ realizations.}
	\label{fig:montecarlo}
\end{figure*}

\begin{figure*}
	\centering
	\includegraphics[scale=0.5]{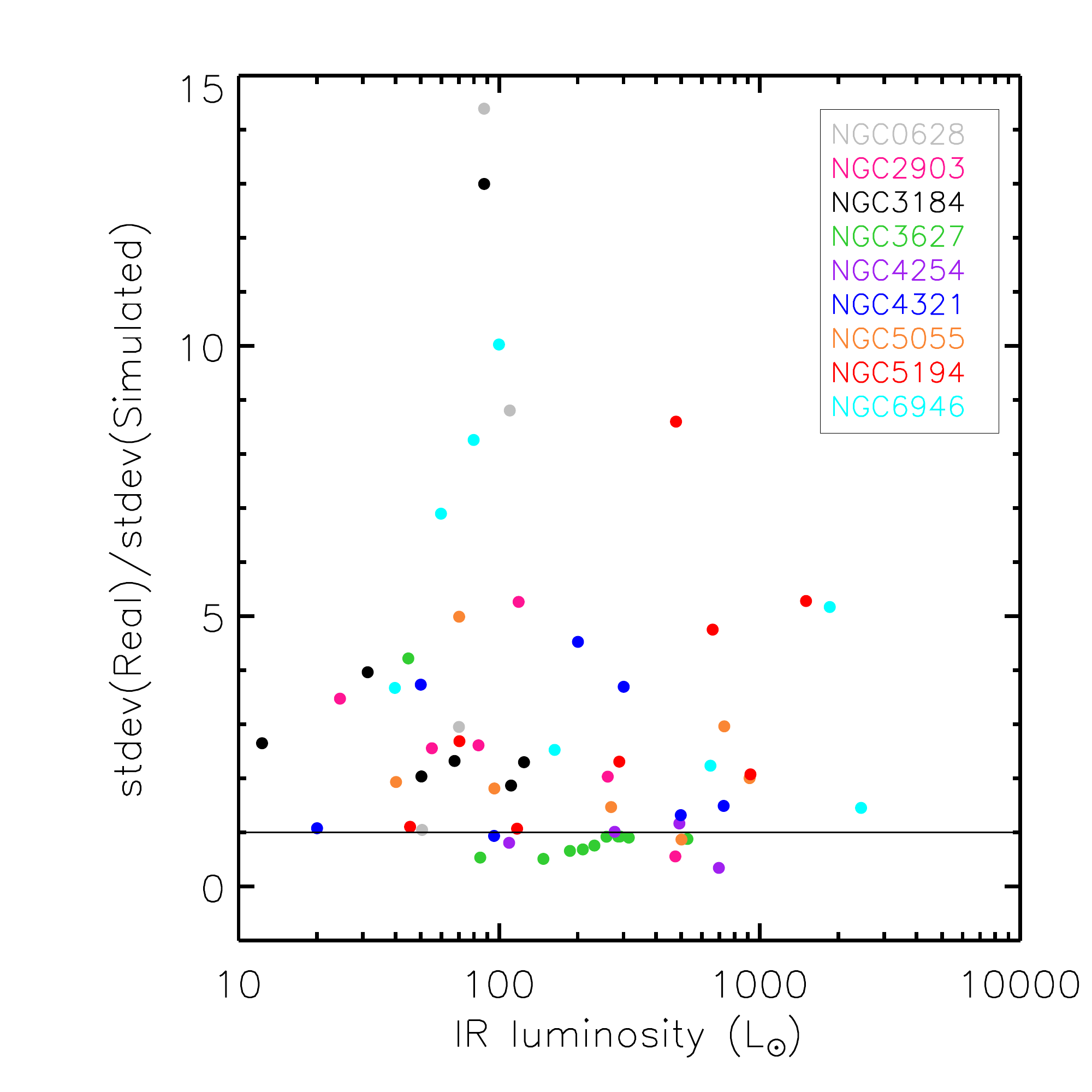}
	\caption{Ratio between standard deviation computed for the EMPIRE data set, and the standard deviation of Montecarlo realizations. Each color represents a different galaxy, and every data point corresponds to a different TIR bin, where the standard deviation was calculated. The horizontal black line marks the unity value, where the scatter in the EMPIRE data set is similar to the one reproduced by the Montecarlo analysis.}
	\label{fig:deviations}
\end{figure*}

Figure \ref{fig:montecarlo} shows the simple Montecarlo test for the EMPIRE data set. It displays the HCN-to-TIR ratio as a function of the TIR emission computed in every individual sampling point for every particular galaxy. The EMPIRE original observations are shown as dark blue points, whereas the light blue points reflect the mean value of a constant HCN/TIR model from $10^5$ realizations. In those models, as described above, the HCN value includes a random, realistic perturbation within the range of the observed uncertainties. The various panels show the different cases obtained depending on the constant chosen for the model, which is initially inferred from our high SNR measurements in the galaxy. The error bars are the (1$\sigma$) standard deviations from the modeled points in $10^5$ realizations. Every panel of Figure \ref{fig:montecarlo}, for all galaxies, shows that there are variations in the HCN-to-TIR ratio for both high and low TIR values, and the amplitude of these variations cannot be explained by our Montecarlo test. We quantify this by calculating the ratio between the typical standard variation of every particular EMPIRE galaxy and its Montecarlo simulated data. This quantity is shown in the $y$-axis of Figure \ref{fig:deviations}, as a function of the TIR luminosity. Every data point corresponds to an independent TIR bin where the standard deviations for the real and the simulated data are computed. The different colors correspond to each EMPIRE galaxy. Figure \ref{fig:deviations} shows that the vast majority of data in EMPIRE has a much larger scatter than the expected one from our Montecarlo realizations. The only exception is NGC\,3627, which shows a comparable scatter in the real and simulated data across its entire disk, except for its very central position. Therefore there are real and systematic variations beyond what we can expect from the noise; there must be physical and chemical processes responsible for the even larger scatter at low TIR values.
	
%\bibliography{empire}

\begin{thebibliography}{apj}

\bibitem[{{Aniano} {et~al.}(2011){Aniano}, {Draine}, {Gordon}, \&
  {Sandstrom}}]{ANIANO11}
{Aniano}, G., {Draine}, B.~T., {Gordon}, K.~D., \& {Sandstrom}, K. 2011, \pasp,
  123, 1218

\bibitem[{{Aravena} {et~al.}(2014){Aravena}, {Hodge}, {Wagg}, {Carilli},
  {Daddi}, {Dannerbauer}, {Lentati}, {Riechers}, {Sargent}, \&
  {Walter}}]{ARAVENA14}
{Aravena}, M., {Hodge}, J.~A., {Wagg}, J., {et~al.} 2014, \mnras, 442, 558

\bibitem[{{Athanassoula}(1992)}]{ATHA92}
{Athanassoula}, E. 1992, \mnras, 259, 345

\bibitem[{{Barnes} {et~al.}(2017){Barnes}, {Longmore}, {Battersby}, {Bally},
  {Kruijssen}, {Henshaw}, \& {Walker}}]{BARNES17}
{Barnes}, A.~T., {Longmore}, S.~N., {Battersby}, C., {et~al.} 2017, \mnras,
  469, 2263

\bibitem[{{Battersby} {et~al.}(2017){Battersby}, {Keto}, {Zhang}, {Longmore},
  {Kruijssen}, {Pillai}, {Kauffmann}, {Walker}, {Lu}, {Ginsburg}, {Bally},
  {Mills}, {Henshaw}, {Immer}, {Patel}, {Tolls}, {Walsh}, {Johnston}, \&
  {Ho}}]{BATTERSBY17}
{Battersby}, C., {Keto}, E., {Zhang}, Q., {et~al.} 2017, in IAU Symposium, Vol.
  322, The Multi-Messenger Astrophysics of the Galactic Centre, ed. R.~M.
  {Crocker}, S.~N. {Longmore}, \& G.~V. {Bicknell}, 90--94

\bibitem[{{Bemis} \& {Wilson}(2019)}]{BEMIS19}
{Bemis}, A., \& {Wilson}, C.~D. 2019, \aj, 157, 131

\bibitem[{{Bendo} {et~al.}(2012){Bendo}, {Boselli}, {Dariush}, {Pohlen},
  {Roussel}, {Sauvage}, {Smith}, {Wilson}, {Baes}, {Cooray}, {Clements},
  {Cortese}, {Foyle}, {Galametz}, {Gomez}, {Lebouteiller}, {Lu}, {Madden},
  {Mentuch}, {O'Halloran}, {Page}, {Remy}, {Schulz}, \& {Spinoglio}}]{BENDO12}
{Bendo}, G.~J., {Boselli}, A., {Dariush}, A., {et~al.} 2012, \mnras, 419, 1833

\bibitem[{{Benincasa} {et~al.}(2016){Benincasa}, {Wadsley}, {Couchman}, \&
  {Keller}}]{BENINCASA16}
{Benincasa}, S.~M., {Wadsley}, J., {Couchman}, H.~M.~P., \& {Keller}, B.~W.
  2016, \mnras, 462, 3053

\bibitem[{{Beuther} {et~al.}(2016){Beuther}, {Bihr}, {Rugel}, {Johnston},
  {Wang}, {Walter}, {Brunthaler}, {Walsh}, {Ott}, {Stil}, {Henning},
  {Schierhuber}, {Kainulainen}, {Heyer}, {Goldsmith}, {Anderson}, {Longmore},
  {Klessen}, {Glover}, {Urquhart}, {Plume}, {Ragan}, {Schneider},
  {McClure-Griffiths}, {Menten}, {Smith}, {Roy}, {Shanahan}, {Nguyen-Luong}, \&
  {Bigiel}}]{BEUTHER16}
{Beuther}, H., {Bihr}, S., {Rugel}, M., {et~al.} 2016, \aap, 595, A32

\bibitem[{{Bigiel} {et~al.}(2008){Bigiel}, {Leroy}, {Walter}, {Brinks}, {de
  Blok}, {Madore}, \& {Thornley}}]{BIGIEL08}
{Bigiel}, F., {Leroy}, A., {Walter}, F., {et~al.} 2008, \aj, 136, 2846

\bibitem[{{Bigiel} {et~al.}(2015){Bigiel}, {Leroy}, {Blitz}, {Bolatto}, {da
  Cunha}, {Rosolowsky}, {Sandstrom}, \& {Usero}}]{BIGIEL15}
{Bigiel}, F., {Leroy}, A.~K., {Blitz}, L., {et~al.} 2015, \apj, 815, 103

\bibitem[{{Bigiel} {et~al.}(2016){Bigiel}, {Leroy}, {Jim{\'e}nez-Donaire},
  {Pety}, {Usero}, {Cormier}, {Bolatto}, {Garcia-Burillo}, {Colombo},
  {Gonz{\'a}lez-Garc{\'{\i}}a}, {Hughes}, {Kepley}, {Kramer}, {Sandstrom},
  {Schinnerer}, {Schruba}, {Schuster}, {Tomicic}, \& {Zschaechner}}]{BIGIEL16}
{Bigiel}, F., {Leroy}, A.~K., {Jim{\'e}nez-Donaire}, M.~J., {et~al.} 2016,
  \apjl, 822, L26

\bibitem[{{Bisbas} {et~al.}(2015){Bisbas}, {Papadopoulos}, \&
  {Viti}}]{BISBAS15}
{Bisbas}, T.~G., {Papadopoulos}, P.~P., \& {Viti}, S. 2015, \apj, 803, 37

\bibitem[{{Bisbas} {et~al.}(2019){Bisbas}, {Schruba}, \& {van
  Dishoeck}}]{BISBAS19}
{Bisbas}, T.~G., {Schruba}, A., \& {van Dishoeck}, E.~F. 2019, \mnras, 412

\bibitem[{{Blitz} \& {Rosolowsky}(2006)}]{BLITZ06}
{Blitz}, L., \& {Rosolowsky}, E. 2006, \apj, 650, 933

\bibitem[{{Bolatto} {et~al.}(2013){Bolatto}, {Wolfire}, \& {Leroy}}]{BOLATTO13}
{Bolatto}, A.~D., {Wolfire}, M., \& {Leroy}, A.~K. 2013, \araa, 51, 207

\bibitem[{{Braine} {et~al.}(2017){Braine}, {Shimajiri}, {Andr{\'e}},
  {Bontemps}, {Gao}, {Chen}, \& {Kramer}}]{BRAINE17}
{Braine}, J., {Shimajiri}, Y., {Andr{\'e}}, P., {et~al.} 2017, \aap, 597, A44

\bibitem[{{Brouillet} {et~al.}(2005){Brouillet}, {Muller}, {Herpin}, {Braine},
  \& {Jacq}}]{BROUILLET05}
{Brouillet}, N., {Muller}, S., {Herpin}, F., {Braine}, J., \& {Jacq}, T. 2005,
  \aap, 429, 153

\bibitem[{{Buchbender} {et~al.}(2013){Buchbender}, {Kramer}, {Gonzalez-Garcia},
  {Israel}, {Garc{\'{\i}}a-Burillo}, {van der Werf}, {Braine}, {Rosolowsky},
  {Mookerjea}, {Aalto}, {Boquien}, {Gratier}, {Henkel}, {Quintana-Lacaci},
  {Verley}, \& {van der Tak}}]{BUCHBENDER13}
{Buchbender}, C., {Kramer}, C., {Gonzalez-Garcia}, M., {et~al.} 2013, \aap,
  549, A17

\bibitem[{{Cald{\'u}-Primo} {et~al.}(2013){Cald{\'u}-Primo}, {Schruba},
  {Walter}, {Leroy}, {Sandstrom}, {de Blok}, {Ianjamasimanana}, \&
  {Mogotsi}}]{CALDUPRIMO13}
{Cald{\'u}-Primo}, A., {Schruba}, A., {Walter}, F., {et~al.} 2013, \aj, 146,
  150

\bibitem[{{Calzetti} {et~al.}(2007){Calzetti}, {Kennicutt}, {Engelbracht},
  {Leitherer}, {Draine}, {Kewley}, {Moustakas}, {Sosey}, {Dale}, {Gordon},
  {Helou}, {Hollenbach}, {Armus}, {Bendo}, {Bot}, {Buckalew}, {Jarrett}, {Li},
  {Meyer}, {Murphy}, {Prescott}, {Regan}, {Rieke}, {Roussel}, {Sheth}, {Smith},
  {Thornley}, \& {Walter}}]{CALZETTI07}
{Calzetti}, D., {Kennicutt}, R.~C., {Engelbracht}, C.~W., {et~al.} 2007, \apj,
  666, 870

\bibitem[{{Carter} {et~al.}(2012){Carter}, {Lazareff}, {Maier}, {Chenu},
  {Fontana}, {Bortolotti}, {Boucher}, {Navarrini}, {Blanchet}, {Greve}, {John},
  {Kramer}, {Morel}, {Navarro}, {Pe{\~n}alver}, {Schuster}, \&
  {Thum}}]{CARTER12}
{Carter}, M., {Lazareff}, B., {Maier}, D., {et~al.} 2012, \aap, 538, A89

\bibitem[{{Chen} {et~al.}(2017){Chen}, {Braine}, {Gao}, {Koda}, \&
  {Gu}}]{CHEN17}
{Chen}, H., {Braine}, J., {Gao}, Y., {Koda}, J., \& {Gu}, Q. 2017, \apj, 836,
  101

\bibitem[{{Chen} {et~al.}(2015){Chen}, {Gao}, {Braine}, \& {Gu}}]{CHEN15}
{Chen}, H., {Gao}, Y., {Braine}, J., \& {Gu}, Q. 2015, \apj, 810, 140

\bibitem[{{Chin} {et~al.}(1998){Chin}, {Henkel}, {Millar}, {Whiteoak}, \&
  {Marx-Zimmer}}]{CHIN98}
{Chin}, Y.-N., {Henkel}, C., {Millar}, T.~J., {Whiteoak}, J.~B., \&
  {Marx-Zimmer}, M. 1998, \aap, 330, 901

\bibitem[{{Chin} {et~al.}(1997){Chin}, {Henkel}, {Whiteoak}, {Millar}, {Hunt},
  \& {Lemme}}]{CHIN97}
{Chin}, Y.-N., {Henkel}, C., {Whiteoak}, J.~B., {et~al.} 1997, \aap, 317, 548

\bibitem[{{Colombo} {et~al.}(2014){Colombo}, {Meidt}, {Schinnerer},
  {Garc{\'{\i}}a-Burillo}, {Hughes}, {Pety}, {Leroy}, {Dobbs}, {Dumas},
  {Thompson}, {Schuster}, \& {Kramer}}]{COLOMBO14B}
{Colombo}, D., {Meidt}, S.~E., {Schinnerer}, E., {et~al.} 2014, \apj, 784, 4

\bibitem[{{Cormier} {et~al.}(2018){Cormier}, {Bigiel}, {Jim{\'e}nez-Donaire},
  {Leroy}, {Gallagher}, {Usero}, {Sandstrom}, {Bolatto}, {Hughes}, {Kramer},
  {Krumholz}, {Meier}, {Murphy}, {Pety}, {Rosolowsky}, {Schinnerer}, {Schruba},
  {Sliwa}, \& {Walter}}]{CORMIER18}
{Cormier}, D., {Bigiel}, F., {Jim{\'e}nez-Donaire}, M.~J., {et~al.} 2018,
  \mnras, 475, 3909

\bibitem[{{Crocker} {et~al.}(2012){Crocker}, {Krips}, {Bureau}, {Young},
  {Davis}, {Bayet}, {Alatalo}, {Blitz}, {Bois}, {Bournaud}, {Cappellari},
  {Davies}, {de Zeeuw}, {Duc}, {Emsellem}, {Khochfar}, {Krajnovi{\'c}},
  {Kuntschner}, {Lablanche}, {McDermid}, {Morganti}, {Naab}, {Oosterloo},
  {Sarzi}, {Scott}, {Serra}, \& {Weijmans}}]{CROCKER12}
{Crocker}, A., {Krips}, M., {Bureau}, M., {et~al.} 2012, \mnras, 421, 1298

\bibitem[{{Dale} {et~al.}(2007){Dale}, {Gil de Paz}, {Gordon}, {Hanson},
  {Armus}, {Bendo}, {Bianchi}, {Block}, {Boissier}, {Boselli}, {Buckalew},
  {Buat}, {Burgarella}, {Calzetti}, {Cannon}, {Engelbracht}, {Helou},
  {Hollenbach}, {Jarrett}, {Kennicutt}, {Leitherer}, {Li}, {Madore}, {Martin},
  {Meyer}, {Murphy}, {Regan}, {Roussel}, {Smith}, {Sosey}, {Thilker}, \&
  {Walter}}]{DALE07}
{Dale}, D.~A., {Gil de Paz}, A., {Gordon}, K.~D., {et~al.} 2007, \apj, 655, 863

\bibitem[{{Dale} {et~al.}(2009){Dale}, {Cohen}, {Johnson}, {Schuster},
  {Calzetti}, {Engelbracht}, {Gil de Paz}, {Kennicutt}, {Lee}, {Begum},
  {Block}, {Dalcanton}, {Funes}, {Gordon}, {Johnson}, {Marble}, {Sakai},
  {Skillman}, {van Zee}, {Walter}, {Weisz}, {Williams}, {Wu}, \& {Wu}}]{DALE09}
{Dale}, D.~A., {Cohen}, S.~A., {Johnson}, L.~C., {et~al.} 2009, \apj, 703, 517

\bibitem[{{Dame} {et~al.}(2001){Dame}, {Hartmann}, \& {Thaddeus}}]{DAME01}
{Dame}, T.~M., {Hartmann}, D., \& {Thaddeus}, P. 2001, \apj, 547, 792

\bibitem[{{Downes} {et~al.}(1996){Downes}, {Reynaud}, {Solomon}, \&
  {Radford}}]{DOWNES96}
{Downes}, D., {Reynaud}, D., {Solomon}, P.~M., \& {Radford}, S.~J.~E. 1996,
  \apj, 461, 186

\bibitem[{{Elmegreen}(1989)}]{ELMEGREEN89}
{Elmegreen}, B.~G. 1989, \apj, 344, 306

\bibitem[{{Elmegreen} \& {Parravano}(1994)}]{ELMEGREEN94}
{Elmegreen}, B.~G., \& {Parravano}, A. 1994, \apjl, 435, L121

\bibitem[{{Engelbracht} {et~al.}(2008){Engelbracht}, {Rieke}, {Gordon},
  {Smith}, {Werner}, {Moustakas}, {Willmer}, \& {Vanzi}}]{ENGELBRACHT08}
{Engelbracht}, C.~W., {Rieke}, G.~H., {Gordon}, K.~D., {et~al.} 2008, \apj,
  678, 804

\bibitem[{{Evans} {et~al.}(2014){Evans}, {Heiderman}, \&
  {Vutisalchavakul}}]{EVANS14}
{Evans}, II, N.~J., {Heiderman}, A., \& {Vutisalchavakul}, N. 2014, \apj, 782,
  114

\bibitem[{{Federrath} \& {Klessen}(2012)}]{FEDERRATH12}
{Federrath}, C., \& {Klessen}, R.~S. 2012, \apj, 761, 156

\bibitem[{{Galametz} {et~al.}(2013){Galametz}, {Kennicutt}, {Calzetti},
  {Aniano}, {Draine}, {Boquien}, {Brandl}, {Croxall}, {Dale}, {Engelbracht},
  {Gordon}, {Groves}, {Hao}, {Helou}, {Hinz}, {Hunt}, {Johnson}, {Li},
  {Murphy}, {Roussel}, {Sandstrom}, {Skibba}, \& {Tabatabaei}}]{GALAMETZ13}
{Galametz}, M., {Kennicutt}, R.~C., {Calzetti}, D., {et~al.} 2013, \mnras, 431,
  1956

\bibitem[{{Gallagher} {et~al.}(2018{\natexlab{a}}){Gallagher}, {Leroy},
  {Bigiel}, {Cormier}, {Jim{\'e}nez-Donaire}, {Ostriker}, {Usero}, {Bolatto},
  {Garc{\'\i}a-Burillo}, \& {Hughes}}]{GALLAGHER18}
{Gallagher}, M.~J., {Leroy}, A.~K., {Bigiel}, F., {et~al.} 2018{\natexlab{a}},
  \apj, 858, 90

\bibitem[{{Gallagher} {et~al.}(2018{\natexlab{b}}){Gallagher}, {Leroy},
  {Bigiel}, {Cormier}, {Jim{\'e}nez-Donaire}, {Hughes}, {Pety}, {Schinnerer},
  {Sun}, \& {Usero}}]{GALLAGHER18b}
---. 2018{\natexlab{b}}, \apj, 868, L38

\bibitem[{{Gao} {et~al.}(2007){Gao}, {Carilli}, {Solomon}, \& {Vanden
  Bout}}]{GAO07}
{Gao}, Y., {Carilli}, C.~L., {Solomon}, P.~M., \& {Vanden Bout}, P.~A. 2007,
  \apjl, 660, L93

\bibitem[{{Gao} \& {Solomon}(2004)}]{GAO04}
{Gao}, Y., \& {Solomon}, P.~M. 2004, \apj, 606, 271

\bibitem[{{Garcia-Burillo} {et~al.}(1998){Garcia-Burillo}, {Sempere}, {Combes},
  \& {Neri}}]{GARCIABURILLO98}
{Garcia-Burillo}, S., {Sempere}, M.~J., {Combes}, F., \& {Neri}, R. 1998, \aap,
  333, 864

\bibitem[{{Garc{\'{\i}}a-Burillo} {et~al.}(2012){Garc{\'{\i}}a-Burillo},
  {Usero}, {Alonso-Herrero}, {Graci{\'a}-Carpio}, {Pereira-Santaella},
  {Colina}, {Planesas}, \& {Arribas}}]{GARCIABURILLO12}
{Garc{\'{\i}}a-Burillo}, S., {Usero}, A., {Alonso-Herrero}, A., {et~al.} 2012,
  \aap, 539, A8

\bibitem[{{Goldsmith} \& {Kauffmann}(2017)}]{GOLDSMITH17}
{Goldsmith}, P.~F., \& {Kauffmann}, J. 2017, \apj, 841, 25

\bibitem[{{Graci{\'a}-Carpio} {et~al.}(2006){Graci{\'a}-Carpio},
  {Garc{\'{\i}}a-Burillo}, {Planesas}, \& {Colina}}]{GARCIACARPIO06}
{Graci{\'a}-Carpio}, J., {Garc{\'{\i}}a-Burillo}, S., {Planesas}, P., \&
  {Colina}, L. 2006, \apjl, 640, L135

\bibitem[{{Graci{\'a}-Carpio} {et~al.}(2008){Graci{\'a}-Carpio},
  {Garc{\'{\i}}a-Burillo}, {Planesas}, {Fuente}, \& {Usero}}]{GARCIACARPIO08}
{Graci{\'a}-Carpio}, J., {Garc{\'{\i}}a-Burillo}, S., {Planesas}, P., {Fuente},
  A., \& {Usero}, A. 2008, \aap, 479, 703

\bibitem[{{Grudi{\'c}} {et~al.}(2018){Grudi{\'c}}, {Hopkins}, {Lee}, {Murray},
  {Faucher-Gigu{\`e}re}, \& {Johnson}}]{GRUDRIC18}
{Grudi{\'c}}, M.~Y., {Hopkins}, P.~F., {Lee}, E.~J., {et~al.} 2018, ArXiv
  e-prints, arXiv:1809.08348

\bibitem[{{Hacar} {et~al.}(2018){Hacar}, {Tafalla}, {Forbrich}, {Alves},
  {Meingast}, {Grossschedl}, \& {Teixeira}}]{HACAR18}
{Hacar}, A., {Tafalla}, M., {Forbrich}, J., {et~al.} 2018, \aap, 610, A77

\bibitem[{{Heiderman} {et~al.}(2010){Heiderman}, {Evans}, {Allen}, {Huard}, \&
  {Heyer}}]{HEIDERMAN10}
{Heiderman}, A., {Evans}, II, N.~J., {Allen}, L.~E., {Huard}, T., \& {Heyer},
  M. 2010, \apj, 723, 1019

\bibitem[{{Helfer} \& {Blitz}(1997)}]{HELFER97}
{Helfer}, T.~T., \& {Blitz}, L. 1997, \apj, 478, 162

\bibitem[{{Hughes} {et~al.}(2013){Hughes}, {Meidt}, {Colombo}, {Schinnerer},
  {Pety}, {Leroy}, {Dobbs}, {Garc{\'{\i}}a-Burillo}, {Thompson}, {Dumas},
  {Schuster}, \& {Kramer}}]{HUGHES13}
{Hughes}, A., {Meidt}, S.~E., {Colombo}, D., {et~al.} 2013, \apj, 779, 46

\bibitem[{{Izumi} {et~al.}(2013){Izumi}, {Kohno}, {Mart{\'{\i}}n}, {Espada},
  {Harada}, {Matsushita}, {Hsieh}, {Turner}, {Meier}, {Schinnerer}, {Imanishi},
  {Tamura}, {Curran}, {Doi}, {Fathi}, {Krips}, {Lundgren}, {Nakai}, {Nakajima},
  {Regan}, {Sheth}, {Takano}, {Taniguchi}, {Terashima}, {Tosaki}, \&
  {Wiklind}}]{IZUMI13}
{Izumi}, T., {Kohno}, K., {Mart{\'{\i}}n}, S., {et~al.} 2013, \pasj, 65, 100

\bibitem[{{Jim{\'e}nez-Donaire}
  {et~al.}(2017{\natexlab{a}}){Jim{\'e}nez-Donaire}, {Bigiel}, {Leroy},
  {Cormier}, {Gallagher}, {Usero}, {Bolatto}, {Colombo},
  {Garc{\'{\i}}a-Burillo}, {Hughes}, {Kramer}, {Krumholz}, {Meier}, {Murphy},
  {Pety}, {Rosolowsky}, {Schinnerer}, {Schruba}, {Tomi{\v c}i{\'c}}, \&
  {Zschaechner}}]{JIMENEZDONAIRE17}
{Jim{\'e}nez-Donaire}, M.~J., {Bigiel}, F., {Leroy}, A.~K., {et~al.}
  2017{\natexlab{a}}, \mnras, 466, 49

\bibitem[{{Jim{\'e}nez-Donaire}
  {et~al.}(2017{\natexlab{b}}){Jim{\'e}nez-Donaire}, {Cormier}, {Bigiel},
  {Leroy}, {Gallagher}, {Krumholz}, {Usero}, {Hughes}, {Kramer}, {Meier},
  {Murphy}, {Pety}, {Schinnerer}, {Schruba}, {Schuster}, {Sliwa}, \&
  {Tomicic}}]{JIMENEZDONAIRE17B}
{Jim{\'e}nez-Donaire}, M.~J., {Cormier}, D., {Bigiel}, F., {et~al.}
  2017{\natexlab{b}}, \apjl, 836, L29

\bibitem[{{Jones} {et~al.}(2012){Jones}, {Burton}, {Cunningham},
  {Requena-Torres}, {Menten}, {Schilke}, {Belloche}, {Leurini},
  {Mart{\'{\i}}n-Pintado}, {Ott}, \& {Walsh}}]{JONES12}
{Jones}, P.~A., {Burton}, M.~G., {Cunningham}, M.~R., {et~al.} 2012, \mnras,
  419, 2961

\bibitem[{{Juneau} {et~al.}(2009){Juneau}, {Narayanan}, {Moustakas}, {Shirley},
  {Bussmann}, {Kennicutt}, \& {Vanden Bout}}]{JUNEAU09}
{Juneau}, S., {Narayanan}, D.~T., {Moustakas}, J., {et~al.} 2009, \apj, 707,
  1217

\bibitem[{{Kauffmann} {et~al.}(2017){Kauffmann}, {Goldsmith}, {Melnick},
  {Tolls}, {Guzman}, \& {Menten}}]{KAUFFMANN17}
{Kauffmann}, J., {Goldsmith}, P.~F., {Melnick}, G., {et~al.} 2017, \aap, 605,
  L5

\bibitem[{{Kennicutt} \& {Evans}(2012)}]{KENNICUTT12}
{Kennicutt}, R.~C., \& {Evans}, N.~J. 2012, \araa, 50, 531

\bibitem[{{Kennicutt} {et~al.}(2011){Kennicutt}, {Calzetti}, {Aniano},
  {Appleton}, {Armus}, {Beir{\~a}o}, {Bolatto}, {Brandl}, {Crocker}, {Croxall},
  {Dale}, {Donovan Meyer}, {Draine}, {Engelbracht}, {Galametz}, {Gordon},
  {Groves}, {Hao}, {Helou}, {Hinz}, {Hunt}, {Johnson}, {Koda}, {Krause},
  {Leroy}, {Li}, {Meidt}, {Montiel}, {Murphy}, {Rahman}, {Rix}, {Roussel},
  {Sandstrom}, {Sauvage}, {Schinnerer}, {Skibba}, {Smith}, {Srinivasan},
  {Vigroux}, {Walter}, {Wilson}, {Wolfire}, \& {Zibetti}}]{KENNICUTT11}
{Kennicutt}, R.~C., {Calzetti}, D., {Aniano}, G., {et~al.} 2011, \pasp, 123,
  1347

\bibitem[{{Kennicutt} {et~al.}(2003){Kennicutt}, {Armus}, {Bendo}, {Calzetti},
  {Dale}, {Draine}, {Engelbracht}, {Gordon}, {Grauer}, {Helou}, {Hollenbach},
  {Jarrett}, {Kewley}, {Leitherer}, {Li}, {Malhotra}, {Regan}, {Rieke},
  {Rieke}, {Roussel}, {Smith}, {Thornley}, \& {Walter}}]{KENNICUTT03}
{Kennicutt}, Jr., R.~C., {Armus}, L., {Bendo}, G., {et~al.} 2003, \pasp, 115,
  928

\bibitem[{{Kepley} {et~al.}(2014){Kepley}, {Leroy}, {Frayer}, {Usero},
  {Marvil}, \& {Walter}}]{KEPLEY14}
{Kepley}, A.~A., {Leroy}, A.~K., {Frayer}, D., {et~al.} 2014, \apjl, 780, L13

\bibitem[{{Kepley} {et~al.}(2018){Kepley}, {Bittle}, {Leroy},
  {Jim{\'e}nez-Donaire}, {Schruba}, {Bigiel}, {Gallagher}, {Johnson}, \&
  {Usero}}]{KEPLEY18}
{Kepley}, A.~A., {Bittle}, L., {Leroy}, A.~K., {et~al.} 2018, \apj, 862, 120

\bibitem[{{Knudsen} {et~al.}(2007){Knudsen}, {Walter}, {Weiss}, {Bolatto},
  {Riechers}, \& {Menten}}]{KNUDSEN07}
{Knudsen}, K.~K., {Walter}, F., {Weiss}, A., {et~al.} 2007, \apj, 666, 156

\bibitem[{{Kohno} {et~al.}(2001){Kohno}, {Matsushita}, {Vila-Vilar{\'o}},
  {Okumura}, {Shibatsuka}, {Okiura}, {Ishizuki}, \& {Kawabe}}]{KOHNO01}
{Kohno}, K., {Matsushita}, S., {Vila-Vilar{\'o}}, B., {et~al.} 2001, in
  Astronomical Society of the Pacific Conference Series, Vol. 249, The Central
  Kiloparsec of Starbursts and AGN: The La Palma Connection, ed. J.~H.
  {Knapen}, J.~E. {Beckman}, I.~{Shlosman}, \& T.~J. {Mahoney}, 672

\bibitem[{{Kormendy} \& {Kennicutt}(2004)}]{KORMENDY04}
{Kormendy}, J., \& {Kennicutt}, Jr., R.~C. 2004, \araa, 42, 603

\bibitem[{{Kregel} {et~al.}(2002){Kregel}, {van der Kruit}, \& {de
  Grijs}}]{KREGEL02}
{Kregel}, M., {van der Kruit}, P.~C., \& {de Grijs}, R. 2002, \mnras, 334, 646

\bibitem[{{Krips} {et~al.}(2008){Krips}, {Neri}, {Garc{\'{\i}}a-Burillo},
  {Mart{\'{\i}}n}, {Combes}, {Graci{\'a}-Carpio}, \& {Eckart}}]{KRIPS08}
{Krips}, M., {Neri}, R., {Garc{\'{\i}}a-Burillo}, S., {et~al.} 2008, \apj, 677,
  262

\bibitem[{{Krumholz} \& {McKee}(2005)}]{KRUMHOLZ05}
{Krumholz}, M.~R., \& {McKee}, C.~F. 2005, \apj, 630, 250

\bibitem[{{Krumholz} \& {Tan}(2007)}]{KRUMHOLZTAN07}
{Krumholz}, M.~R., \& {Tan}, J.~C. 2007, \apj, 654, 304

\bibitem[{{Krumholz} \& {Thompson}(2007)}]{KRUMHOLZ07}
{Krumholz}, M.~R., \& {Thompson}, T.~A. 2007, \apj, 669, 289

\bibitem[{{Lada} {et~al.}(2012){Lada}, {Forbrich}, {Lombardi}, \&
  {Alves}}]{LADA12}
{Lada}, C.~J., {Forbrich}, J., {Lombardi}, M., \& {Alves}, J.~F. 2012, \apj,
  745, 190

\bibitem[{{Lada} {et~al.}(2010){Lada}, {Lombardi}, \& {Alves}}]{LADA10}
{Lada}, C.~J., {Lombardi}, M., \& {Alves}, J.~F. 2010, \apj, 724, 687

\bibitem[{{Lada} {et~al.}(2013){Lada}, {Lombardi}, {Roman-Zuniga}, {Forbrich},
  \& {Alves}}]{LADA13}
{Lada}, C.~J., {Lombardi}, M., {Roman-Zuniga}, C., {Forbrich}, J., \& {Alves},
  J.~F. 2013, \apj, 778, 133

\bibitem[{{Lee} {et~al.}(2009){Lee}, {Gil de Paz}, {Tremonti}, {Kennicutt},
  {Salim}, {Bothwell}, {Calzetti}, {Dalcanton}, {Dale}, {Engelbracht}, {Funes},
  {Johnson}, {Sakai}, {Skillman}, {van Zee}, {Walter}, \& {Weisz}}]{LEE09}
{Lee}, J.~C., {Gil de Paz}, A., {Tremonti}, C., {et~al.} 2009, \apj, 706, 599

\bibitem[{{Leroy} {et~al.}(2008){Leroy}, {Walter}, {Brinks}, {Bigiel}, {de
  Blok}, {Madore}, \& {Thornley}}]{LEROY08}
{Leroy}, A.~K., {Walter}, F., {Brinks}, E., {et~al.} 2008, \aj, 136, 2782

\bibitem[{{Leroy} {et~al.}(2009){Leroy}, {Walter}, {Bigiel}, {Usero}, {Weiss},
  {Brinks}, {de Blok}, {Kennicutt}, {Schuster}, {Kramer}, {Wiesemeyer}, \&
  {Roussel}}]{LEROY09}
{Leroy}, A.~K., {Walter}, F., {Bigiel}, F., {et~al.} 2009, \aj, 137, 4670

\bibitem[{{Leroy} {et~al.}(2012){Leroy}, {Bigiel}, {de Blok}, {Boissier},
  {Bolatto}, {Brinks}, {Madore}, {Munoz-Mateos}, {Murphy}, {Sandstrom},
  {Schruba}, \& {Walter}}]{LEROY12}
{Leroy}, A.~K., {Bigiel}, F., {de Blok}, W.~J.~G., {et~al.} 2012, \aj, 144, 3

\bibitem[{{Leroy} {et~al.}(2013){Leroy}, {Walter}, {Sandstrom}, {Schruba},
  {Munoz-Mateos}, {Bigiel}, {Bolatto}, {Brinks}, {de Blok}, {Meidt}, {Rix},
  {Rosolowsky}, {Schinnerer}, {Schuster}, \& {Usero}}]{LEROY13}
{Leroy}, A.~K., {Walter}, F., {Sandstrom}, K., {et~al.} 2013, \aj, 146, 19

\bibitem[{{Leroy} {et~al.}(2016){Leroy}, {Hughes}, {Schruba}, {Rosolowsky},
  {Blanc}, {Bolatto}, {Colombo}, {Escala}, {Kramer}, {Kruijssen}, {Meidt},
  {Pety}, {Querejeta}, {Sandstrom}, {Schinnerer}, {Sliwa}, \&
  {Usero}}]{LEROY16}
{Leroy}, A.~K., {Hughes}, A., {Schruba}, A., {et~al.} 2016, \apj, 831, 16

\bibitem[{{Leroy} {et~al.}(2017{\natexlab{a}}){Leroy}, {Usero}, {Schruba},
  {Bigiel}, {Kruijssen}, {Kepley}, {Blanc}, {Bolatto}, {Cormier}, {Gallagher},
  {Hughes}, {Jim{\'e}nez-Donaire}, {Rosolowsky}, \& {Schinnerer}}]{LEROY17}
{Leroy}, A.~K., {Usero}, A., {Schruba}, A., {et~al.} 2017{\natexlab{a}}, \apj,
  835, 217

\bibitem[{{Leroy} {et~al.}(2017{\natexlab{b}}){Leroy}, {Schinnerer}, {Hughes},
  {Kruijssen}, {Meidt}, {Schruba}, {Sun}, {Bigiel}, {Aniano}, \&
  {Blanc}}]{LEROY17B}
{Leroy}, A.~K., {Schinnerer}, E., {Hughes}, A., {et~al.} 2017{\natexlab{b}},
  \apj, 846, 71

\bibitem[{{Liszt} \& {Pety}(2016)}]{LISZT16}
{Liszt}, H.~S., \& {Pety}, J. 2016, \apj, 823, 124

\bibitem[{{Liu} {et~al.}(2015){Liu}, {Gao}, \& {Greve}}]{LIU15}
{Liu}, L., {Gao}, Y., \& {Greve}, T.~R. 2015, \apj, 805, 31

\bibitem[{{Loenen} {et~al.}(2008){Loenen}, {Spaans}, {Baan}, \&
  {Meijerink}}]{LOENEN08}
{Loenen}, A.~F., {Spaans}, M., {Baan}, W.~A., \& {Meijerink}, R. 2008, \aap,
  488, L5

\bibitem[{{Longmore} {et~al.}(2013){Longmore}, {Bally}, {Testi}, {Purcell},
  {Walsh}, {Bressert}, {Pestalozzi}, {Molinari}, {Ott}, {Cortese}, {Battersby},
  {Murray}, {Lee}, {Kruijssen}, {Schisano}, \& {Elia}}]{LONGMORE13}
{Longmore}, S.~N., {Bally}, J., {Testi}, L., {et~al.} 2013, \mnras, 429, 987

\bibitem[{{Makarov} {et~al.}(2014){Makarov}, {Prugniel}, {Terekhova},
  {Courtois}, \& {Vauglin}}]{MAKAROV14}
{Makarov}, D., {Prugniel}, P., {Terekhova}, N., {Courtois}, H., \& {Vauglin},
  I. 2014, \aap, 570, A13

\bibitem[{{Meidt}(2016)}]{MEIDT16}
{Meidt}, S.~E. 2016, \apj, 818, 69

\bibitem[{{Meidt} {et~al.}(2012){Meidt}, {Schinnerer}, {Knapen}, {Bosma},
  {Athanassoula}, {Sheth}, {Buta}, {Zaritsky}, {Laurikainen}, {Elmegreen},
  {Elmegreen}, {Gadotti}, {Salo}, {Regan}, {Ho}, {Madore}, {Hinz}, {Skibba},
  {Gil de Paz}, {Mu{\~n}oz-Mateos}, {Men{\'e}ndez-Delmestre}, {Seibert}, {Kim},
  {Mizusawa}, {Laine}, \& {Comer{\'o}n}}]{MEIDT12}
{Meidt}, S.~E., {Schinnerer}, E., {Knapen}, J.~H., {et~al.} 2012, \apj, 744, 17

\bibitem[{{Meidt} {et~al.}(2013){Meidt}, {Schinnerer}, {Hughes}, {Colombo},
  {Pety}, {Garcia-Burillo}, {Leroy}, {Dobbs}, {Schuster}, {Kramer}, {Dumas}, \&
  {Thompson}}]{MEIDT13}
{Meidt}, S.~E., {Schinnerer}, E., {Hughes}, A., {et~al.} 2013, in IAU
  Symposium, Vol. 292, Molecular Gas, Dust, and Star Formation in Galaxies, ed.
  T.~{Wong} \& J.~{Ott}, 139--142

\bibitem[{{Meidt} {et~al.}(2014){Meidt}, {Schinnerer}, {van de Ven},
  {Zaritsky}, {Peletier}, {Knapen}, {Sheth}, {Regan}, {Querejeta},
  {Mu{\~n}oz-Mateos}, {Kim}, {Hinz}, {Gil de Paz}, {Athanassoula}, {Bosma},
  {Buta}, {Cisternas}, {Ho}, {Holwerda}, {Skibba}, {Laurikainen}, {Salo},
  {Gadotti}, {Laine}, {Erroz-Ferrer}, {Comer{\'o}n}, {Men{\'e}ndez-Delmestre},
  {Seibert}, \& {Mizusawa}}]{MEIDT14}
{Meidt}, S.~E., {Schinnerer}, E., {van de Ven}, G., {et~al.} 2014, \apj, 788,
  144

\bibitem[{{Meidt} {et~al.}(2018){Meidt}, {Leroy}, {Rosolowsky}, {Kruijssen},
  {Schinnerer}, {Schruba}, {Pety}, {Blanc}, {Bigiel}, {Chevance}, {Hughes},
  {Querejeta}, \& {Usero}}]{MEIDT18}
{Meidt}, S.~E., {Leroy}, A.~K., {Rosolowsky}, E., {et~al.} 2018, \apj, 854, 100

\bibitem[{{Meier} {et~al.}(2014){Meier}, {Turner}, \& {Beck}}]{MEIER14}
{Meier}, D.~S., {Turner}, J.~L., \& {Beck}, S.~C. 2014, \apj, 795, 107

\bibitem[{{Meier} {et~al.}(2015){Meier}, {Walter}, {Bolatto}, {Leroy}, {Ott},
  {Rosolowsky}, {Veilleux}, {Warren}, {Wei{\ss}}, {Zwaan}, \&
  {Zschaechner}}]{MEIER15}
{Meier}, D.~S., {Walter}, F., {Bolatto}, A.~D., {et~al.} 2015, \apj, 801, 63

\bibitem[{{Mills} \& {Battersby}(2017)}]{MILLS17}
{Mills}, E. A.~C., \& {Battersby}, C. 2017, \apj, 835, 76

\bibitem[{{Moustakas} {et~al.}(2010){Moustakas}, {Kennicutt}, {Tremonti},
  {Dale}, {Smith}, \& {Calzetti}}]{MOUSTAKAS10}
{Moustakas}, J., {Kennicutt}, Jr., R.~C., {Tremonti}, C.~A., {et~al.} 2010,
  \apjs, 190, 233

\bibitem[{{Murphy} {et~al.}(2011){Murphy}, {Condon}, {Schinnerer}, {Kennicutt},
  {Calzetti}, {Armus}, {Helou}, {Turner}, {Aniano}, {Beir{\~a}o}, {Bolatto},
  {Brandl}, {Croxall}, {Dale}, {Donovan Meyer}, {Draine}, {Engelbracht},
  {Hunt}, {Hao}, {Koda}, {Roussel}, {Skibba}, \& {Smith}}]{MURPHY11}
{Murphy}, E.~J., {Condon}, J.~J., {Schinnerer}, E., {et~al.} 2011, \apj, 737,
  67

\bibitem[{{Narayanan} {et~al.}(2008){Narayanan}, {Cox}, {Shirley}, {Dav{\'e}},
  {Hernquist}, \& {Walker}}]{NARAYANAN08}
{Narayanan}, D., {Cox}, T.~J., {Shirley}, Y., {et~al.} 2008, \apj, 684, 996

\bibitem[{{Nishimura} {et~al.}(2016){Nishimura}, {Shimonishi}, {Watanabe},
  {Sakai}, {Aikawa}, {Kawamura}, \& {Yamamoto}}]{Nishimura2016SpectralIC10}
{Nishimura}, Y., {Shimonishi}, T., {Watanabe}, Y., {et~al.} 2016, \apj, 829, 94

\bibitem[{{Onus} {et~al.}(2018){Onus}, {Krumholz}, \& {Federrath}}]{ONUS18}
{Onus}, A., {Krumholz}, M.~R., \& {Federrath}, C. 2018, \mnras, 479, 1702

\bibitem[{{Orr} {et~al.}(2019){Orr}, {Hayward}, \& {Hopkins}}]{ORR18}
{Orr}, M.~E., {Hayward}, C.~C., \& {Hopkins}, P.~F. 2019, \mnras, 486, 4724

\bibitem[{{Ostriker} {et~al.}(2010){Ostriker}, {McKee}, \&
  {Leroy}}]{OSTRIKER10}
{Ostriker}, E.~C., {McKee}, C.~F., \& {Leroy}, A.~K. 2010, \apj, 721, 975

\bibitem[{{Padoan} \& {Nordlund}(2002)}]{PADOAN02}
{Padoan}, P., \& {Nordlund}, {\AA}. 2002, \apj, 576, 870

\bibitem[{{Papadopoulos}(2007)}]{PAPADOPOULOS07}
{Papadopoulos}, P.~P. 2007, \apj, 656, 792

\bibitem[{{Paturel} {et~al.}(2003){Paturel}, {Petit}, {Prugniel}, {Theureau},
  {Rousseau}, {Brouty}, {Dubois}, \& {Cambr{\'e}sy}}]{PATUREL03}
{Paturel}, G., {Petit}, C., {Prugniel}, P., {et~al.} 2003, \aap, 412, 45

\bibitem[{{Pellegrini} {et~al.}(2009){Pellegrini}, {Baldwin}, {Ferland},
  {Shaw}, \& {Heathcote}}]{PELLEGRINI09}
{Pellegrini}, E.~W., {Baldwin}, J.~A., {Ferland}, G.~J., {Shaw}, G., \&
  {Heathcote}, S. 2009, \apj, 693, 285

\bibitem[{{Pety}(2005)}]{PETY05}
{Pety}, J. 2005, in SF2A-2005: Semaine de l'Astrophysique Francaise, ed.
  F.~{Casoli}, T.~{Contini}, J.~M. {Hameury}, \& L.~{Pagani}, 721

\bibitem[{{Pety} {et~al.}(2017){Pety}, {Guzm{\'a}n}, {Orkisz}, {Liszt},
  {Gerin}, {Bron}, {Bardeau}, {Goicoechea}, {Gratier}, {Le Petit}, {Levrier},
  {{\"O}berg}, {Roueff}, \& {Sievers}}]{PETY17}
{Pety}, J., {Guzm{\'a}n}, V.~V., {Orkisz}, J.~H., {et~al.} 2017, \aap, 599, A98

\bibitem[{{Privon} {et~al.}(2015){Privon}, {Herrero-Illana}, {Evans},
  {Iwasawa}, {Perez-Torres}, {Armus}, {D{\'{\i}}az-Santos}, {Murphy},
  {Stierwalt}, {Aalto}, {Mazzarella}, {Barcos-Mu{\~n}oz}, {Borish}, {Inami},
  {Kim}, {Treister}, {Surace}, {Lord}, {Conway}, {Frayer}, \&
  {Alberdi}}]{PRIVON15}
{Privon}, G.~C., {Herrero-Illana}, R., {Evans}, A.~S., {et~al.} 2015, \apj,
  814, 39

\bibitem[{{Querejeta} {et~al.}(2015){Querejeta}, {Meidt}, {Schinnerer},
  {Cisternas}, {Mu{\~n}oz-Mateos}, {Sheth}, {Knapen}, {van de Ven}, {Norris},
  {Peletier}, {Laurikainen}, {Salo}, {Holwerda}, {Athanassoula}, {Bosma},
  {Groves}, {Ho}, {Gadotti}, {Zaritsky}, {Regan}, {Hinz}, {Gil de Paz},
  {Menendez-Delmestre}, {Seibert}, {Mizusawa}, {Kim}, {Erroz-Ferrer}, {Laine},
  \& {Comer{\'o}n}}]{QUEREJETA15}
{Querejeta}, M., {Meidt}, S.~E., {Schinnerer}, E., {et~al.} 2015, \apjs, 219, 5

\bibitem[{{Querejeta} {et~al.}(2019){Querejeta}, {Schinnerer}, {Schruba},
  {Murphy}, {Meidt}, {Usero}, {Leroy}, {Pety}, {Bigiel}, \&
  {Chevance}}]{QUEREJETA19}
{Querejeta}, M., {Schinnerer}, E., {Schruba}, A., {et~al.} 2019, \aap, 625, A19

\bibitem[{{Radford} {et~al.}(1991){Radford}, {Solomon}, \&
  {Downes}}]{RADFORD91}
{Radford}, S. J.~E., {Solomon}, P.~M., \& {Downes}, D. 1991, \apj, 368, L15

\bibitem[{{Rahner} {et~al.}(2017){Rahner}, {Pellegrini}, {Glover}, \&
  {Klessen}}]{RAHNER17}
{Rahner}, D., {Pellegrini}, E.~W., {Glover}, S. C.~O., \& {Klessen}, R.~S.
  2017, \mnras, 470, 4453

\bibitem[{{Rahner} {et~al.}(2019){Rahner}, {Pellegrini}, {Glover}, \&
  {Klessen}}]{RAHNER18}
---. 2019, \mnras, 483, 2547

\bibitem[{{Sakamoto} {et~al.}(1995){Sakamoto}, {Okumura}, {Minezaki},
  {Kobayashi}, \& {Wada}}]{SAKAMOTO95}
{Sakamoto}, K., {Okumura}, S., {Minezaki}, T., {Kobayashi}, Y., \& {Wada}, K.
  1995, \aj, 110, 2075

\bibitem[{{Sandstrom} {et~al.}(2013){Sandstrom}, {Leroy}, {Walter}, {Bolatto},
  {Croxall}, {Draine}, {Wilson}, {Wolfire}, {Calzetti}, {Kennicutt}, {Aniano},
  {Donovan Meyer}, {Usero}, {Bigiel}, {Brinks}, {de Blok}, {Crocker}, {Dale},
  {Engelbracht}, {Galametz}, {Groves}, {Hunt}, {Koda}, {Kreckel}, {Linz},
  {Meidt}, {Pellegrini}, {Rix}, {Roussel}, {Schinnerer}, {Schruba}, {Schuster},
  {Skibba}, {van der Laan}, {Appleton}, {Armus}, {Brandl}, {Gordon}, {Hinz},
  {Krause}, {Montiel}, {Sauvage}, {Schmiedeke}, {Smith}, \&
  {Vigroux}}]{SANDSTROM13}
{Sandstrom}, K.~M., {Leroy}, A.~K., {Walter}, F., {et~al.} 2013, \apj, 777, 5

\bibitem[{{Schinnerer} {et~al.}(2013){Schinnerer}, {Meidt}, {Pety}, {Hughes},
  {Colombo}, {Garc{\'{\i}}a-Burillo}, {Schuster}, {Dumas}, {Dobbs}, {Leroy},
  {Kramer}, {Thompson}, \& {Regan}}]{SCHINNERER13}
{Schinnerer}, E., {Meidt}, S.~E., {Pety}, J., {et~al.} 2013, \apj, 779, 42

\bibitem[{{Schruba} {et~al.}(2011){Schruba}, {Leroy}, {Walter}, {Bigiel},
  {Brinks}, {de Blok}, {Dumas}, {Kramer}, {Rosolowsky}, {Sandstrom},
  {Schuster}, {Usero}, {Weiss}, \& {Wiesemeyer}}]{SCHRUBA11}
{Schruba}, A., {Leroy}, A.~K., {Walter}, F., {et~al.} 2011, \aj, 142, 37

\bibitem[{{Seifried} {et~al.}(2019){Seifried}, {Walch}, {Reissl}, \&
  {Ib{\'a}{\~n}ez- Mej{\'\i}a}}]{SEIFRIED19}
{Seifried}, D., {Walch}, S., {Reissl}, S., \& {Ib{\'a}{\~n}ez- Mej{\'\i}a},
  J.~C. 2019, \mnras, 482, 2697

\bibitem[{{Sheth} {et~al.}(2005){Sheth}, {Vogel}, {Regan}, {Thornley}, \&
  {Teuben}}]{SHETH05}
{Sheth}, K., {Vogel}, S.~N., {Regan}, M.~W., {Thornley}, M.~D., \& {Teuben},
  P.~J. 2005, \apj, 632, 217

\bibitem[{{Sheth} {et~al.}(2010){Sheth}, {Regan}, {Hinz}, {Gil de Paz},
  {Men{\'e}ndez-Delmestre}, {Mu{\~n}oz-Mateos}, {Seibert}, {Kim},
  {Laurikainen}, {Salo}, {Gadotti}, {Laine}, {Mizusawa}, {Armus},
  {Athanassoula}, {Bosma}, {Buta}, {Capak}, {Jarrett}, {Elmegreen},
  {Elmegreen}, {Knapen}, {Koda}, {Helou}, {Ho}, {Madore}, {Masters},
  {Mobasher}, {Ogle}, {Peng}, {Schinnerer}, {Surace}, {Zaritsky},
  {Comer{\'o}n}, {de Swardt}, {Meidt}, {Kasliwal}, \& {Aravena}}]{SHETH10}
{Sheth}, K., {Regan}, M., {Hinz}, J.~L., {et~al.} 2010, \pasp, 122, 1397

\bibitem[{{Shimajiri} {et~al.}(2017){Shimajiri}, {Andr{\'e}}, {Braine},
  {K{\"o}nyves}, {Schneider}, {Bontemps}, {Ladjelate}, {Roy}, {Gao}, \&
  {Chen}}]{SHIMAJIRI17}
{Shimajiri}, Y., {Andr{\'e}}, P., {Braine}, J., {et~al.} 2017, \aap, 604, A74

\bibitem[{{Shirley}(2015)}]{SHIRLEY15}
{Shirley}, Y.~L. 2015, \pasp, 127, 299

\bibitem[{{Sormani} {et~al.}(2018){Sormani}, {Sobacchi}, {Fragkoudi}, {Ridley},
  {Tre{\ss}}, {Glover}, \& {Klessen}}]{SORMANI18}
{Sormani}, M.~C., {Sobacchi}, E., {Fragkoudi}, F., {et~al.} 2018, \mnras, 481,
  2

\bibitem[{{Stephens} {et~al.}(2016){Stephens}, {Jackson}, {Whitaker},
  {Contreras}, {Guzm{\'a}n}, {Sanhueza}, {Foster}, \& {Rathborne}}]{STEPHENS16}
{Stephens}, I.~W., {Jackson}, J.~M., {Whitaker}, J.~S., {et~al.} 2016, \apj,
  824, 29

\bibitem[{{Sternberg} {et~al.}(2014){Sternberg}, {Le Petit}, {Roueff}, \& {Le
  Bourlot}}]{STERNBERG14}
{Sternberg}, A., {Le Petit}, F., {Roueff}, E., \& {Le Bourlot}, J. 2014, \apj,
  790, 10

\bibitem[{{Sun} {et~al.}(2018){Sun}, {Leroy}, {Schruba}, {Rosolowsky},
  {Hughes}, {Kruijssen}, {Meidt}, {Schinnerer}, {Blanc}, \& {Bigiel}}]{SUN18}
{Sun}, J., {Leroy}, A.~K., {Schruba}, A., {et~al.} 2018, \apj, 860, 172

\bibitem[{{Tamburro} {et~al.}(2009){Tamburro}, {Rix}, {Leroy}, {Mac Low},
  {Walter}, {Kennicutt}, {Brinks}, \& {de Blok}}]{TAMBURRO09}
{Tamburro}, D., {Rix}, H.-W., {Leroy}, A.~K., {et~al.} 2009, \aj, 137, 4424

\bibitem[{{Tully} {et~al.}(2009){Tully}, {Rizzi}, {Shaya}, {Courtois},
  {Makarov}, \& {Jacobs}}]{TULLY09}
{Tully}, R.~B., {Rizzi}, L., {Shaya}, E.~J., {et~al.} 2009, \aj, 138, 323

\bibitem[{{Usero} {et~al.}(2015){Usero}, {Leroy}, {Walter}, {Schruba},
  {Garc{\'{\i}}a-Burillo}, {Sandstrom}, {Bigiel}, {Brinks}, {Kramer},
  {Rosolowsky}, {Schuster}, \& {de Blok}}]{USERO15}
{Usero}, A., {Leroy}, A.~K., {Walter}, F., {et~al.} 2015, \aj, 150, 115

\bibitem[{{Utomo} {et~al.}(2018){Utomo}, {Sun}, {Leroy}, {Kruijssen},
  {Schinnerer}, {Schruba}, {Bigiel}, {Blanc}, {Chevance}, {Emsellem},
  {Herrera}, {Hygate}, {Kreckel}, {Ostriker}, {Pety}, {Querejeta},
  {Rosolowsky}, {Sandstrom}, \& {Usero}}]{UTOMO18}
{Utomo}, D., {Sun}, J., {Leroy}, A.~K., {et~al.} 2018, \apj, 861, L18

\bibitem[{{van der Kruit}(1988)}]{VANDERKRUIT88}
{van der Kruit}, P.~C. 1988, \aap, 192, 117

\bibitem[{{van der Kruit} \& {Freeman}(2011)}]{VANDERKRUIT11}
{van der Kruit}, P.~C., \& {Freeman}, K.~C. 2011, Annual Review of Astronomy
  and Astrophysics, 49, 301

\bibitem[{{van der Tak} {et~al.}(2007){van der Tak}, {Black}, {Sch{\"o}ier},
  {Jansen}, \& {van Dishoeck}}]{TAK07}
{van der Tak}, F.~F.~S., {Black}, J.~H., {Sch{\"o}ier}, F.~L., {Jansen}, D.~J.,
  \& {van Dishoeck}, E.~F. 2007, \aap, 468, 627

\bibitem[{{Vazquez-Semadeni}(1994)}]{VAZQUEZSEMADENI94}
{Vazquez-Semadeni}, E. 1994, \apj, 423, 681

\bibitem[{Vincenzo {et~al.}(2016)Vincenzo, Belfiore, Maiolino, Matteucci, \&
  Ventura}]{Vincenzo2016NitrogenUniverseb}
Vincenzo, F., Belfiore, F., Maiolino, R., Matteucci, F., \& Ventura, P. 2016,
  MNRAS, 3477, 3466

\bibitem[{{Vollmer} {et~al.}(2017){Vollmer}, {Gratier}, {Braine}, \&
  {Bot}}]{VOLLMER17}
{Vollmer}, B., {Gratier}, P., {Braine}, J., \& {Bot}, C. 2017, \aap, 602, A51

\bibitem[{{Vutisalchavakul} {et~al.}(2016){Vutisalchavakul}, {Evans}, \&
  {Heyer}}]{VUTI16}
{Vutisalchavakul}, N., {Evans}, Neal~J., I., \& {Heyer}, M. 2016, \apj, 831, 73

\bibitem[{{Walker} {et~al.}(2018){Walker}, {Longmore}, {Zhang}, {Battersby},
  {Keto}, {Kruijssen}, {Ginsburg}, {Lu}, {Henshaw}, {Kauffmann}, {Pillai},
  {Mills}, {Walsh}, {Bally}, {Ho}, {Immer}, \& {Johnston}}]{WALKER18}
{Walker}, D.~L., {Longmore}, S.~N., {Zhang}, Q., {et~al.} 2018, \mnras, 474,
  2373

\bibitem[{{Walter} {et~al.}(2008){Walter}, {Brinks}, {de Blok}, {Bigiel},
  {Kennicutt}, {Thornley}, \& {Leroy}}]{WALTER08}
{Walter}, F., {Brinks}, E., {de Blok}, W.~J.~G., {et~al.} 2008, \aj, 136, 2563

\bibitem[{{Wolfire} {et~al.}(2010){Wolfire}, {Hollenbach}, \&
  {McKee}}]{WOLFIRE10}
{Wolfire}, M.~G., {Hollenbach}, D., \& {McKee}, C.~F. 2010, \apj, 716, 1191

\bibitem[{{Wong} \& {Blitz}(2002)}]{WONG02}
{Wong}, T., \& {Blitz}, L. 2002, \apj, 569, 157

\bibitem[{{Wu} {et~al.}(2005){Wu}, {Evans}, {Gao}, {Solomon}, {Shirley}, \&
  {Vanden Bout}}]{WU05}
{Wu}, J., {Evans}, II, N.~J., {Gao}, Y., {et~al.} 2005, \apjl, 635, L173

\bibitem[{{Wu} {et~al.}(2010){Wu}, {Evans}, {Shirley}, \& {Knez}}]{WU10}
{Wu}, J., {Evans}, II, N.~J., {Shirley}, Y.~L., \& {Knez}, C. 2010, \apjs, 188,
  313

\end{thebibliography}

\end{document}